\documentclass[twocolumn,secnumarabic,amssymb,amsmath, nobibnotes, aps,showpacs,superscriptaddress,prd]{revtex4-1}
\usepackage{color}
\usepackage{amssymb,amsmath,graphicx}
\usepackage{lineno}

\usepackage{enumitem}
\usepackage{threeparttable}

\begin{document}

%\begin{frontmatter}

\title{Final results of Borexino Phase-I on low-energy solar-neutrino spectroscopy}
\author{G.~Bellini}
\affiliation{Dipartimento di Fisica, Universit\`{a} degli Studi e INFN, Milano 20133, Italy}
\author{J.~Benziger}
\affiliation{Physics Department, Princeton University, Princeton, NJ 08544, USA}
\author{D.~Bick}
\affiliation{Institut f\"ur Experimentalphysik, Universit\"at Hamburg, Germany}
\author{G.~Bonfini}
\affiliation{INFN Laboratori Nazionali del Gran Sasso, Assergi 67010, Italy}
\author{D.~Bravo}
\affiliation{Physics Department, Virginia Polytechnic Institute and State University, Blacksburg, VA 24061, USA}
\author{M.~Buizza Avanzini}
\affiliation{Dipartimento di Fisica, Universit\`{a} degli Studi e INFN, Milano 20133, Italy}
\author{B.~Caccianiga}
\affiliation{Dipartimento di Fisica, Universit\`{a} degli Studi e INFN, Milano 20133, Italy}
\author{L.~Cadonati}
\affiliation{Physics Department, University of Massachusetts, Amherst MA 01003, USA}
\author{F.~Calaprice}
\affiliation{Physics Department, Princeton University, Princeton, NJ 08544, USA}
\author{P.~Cavalcante}
\affiliation{INFN Laboratori Nazionali del Gran Sasso, Assergi 67010, Italy}
\author{A.~Chavarria}
\affiliation{Physics Department, Princeton University, Princeton, NJ 08544, USA}
\author{A.~Chepurnov}
\affiliation{Lomonosov Moscow State University Skobeltsyn Institute of Nuclear  Physics, Moscow 119234, Russia}
\author{D.~D{\textquoteright}Angelo}
\affiliation{Dipartimento di Fisica, Universit\`{a} degli Studi e INFN, Milano 20133, Italy}
\author{S.~Davini}
\affiliation{Department of Physics, University of Houston, Houston,  TX 77204, USA}
\author{A.~Derbin}
\affiliation{St. Petersburg Nuclear Physics Institute, Gatchina 188350, Russia}
\author{A.~Empl}
\affiliation{Department of Physics, University of Houston, Houston,  TX 77204, USA}
\author{A.~Etenko}
\affiliation{NRC Kurchatov Institute, Moscow 123182, Russia}
\author{K.~Fomenko}
\affiliation{Joint Institute for Nuclear Research, Dubna 141980, Russia}
\affiliation{INFN Laboratori Nazionali del Gran Sasso, Assergi 67010, Italy}
\author{D.~Franco}
\affiliation{APC, Univ. Paris Diderot, CNRS/IN2P3, CEA/Irfu, Obs. de Paris, Sorbonne Paris Cit\'e, France}
\author{F.~Gabriele}
\affiliation{INFN Laboratori Nazionali del Gran Sasso, Assergi 67010, Italy}
\author{C.~Galbiati}
\affiliation{Physics Department, Princeton University, Princeton, NJ 08544, USA}
\author{S.~Gazzana}
\affiliation{INFN Laboratori Nazionali del Gran Sasso, Assergi 67010, Italy}
\author{C.~Ghiano}
\affiliation{APC, Univ. Paris Diderot, CNRS/IN2P3, CEA/Irfu, Obs. de Paris, Sorbonne Paris Cit\'e, France}
\author{M.~Giammarchi}
\affiliation{Dipartimento di Fisica, Universit\`{a} degli Studi e INFN, Milano 20133, Italy}
\author{M.~G\"{o}ger-Neff}
\affiliation{Physik Department, Technische Universit\"{a}t M\"{u}nchen, Garching 85747, Germany}
\author{A.~Goretti}
\affiliation{Physics Department, Princeton University, Princeton, NJ 08544, USA}
\author{L.~Grandi}
\affiliation{Physics Department, Princeton University, Princeton, NJ 08544, USA}
\author{M.~Gromov}
\affiliation{Lomonosov Moscow State University Skobeltsyn Institute of Nuclear  Physics, Moscow 119234, Russia}
\author{C.~Hagner}
\affiliation{Institut f\"ur Experimentalphysik, Universit\"at Hamburg, Germany}
\author{E.~Hungerford}
\affiliation{Department of Physics, University of Houston, Houston,  TX 77204, USA}
\author{Aldo Ianni}
\affiliation{INFN Laboratori Nazionali del Gran Sasso, Assergi 67010, Italy}
\author{Andrea Ianni}
\affiliation{Physics Department, Princeton University, Princeton, NJ 08544, USA}
\author{V.~Kobychev}
\affiliation{Kiev Institute for Nuclear Research, Kiev 06380, Ukraine}
\author{D.~Korablev}
\affiliation{Joint Institute for Nuclear Research, Dubna 141980, Russia}
\author{G.~Korga} 
\affiliation{Department of Physics, University of Houston, Houston,  TX 77204, USA}
\author{D.~Kryn}
\affiliation{APC, Univ. Paris Diderot, CNRS/IN2P3, CEA/Irfu, Obs. de Paris, Sorbonne Paris Cit\'e, France}
\author{M.~Laubenstein}
\affiliation{INFN Laboratori Nazionali del Gran Sasso, Assergi 67010, Italy}
\author{T.~Lewke}
\affiliation{Physik Department, Technische Universit\"{a}t M\"{u}nchen, Garching 85747, Germany}
\author{E.~Litvinovich}
\affiliation{NRC Kurchatov Institute, Moscow 123182, Russia}
\affiliation{National Nuclear Research University "MEPhI", 31 Kashirskoe Shosse, Moscow, Russia}
\author{B.~Loer}
\affiliation{Physics Department, Princeton University, Princeton, NJ 08544, USA}
\author{F.~Lombardi}
\affiliation{INFN Laboratori Nazionali del Gran Sasso, Assergi 67010, Italy}
\author{P.~Lombardi}
\affiliation{Dipartimento di Fisica, Universit\`{a} degli Studi e INFN, Milano 20133, Italy}
\author{L.~Ludhova}
\affiliation{Dipartimento di Fisica, Universit\`{a} degli Studi e INFN, Milano 20133, Italy}
\author{G.~Lukyanchenko}
\affiliation{NRC Kurchatov Institute, Moscow 123182, Russia}
\author{I.~Machulin}
\affiliation{NRC Kurchatov Institute, Moscow 123182, Russia}
\affiliation{National Nuclear Research University "MEPhI", 31 Kashirskoe Shosse, Moscow, Russia}
\author{S.~Manecki}
\affiliation{Physics Department, Virginia Polytechnic Institute and State University, Blacksburg, VA 24061, USA}
\author{W.~Maneschg}
\affiliation{Max-Planck-Institut f\"{u}r Kernphysik, Saupfercheckweg 1, 69117 Heidelberg, Germany}
\author{G.~Manuzio}
\affiliation{Dipartimento di Fisica, Universit\`{a} e INFN, Genova 16146, Italy}
\author{Q.~Meindl}
\affiliation{Physik Department, Technische Universit\"{a}t M\"{u}nchen, Garching 85747, Germany}
\author{E.~Meroni}
\affiliation{Dipartimento di Fisica, Universit\`{a} degli Studi e INFN, Milano 20133, Italy}
\author{L.~Miramonti}
\affiliation{Dipartimento di Fisica, Universit\`{a} degli Studi e INFN, Milano 20133, Italy}
\author{M.~Misiaszek}
\affiliation{M. Smoluchowski Institute of Physics, Jagiellonian University, Cracow, 30059, Poland}
\author{P.~Mosteiro}
\affiliation{Physics Department, Princeton University, Princeton, NJ 08544, USA}
\author{V.~Muratova}
\affiliation{St. Petersburg Nuclear Physics Institute, Gatchina 188350, Russia}
\author{L.~Oberauer}
\affiliation{Physik Department, Technische Universit\"{a}t M\"{u}nchen, Garching 85747, Germany}
\author{M.~Obolensky}
\affiliation{APC, Univ. Paris Diderot, CNRS/IN2P3, CEA/Irfu, Obs. de Paris, Sorbonne Paris Cit\'e, France}
\author{F.~Ortica}
\affiliation{Dipartimento di Chimica, Biologia e Biotecnologie, Universit\`{a} e INFN, Perugia 06123, Italy}
\author{K.~Otis}
\affiliation{Physics Department, University of Massachusetts, Amherst MA 01003, USA}
\author{M.~Pallavicini}
\affiliation{Dipartimento di Fisica, Universit\`{a} e INFN, Genova 16146, Italy}
\author{L.~Papp}
\affiliation{Physics Department, Virginia Polytechnic Institute and State University, Blacksburg, VA 24061, USA}
\author{C.~Pena--Garay}
\affiliation{Istituto de Fisica Corpuscular, Valencia, E-46071, Spain}
\author{L.~Perasso}
\affiliation{Dipartimento di Fisica, Universit\`{a} e INFN, Genova 16146, Italy}
\author{S.~Perasso}
\affiliation{Dipartimento di Fisica, Universit\`{a} e INFN, Genova 16146, Italy}
\author{A.~Pocar}
\affiliation{Physics Department, University of Massachusetts, Amherst MA 01003, USA}
\author{G.~Ranucci}
\affiliation{Dipartimento di Fisica, Universit\`{a} degli Studi e INFN, Milano 20133, Italy}
\author{A.~Razeto}
\affiliation{INFN Laboratori Nazionali del Gran Sasso, Assergi 67010, Italy}
\author{A.~Re}
\affiliation{Dipartimento di Fisica, Universit\`{a} degli Studi e INFN, Milano 20133, Italy}
\author{A.~Romani}
\affiliation{Dipartimento di Chimica, Biologia e Biotecnologie, Universit\`{a} e INFN, Perugia 06123, Italy}
\author{N.~Rossi}
\affiliation{Physics ans Astronomy Department, University of California Los Angeles (UCLA), Los Angeles, CA 90095, USA}
\author{R.~Saldanha}
\affiliation{Physics Department, Princeton University, Princeton, NJ 08544, USA}
\author{C.~Salvo}
\affiliation{Dipartimento di Fisica, Universit\`{a} e INFN, Genova 16146, Italy}
\author{S.~Sch\"onert}
\affiliation{Physik Department, Technische Universit\"{a}t M\"{u}nchen, Garching 85747, Germany}
\author{H.~Simgen}
\affiliation{Max-Planck-Institut f\"{u}r Kernphysik, Saupfercheckweg 1, 69117 Heidelberg, Germany}
\author{M.~Skorokhvatov}
\affiliation{NRC Kurchatov Institute, Moscow 123182, Russia}
\affiliation{National Nuclear Research University "MEPhI", 31 Kashirskoe Shosse, Moscow, Russia}
\author{O.~Smirnov}
\affiliation{Joint Institute for Nuclear Research, Dubna 141980, Russia}
\author{A.~Sotnikov}
\affiliation{Joint Institute for Nuclear Research, Dubna 141980, Russia}
\author{S.~Sukhotin}
\affiliation{NRC Kurchatov Institute, Moscow 123182, Russia}
\author{Y.~Suvorov}
\affiliation{Physics ans Astronomy Department, University of California Los Angeles (UCLA), Los Angeles, CA 90095, USA}
\affiliation{NRC Kurchatov Institute, Moscow 123182, Russia}
\author{R.~Tartaglia}
\affiliation{Physics ans Astronomy Department, University of California Los Angeles (UCLA), Los Angeles, CA 90095, USA}
\author{G.~Testera}
\affiliation{Dipartimento di Fisica, Universit\`{a} e INFN, Genova 16146, Italy}
\author{D.~Vignaud}
\affiliation{APC, Univ. Paris Diderot, CNRS/IN2P3, CEA/Irfu, Obs. de Paris, Sorbonne Paris Cit\'e, France}
\author{R.B.~Vogelaar}
\affiliation{Physics Department, Virginia Polytechnic Institute and State University, Blacksburg, VA 24061, USA}
\author{F.~von Feilitzsch}
\affiliation{Physik Department, Technische Universit\"{a}t M\"{u}nchen, Garching 85747, Germany}
\author{J.~Winter}
\affiliation{Physik Department, Technische Universit\"{a}t M\"{u}nchen, Garching 85747, Germany}
\author{M. Wojcik}
\affiliation{M. Smoluchowski Institute of Physics, Jagiellonian University, Cracow, 30059, Poland}
\author{A.~Wright}
\affiliation{Physics Department, Princeton University, Princeton, NJ 08544, USA}
\author{M.~Wurm}
\affiliation{Institut f\"ur Experimentalphysik, Universit\"at Hamburg, Germany}
\author{J.~Xu}
\affiliation{Physics Department, Princeton University, Princeton, NJ 08544, USA}
\author{O.~Zaimidoroga}
\affiliation{Joint Institute for Nuclear Research, Dubna 141980, Russia}
\author{S.~Zavatarelli}
\affiliation{Dipartimento di Fisica, Universit\`{a} e INFN, Genova 16146, Italy}
\author{G.~Zuzel}
\affiliation{M. Smoluchowski Institute of Physics, Jagiellonian University, Cracow, 30059, Poland}

\collaboration{Borexino Collaboration} 
\noaffiliation

\begin{abstract}

Borexino has been running since May 2007 at the LNGS laboratory in Italy with the primary goal of detecting solar neutrinos.
The detector, a large, unsegmented 
liquid scintillator calorimeter characterized by unprecedented low levels of
intrinsic radioactivity, is optimized for the study of the lower 
energy part of the spectrum.  During the Phase-I (2007 - 2010), Borexino first detected and then precisely measured 
the flux of the $^7$Be solar neutrinos, ruled out 
any significant day--night asymmetry of their interaction rate, made the first 
direct observation of the $pep$ neutrinos, and set the 
tightest upper limit on the flux of CNO solar neutrinos.
In this paper we discuss the signal signature and provide a comprehensive description of the backgrounds, quantify their event rates, describe the methods for
their identification, selection or subtraction, and describe data analysis.
Key features are an extensive in 
situ calibration program using radioactive sources, the detailed modeling of 
the detector response, the ability to define an innermost fiducial volume with 
extremely low background via software cuts, and the excellent
pulse--shape discrimination capability of the scintillator that allows particle
identification.
We report a measurement of the annual
modulation of the $^7$Be neutrino interaction rate.
The period, the amplitude, and the phase of the observed modulation are 
consistent with the solar origin of these events, and the absence of 
their annual modulation is rejected with higher than 99$\%$ C.L.
The physics implications of Phase-I results  in the context of the neutrino oscillation physics and solar models are presented.

\end{abstract}



\maketitle

\tableofcontents

%\end{frontmatter}

%%
%% Start line numbering here if you want
%%
%\linenumbers

\section{Introduction}
\label{sec:intro}

The study of neutrinos emitted by the Sun with energies
below $\sim$3000\,keV (low-energy solar neutrinos) is a science at the intersection of elementary particle physics
and astrophysics:  on one hand these neutrinos allow for the study of neutrino oscillations, and on the other they provide key information for accurate solar modeling.

\begin{figure}[t]
\begin{center}
\hspace{-5 mm}
\centering{\includegraphics[width = 0.48 \textwidth]{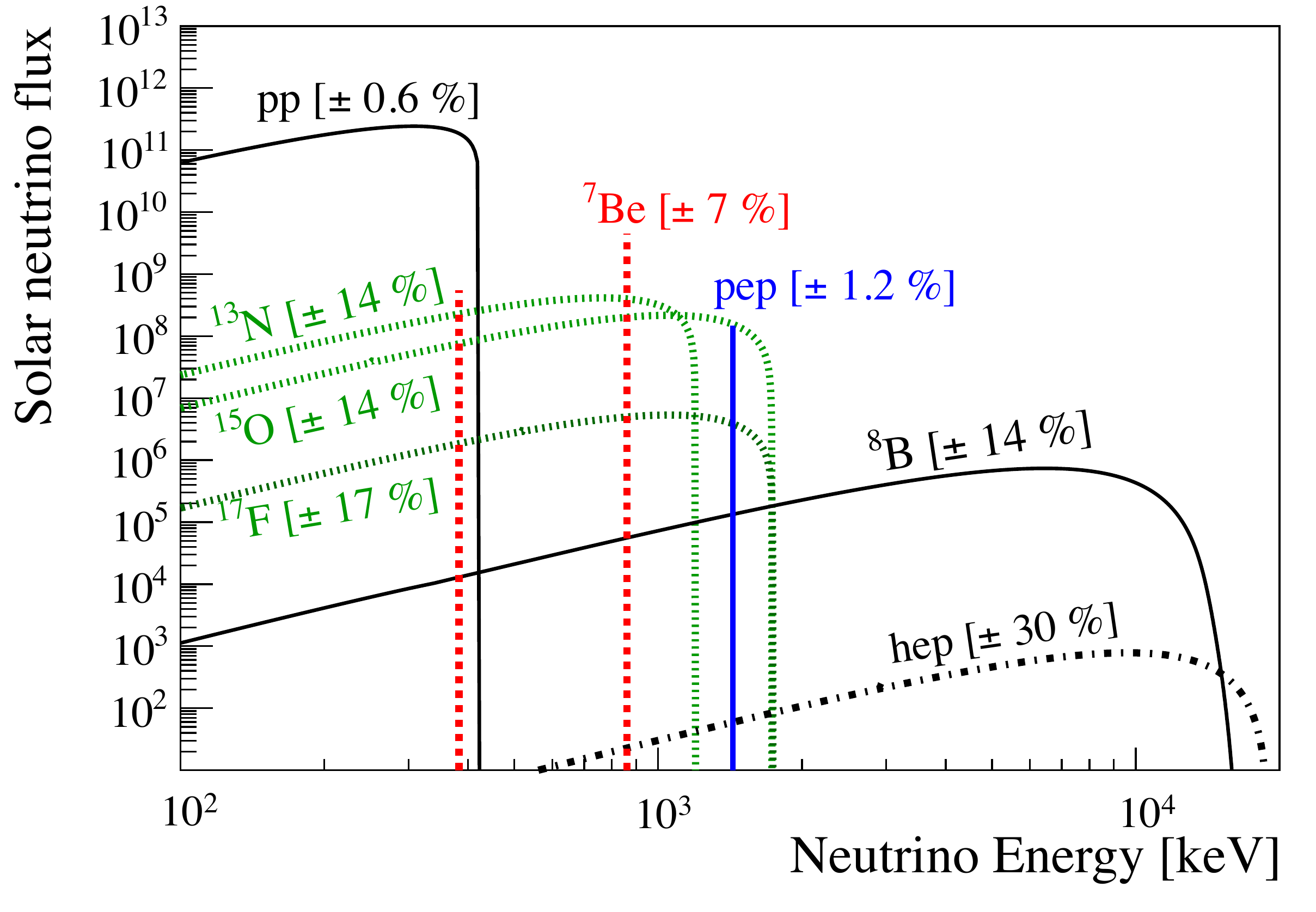}}
\caption{Energy spectrum of solar neutrinos.  The spectral shapes are taken from~\cite{BahcallPage} while the flux normalisation from~\cite{HighMet_GS98SSM}. The vertical axis report the flux in cm$^{-2}$ s$^{-1}$\,(10$^3$ keV)$^{-1}$ for the continuous neutrino spectra, while in cm$^{-2}$ s$^{-1}$ for the mono--chromatic lines  ($^7$Be--$\nu$ at 384 and 862\,keV, shown as red dotted lines, and $pep$--$\nu$ at 1440\,keV, shown as a blue continuous line). The numbers in parenthesis represent the theoretical uncertainties on the expected fluxes. }
\label{fig:SolarNuSpectrum}
\end{center}
\end{figure}

The spectrum of electron neutrinos ($\nu_e$) generated in the core of the Sun is shown in 
Fig.~\ref{fig:SolarNuSpectrum}. The spectral shapes are taken from~\cite{BahcallPage} while the flux normalisation from~\cite{HighMet_GS98SSM}. 

Borexino is presently the only detector able to measure the solar-neutrino interaction rate 
%event--by--event 
down to energies as low as $\sim$150\,keV
and to reconstruct  the energy spectrum of the events.
Previous radiochemical experiments, with an energy threshold of 233\,keV, could not extract information about the neutrino energy spectrum ~\cite{GALLEX},~\cite{SAGE}.  In particular, Borexino  is the only experiment to date to have measured the interaction rate of the $^7$Be 862\,keV solar neutrinos~\cite{BxBe},~\cite{BxBe2}.  The accuracy of the measurement  has recently reached $5\%$~\cite{be7-2011}, %Borexino 
and any significant day--night asymmetry of the $^7$Be solar neutrino flux has been excluded~\cite{DNLetter}.
Borexino has also made the first direct observation of the mono--energetic 1440\,keV $pep$ solar neutrinos~\cite{PepBorex},
and set the strongest upper limit of the CNO solar neutrinos flux to date.
Furthermore, the experiment has measured the $^8$B solar neutrinos with an energy threshold of 3000\,keV~\cite{bxB8},
lower than that achieved by previous experiments.

Lepton-flavor changing neutrino oscillations have been detected by several experiments covering a wide range of source-to-detector distances and neutrino energies. Numerous experiments, measuring atmospheric and solar neutrinos or using neutrino and antineutrino beams from nuclear reactors and accelerators, contribute to our current understanding of neutrino oscillations ~\cite{SNODN},~\cite{NuOsc}, the phenomenological description of which involves the square of the neutrino mass differences $\Delta m^2_{ij}$ ($i$ and $j$ label mass eigenstates) and their mixing angles $\theta_{ij}$.
Indeed, solar $\nu_e$'s are a very sensitive probe for oscillations.
Because they oscillate, they reach the Earth as a mixture of $\nu_e$, $\nu_\mu$, and $\nu_\tau$. One of the mixing angles, $\theta_{13}$, is small and then only two parameters ($\Delta m^2_{12}$ and $\theta_{12}$) are sufficient to describe well the main features of solar neutrino oscillations.

Neutrino interactions with the electrons inside the Sun play an important role in the oscillation dynamics of solar neutrinos, via the MSW effect~\cite{MSW}.
Additionally, depending on the allowed region of the oscillation parameters and neutrino energy,
neutrino interactions with the Earth electrons may induce
a regeneration effect of the disappeared $\nu_e$. The result could be a different flux of neutrinos reaching the
detector during the night time (when neutrinos cross the
Earth during their path from the Sun to the detector) and
during the day (when neutrinos do not cross the Earth)~\cite{BahcallDN}.
This effect was studied already   for $^8B$ solar neutrinos  ~\cite{SNODN} and it has recently detected with a statistical significance of 2.7 $\sigma$ ~\cite{SKDN2014}
for that solar neutrino component.
Borexino has provided a measurement for the lower energy $^7Be$ solar  neutrinos.

The currently measured solar neutrino oscillation parameters~\cite{OscParam} are $\Delta m^2_{12} = (7.54 ^{+0.26}_{-0.22}) \times 10^{-5} $ eV$^2$ and $\sin^2 \theta_{12} =0.307 ^{+0.018}_{-0.016}$, also known as the MSW--LMA (large mixing angle)~\cite{MSW} solution. These values have been obtained in a global 3 lepton flavor analysis of all available neutrino data, including the recent discovery of non-zero value of $\theta_{13}$ mixing angle~\cite{DayaBay}. The current best value of $\sin^2\theta_{13}$ is $0.0241 \pm 0.0025$, taken as well from \cite{OscParam}.

%these two parameters are obtained combining the measurements of high-energy $^8$B solar neutrino fluxes, of the integrated solar neutrino fluxes from %radiochemical experiments and the electron flavor survival probability measured versus the distance from the source of antineutrinos from nuclear reactors. 

The MSW--LMA model predicts an energy dependent survival probability $P_{ee}$ of electron  neutrinos with two oscillation regimes, in vacuum and in matter, and a transition region in between. Non-standard neutrino interaction models~\cite{NSI} predict $P_{ee}$ curves that deviate significantly from the MSW--LMA, particularly between 1000 and 4000\,keV.  Low--energy solar neutrinos are thus a sensitive tool to test the MSW--LMA paradigm by measuring $P_{ee}$ versus neutrino energy.

The Standard Solar Model (SSM) identifies two distinct nuclear fusion processes occurring in our star. One is the  dominant $pp$ fusion chain and the other the sub--dominant CNO cycle~\cite{HighMet_GS98SSM},~\cite{LowMet_AGSS09}. Together, they yield the neutrino fluxes as in Fig.~\ref{fig:SolarNuSpectrum}. A measurement of solar neutrinos from the CNO cycle has important implications in solar physics and 
astrophysics more generally, as this is believed to be the primary process fueling massive stars ($>$1.5$M_{\rm{Sun}}$).
The CNO solar neutrino flux is sensitive to the abundance of heavy elements in the Sun
(metallicity), an experimental input parameter in solar models.  The CNO flux is 40$\%$ higher in high--metallicity models~\cite{HighMet_GS98SSM} than it is in low--metallicity ones~\cite{LowMet_AGSS09}. A precise CNO solar neutrino flux measurement
has therefore the potential to discriminate between these competing models and to shed light on the inner workings of heavy stars.

This paper provides a detailed description of the
analysis methods used to obtain the aforementioned measurements of $^7$Be,
$pep$, and CNO (upper limit) solar neutrino interaction rates in Borexino. After a brief description of the detector, we discuss the expected neutrino signal,
the backgrounds, the variables used in the analysis, and the procedures
adopted to extract the signal.  We then report on a measurement of the annual modulation of the $^7$Be solar neutrino rate.
Finally, we discuss the physics implications of the Borexino solar neutrino results and we report a global analysis of the
Borexino data combined with that of other solar neutrino experiments and of reactor experiments sensitive to $\Delta m^2_{12}$ and $\theta_{12}$.

This paper reports the final results of the Borexino Phase-I. Phase-II, with an even better radio--purity already obtained after an extensive purification campaign of the scintillator, has already started data taking in 2012 and will continue for several years. The goals of the Phase-II will be reported in a separate paper.

\section{The Borexino detector}
\label{sec:detector}

\begin{figure}[t]
\begin{center}
\includegraphics[width = 8 cm]{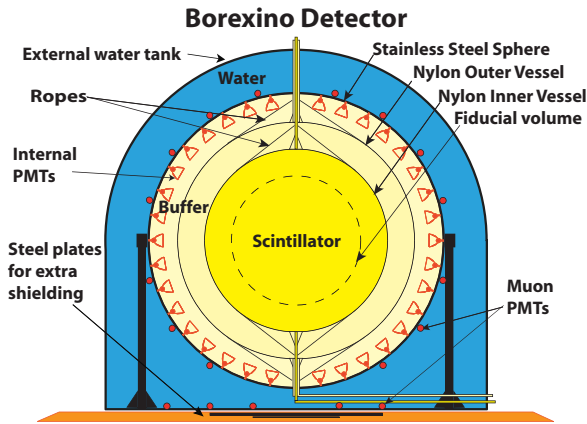}
\caption{The schematic view of the Borexino detector.}
\label{fig:Borexino}
\end{center}
\end{figure}

Borexino is installed in Hall C of the Laboratori Nazionali del
Gran Sasso (LNGS) in Italy.
Its design~\cite{BxNim}  is based on the principle of graded shielding,
with the inner scintillating core at the center of a set of concentric
shells of decreasing radio--purity from inside to outside (see Fig.~\ref{fig:Borexino}). The active  medium is a solution of PPO (2,5-diphenyloxazole, a fluorescent dye) in pseudocumene (PC, 1,2,4-trimethylbenzene) at a concentration of 1.5\,g/l~\cite{quencherPaper}. The scintillator mass ($\sim$278\,ton) is contained in a 125\,$\mu$m thick spherical nylon Inner Vessel (IV)~\cite{Vessel} with 4.25\,m radius surrounded by 2212 photomultipliers (PMTs) labeled as Internal PMTs in
Fig.~\ref{fig:Borexino}. All but 371 PMTs are equipped with aluminum light concentrators designed to increase the light collection efficiency.

Within the IV a fiducial volume (FV) is
software defined through the measured event position, obtained from
the PMTs timing data via a time--of--flight algorithm (see Section~
\ref{sec:position}).  A second 5.5\,m radius nylon Outer Vessel (OV)
surrounds the IV, acting as a barrier against radon and other
background contamination originating from outside. The region between
the IV and the OV contains a passive shield composed of PC and a small quantity of DMP (dimethylphthalate), a material that
quenches the residual scintillation of PC so that scintillation
signals arise dominantly from the interior of the
IV~\cite{quencherPaper}. The concentration of DMP in PC was 5.0\,g/l at the beginning of data taking and 
was later reduced to 3.0\,g/l (and then to 2.0\,g/l) to mitigate the effects of a small leak in the IV (discussed in Subsection~\ref{subsec:leak}).
A 6.85\,m radius Stainless Steel Sphere (SSS)
encloses the central part of the detector and serves also as a support
structure for the 2212 8" (ETL 9351) PMTs.

The region between the OV and the SSS is filled with the same inert
 buffer fluid (PC plus DMP) which is layered between the IV and the
 OV.  The apparatus consisting of the PC and its solvents, the nylon vessels and the Internal PMTs is called Inner Detector (ID). 
 
 The ID  is contained in a tank (9\,m base radius,
 16.9\,m height) filled by ultra--pure water. The total liquid passive
 shielding of the central volume from external radiation (such as that
 originating from the rock) is thus 5.5\,m of water equivalent.
 The water tank (WT) serves also as an active veto (Outer Detector OD)
 allowing the detection of the Cherenkov light induced by muons
 in water. For this purpose 208~PMTs are installed on the
 external side of the SSS and on the WT walls. The walls of the water tank are covered by
 a reflective material to enhance the light collection. Details of the
 OD are described in~\cite{BxMuons}.

All the materials of the detector internal components (stainless
steel, phototubes, cables, light concentrators, nylon) were specially
selected for extremely low
radioactivity. Furthermore, only qualified
ultra--clean processes were employed for their realization, followed by
careful surface cleaning methods.

The final assembly of the elements in the SSS was carried out in clean
room conditions: the entire interior of the sphere was converted into
a class 1000 clean room, while in front of the main entrance of the
sphere itself an on purpose clean room of class 100 - 1000 was used for
all the final cleaning procedures of the equipment.
Key elements determining the success of the experiment were also the
many liquid purification and handling systems~\cite{BxFluidHandling}, which were designed and
installed to ensure the proper fluid manipulation at the exceptional purity 
level demanded by Borexino.

The PC was specially produced for Borexino by Polimeri Europa
(Sarroch-IT), according to a stringent quality control plan jointly
developed. It was shipped to LNGS through custom--built transport tanks
especially cleaned and treated. The first underground operation was
the PC transfer via a dedicated unloading station to four big
reservoir tanks. Taken from this storage area, the PC was first
purified via distillation, then either mixed with PPO for insertion in
the IV or mixed with DMP for the insertion in the buffer
region. Furthermore, the PPO was pre-mixed with a limited quantity of
PC in a dedicated PPO system, originating a concentrated PPO solution
which was then mixed in line with the PC. 

Other important ancillary plants are the N$_2$ systems, which deliver regular, or on site
purified, or specially produced N$_2$. The last one has exceptionally
low content of $^{39}$Ar and $^{85}$Kr, to be used for the crucial
manipulations of the liquid in the IV. Finally, an ultra--pure water system was used to produce the water for
the cleaning operations, for the WT fill, and for the preliminary
water fill of the SSS.

The selection of the low radioactivity materials, the liquid handling procedures, the purification strategies, and many scintillator properties have been tested using a prototype of Borexino called Counting Test Facility (CTF). This detector ($\simeq$5\,m$^3$ vessel filled by an organic liquid scintillator viewed by 100 PMTs) collected data from the year 1995 until the year 2011 in Hall C of the Laboratori Nazionali del Gran Sasso in Italy. The  CTF data have allowed to understand the relevant background expected in Borexino, to setup the correct purification procedure, to select the most suitable scintillator mixture, and to fully demonstrate the feasibility of Borexino itself. Relevant CTF results are reported in~\cite{bib:CTFlist1},\cite{bib:CTFlist2},\cite{bib:CTFlist3},\cite{bib:CTFlist4},\cite{bib:CTFlist5},\cite{bib:CTFlist6},\cite{bib:CTFlist7}.
 
\subsection{Inner--Vessel leak}
\label{subsec:leak}

A leak of scintillator from the IV to the buffer region within the OV started approximately on April 9$^{\rm th}$ 2008, for reasons which we could not exactly determine.

The small hole in the IV was reconstructed to have location as $26^\circ < \theta < 37^\circ$ and $225^\circ < \phi < 270^\circ$.
This leak was detected only in September 2008 based on a large rate of events reconstructed out of the IV.
Its presence was then confirmed by abnormally high PPO concentration in the samples of OV--buffer.

The IV shape and volume can be reconstructed based on the inner--detector pictures taken with the seven CCD cameras~\cite{BxCalibPaper}.
By this technique, the leak rate was estimated to be about 1.33\,m$^3$/month.
In order to minimize the leak rate, the density difference between the scintillator and the buffer fluids, and hence the pressure difference across the leak, was reduced by partial removal of DMP from the buffer by distillation.
Between February 12$^{\rm th}$, 2009 and April 3$^{\rm rd}$, 2009 the buffer liquid was purified and the DMP concentration reduced from 5\,g/l to 3\,g/l, thus reducing the density difference between the scintillator and the buffer and in turn the buoyant forces on the IV.
This reduced the leak rate to about 0.56\,m$^3$/month, and greatly reduced the number of scintillation events occurring in the buffer. 
In December 2009 it was decided to further reduce the DMP concentration to 2~g/l, to  approach neutral buoyancy between buffer and scintillator.  This concentration is still high enough to suppress the PC scintillation in the buffer.
Following this operation, concluded at the end of January 2010, the leak rate was further reduced to $\sim$1.5\,m$^3$/year.  The IV shape appeared to be stabilized.
The lost scintillator volume in the IV was compensated by several refilling operations using PC. 

\section{Solar neutrinos detection in Borexino}
\label{sec:Nudetection}

Solar neutrinos of all flavors are detected by means of their elastic scattering off electrons:
\begin{equation}
\nu_{e,\mu,\tau} +e^- \rightarrow \nu_{e,\mu,\tau} +e^- .
\label{es}
\end{equation}
In the elastic scattering process only a fraction of the neutrino energy $E_\nu$ is transferred to an electron and the interaction of the latter with the medium originates the scintillation signal. The electron recoil spectrum is thus continuous even in the case of mono--energetic neutrinos and it extends up to a maximum energy $T^{max}$ given by 
\begin{equation}
T^{max}= \frac {E_\nu } {1+ \frac {m_e c^2} {2 E_\nu}}, 
\label{Eq:Temax}
\end{equation}
where $m_e c^2$ is the electron rest energy.
For the mono--energetic 862\,keV $^7$Be and 1440\,keV $pep$ solar neutrinos, $T^{max}$ is 665\,keV and 1220\,keV, respectively.

The rate of $\nu_{e,\mu,\tau}$ -- electron elastic scattering interactions in a given target is a product of the incoming neutrino flux, the number of electrons in the target  $N_e$ (in Borexino $\rm{(3.307 \pm 0.003) \times 10^{31}}$ $e^{-}$/100\,ton), and the neutrino--electron elastic scattering cross section. The cross sections $\sigma_e$ and $\sigma_{\mu,\tau}$ are obtained from the electroweak Standard Model (SM). Radiative corrections to the total cross sections for solar--$\nu$s' elastic scattering and thus to the electron--recoil energy spectra are described in~\cite{RadCorr}. Table~\ref{tab:CrossSection} shows the total cross sections for solar neutrinos (weighted for the spectral shape in case of continuous energy spectra) calculated following the procedure of~\cite{RadCorr} with updated values for numerical constants according to~\cite{PDG2010} and with the constant term of (A4) from~\cite{RadCorr} equal to 0.9786 according to~\cite{RadCorrPrivate}.  
The radiative corrections change monotonically the electron recoil spectrum for incident $^8$B solar neutrinos,  with the relative probability of observing recoil electrons being reduced by about 4\% at the highest electron energies. For $pep$ and $^7$Be solar neutrinos, the recoil spectra are not affected significantly.

\begin{figure}[t]
\centerline{\includegraphics[width = 0.5\textwidth]{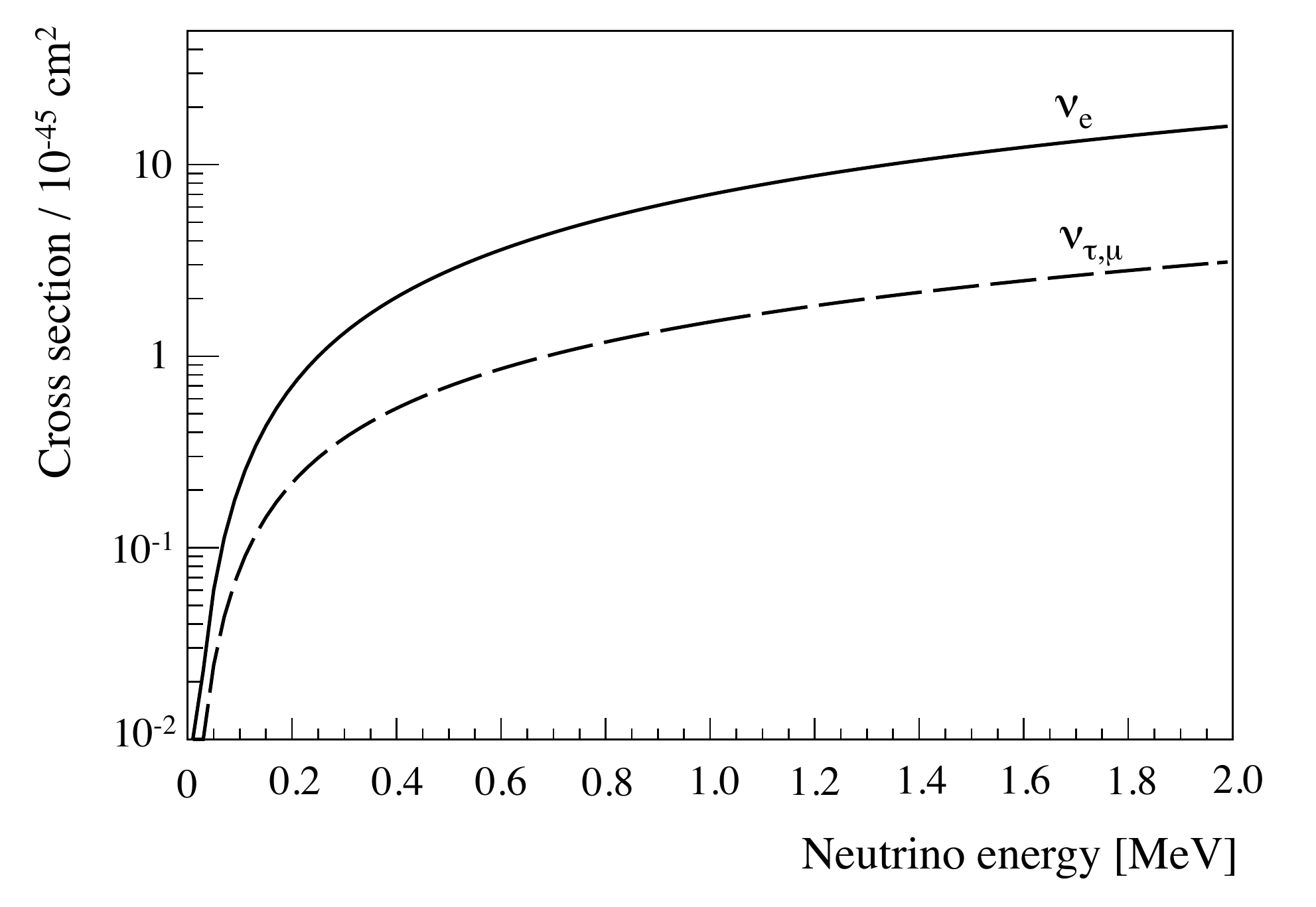}}
\caption{Neutrino -- electron elastic scattering cross section as a function of the neutrino energy for $\nu_e$ (solid line) and for $\nu_\mu$ or $\nu_\tau$ (dashed line).} 
\label{fig:CrossSection}
\end{figure}	

Borexino can detect neutrinos of all flavors, but $\nu_e$ have a larger cross section than $\nu_\mu$ and $\nu_\tau$, because $\nu_e$ interact through both charged current (CC) and neutral current (NC), while $\nu_\mu$ and $\nu_\tau$ interact only via the NC. Figure~\ref{fig:CrossSection} shows the total cross section for neutrino electron elastic scattering as a function of neutrino energy.
The interaction probability increases with energy and it is about 4-5 times larger for $\nu_e$ than for $\nu_\mu$,$\nu_\tau$ in the energy region of our interest.
Electron recoils induced by different neutrino flavors cannot be distinguished event--by--event. In principle, the different recoil energy spectra between $\nu_e$ and $\nu_\mu$,$\nu_\tau$ might allow a statistical separation, but this is practically not possible with the current amount of data. 

\begin{table*}
\begin{center}
\begin{tabular}{lllccc} 
\hline \hline
Solar $\nu$  &  $E_{\nu}$   & $T^{max}$ & $\sigma_e$                            & $\sigma_{\mu,\tau}$ &  $P_{ee}$\\
                   &  [keV]           & [keV]               & $[\times 10^{-46}$\,cm$^2]$ &  $[\times 10^{-46}$\,cm$^2]$ \\
\hline
$pp$         & $\leq$420 & 261 & 11.38  & 3.22 &  $0.542 \pm 0.016$\\
$^7$Be        & 384 & 231     & 19.14 & 5.08  & $0.537 \pm 0.015$ \\
$^7$Be       & 862 & 665      & 57.76 & 12.80 & $0.524 \pm 0.014$ \\
$pep$        & 1440 & 1220   	& 108.49 & 22.08 &  $0.514 \pm 0.012$ \\
$^{13}$N     & $\leq$1199 & 988 	& 45.32 & 10.29 & $0.528 \pm 0.014$\\
$^{15}$O     & $\leq$1732 & 1509 	& 70.07 & 14.96 &  $0.517 \pm 0.013$\\
$^{17}$F     & $\leq$1740 &    1517 & 70.34 & 15.01 & $ 0.517 \pm 0.019$\\
$^8$B        &  $\leq$15000 & 14500 & 596.71 & 106.68 & $0.384 \pm 0.009$ \\
\hline   
\hline   
\end{tabular}
\end{center}
\caption{The total cross sections $\sigma_e$ and $\sigma_{\mu,\tau}$ for solar neutrinos, weighted for the spectral shape in case of continuos energy spectra. $E_{\nu}$ is the neutrino energy (end--point for continous energy spectra) and $T^{max}$ is the maximal energy of the scattered $e^{-}$ according to Eq.~\ref{Eq:Temax}. The last column gives the electron neutrino survival probability $P_{ee}$,  weighted for the spectral shape in case of continuos energy spectra, and calculated according to~\cite{MSWSolar} using the oscillation parameters from~\cite{OscParam}.}
\label{tab:CrossSection}
\end{table*}

Considering solar neutrino oscillations, the expected neutrino interaction rate in Borexino $R_{\nu}$ is: 

\begin{equation}
\begin{split}
&R_{\nu}= N_e \Phi_{\nu} \int dE_{\nu}\, \frac{d\lambda}{dE_{\nu}}  \\
&\int \left\{ \frac{d\sigma_e(E_{\nu}, T)}{dT}\, P_{ee}(E_{\nu}) + \frac{d\sigma_{\mu}(E_{\nu}, T)}{dT}\,[1-P_{ee}(E_{\nu})]\right\} dT,
\end{split}
\label{Eq:Rates}
\end{equation}
where $N_e$ is the number of target electrons, $\Phi_{\nu}$ is the SSM solar neutrino flux, $d\lambda /dE_{\nu}$ is the differential energy spectrum of solar neutrinos, and $P_{ee}$ is the electron neutrino survival probability defined in~\cite{MSWSolar}.
Table~\ref{table:Rate} reports the expected interaction rates of solar neutrinos in Borexino according to the high--metallicity~\cite{HighMet_GS98SSM} and the low--metallicity~\cite{LowMet_AGSS09} hypothesis of the Standard Solar Model, using the oscillation parameters from~\cite{OscParam}. The spectral shapes $d\lambda /dE_{\nu}$ are taken from~\cite{BahcallPage} with the exception of $^8$B--$\nu$ taken from~\cite{B8spectra}. The low count rate between a few and a few tens of counts-per-day (cpd)/100\,ton defines the required background rates and the needed radio--purity of the detector.

\begin{table*}
\begin{center}
\begin{threeparttable}
\begin{tabular}{lllllcc} 
\hline \hline
Solar--$\nu$  &  $\Phi_{\nu}$(GS98)  &$\Phi_{\nu}$(AGSS09)                    & $R_{\nu}$(GS98)       &  $R_{\nu}$(AGSS09)   & Main       \\
               &  High--metallicity           &Low--metallicity             &   High--metallicity       &  Low--metallicity &    \,\,\,\,    background \\
               &$\rm{[cm^{-2}s^{-1}]}$    &$\rm{[ cm^{-2}s^{-1}]}$             &   [cpd/100\,ton]                &      [cpd/100\,ton]        & \\

\hline
$pp$          &5.98 (1$\pm$0.006)               & 6.03 (1$\pm$0.006$)$                 	& 130.8 $\pm$ 2.4  	& 131.9 $\pm$ 2.4  &   $^{14}$C\\
$^7$Be\tnote{$\ast$} (384\,keV)      &0.53 (1$\pm$ 0.07)  & 0.48 (1$\pm$ 0.07$)$	& 1.90 $\pm$ 0.14    	& 1.73 $\pm$ 0.12          &        $^{85}$Kr, $^{210}$Bi\\
$^7$Be\tnote{$\ast$} (862\,keV)       &4.47 (1$\pm$ 0.07)  & 4.08 (1$\pm$ 0.07$)$	& 46.48 $\pm$ 3.35    	& 42.39 $\pm$ 3.05          &        $^{85}$Kr, $^{210}$Bi\\
$pep$         &1.44 (1$\pm$ 0.012)              & 1.47 (1$\pm$ 0.012$)$                  	& 2.73 $\pm$ 0.05 	& 2.79 $\pm$ 0.06        &   $^{11}$C, $^{210}$Bi\\
$^{13}$N    &2.96 (1$\pm$ 0.14)               & 2.17 (1$\pm$ 0.14)                      	& 2.42 $\pm$ 0.34   	& 1.78 $\pm$ 0.23        &                       \\
$^{15}$O       &2.23 (1$\pm$ 0.15)               & 1.56 (1$\pm$ 0.15)                  	& 2.75 $\pm$ 0.42   	& 1.92 $\pm$ 0.29 	    &                       \\
$^{17}$F     &5.52 (1$\pm$ 0.17)               & 3.40 (1$\pm$ 0.16)                      	& 0.068 $\pm$ 0.012  & 0.042 $\pm$ 0.007 &                       \\
CNO           &5.24 (1$\pm$ 0.21)               & 3.76 (1$\pm$ 0.21)                      	& 5.24 $\pm$ 0.54   	&  3.74 $\pm$ 0.37  &   $^{11}$C, $^{210}$Bi\\
$^8$B            &5.58 (1$\pm$ 0.14)               & 4.59 (1$\pm$ 0.14)              	& 0.44 $\pm$ 0.07   	&   0.37 $\pm$ 0.05     &   $^{208}$Tl, ext $\gamma$ \\
\hline   
\hline   
\end{tabular}
\begin{tablenotes}
\item[$\ast$] \footnotesize{The production branching ratios of the 384 and 862\,keV $^7$Be--$\nu$ lines are 0.1052 and 0.8948, respectively. The respective ratio of interaction rates in Borexino is 3.9 : 96.1.}
\end{tablenotes}

\end{threeparttable}
\end{center}
\caption{The solar--neutrino fluxes $\Phi_{\nu}$ calculated with the high--metallicity Standard Solar Model (GS98)~\cite{HighMet_GS98SSM}, the ones obtained with the low--metallicity model (AGSS09)~\cite{LowMet_AGSS09}, and the corresponding expected $\nu$--interaction rates $R_{\nu}$ in Borexino. The fluxes are given in units of $10^{10} $($pp$), $10^9$ ($^7$Be), $10^8$ ($pep$, $^{13}$N, $^{15}$O), and  $10^6$ ($^8$B, $^{17}$F). The CNO flux is the sum of the $^{13}$N, $^{15}$O, and $^{17}$F fluxes.
The rate calculations are based on Eq.~\ref{Eq:Rates}.    
The last column lists some of the relevant background components  in the energy region of interest of a given $\nu$--species: see Section~\ref{subsec:InternalBack} for a discussion about them.}
\label{table:Rate}
\end{table*}

\section{The data set}
\label{sec:DataSet}

Borexino is collecting data in its final configuration since May 16$^{\rm {th}}$, 2007. 
For the precision measurement of the interaction rate of the $^7$Be neutrinos~\cite{be7-2011} we have used all the available data until
May~8$^{\rm{th}}$, 2010. The live--time after the analysis cuts is
740.7\,days which corresponds to the 153.6\,ton\,$\times$\,year fiducial exposure.
For the measurement of the interaction rate of the $pep$ and CNO neutrinos~\cite{PepBorex} we have used the data collected from
January 13$^{\rm{th}}$, 2008 to May 9$^{\rm{th}}$, 2010. The total live--time
after the cuts  but before the subtraction of the background signal due  to the cosmogenic $^{11}$C (see Section~\ref{sec:c11}) is 598.3\,days.
The final spectrum obtained after  the $^{11}$C subtraction corresponds to 55.9\,ton\,$\times$\,year and it preserves 48.5\% of the total exposure.

The data have been collected almost continuously over time with some interruptions due to
maintenance or calibrations with radioactive sources described in Section~\ref{sec:calibration} and in~\cite{BxCalibPaper}.
The data taking is organized in periods called "runs" with a typical duration of few hours.  

\section{The analysis methods}
\label{sec:method}

The emission of scintillation light is isotropic and any information about the
initial direction of solar neutrinos is lost. This is a weak point compared to Cherenkov detectors, which can measure the incoming neutrino direction and have been widely employed to study the high--energy part of the solar--neutrinos spectrum~\cite{Cherenkov}. However, the light yield of
Cherenkov emitters is too small (about 50 times smaller than the scintillation one) to allow their use for detecting the low-energy
part of the solar--neutrinos spectrum, so liquid scintillators are the only practical possibility for real--time detection. 

Neutrino--induced events in liquid scintillator are thus intrinsically indistinguishable on an event--by--event basis from the background due to $\beta$ or $\gamma$ decays. The analysis procedure begins removing from the available data 
single events due to taggable background (radioactive decays from
delayed coincidences, muons and events following muons within a given time
window) or due to electronics noise. The set of these event--by--event
based cuts (standard cuts) is described in Section~\ref{sec:cuts}.
Additional background components can be eventually suppressed by removing all the events detected within a given volume during a proper time window: this is the case of the cosmogenic $^{11}$C suppression (see Section~\ref{sec:c11}) applied in the $pep$ and CNO neutrino analysis.

In general, the majority of the background types cannot be eliminated
by these methods.
The analysis procedure continues by building the distributions of the quantities of interest (energy estimators, radial position of events,
particular shape parameters built to distinguish between signal and
background) and fitting them by means of analytical models or Monte Carlo (MC) spectra to extract the contribution of the signal
and background. When possible, some background is removed from these
distributions by applying statistical subtraction techniques based on
the particle pulse--shape identification.
The ability to  define   a fiducial volume through the reconstruction of the position of the scintillation events is a crucial feature made possible by the fast
time response of the scintillator and of the PMTs: this handle allows to strongly suppress any external background. 
For the $^7$Be-neutrino analysis we fit only the energy spectrum of the events
surviving the standard cuts with and without the application of a
statistical subtraction procedure aiming to remove background due to
the $\alpha$ decay of $^{210}$Po. For the $pep$ and CNO neutrino analysis we
developed a multivariate likelihood fit including distributions of the energy estimator,
the radial position, and the shape parameter able to separate scintillation induced by the $\beta^+$ decay of $^{11}$C from the scintillation due to electrons. 

The results achieved by Borexino have been made possible by the extremely low background of the detector obtained after many years of tests and research.
This high radio--purity is the element making the performances of Borexino unique.
In addition, the accuracy of all the analysis is related to a careful modeling of the detector response function
achieved through a calibration campaign of the detector. The detector response function is in general the probability distribution function of a physical quantity of interest, like the energy deposit of $\alpha$, $\beta$ or $\gamma$ and/or the interaction position inside the scintillator volume.
It allows the link between the physical information 
and the measured quantities. They are ideally the list of the PMTs
detecting one or more photoelectrons (p.e.), the times $t_i^j$ when the hit $i$ is detected by the PMT $j$ and its associated charge $q_i^j$.
The number of detected hits and the corresponding charge allow to measure the energy released in the scintillator while the list of the times $t_i^j$ permits to reconstruct the 
interaction position and is the base for the construction of several pulse--shape variables. 
Some remarks about the detector
(Section~\ref{sec:detector}), the electronics, and the data acquisition system (Section~\ref{sec:electronics}),
and about the scintillator (Section~\ref{sec:scintillator}) are necessary to understand
the observables and the details of the analysis.

\section{Electronics and triggers }
\label{sec:electronics}

The quantities recorded for each event by the Borexino detector are the
amount of light collected by each PMT and the relative detection times of 
the photons.

Every PMT is AC--coupled to an electronic chain made by an analogue
front--end followed by a digital circuit. The analogue front--end performs
two tasks: it amplifies the PMT pulse, thus providing a fast input
signal for a threshold discriminator mounted on the digital board, and
it continuously integrates the PMT current using a gate--less
integrator~\cite{gateless}. The integrator output rises when a pulse
is generated on the PMT output and it stays constant for 80\,ns; then
it exponentially decays due to the AC coupling between the PMT and the
front--end circuit with a time constant of 500\,ns. The firing of the
discriminator defines the hit time $t_i^j$ introduced in Section~\ref{sec:method}.

The time of multiple hits on the same PMT is not detected by the digital board 
when the time delay between two consecutive pulses is less than 140\,ns (channel dead
time). This dead time is software extended to 180\,ns.  When the
discriminator fires, then the output of the integrator is sampled and
digitized by an 8 bit flash ADC; it provides the charge $q_i^j$ of all those
pulses reaching the PMT $j$ within 80\,ns from the time of the discriminator firing at time $t_i^j$.
Details of the charge measurement including the managing of the multiple hits on the same PMT during the 500\,ns
decay time are discussed in~\cite{gateless}. More details concerning the digital 
electronics and the triggering system are in~\cite{laben} and~\cite{BxNim}.

\begin{figure}[t]
\begin{center}
\includegraphics[width = 0.45\textwidth]{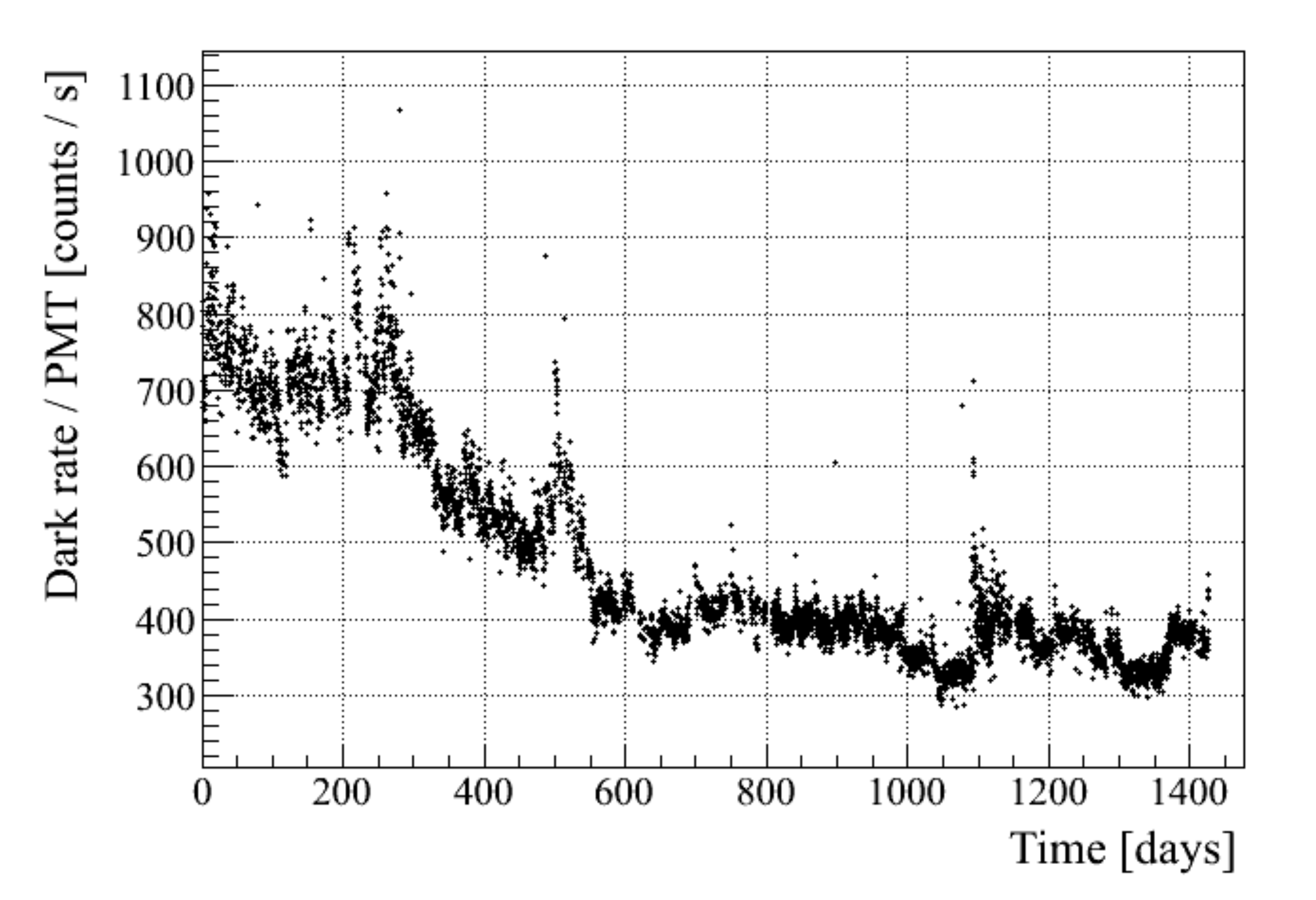}
\caption{Mean dark--noise rate per PMT in counts per second as a
  function of time starting from May 16$^{\rm{th}}$, 2007 (day 0).}
\label{fig:dark}
\end{center}
\end{figure}

Borexino is a self--triggering multiplicity detector, and thus the main
trigger fires when a minimum of $K$ Inner-Detector PMTs detect at least one
photoelectron within a selected time window, normally set to 99\,ns. The $K$-threshold was set in
the range 25 to 30 hits PMTs in the data runs considered in this paper,
corresponding approximately to an energy threshold ranging between 50
and 60\,keV. When a trigger occurs, the time $t_i^j$ and the charge $q_i^j$ of each hit detected in a
time gate of predefined length are acquired and stored. The gate length was initially 6.9\,$\mu$s and was  enlarged to 16.5\,$\mu$s in December 2007, with dead time between two consecutive gates of 6.1\,$\mu$s and 2.5\,$\mu$s, respectively.

The hit time is measured by a Time--to--Digital Converter (TDC) with
about 0.5\,ns resolution which is smaller than the intrinsic 1.2\,ns time
jitter of the PMTs. A dedicated sub--ns 394\,nm pulsed laser system is
used to measure and then to software equalize the time response of all the PMTs~\cite{BxNim} via a
set of optical fibers that reach every PMT. The typical accuracy of
the time equalization is better than 0.5\,ns and the time calibration
procedure is performed at least once a week. A similar system based on
a set of LEDs is employed for the Outer Detector. 

The typical dark rate of internal PMT is about 400 - 500~counts/s (see Fig.~\ref{fig:dark}), which yields, on average, 15~random hits within the
16\,$\mu$s acquisition gate and less than 0.5 hits on average in a typical scintillation pulse (considering 500\,ns duration of the signal). 

\begin{figure}[t]
\includegraphics[width = 0.5\textwidth]{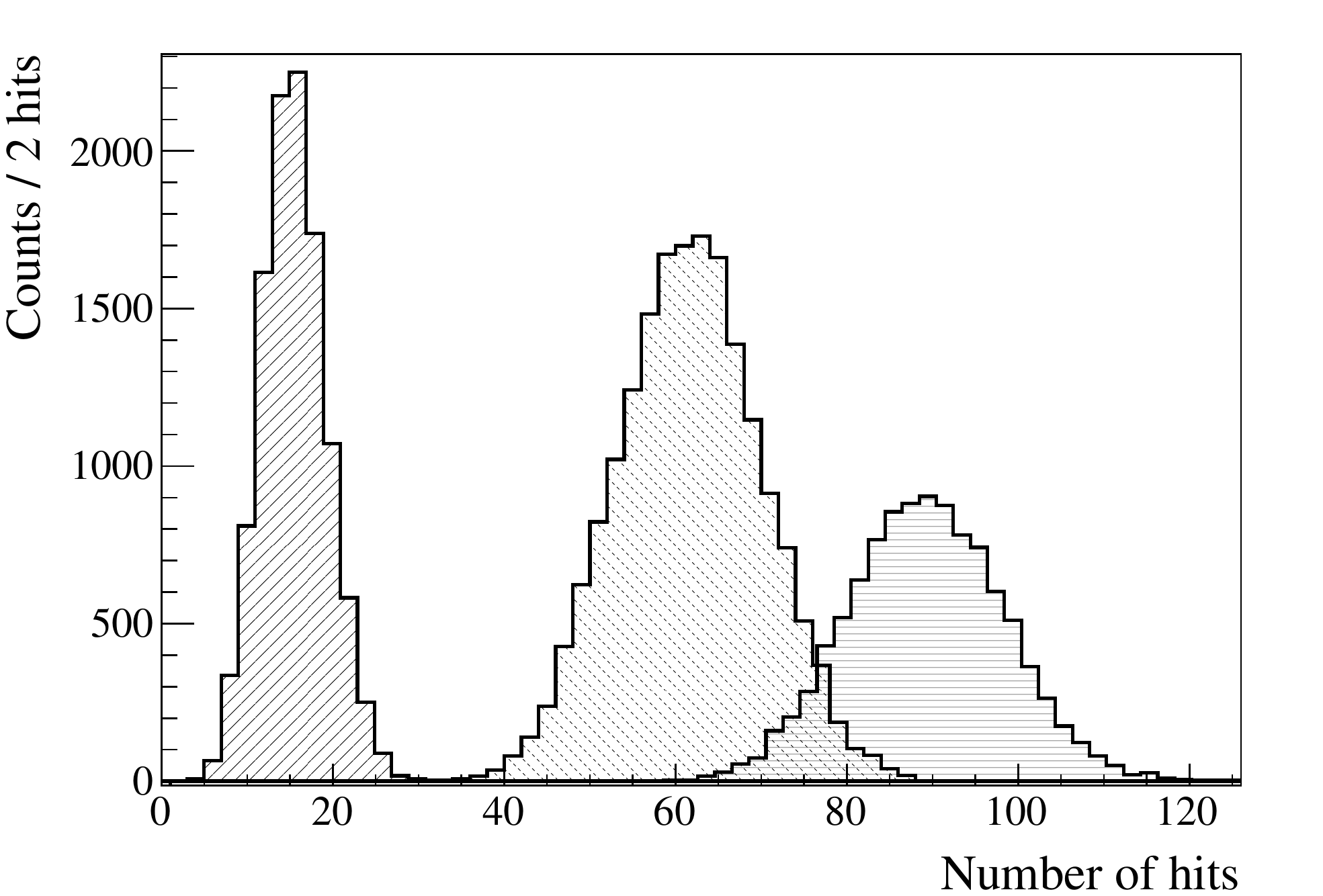} \\
\includegraphics[width=0.5\textwidth]{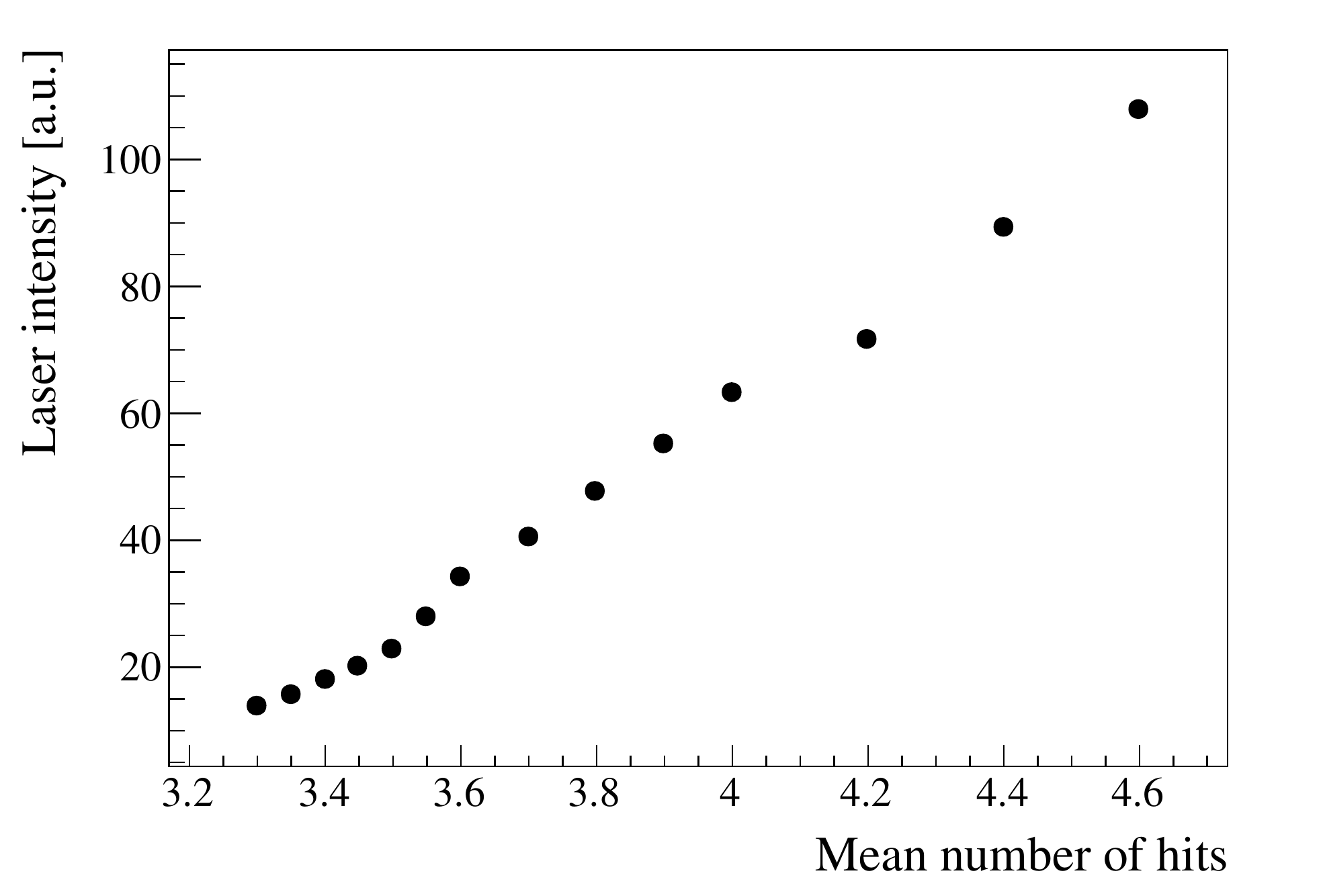}
\caption{\label{fig:trigger-laser-calibration1} Top: examples of the distributions of the number of PMT hits for three laser intensities; Bottom: relation between the average number of PMT hits detected by Borexino Inner  Detector and the laser intensity. These calibration values, which are approximately, but not exactly, linear, were used to compute the detection efficiency shown in Fig.~\ref{fig:trigger-laser-calibration2}.}
\end{figure}

The efficiency of the triggering system was measured by means of the following procedure: by using the 394\,nm laser and the optical fibers that deliver the laser pulse to each internal PMT, we have first calibrated the laser intensity. In a set of runs with variable laser intensity a pulse was sent both to the laser and to the Borexino Trigger Board (BTB), yielding a precise measurement of the average number of PMT hits as a function of the laser intensity.
The pulse sent to the BTB guarantees that data is acquired regardless of the number of PMTs fired, particularly important at the lowest laser intensities. 
We scanned 14~different laser intensities, ranging from 14 up to 120 average PMT hits per event, which roughly corresponds to the energy region between 30\,keV and 240\,keV. 
The result of this calibration scan is shown in Fig.~\ref{fig:trigger-laser-calibration1}.

\begin{figure}[t]
\begin{center}
\includegraphics[width = 0.5\textwidth]{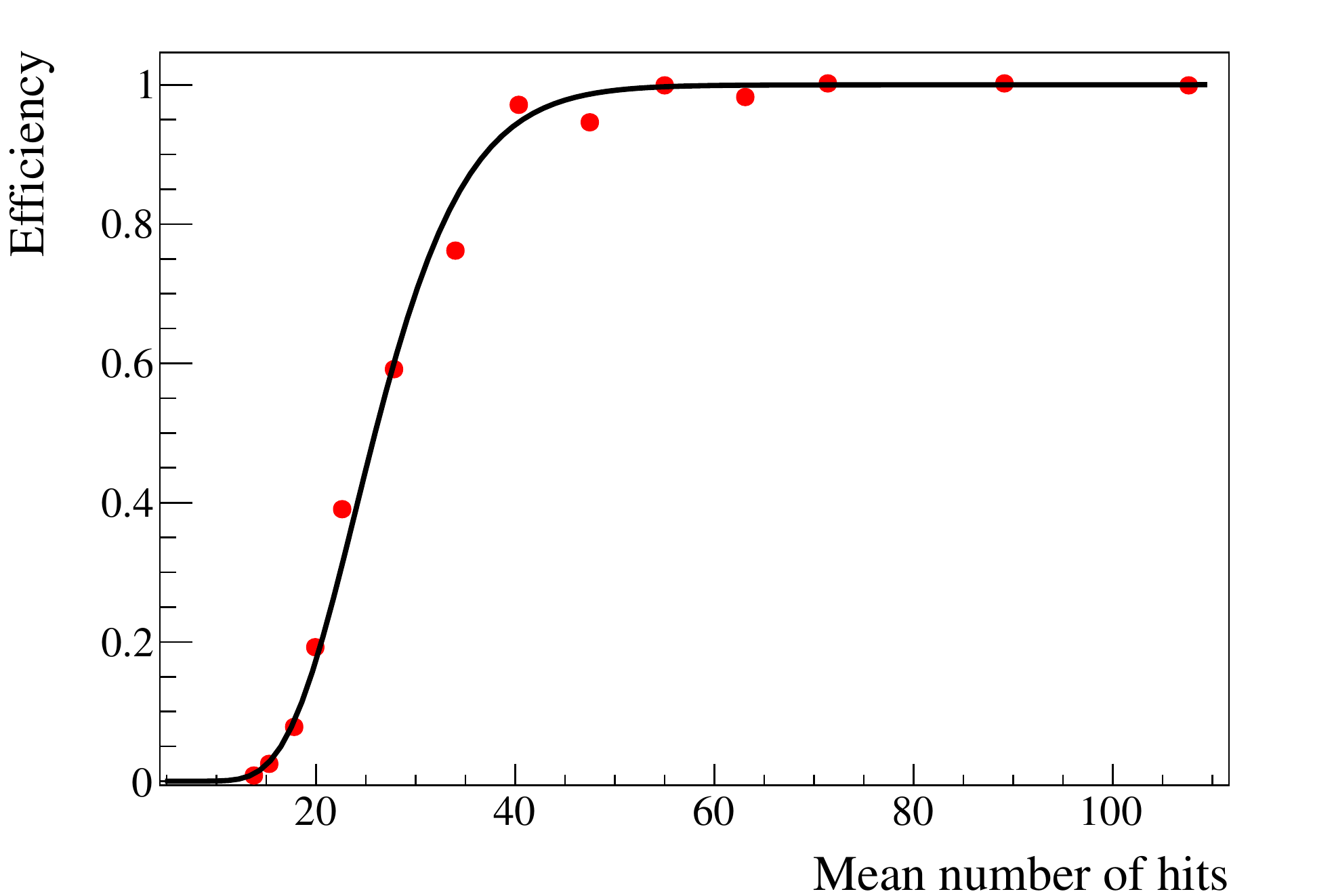}
\caption{The trigger efficiency as a function of the average number of detected PMT hits. For each point, the average number of hits was obtained from the calibration shown in Fig.~\ref{fig:trigger-laser-calibration1}. The fit function is an error function with standard deviation computed assuming exact Poisson statistics. The fit mean value is 25.7, in a very good agreement with the nominal triggering threshold of $K$ = 25. }
\label{fig:trigger-laser-calibration2}
\end{center}
\end{figure}

We then performed again a similar scan, this time avoiding to trigger Borexino 
with the pulse sent to the BTB but just leaving the standard Borexino trigger function as in normal 
physics run.
%the 
The detection efficiency for a given laser intensity is then defined as a fraction of the fired laser pulses (counted by a scaler) which actually gave a DAQ trigger.
For each laser intensity, the average number of PMT hits was obtained from the calibration shown in 
Fig.~\ref{fig:trigger-laser-calibration1}.
The resulting trigger efficiency as a function of the mean number of PMT hits is shown in Fig.~\ref{fig:trigger-laser-calibration2}. 
The fit function is an error function with standard deviation computed assuming exact Poisson statistics.
The fit mean value is 25.7, in a very good agreement with the nominal triggering threshold set during the measurement to $K$ = 25.
The curve clearly shows that the triggering efficiency is effectively one when the number of
fired PMTs is above 40, corresponding approximately to a deposited energy of 80\,keV.

The trigger efficiency measurement shown in Fig.~\ref{fig:trigger-laser-calibration2} was done without applying any correction due to the number of dead channels. The correction is time--dependent and is normally done for data analysis, as later described in Section~\ref{sec:estimators}. This correction is not relevant here, being the purpose of this test to show that the triggering logic was working properly, and that the triggering efficiency can be safely assumed to be 1 for all energies of interest for this paper. The number of live PMTs in a run is always at least 80\% of the total, so even applying a correction, the effective threshold raises from about 40 to about 50, well below the physics region of interest to this paper. 

The trigger efficiency at higher energy (514\,keV) was also studied with the $^{85}$Sr calibration source as reported in ~\cite{BxCalibPaper} and was again found to be well compatible with 1. However, the uncertainty in the activity of the calibration sources was too large to use those tests as a definitive proof of the good behavior of the triggering system. 

\begin{figure}[b]
\begin{center}
\includegraphics[width = 0.5\textwidth]{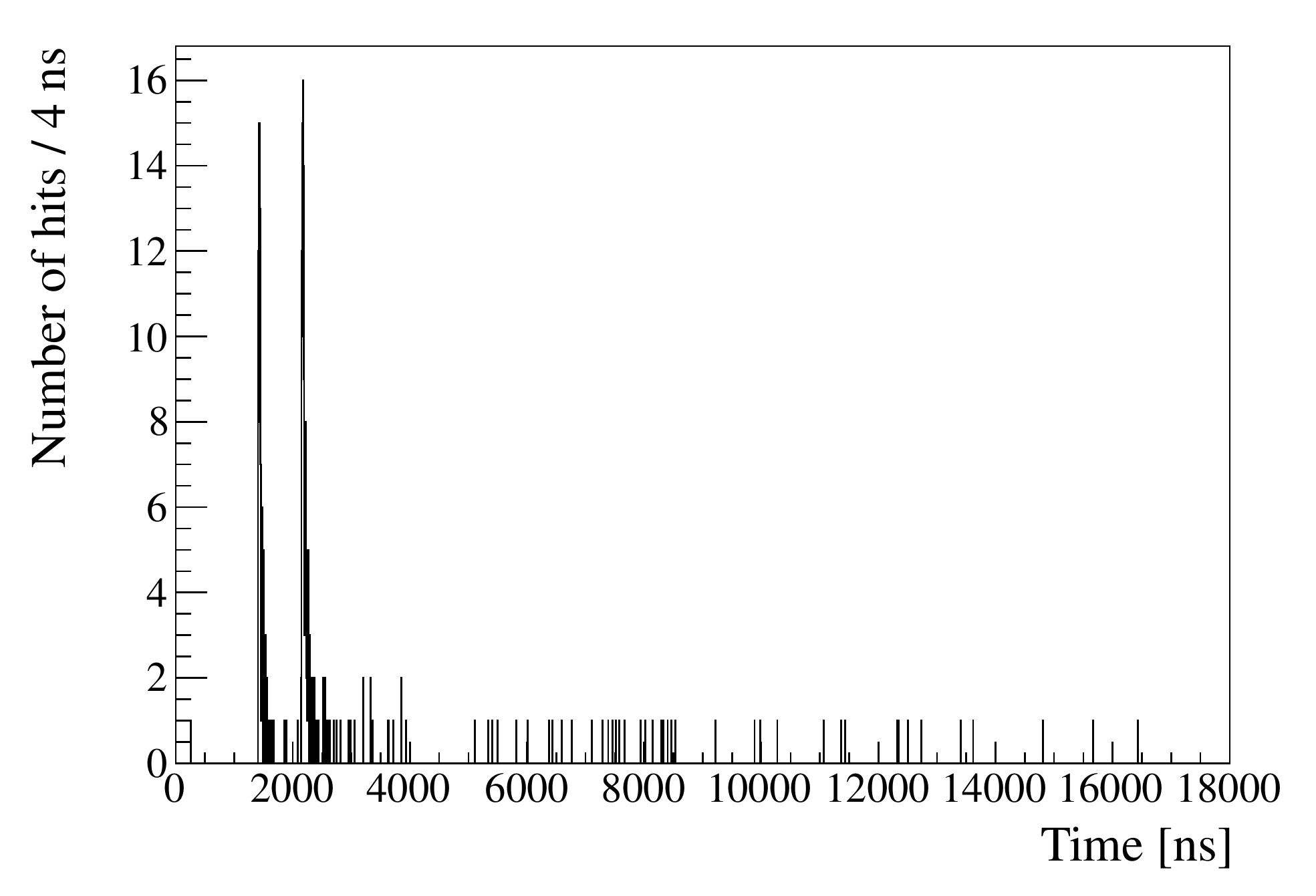}
\caption{An example of a single data acquisition gate  (so called event) containing two well separated clusters, which are due to two different interactions inside the scintillator.} 
\label{fig:cluster}
\end{center}
\end{figure}

A software code (called clustering algorithm) identifies within the
acquisition  gate the group of hits that belong to a single scintillation event
(here called cluster). The cluster duration is typically 1.5\,$\mu$s long, although different values have been used for some analysis.
Fast radioactive decays or random coincidence events detected in a single trigger gate are separated by this clustering
algorithm. Delayed coincidences separated by more than the gate width are detected in two separate events (DAQ triggers).  Figure~\ref{fig:cluster} shows an event
with two clusters.

The readout sequence can also be activated by the OD through a
dedicated triggering system firing when at least six Outer-Detector PMTs detect light
within a time window of 150\,ns. Regardless of the trigger type, the
data from both the Inner and Outer Detectors are always recorded.

A dedicated trigger was developed for cosmogenic neutron detection. After each muon passing and triggering both the OD and the ID, a 1.6\,ms wide acquisition gate is opened. This duration is sufficient since it corresponds to more than six times the neutron capture time. Neutrons are searched for as clusters in this dedicated long trigger as well as clusters within the muon gate itself. The dead--time between the muon and neutron trigger is (150 $\pm$ 50)\,ns. To test the neutron detection efficiency a 500\,MHz waveform digitizer (CAEN v 1731) is fed with the Analogue Sum of all
the signals of the ID PMTs properly attenuated. The data from this single channel Analogue Sum system
are acquired every time OD triggers, regardless if the ID did or did not trigger. More details about the neutron detection can be found in~\cite{BxMuons}. The single channel sum  provides a high-resolution copy of the total
detector signal after a muon shower, which is used to cross check  the performances of the electronics and data acquisition  processing the signal of each PMT. 

For detector monitoring and calibration purposes, every two seconds three
calibration events of three different types are acquired: $i)$ \emph{laser}
events in which the ID--PMTs are synchronously illuminated by
the 394\,nm laser pulse; $ii)$ \emph{pulser} events with a
calibration pulse used for testing the electronics chain independently
of the PMT status; $iii)$ \emph{random triggers} are acquired
without any calibration signal in order to follow the PMT dark
rate. The typical triggering rate during the runs analyzed in this
paper was in the range 25 - 30\,s$^{-1}$, including all trigger types. The triggering rate is largely dominated by the $\beta$-decay of the $^{14}$C isotope with 156\,keV
end point energy.

The gain of the PMTs is checked for every run by fitting the $^{14}$C charge spectrum of each photomultiplier with the sum of
two Gaussian curves representing the single and double photoelectron
response. 
The $^{14}$C data provide a natural calibration
source giving single photoelectrons on the hit PMTs with very good approximation, since less than 100 PMTs from 2200 are hit in single event.
The gain of the PMT $j$ is measured through the ADC position $P^j_{ADC}$ of its first
photoelectron peak. After such calibration, the charge $q_{iADC}^j$ associated to the  hits i and  measured in ADC channels can be converted  in the number of photoelectrons $q_i^j$ (often called p.e.):
\begin{equation}
q_i^j = \frac{q^j_{iADC}}{P^j_{ADC}} 
\label{Eq:charge_calib}
\end{equation} 

\section{Scintillator properties}
\label{sec:scintillator}

The scintillator properties and the processes dominating the light
propagation (absorption, re-emission, Rayleigh scattering) are largely
discussed in~\cite{SciProp1},~\cite{SciProp2}, and~\cite{CTFLightPropPaper}. Particularly relevant
for the measurements under discussion are the high light yield (about
$10^4$ photons/MeV) and transparency (the attenuation length is close
to 10\,m at 430\,nm), the fast time response, the ionization quenching
effect, and the pulse--shape discrimination capability.

Charged particles loosing energy in organic liquid scintillators excite the scintillator
molecules, which then de--excite emitting fluorescence light. 
The amount of emitted light is not simply related to the total energy lost by the
particle but to details of the energy--loss mechanism. 
The ionization quenching effect~\cite{Birks} introduces an intrinsic non--linear
relation between the deposited energy and the emitted light which, for a fixed energy, depends on the particle type.
This non--linearity must be known and taken into account in the detector energy response function.
In general, the higher is the specific energy loss $dE/dx$, the lower is the number of scintillation photons $dY^{\rm ph}$ emitted per unit 
of path length $dx$; different semi--empirical relations between $dE/dx$ and $dY^{\rm ph}/dx$ can be found in the literature~\cite{VariousBirks}. For $\beta^+$ and $\beta^-$ we are using the following, so called Birk's relation: 
\begin{equation}
\frac{dY^{\rm ph}}{dx} = \frac {Y_0^{\rm ph} \cdot dE/dx} {1+kB \cdot dE/dx},
\label{eq:Birks}
\end{equation}
where  $kB$ is called the quenching parameter, and $Y_0^{\rm ph}$\,($\simeq$$10^{4}$ photons/MeV) is the
scintillation light yield in absence of quenching ($kB = 0$).

The quenching parameter $kB$ is of the order of $10^{-2}$\,cm/MeV, but its
precise value has to be experimentally determined for specific scintillator mixtures.
The $kB$ of the Borexino scintillator was determined based on the calibration with $\gamma$--sources.
Two independent methods give consistent results: $kB$ = (0.0109 $\pm$ 0.0006)\,cm/MeV obtained with the Monte Carlo-based procedure (see Section~\ref{sec:MC}) and $kB$ = (0.0115 $\pm$ 0.0007)\,cm/MeV based on the analytical method (Section~\ref{sec:analytical}).

 \begin{figure}[t]
\centerline{\includegraphics[width = 0.5\textwidth]{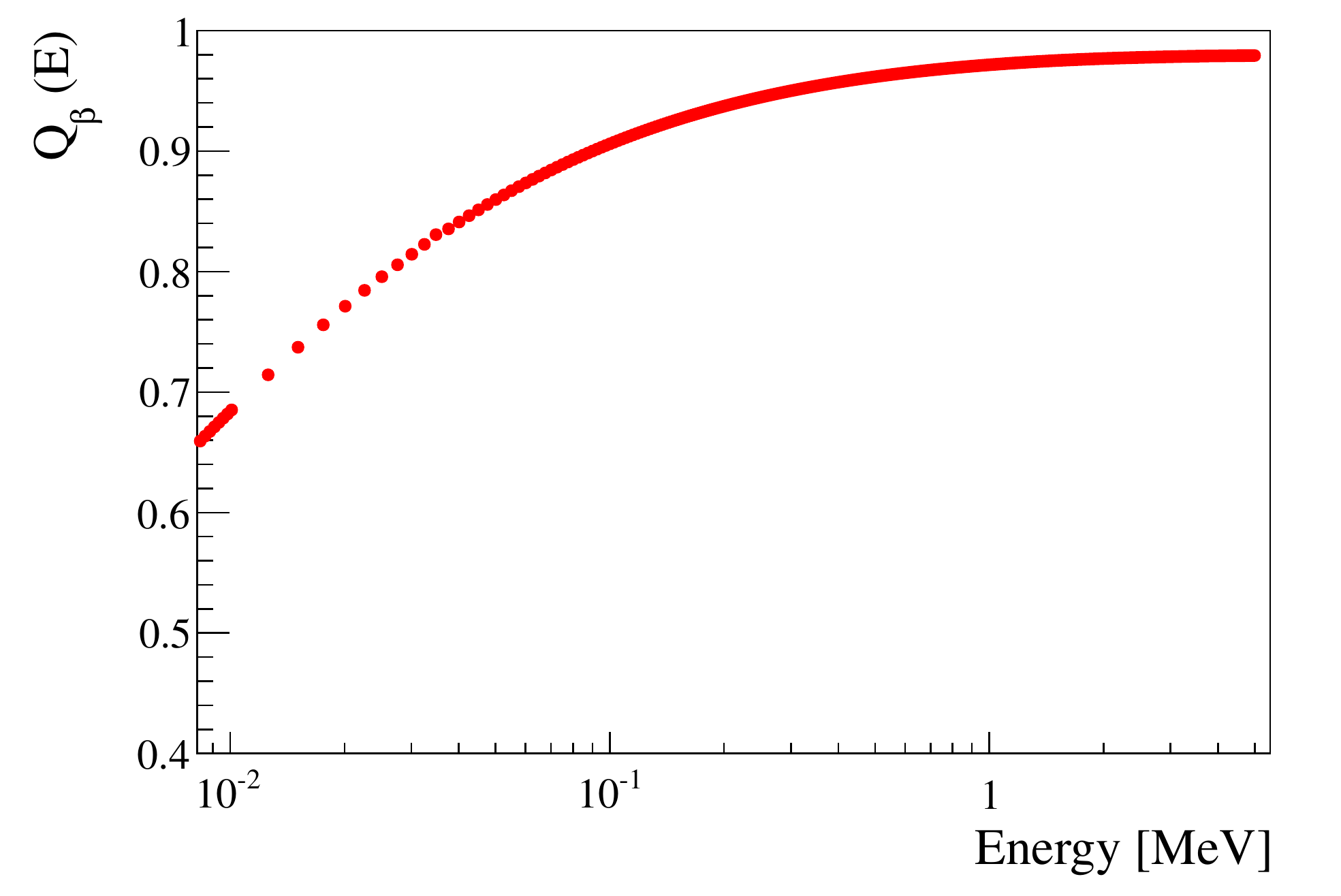}}
\caption{The quenching factor $Q_\beta(E)$ calculated from
Eq.~\ref{Qdef} with the Borexino quenching parameter $kB$ = 0.011\,cm/MeV.}
\label{fig:Qe}
\end{figure}
  
Regardless of the particular functional shape that links $dE/dx$ to $dY^{\rm ph}/dx$, the total number of emitted photons $Y_p^{\rm ph}$ is then related to the amount of deposited energy $E$ through a non--linear relation:
\begin{equation}
Y_p^{\rm ph}= Y_0^{\rm ph} \cdot Q_{p}(E) \cdot E,
\label{Eq:Y_p_ph}
\end{equation}
where $Q_p (E) <1$ is called quenching factor. 
The suffix $p$ recalls that
$Q_{p}(E)$ and $Y_p^{\rm ph}$ depend on the particle type $p$ ($\alpha$, $\beta$, or
$\gamma$).
Considering also the relation of Eq.~\ref{eq:Birks}, the quenching factor for  $\beta$ particles $Q_{\beta}(E)$ can be obtained by integrating $dY^{\rm ph}/dx$; 
\begin{equation}
Q_{\beta}(E) = \frac {1} {E} \int_0^E   \frac {dE} {1+kB \cdot dE/dx} \cdot
\label{Qdef}
\end{equation} 
The energy dependence of $Q_{\beta}(E)$ calculated with $kB$ = 0.011\,cm/MeV 
is reported in Fig.~\ref{fig:Qe}.
The non--linear effect is more and more relevant as long as the energy deposit is below a few hundreds keV.

The quenching effect for $\alpha$ particles with a few MeV of energy (as those from radioactive decays of nuclides at rest) is higher, 
and consequently the amount of emitted light is reduced, 
by a factor of the order of ten with respect to an electron with the same energy~\cite{BxNim}. 
For instance, the 5400\,keV $\alpha$'s emitted by the $^{210}$Po populate the range around 420\,keV (only about 100\,keV is lost in the nucleus recoil).
The determination of the Q$_\alpha$ is discussed in section \ref{sec:alphaQ}.
Quenching is also important for protons but it is not relevant for the results discussed in this paper.

Finally, the quenching effect influences also the detection of $\gamma$ rays.
In general, the amount of scintillation light emitted when a $\gamma$ with energy $E$ is fully
absorbed by the scintillator is significantly lower than the amount of light emitted by
an electron with the same energy $E$.
This effect originates from the fact that $\gamma$ rays cannot directly excite the
molecules of the scintillator.
In fact, the interactions of $\gamma$ rays in the scintillator are observed by detecting the scintillation light
emitted due to the various electrons (and positrons) scattered (or
produced)  by the parent
$\gamma$'s.
Every electron deposits in the scintillator an amount of
energy $E_i$ which is a fraction of the initial energy of the
$\gamma$ ray.
The amount of scintillation light $Y_{\gamma}^{\rm ph}$ generated by the $\gamma$ is then obtained by summing
over all the electron contributions $i$ obtaining the following relation:
\begin{equation}
Y_{\gamma}^{\rm ph} = Y_0^{\rm ph} \sum_i E_i Q_\beta(E_i) \equiv Y_0^{\rm ph} \cdot Q_\gamma(E) \cdot E,
\label{eqn:Qgamma}
\end{equation}
which defines $Q_\gamma(E)$.
Since $Q_\beta(E)$ decreases as a function of the energy, it results that  $Q_\gamma(E)$ is smaller than $Q_\beta (E)$ for the same energy $E$.
As a result, the quenching factor is not negligible for $\gamma$ rays with $E$ in the MeV range.

The amount of Cherenkov light produced is expected to be at the level of
few percent of the scintillation light yield and therefore not
negligible. The number of Cherenkov photons $N_{\rm Ch}^{\rm ph}$ radiated per unit length and wavelength
is~\cite{Tamm}:
 \begin{equation}
 \left( \frac{d^2N^{\rm ph}_{\rm Ch}}{dx d\lambda}\right)_{Ch} \propto
 \frac{1}{\lambda^2} \left( 1- \frac{c^2}{\rm{v}^2 \cdot n^2(\lambda)} \right),
\label{Eq:Cherenkov}
 \end{equation}
 where $n(\lambda)$ is the wavelength dependent refraction index in
 the scintillator and v  is the particle velocity in the
 scintillator.  The dependency of the refraction index of the
 scintillator $n$ on the wavelength $\lambda$ makes the primary spectrum of the
 Cherenkov light energy dependent: in fact, the condition
\begin{equation}
 \left( 1- \frac{c^2}{\rm{v}^2 \cdot n^2(\lambda)}  \right) >0 
 \end{equation}
 must be satisfied. The primary spectrum of
 the Cherenkov light extends into the ultraviolet region which is not
 directly detectable by the PMTs. The mean free path of this
 ultraviolet light in the scintillator is very short (sub-mm) and then
 this light is almost totally absorbed by the scintillator. However,
 the scintillator re--emits a fraction of this absorbed light with
 probability of $\sim$80\% according to its emission
 spectrum. In this way the ultraviolet light invisible to the PMTs is
 transformed into detectable light.
    
The emission times of all types of produced light depend on details of the
charged particle energy loss.
This is the base for the particle discrimination capability of the scintillator. 
We describe the probability $P(t)$ of the light emission time according to
\begin{equation}
P(t) = \sum\limits_{i = 1}^4 \frac{w_i}{\tau_i} \exp^{-t/\tau_i},
\label{eq:P(t)}
\end{equation}
assuming that the energy deposit happens at the time $t$~=~0. The values
of $\tau_{1,2,3,4}$ and $w_{1,2,3,4}$, reported in Table~\ref{table:prop}, have been obtained fitting the experimental data measured in a dedicated 
setup \cite{SciTimeMeas}.

\begin{table}[b]
\begin{center}
\begin{tabular}{l c c c c c } \hline \hline
              & i =    &      1  & 2 & 3   &4           \\
\hline 
$\tau_i$~[ns] & $\beta $    & 3.2& 25& 73.4& 500          \\ 
$w_i$         & $\beta$     & 0.86& 0.05& 0.06& 0.02      \\
\hline
$\tau_i$~[ns] & $ \alpha$    & 3.2& 13.5& 63.9& 480    \\ 
$w_i$         & $\alpha   $    & 0.58& 0.18& 0.14& 0.09    \\

\hline \hline
\end{tabular}
\end{center}
\caption{Parameters used in Eq.~\ref{eq:P(t)} for the calculation of
of the light-emission time probability given separately for $\alpha$
and $\beta$ particles.}
\label{table:prop}
\end{table}

As it results from the Table~\ref{table:prop}, the time distribution of
light generated by $\alpha$ particles has a tail longer than that of
the $\beta$'s. This feature is used to statistically subtract the
$\alpha$--background mainly from $^{210}$Po, as described in detail in Subsection~\ref{subsec:softAB} and in Section~\ref{sec:ab}.

The emission spectrum of the scintillator, the attenuation length, and
the index of refraction as functions of the wavelength have been
extensively measured ~\cite{SciProp1},~\cite{SciProp2},~\cite{CTFLightPropPaper}.
Their values influence the light propagation
inside the detector and the number of photoelectrons collected by the
PMTs.

\section{The calibration with radioactive sources}
\label{sec:calibration}

\begin{table}
\begin{center}
\begin{tabular}{l  c c} \hline \hline
%\begin{tabular}{|l|c|c|} \hline \hline
Isotope           & Type       & Energy [keV]    \\ \hline
Inner Vessel &  &  \\
$^{57}$Co     & $\gamma$   &  122+14 (89$\%$) \\
$^{57}$Co     & $\gamma$   &  136 (11$\%$) \\
$^{139}$Ce    & $\gamma$   &  165             \\ 
$^{203}$Hg    & $\gamma$   &  279             \\ 
$^{85}$Sr       & $\gamma$   &  514             \\ 
$^{54}$Mn     & $\gamma$   &  834             \\ 
$^{65}$Zn      & $\gamma$   &  1115            \\
$^{60}$Co     & $\gamma$   &  1173, 1332     \\ 
$^{40}$K       & $\gamma$   &  1460            \\ 
$^{222}$Rn   &  $\alpha$ \ $\beta$ & 0 - 3200   \\ 
$^{14}$C      &  $\beta$     &  0 - 156         \\ 
$^{241}$Am -- $^9$Be  &  neutrons    & $<$ 11000 \\
                                      &  $\gamma$ (H)  & 2233   \\  
                                      & $\gamma$   $(^{12}$C)   &   4946  \\ \hline
Outer Buffer  &  & \\
$^{228}$Th ($^{208}$Tl)   &   $\gamma$  & 2615 \\
 \hline \hline
\end{tabular}
\end{center}
\caption{Isotopes used in the calibration campaign. The last two rows of the sources deployed in the IV give the two $\gamma$ lines obtained when neutrons are captured on H or by $^{12}$C.}
\label{table:source}
\end{table}

\begin{figure}[b]
\begin{center}
\includegraphics[width = 0.5\textwidth]{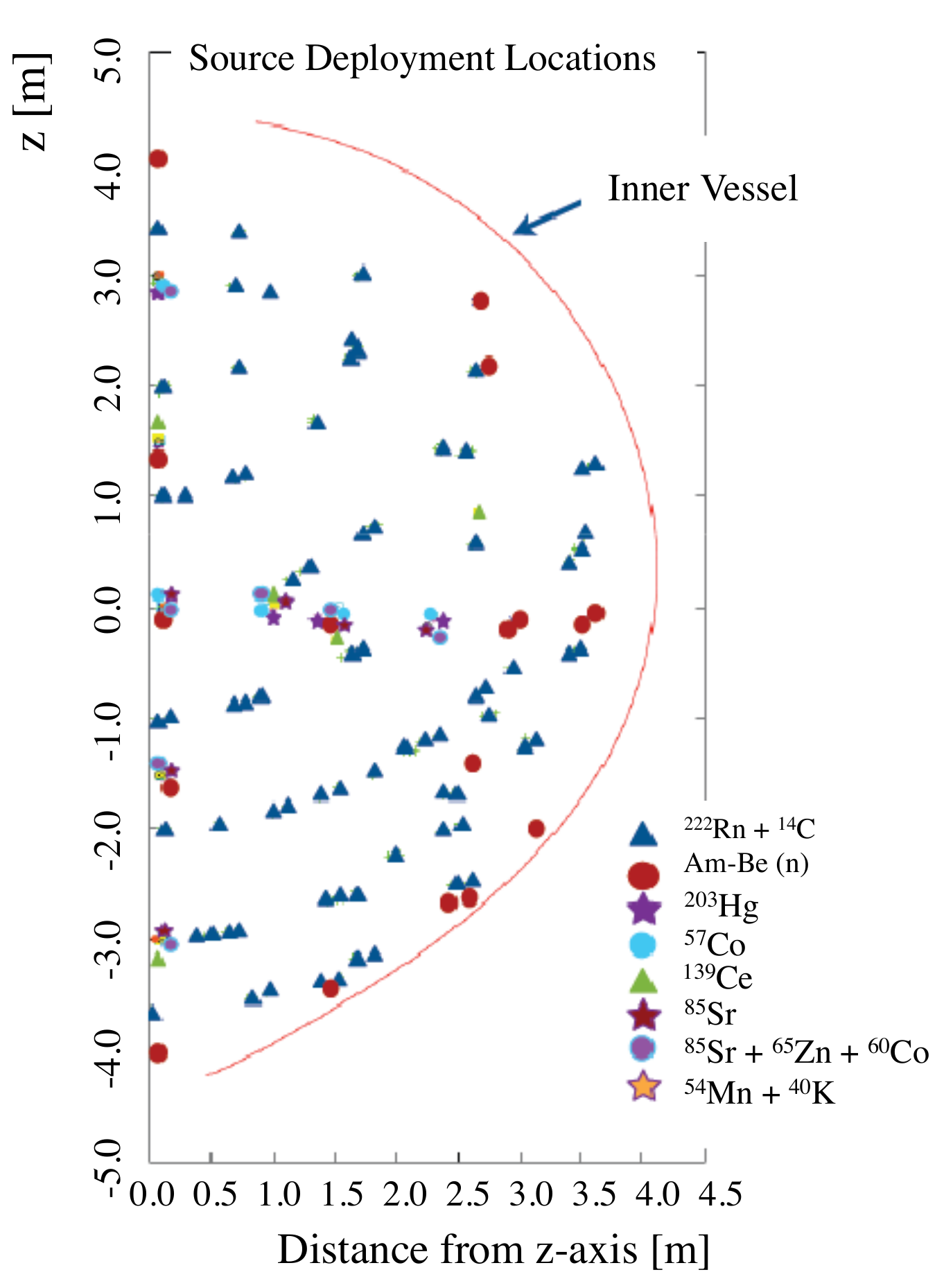}
\caption{Position of the various radioactive sources deployed in the scintillator and used to calibrate the Borexino detector.}
\label{fig:source}
\end{center}
\end{figure}

The detector response function has been modeled in two ways: one is
based on the use of a Monte Carlo code and another one relies on
analytical models. Both approaches benefit from dedicated
calibration campaigns performed with radioactive
sources inserted in the detector.
The campaigns with internal radioactive sources inserted in the scintillator have been performed in October 2008, January, June, and July 2009, while that with an external $\gamma$ source located in the outer
buffer region has been performed in July 2010 and December 2011.
Table~\ref{table:source} lists the sources deployed in the IV and in the outer buffer and Fig.~\ref{fig:source} shows the location  of the various sources in the IV.

All the hardware and details of the calibration systems, as well as demonstration that it preserved the detector radio--purity can be found in~\cite{BxCalibPaper}.

The $\gamma$ sources have been realized by dissolving the radioisotope in
an aqueous solution inside a 1\,cm radius quartz vial. They have been placed
in the detector center and in some off--center positions.  The $\gamma$
particles loose basically all their energy in the scintillator and
they allow to calibrate the absolute energy scale. 

The radon source (a scintillator vial loaded by radon) has been deployed in about 200
positions. This source allowed to study the accuracy of the position
reconstruction and the uniformity of the energy response of the
detector, namely the change of the amount of collected light when a
given energy deposit happens in various positions in the scintillator
volume.

A 10\,Bq $^{241}$Am - $^9$Be neutron source was inserted into the detector in order to study the detector response to neutrons with energies up to 9\,MeV. The neutron interactions in the scintillator made it possible to study also the recoil protons from neutron scattering in the medium and to extend the calibration in the energy range above 2\,MeV. The latter, thanks to the $\gamma$ lines generated by neutron capture on Hydrogen (2233\,keV) and, with a probability at the \% level, on $^{12}$C nucleus (4946\,keV). The neutron captures on the stainless steel insertion arm additionally produced gamma lines up to 8 - 9\,MeV.

A custom made 5.41\,Bq $^{228}$Th source has been placed in the outer buffer in ten different positions. The main purpose of this calibration was to study the external background (Section~\ref{sec:FV}), mainly the energy and radial distributions of the $\gamma$'s from the $^{208}$Tl decay. This isotope is one of the daughter isotopes of $^{228}$Th ($\tau$ = 2.76~years) and the emission probability of the 2615\,keV $\gamma$ ray is 35.6\%.

\section{Energy reconstruction }
\label{sec:estimators}

Borexino works mainly in single photoelectron regime, which means that each
PMT detects on average much less than one photon hit per event. We
define four energy estimators called $N_p$, $N_h$, $N_{pe}$, and
$N_{pe}^d$ counting the number of measured quantities such as number of hit PMTs, hits or photoelectrons during the duration of the cluster defined in Section~\ref{sec:electronics}.

$N_p$ is the number of PMTs that have detected at
least one hit. $N_h$ is the total number of detected hits. $N_h$ generally differs from $N_p$ because a PMT can collect more than one hit if their time separation is more than the dead time discussed in
Section~\ref{sec:electronics}. Both $N_p$ and $N_h$ are computed starting from
the measured values $N_p^{m}$ and $N_h^{m}$:
\begin{equation}
N_p^{m}=  \sum_{j=1}^{N^{\prime}} p^j 
\end{equation}
\begin{equation}
N_h^{m}=  \sum_{j=1}^{N^{\prime}} h^j, 
\end{equation}
where $p^j$ = 1 when at least one photon is detected by PMT $j$ and $p^j$ = 0 otherwise,  $h^j$ can assume the values 0 or
1, 2,...$n$ if the PMT $j$ detects 0, 1, 2,...$n$ hits, and
$N^{\prime}$ is the number of correctly working channels. $N^{\prime}$
is smaller than the total number of installed PMTs because of
temporary electronics problems and due to the PMT failures.
$N^{\prime}$ is evaluated on a nearly event--by--event basis using
calibration events acquired during the run, as described in
Section~\ref{sec:electronics}. The $N_p$ and $N_h$ variables are then
obtained after normalizing the measured values to $N_{tot}= 2000$
working channels through the relations:
\begin{equation}
N_p= \frac{N_{tot}}{N^{\prime}(t)} N_p^{m} = f_{eq}(t) N_p^{m}
\label{eq:equa}
\end{equation}
\begin{equation}
N_h= \frac{N_{tot}}{N^{\prime}(t)} N_h^{m} = f_{eq}(t) N_h^{m},
\end{equation}
in which the time dependent equalization function $f_{eq}(t)$ is
defined as $f_{eq}(t)=\frac{N_{tot}}{N^{\prime}(t)}$.

The third energy variable, $N_{pe}$, is the total number of
collected photoelectrons (p.e.) normalized to $N_{tot}$ channels. 
First, the measured charge $N_{pe}^{m}$ of an event is calculated by summing the hit charges $q_i^j$ expressed in p.e. (based on the charge calibration of single channel given in Eq.~\ref{Eq:charge_calib}):  
\begin{equation}
N_{pe}^{m}= \sum_{i=1}^{N_h^{m}} q_i^j.
\end{equation} 

The number of channels with working ADC's and charge readout, $N^{\prime\prime}$, 
is normally fewer than $N^{\prime}$ by a few tens of channels.
In this charge calculation only hits from such correctly working channels are considered.
The charge $N_{pe}^{m}$ is then normalized to $N_{tot}$=2000 working channels:
\begin{equation}
N_{pe} = \frac{N_{tot}} {N^{\prime\prime}(t)} N_{pe}^{m} = c_{eq}(t) N_{pe}^{m},
\label{eq:Npe}
\end{equation} 
which also defines $c_{eq}(t)$, a time dependent charge normalization function.

An additional variable $N_{pe}^d$,  similar to $N_{pe}$, is calculated by subtracting
the expected number of photoelectrons $q_d$ due to dark noise during the signal duration. It is an estimate of the true number of photoelectrons produced during each event, defined as:
\begin{equation}
N_{pe}^d= c_{eq}(t) \left( \sum_{i=1}^{N_h^{m}} q_i^j -q_d \right).
\end{equation} 
We note that for the purposes of noise reduction, described in
Section~\ref{sec:cuts}, we also define an additional variable
$N_{pe-avg}$ which only differs from $N_{pe}$ in that the sum is
carried over all usable channels $N^{\prime}$:
for those channels that do not
have a working ADC or charge readout, the charge is estimated as the
average charge of all other (valid) hits in a 15\,ns window around the hit.
 
\begin{figure}[th]
\begin{center}
\includegraphics[width = 0.5\textwidth]{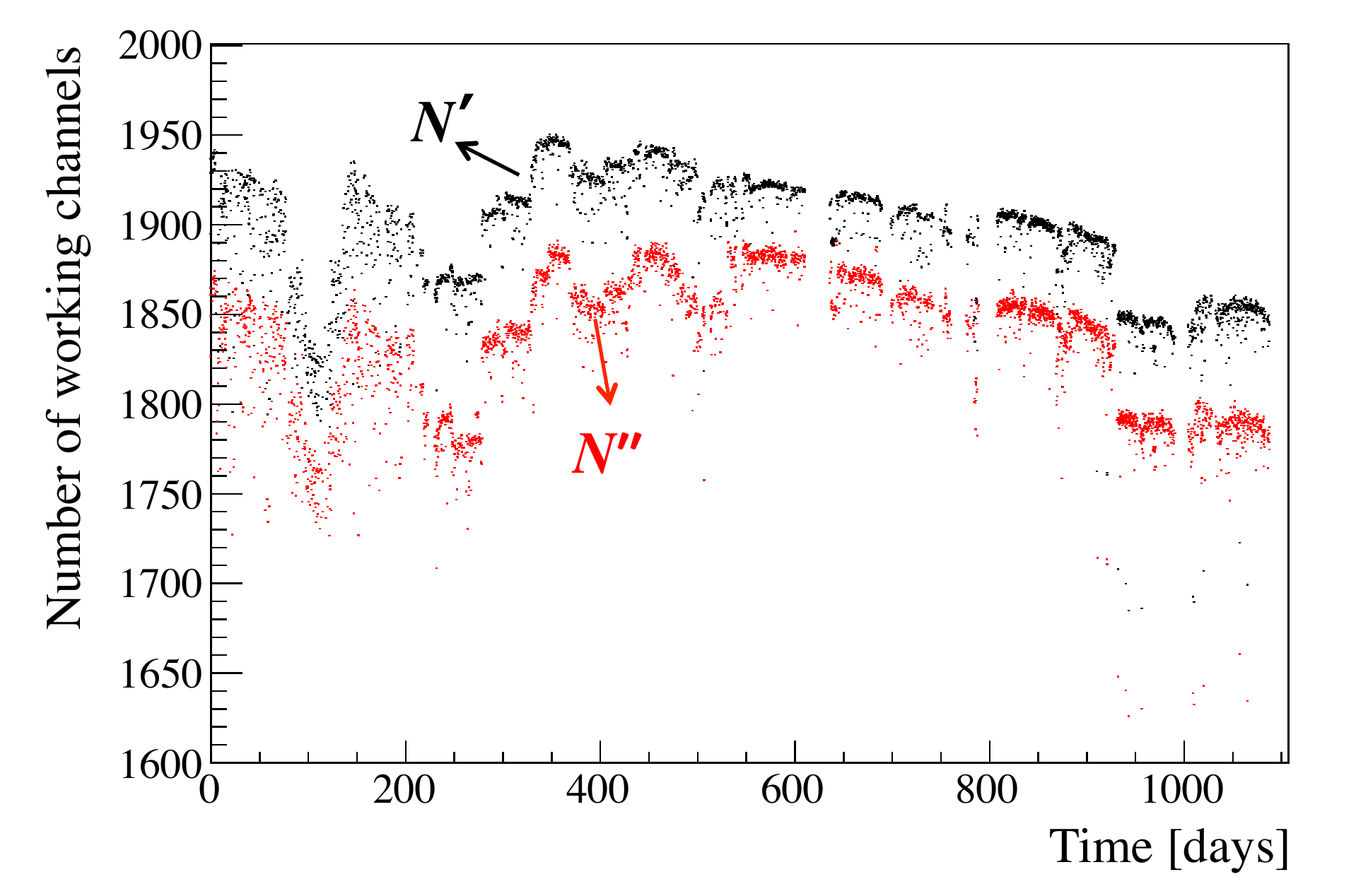}
\caption{The number $N^{\prime}$ (black) and $N^{\prime\prime}$ (red)
of available channels for the computation of $N_{h}$ / $N_p$ and
$N_{pe}$ energy estimators, respectively, as a function of time starting from May 16$^{\rm{th}}$, 2007 (day 0). The slow decrease is due to the PMT mortality, while the sudden changes are due to the failure/repair of the electronics.}
\label{fig:LivePmts}
\end{center}
\end{figure}

Figure~\ref{fig:LivePmts} shows $N^{\prime}$ and $N^{\prime\prime}$ as
a function of time during the data taking.

The different estimators are not independent.
The precise relation between them and the true energy deposit inside the scintillator is
one of the key elements determining the accuracy of the solar--neutrino measurement in Borexino.
Note that in the energy region of interest in this paper, the two estimators 
$N_p$ and $N_h$ are very similar. In addition, the difference between $N_{pe}$ and $N_{pe}^d$ is also very small.
 While details are discussed in Sections~\ref{sec:analytical} and \ref{sec:MC}, here it is useful to point out that an energy deposit of 1\,MeV corresponds to about $N_{pe} \simeq N_{pe}^d \simeq$ 500.

\section{Position reconstruction}
\label{sec:position}

\begin{figure}[t]
\begin{center}
\includegraphics[width = 0.5 \textwidth]{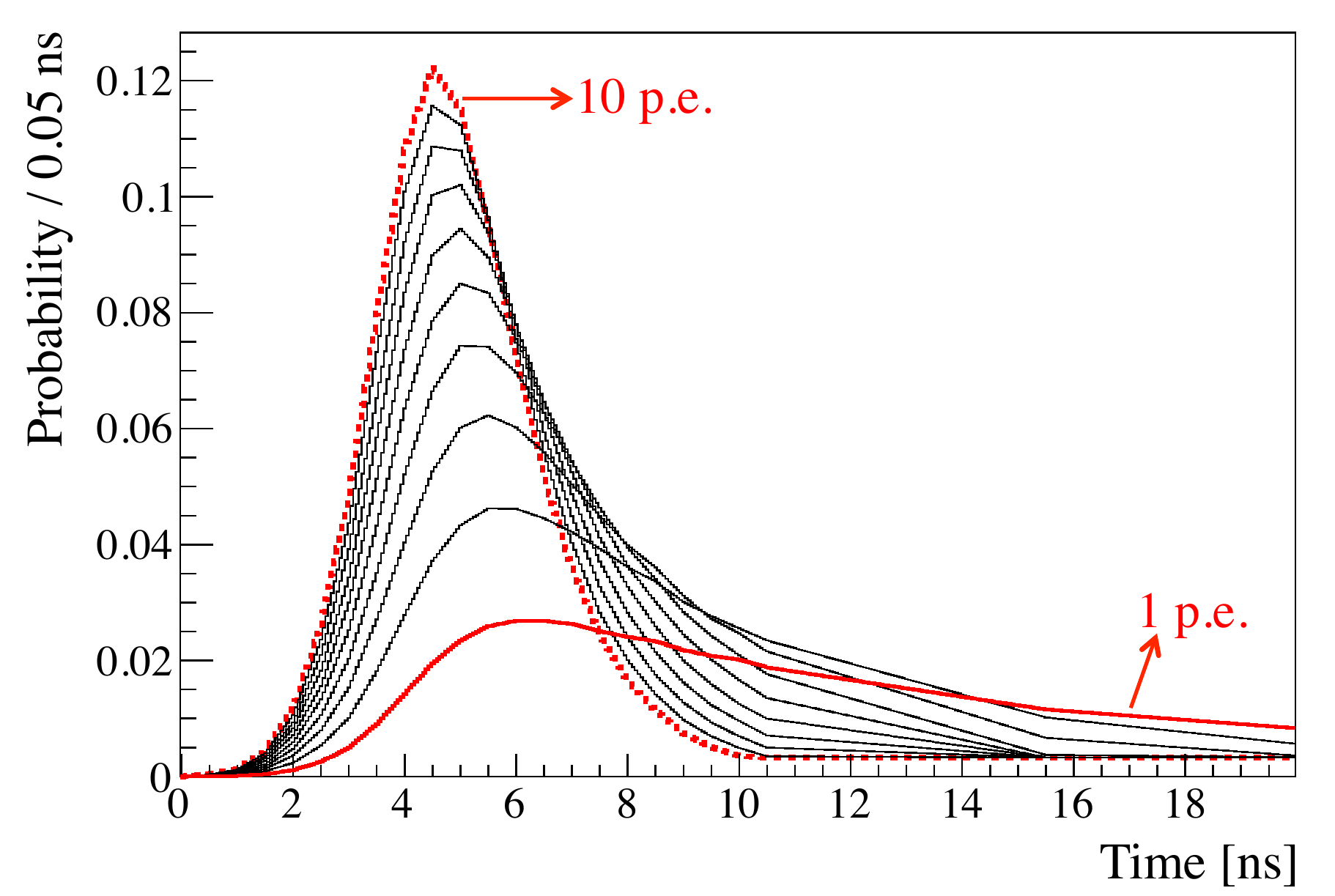}
\end{center}
\caption{Probability density functions for detected hits as a function of time elapsed from the emission of scintillation light. Different curves are for increasing values of the hit charge $q_i^j$ collected at the phototube: from 1 p.e. (red solid curve) to 10 p.e. (red dashed curve).}
\label{fig:pdf}
\end{figure}

The position reconstruction algorithm determines the most likely
vertex position $\vec r_0$ of the interaction using the arrival times $t_i^j$ of the detected photons on each PMT (defined in Section~\ref{sec:electronics}) and the position vectors $\vec r^j$ of these PMTs.
The algorithm subtracts from each measured time $t_i^j$ a position dependent time--of--flight $T_{flight}^j$ from the interaction point to the PMT $j$:
\begin{equation}
T_{flight}^j(\vec r_0, \vec r^j)  = \mid \vec r_0 -\vec r^j \mid \frac{n_{\rm eff}}{c}
\label{eq:tof}
\end{equation}
and then it maximizes the likelihood $L_E(\vec r_0, t_0 \mid (\vec
r^j, t_i^j))$ that the event occurs at the time $t_0$ in the position
$\vec r_0$ given the measured hit space--time pattern $(\vec
r^j,t_i^j)$.
The maximization uses the probability density functions (p.d.f.) of the hit detection, as a function of time elapsed from the emission of scintillation light, which are shown in Fig.~\ref{fig:pdf}.
As it can be seen, the exact shape of the p.d.f. used for every hit depends on its charge $q_i^j$.

The quantity $n_{\rm eff}$ appearing in Eq.~\ref{eq:tof}  is called {\it "effective refraction index"} and it is used to define
an effective velocity of the photons: it is a single parameter  that globally takes
 into account the fact that photons with different wavelengths travel with 
different group velocity and that photons do not go straight from the emission to the detection points but they can be scattered or reflected. 
The value $n_{\rm eff}$~=~1.68 was determined using the calibration data with radioactive sources described in Section~\ref{sec:calibration}.
More details on this parameter can be found in~\cite{BxCalibPaper}.
Note that $n_{\rm eff}$ is larger than the actual index of
refraction of pseudocumene measured at 600\,nm to be $n_{\rm
PC}$~=~1.50.

\begin{figure}[h]
\begin{center}
\includegraphics[width = 0.5\textwidth]{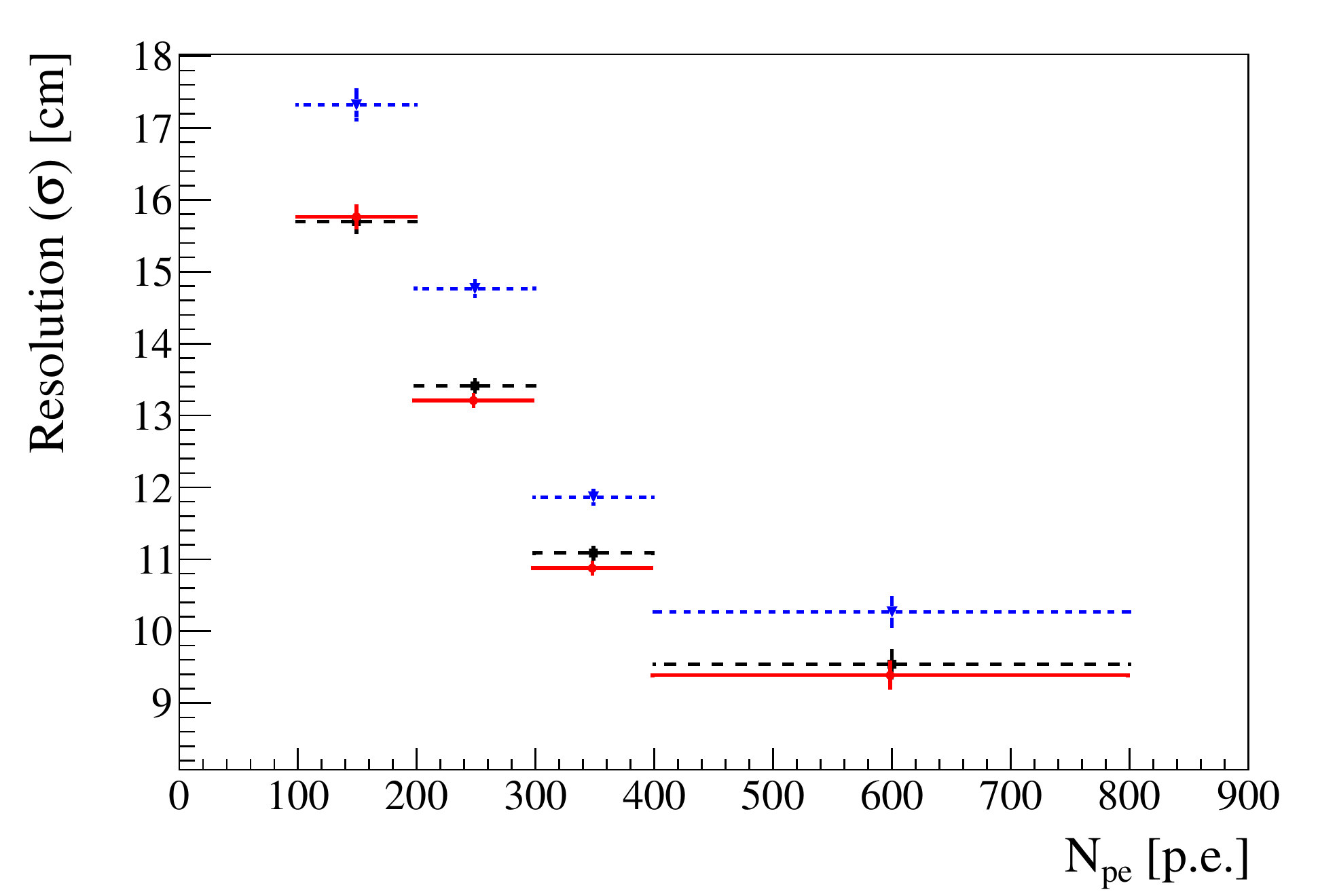}
\caption{Resolution ($\sigma$) of the reconstructed $x$ (solid red), $y$
(dashed black), and $z$ (dotted blue) coordinates as a function of energy (in
number of photoelectrons) for events from calibration sources placed in the center of the detector. }
\label{fig:Radon_resol}
\end{center}
\end{figure}

The data collected during the calibration campaign allowed to map thoroughly
the performance of the position reconstruction code
as a function of energy and position. In particular, the position
reconstruction resolution ($\sigma_{x,y,z}$) has been studied for
different energies and positions. As an example,
Fig.~\ref{fig:Radon_resol} shows the dependency of $\sigma_x$, $\sigma_y$, $\sigma_z$
on energy for events in the center: the resolution for coordinates $x$
and $y$ ranges from 15\,cm at lower energies ($N_{pe}\simeq$ 150\, corresponding to about 300\,keV) to 9\,cm at
higher energies ($N_{pe}\simeq$ 500\,corresponding to about 1\,MeV). The $z$ coordinate is reconstructed with a
slightly worse resolution as expected, since the PMT coverage in
$z$ has a larger granularity.

\begin{figure}[t]
\begin{center}
\centerline{\includegraphics[width = 0.5\textwidth ]{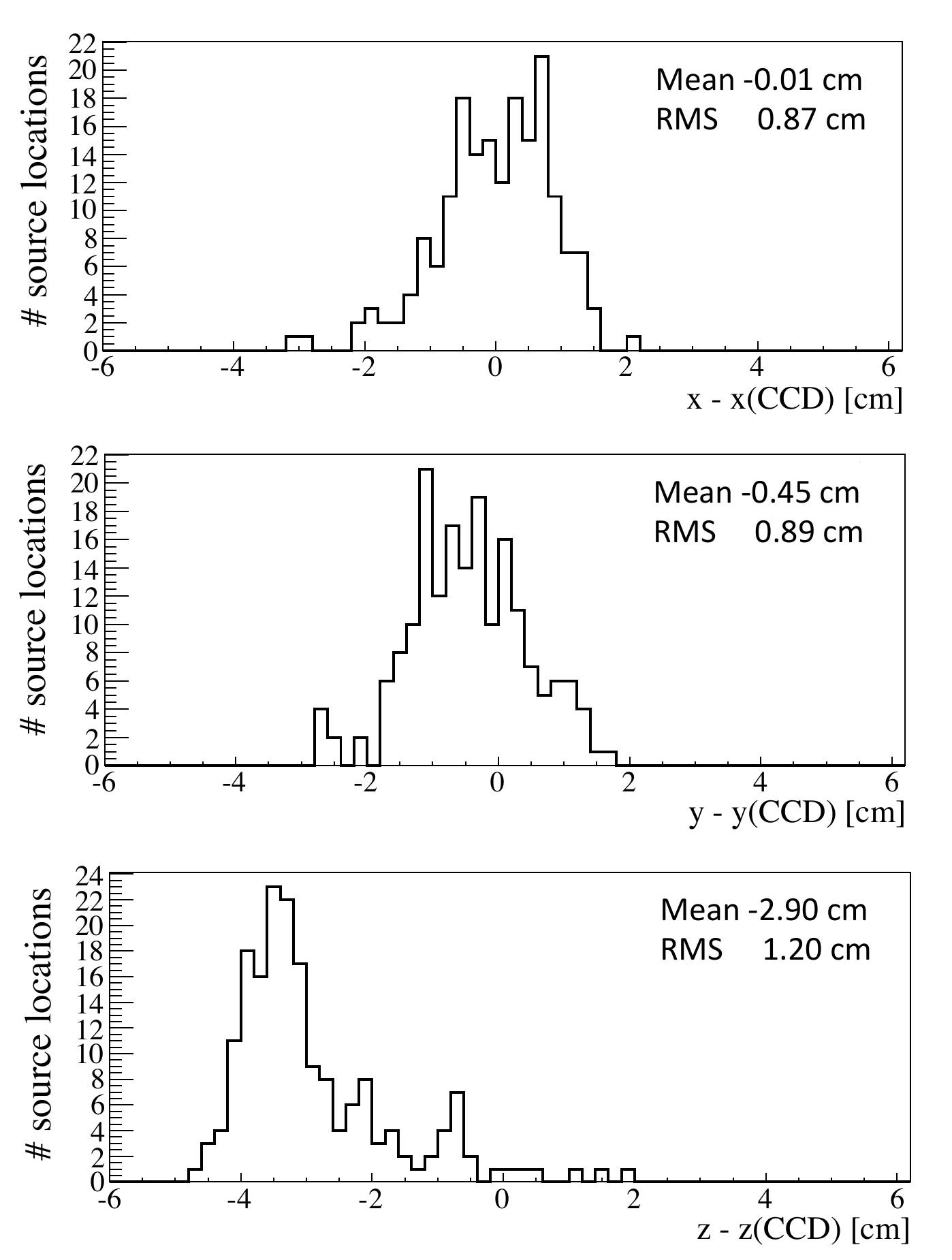}}
\end{center}
\vspace{-1 cm}
\caption{Distribution of differences between the reconstructed and nominal (CCD) coordinates for the radon source data ($^{214}$Po) measured in 182 different positions in the scintillator: $x$ (top), $y$ (center), and $z$ (bottom)   }
\label{fig:xyz_rn}
\end{figure}

The nominal and reconstructed source positions have been compared for
all the collected calibration data. The nominal position of the source
is obtained independently by a system of 7~CCD cameras mounted on the
Stainless Steel Sphere.  Figure~\ref{fig:xyz_rn} shows, as an example,
the difference between the mean value of the reconstructed coordinates
$x$, $y$, and $z$ and the corresponding nominal values for events due
to $^{214}$Po alphas from the $^{222}$Rn chain.  The coordinates $x$ and $y$
are well reconstructed: the sigma of the distribution is $\sim$0.8\,cm
with tails that extend up to 3\,cm (note that the contribution of the 
uncertainty on the CCD position is not disentangled). Instead, the $z$--coordinate shows a systematic shift of
$\sim$3\,cm downwards with respect to the nominal position. The origin
of this shift is unclear: it is not related to the algorithm itself,
since the reconstruction of Monte Carlo simulated data does not show
this effect. It could be due to a small variation of the index of
refraction as a function of $z$ due to the gradient in temperature (and
therefore in density) of the scintillator. In any case, the
contribution of this shift to the systematics of the fiducial volume
determination is small (less than 0.2\%).

\section{Backgrounds  and choice of fiducial volume}
\label{sec:FV}

The achievement of extremely low background levels  in Borexino represents  the essential milestone
that has allowed   to obtain the  solar neutrino results.
In this section we describe different background components classified as:
\begin{itemize}[leftmargin = *]
\item{\it  External and surface background}: events generated outside the
scintillator are referred to as external background, while events generated by the radioactive contaminants of
the nylon IV are referred to as surface background. These background components, described in Subsection~\ref{sub:extBk}, determine the shape of the wall--less region in the Borexino scintillator (Fiducial Volume, FV) used in different analysis, as explained in Subsection~\ref{sub:FV}.

\item {\it Internal background}: events generated by the decay of radioactive isotopes contaminating the scintillator are described in Subsection~\ref{subsec:InternalBack}.
\item{\it Cosmic muons and cosmogenic background}: 
subsection~\ref{subsec:CosmogenicBack} is dedicated to the discussion of the residual muon flux and to the muon--induced radioisotopes. More details about the cosmogenic background in Borexino at 3800\,m water--equivalent depth can be found in~\cite{cosmogenics}.
\end{itemize} 

\subsection{External and  surface background}
\label{sub:extBk}

\begin{table}[b]
\begin{center}
\begin{tabular}{l c}
\hline \hline
Source &  cpd in the $pep$--FV  \\
       &  250\,keV $<$ E $<$ 1300\,keV   \\
\hline
$^{208}$Tl from PMTs  &$\sim$0.2  \\  
$^{214}$Bi from PMTs  &$\sim$0.9  \\  
$^{40}$K from PMTs  &$\sim$0.2 \\
Light cones and SSS & 0.6 -- 1.8  \\
Nylon vessels & $<0.05$     \\
End--cap regions & $<0.06$    \\
Buffer & $<0.02$  \\
\hline \hline
\end{tabular}\end{center}
\caption{Expected rates of external $\gamma$--ray backgrounds relevant
for the $pep$ and CNO analysis. The estimates for $^{208}$Tl, $^{214}$Bi, and
$^{40}$K from the PMTs were made using the Geant-4 Monte Carlo code, starting
from the measured contamination of the PMTs~\cite{BxBack}. The lower and upper
limits for the rates from the SSS and light cones were scaled to
the $pep$--FV from previous estimates~\cite{Vessel}, assuming a
$\gamma$--ray absorption length of 25\,cm. The upper limit for the background originating in the buffer
fluid has been estimated with the Monte Carlo assuming contamination of
$^{238}$U and $^{232}$Th in the buffer that are 100 times greater than
those in the scintillator.}
\label{tab:residuals_ext}
\end{table}

The main source of external background is the radioactivity of the materials that contain and surround the scintillator: examples are  the
vessel support structure, PMTs, light cones, and other
hardware mounted on the Stainless Steel Sphere. Since the radioactive
decays occur outside the scintillator, the only background that can
reach the inner volume and deposit energy are $\gamma$ rays.
The position reconstruction (Section~\ref{sec:position}) allows to select the FV where the event rate due
to external background is negligible. The $\beta$ or $\gamma$ decays due
to surface background events may be reconstructed at some distance
from the IV and the FV definition aims to exclude also these events.

\begin {figure*}
\begin {minipage}{0.45\linewidth}
\center
\includegraphics [width=3.2in] {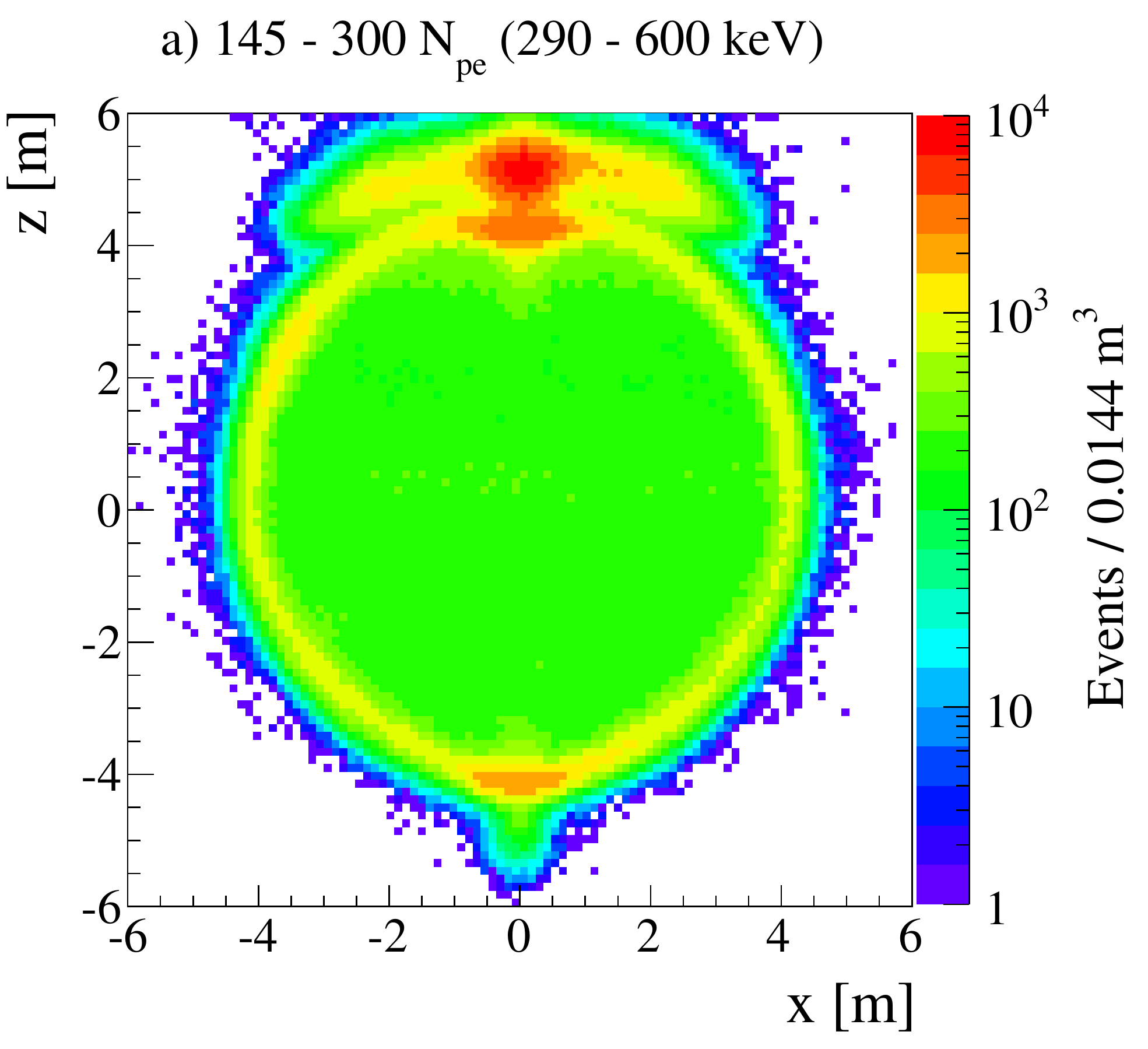}
\end {minipage}
\begin{minipage}{0.45\linewidth}
\center
\includegraphics [width=3.2in] {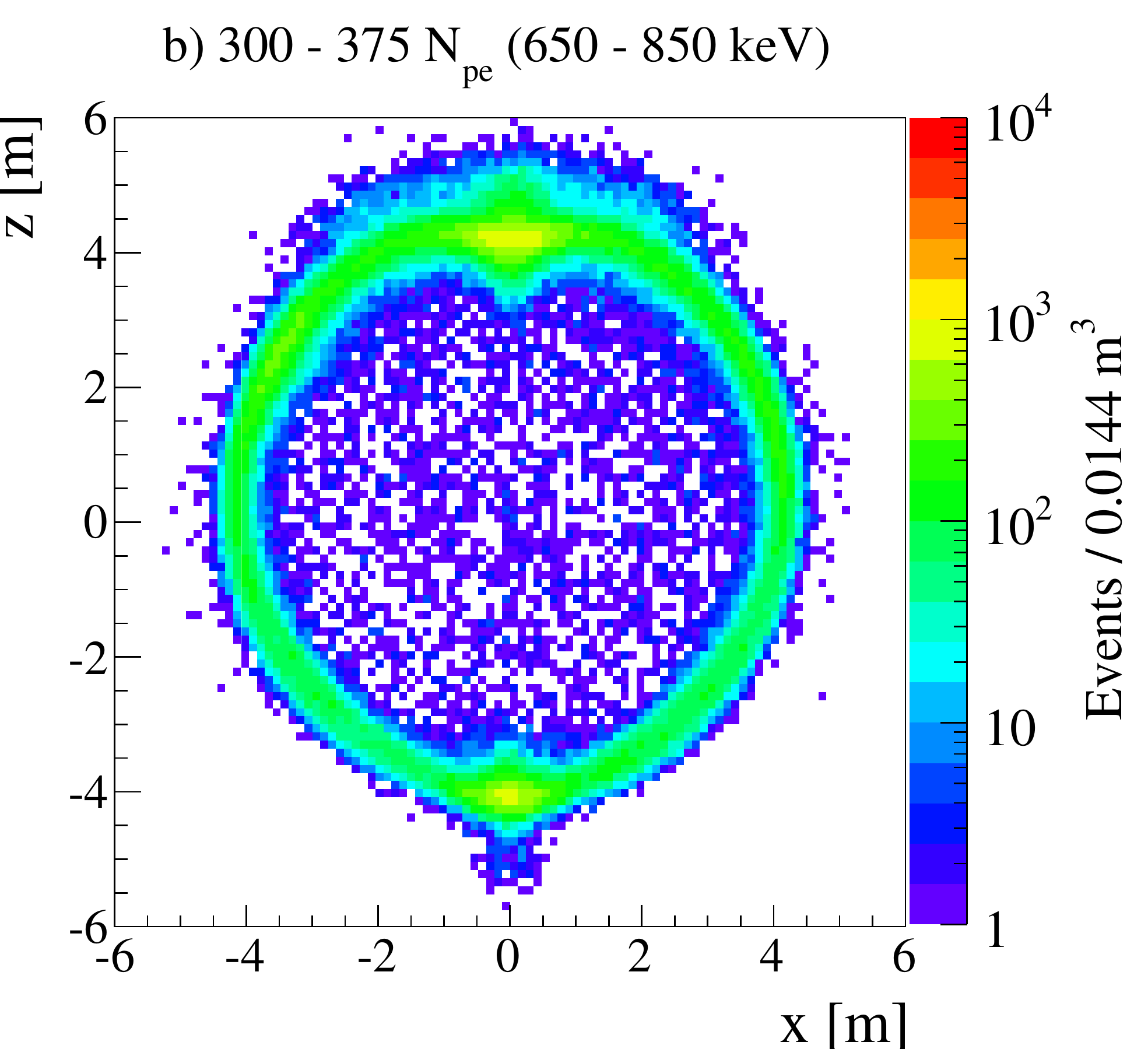}
\end {minipage}
\begin {minipage}{0.45\linewidth}
\vspace{3 mm}
\center
\includegraphics [width=3.2in] {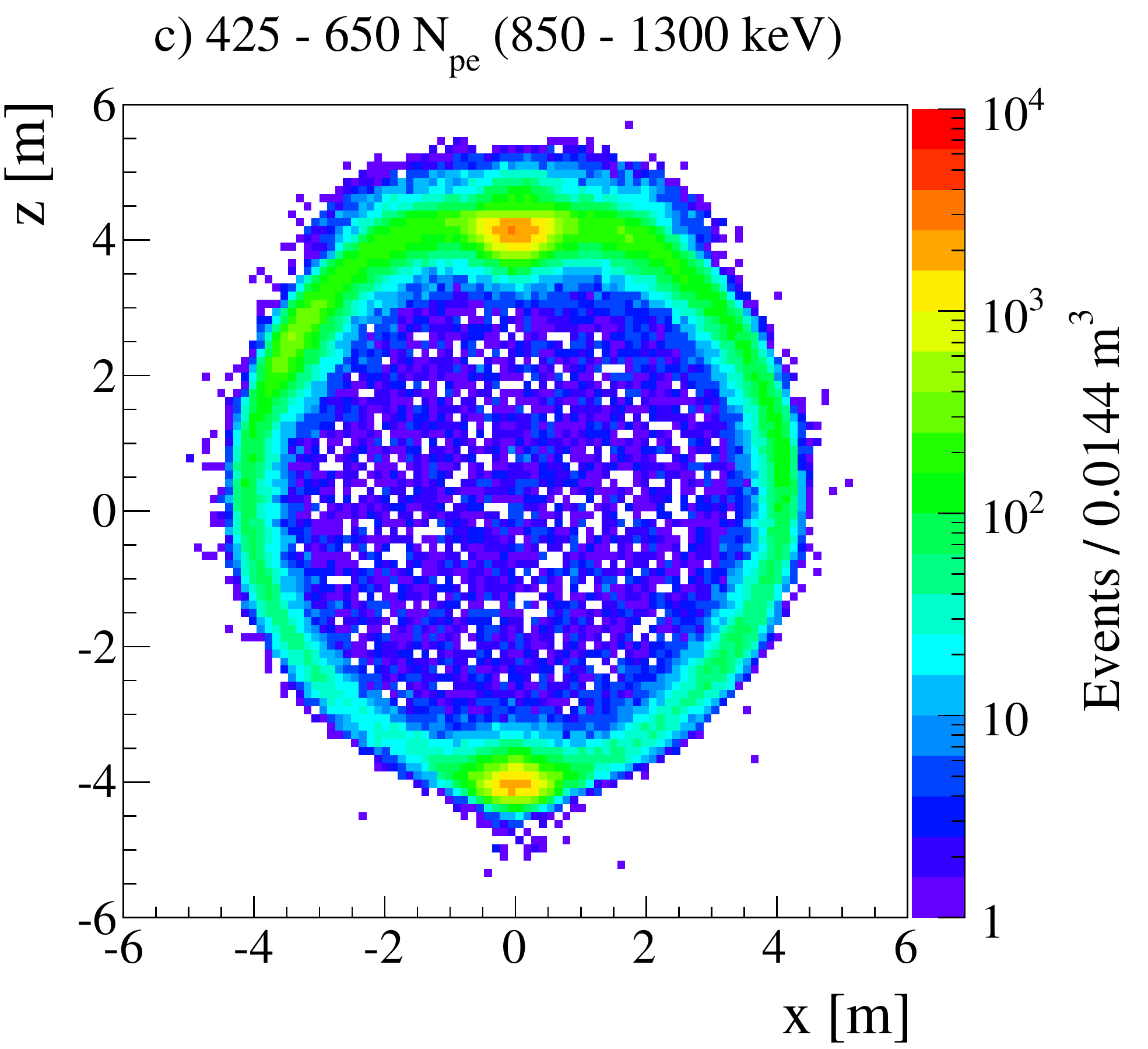}
\end {minipage}
\begin{minipage}{0.45\linewidth}
\vspace{3 mm}
\center
\includegraphics [width=3.2in] {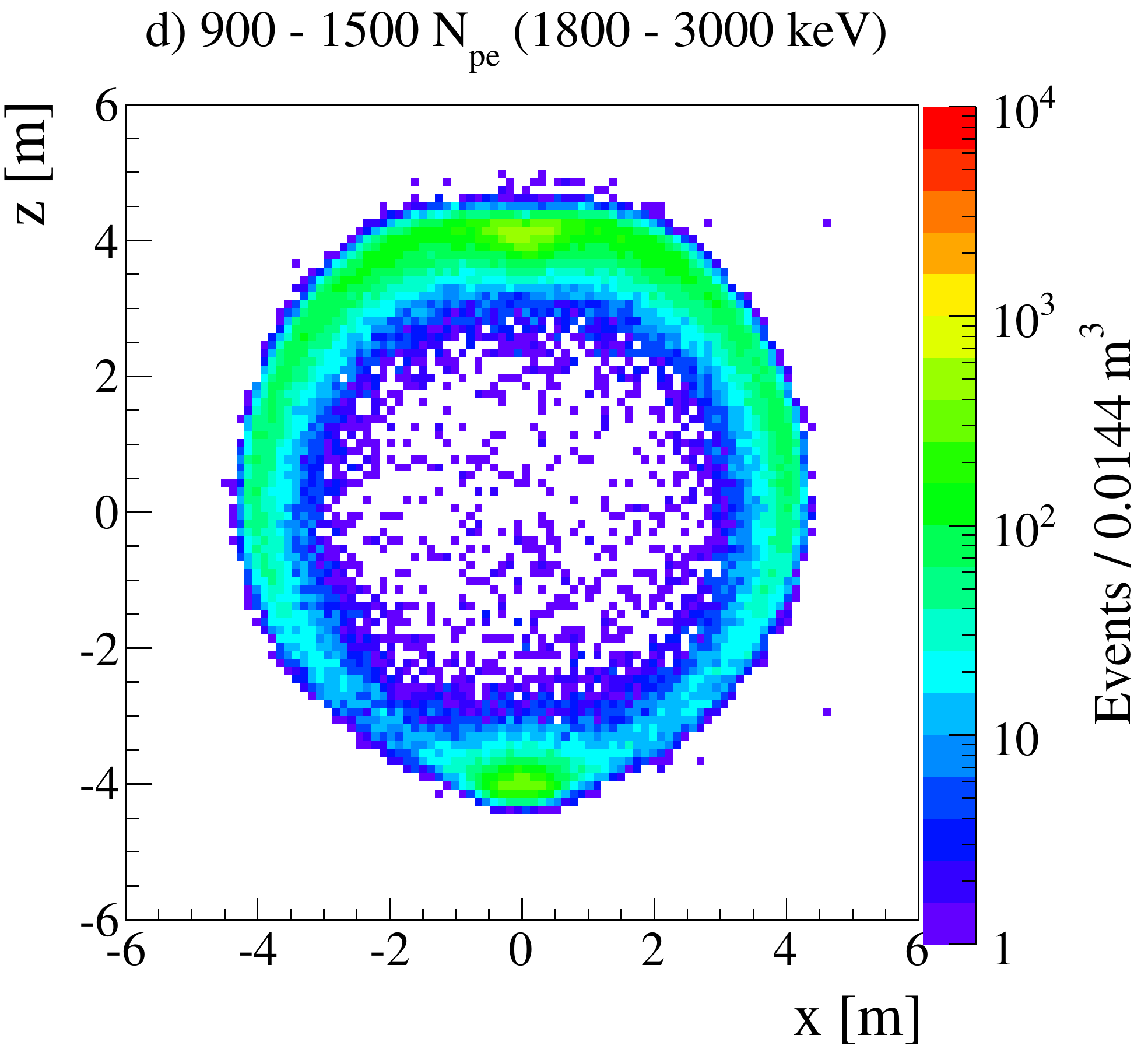}
\end {minipage}
\caption {Spatial distribution (in the $x-z$ plane, $|y| < 0.5$ m) of all reconstructed events (besides muons) in
different energy regions. The color axis represents the number of events per 0.0144 m$^3$ in a pixel of 0.12\,m x 1.00\,m x 0.12\,m ($x$ x $y$ x $z$). a) $^{210}$Po peak region: 145 - 300\,$N_{pe}$ (290 - 600\,keV),  b) $^7$Be shoulder: 300 - 375\,$N_{pe}$ (650 - 850\,keV), c) $^{11}$C energy region: 425 - 650\,$N_{pe}$ (850 - 1300\,keV), d) $^{208}$Tl peak region: 900 - 1500\,$N_{pe}$ (1800 - 3000\,keV). Events occurring outside the IV ($R$ = 4.25\,m), near the top end-cap ($z$ = 4.25\,m), most clearly seen in the low--energy region shown in a) plot, are due to the small leak of scintillator from the IV to the buffer region (Section~\ref{subsec:leak}).}
\label {fig:BackStudy} 
\end {figure*}

Figure~\ref{fig:BackStudy} shows the distribution of all detected
events in the scintillator for different energy ranges. The IV is clearly visible as a ring of
higher activity at a radius $R \simeq 4.25$\,m. Also distinctly visible are the vessel end--caps (IV
end--caps at $z$~=~$\pm$4.25\,m, OV end--caps at $z$~=~$\pm$5.5\,m), the regions of 
the highest activity in the scintillator. These events mainly populate the energy region between the $^{14}$C end--point and the $^{210}$Po peak, shown in a) panel of Fig.~\ref{fig:BackStudy}. The high rate of events occurring outside the IV, above the top end--cap, most prominently seen in the 145 - 300 $N_{pe}$ region, is due to a small leak in the IV (see Subsection~\ref{subsec:leak}).

The spatial distribution of the external background is shown in panel d) of Fig.~\ref{fig:BackStudy},
reporting the reconstructed position of events with  $N_{pe}$ between
900 and 1500. The higher rate of external background in the top hemisphere, compared to the bottom one, is due to the nylon vessel being shifted slightly upwards and therefore closer to the PMTs. As can be seen, the number of
events decreases as one moves radially away from the SSS towards the
IV center. This energy region is dominated by $^{208}$Tl and
$^{214}$Bi $\gamma$--ray interactions, with a smaller contribution
($\sim$25\%) from muon--induced $^{10}$C and $^{11}$C decays, dominating the energy region 425 - 650 $N_{pe}$, shown in panel c) of Fig.~\ref{fig:BackStudy}. 

\begin{figure}[h]
\centering{\includegraphics[width = 0.5\textwidth]{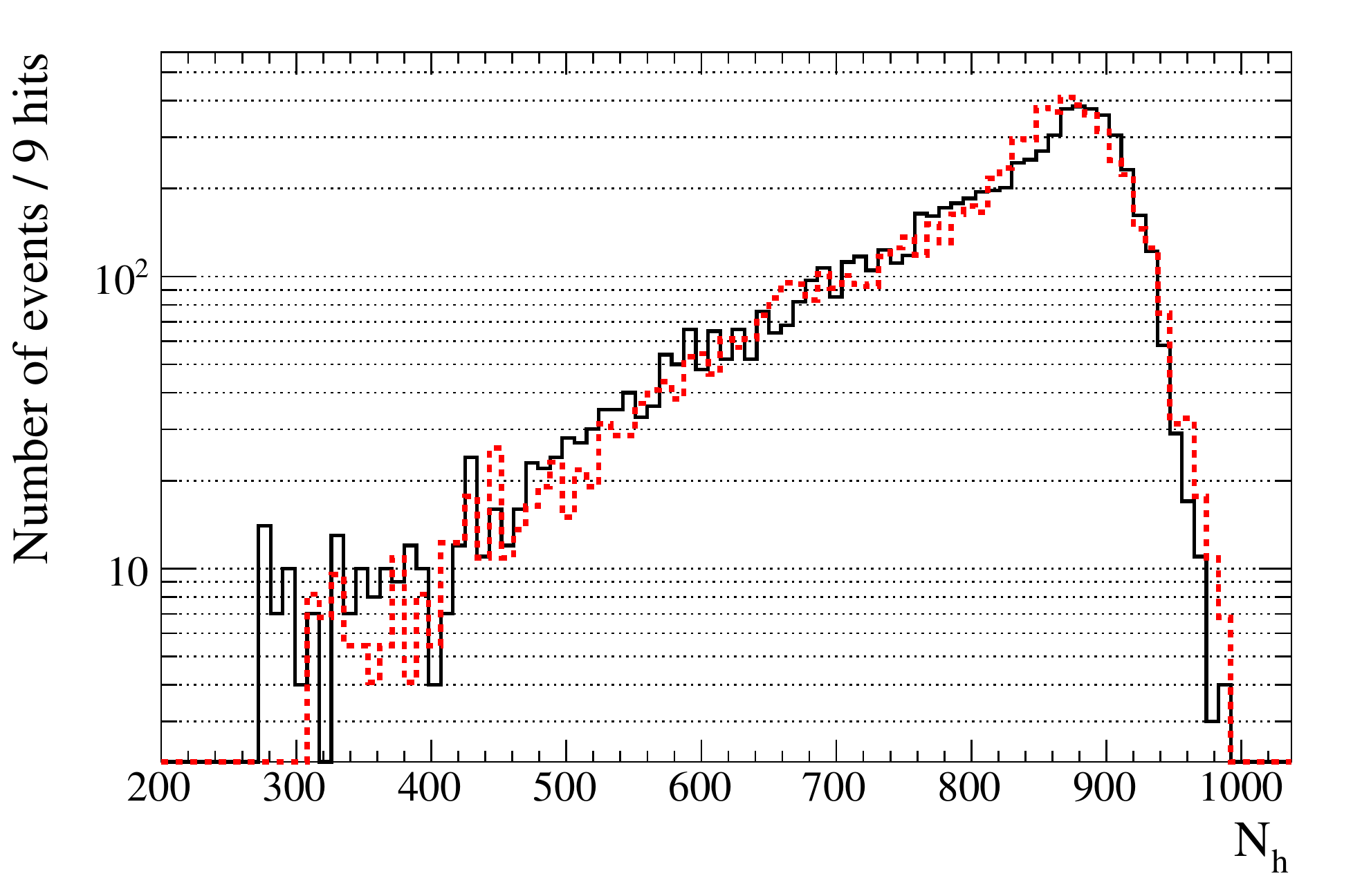}}
\caption{Comparison between the energy spectra ($N_h$ energy estimator) for events from the external $^{228}$Th / $^{208}$Tl source calibration data (black solid line) and from the 2610\,keV simulated external $\gamma$ rays (red dotted line), reconstructed within 3\,m from the detector center. The $\chi^2$/{\rm NDF} between the two histograms is $\simeq$1.2. The source is positioned in the upper hemisphere; a similar result is obtained for the lower one.}
\label{fig:MCThSourceE}
\end{figure}

The contribution of the external background is small in the $^7$Be-neutrino measurement with the 75\,ton FV but it is important for the $pep$ and CNO neutrino detection. Table~\ref{tab:residuals_ext} presents the expected count rates for $\gamma$ rays of different isotopes from different external sources in the FV used in latter analysis. The exact shape of this FV is defined in Table~\ref{tab:FV}. 
The external background contribution has been included in the multivariate--fit approach (see Section~\ref{sec:Multivariate}) exploiting the different radial dependencies of the external background (which exponentially decreases inside the IV) anf of the internal background and the signal (which are both assumed to be uniformly distributed in the FV).

\begin{figure*}
$
\begin{array}{cc}
\centering{\includegraphics[width = 0.5 \textwidth]{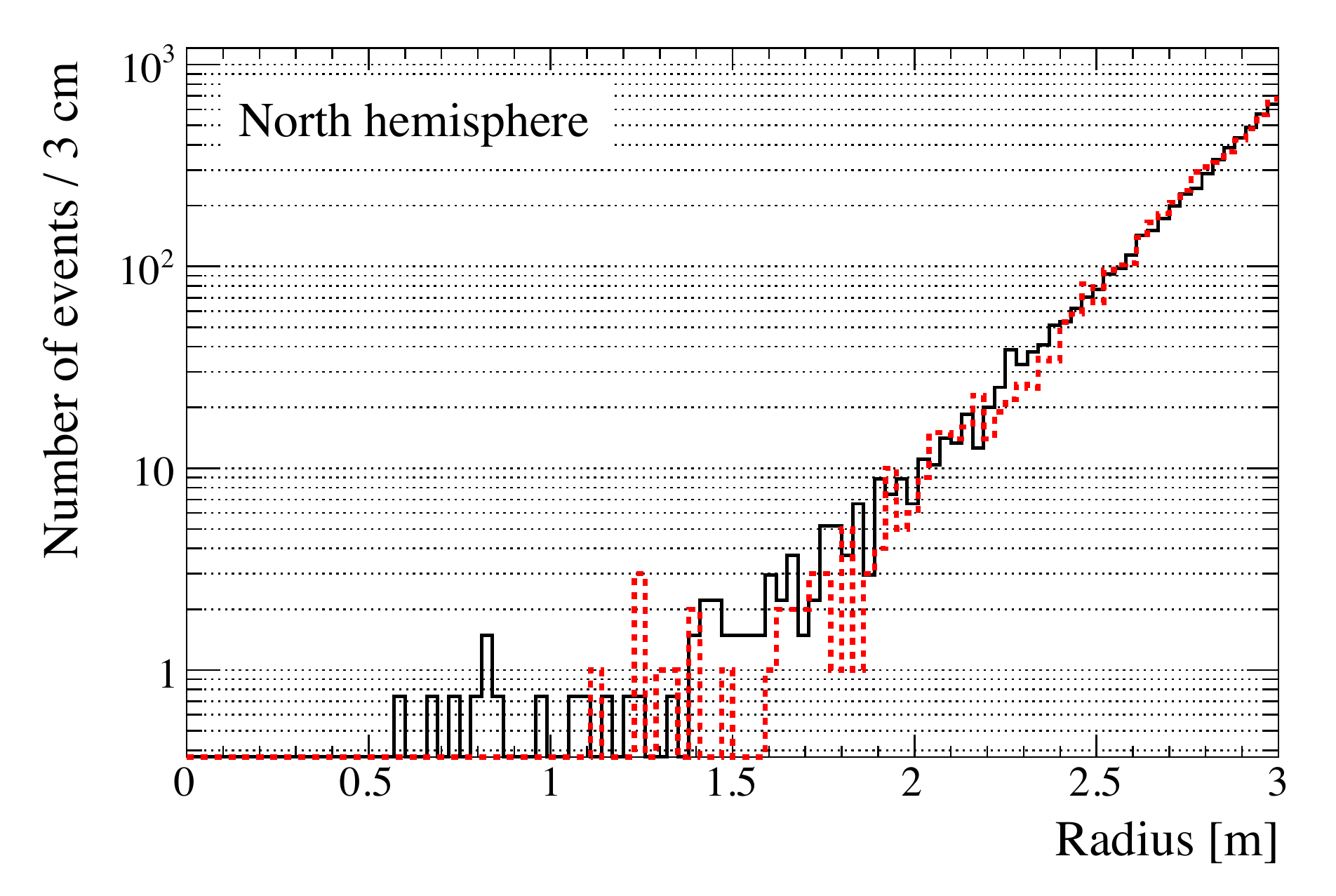}} &
\centering{\includegraphics[width = 0.5 \textwidth]{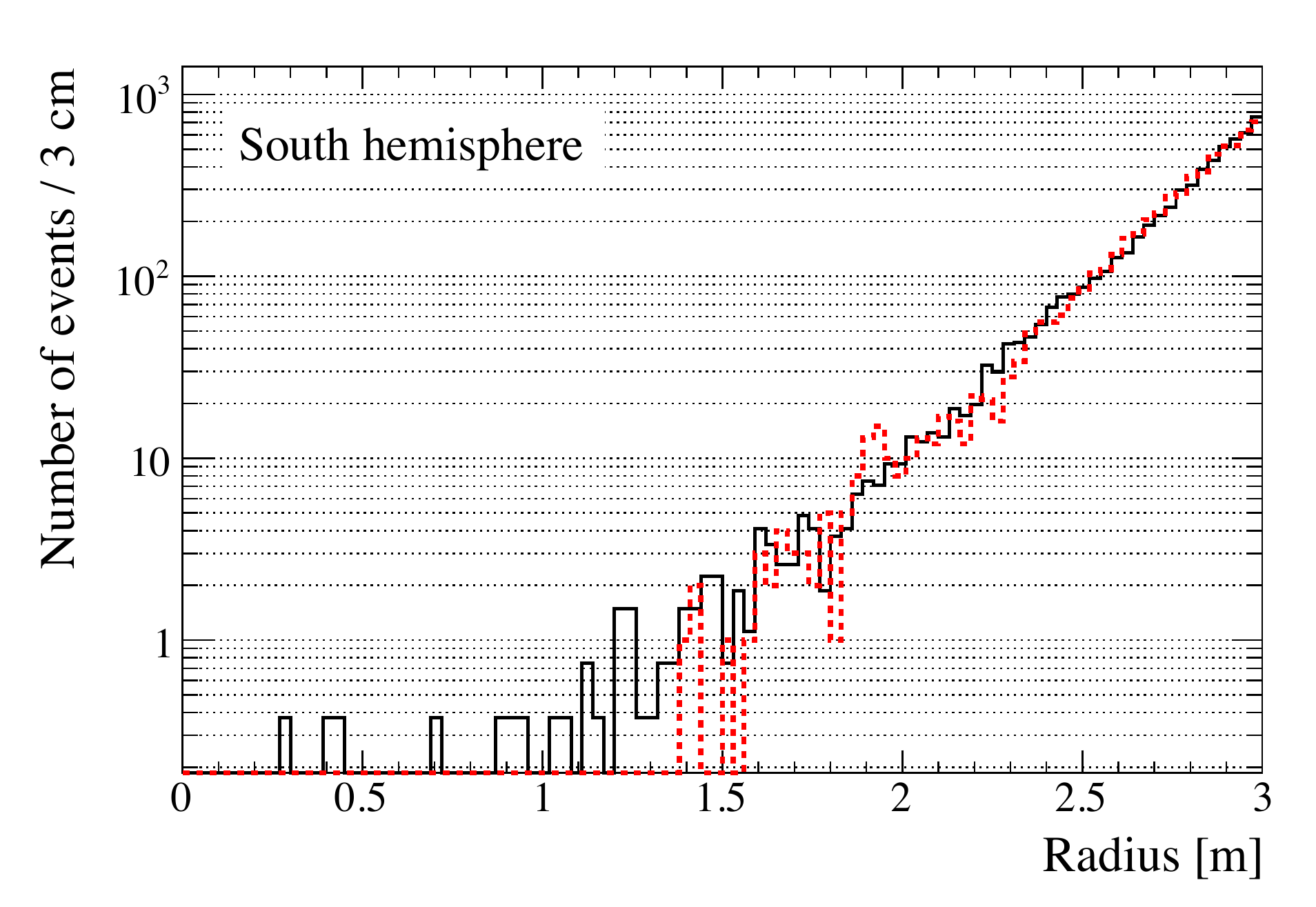}}
\end{array}
$
\caption{Comparison between the reconstructed radius of events with energy $N_h> 400$ from the external $^{228}$Th / $^{208}$Tl source calibration data (black solid line) and from the 2610\,keV simulated external $\gamma$ rays (red dotted line), reconstructed within 3\,m from the detector center. The left and right plots show examples for a source position in the upper/north and lower/south hemispheres, respectively. The $\chi^2$/NDF between the two histograms (Monte Carlo and data) is in the range 0.8 -- 0.9.}
\label{fig:MCThSourceR}
\end{figure*}

The Monte Carlo code has been used to obtain the energy spectrum and the radial distribution of the external background. The background originated from the radioactive contamination ($^{208}$Tl and $^{214}$Bi) of all the PMTs is simulated adopting particular software procedures to reduce the amount of necessary computation time: in fact, obtaining at the end of the simulation a spectrum of about $10^4$ events requires to generate and track more than $10^{12}$ events. The resulting CPU time needed for a single event is about $6 \times 10^{-5}$ s leading to an acceptable  total computation time. The validity of the simulation has been established by comparing the radial and energy distributions of the events measured with the $^{228}$Th external calibration source (having $^{208}$Tl as one of its daughters, see Section~\ref{sec:calibration}) and their simulation. Figure~\ref{fig:MCThSourceE} compares the energy spectra of the simulated and measured events while Fig.~\ref{fig:MCThSourceR} shows the agreement between the measured and simulated radial distributions. The attenuation length of 2610\,keV $\gamma$ rays was measured to be 25\,cm.

\subsection{Fiducial volume in different analyses}
\label{sub:FV}

The optimal choice of FV depends on the type of analysis to be performed. For the measurement of the interaction
 rate of $^7$Be, $pep$, and CNO neutrinos, the FV was defined searching for a volume where all events are almost uniformly distributed and thus the external background is negligible or, at least, is  strongly suppressed. Additional requirements have been introduced in the $pep$ and CNO neutrino analysis (Section~\ref{sec:pepCNOResults}) and are related to the best efficiency of the $^{11}$C subtraction (Section~\ref{sec:c11}).
 We have therefore chosen the FV to lie within a sphere of radius $R_{max}$  with a cut in the  $z$ coordinate ($z<z_{min}$ and $z>z_{max}$)
  to remove the end--cap events for  the $^7$Be, $pep$ and CNO analysis. Table~\ref{tab:FV} summarizes these values  and some additional relevant quantities for each choice of FV (dimensions, total volume and mass, number of target electrons). 

In the context of the search for a possible day--night effect in the $^7$Be neutrinos interaction rate
(Section~\ref{sec:DN}), the spectra of the events collected during the day and night times have been subtracted and analyzed without performing a spectral fit of all components. An enlarged FV, a sphere of 3.3\,m radius, including a contribution from the external background but with a larger number of neutrino--induced events results to be convenient.

\begin{figure}[h]
\begin{center}
\vspace{2 mm}
\centerline {\includegraphics[width = 0.465\textwidth]{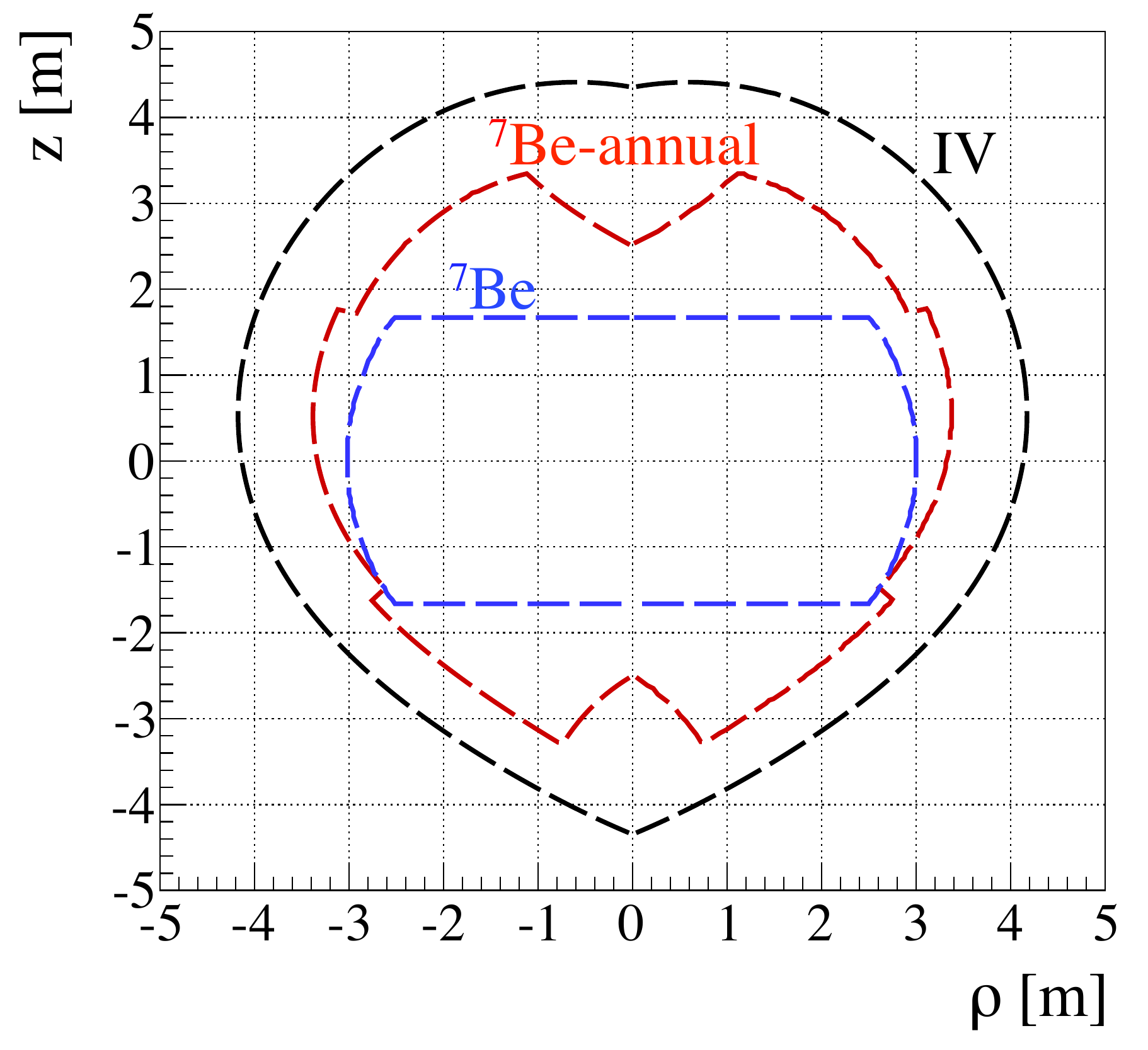}}
\caption{$\rho$--$z$ projection of the IV as of May 3$^{\rm rd}$, 2009 (black). The blue curve shows the shape of the 75\,ton FV used for the measurement of the $^7$Be--$\nu$ interaction rate.  The red curve illustrates the profile of the FV used for the $^7$Be--$\nu$ annual modulation analysis.}
\label{fig:fvshape}
\end{center}
\end{figure}

A search for the optimal choice of an enlarged FV has been done also in the framework of the analysis of the annual modulation of $^7$Be--$\nu$ interaction rate due to the annual variation of the Earth -- Sun distance (Section~\ref{sec:annmod}). In this context, particular attention has been devoted to the IV shape and position. In fact, these are slowly changing in time since the IV is not mechanically fixed; a stronger deformation has developed during the formation of a small leak in the IV (Section~\ref{subsec:leak}). Thus, an enlarged FV may contain a surface background contribution variable in time. We have developed an algorithm to continuously monitor the IV position and shape (Subsection~\ref{sub:VesShape}).
The FV used in this analysis is defined as the volume including all the events for which the reconstructed position
is at a distance larger than a given $d$ from the IV surface. Due to the asymmetric vessel deformation, the selection of $d$ is angle dependent such that for: $\theta$(0, $\pi$/3) $d$ = 100\,cm, $\theta$($\pi$/3, 2$\pi$/3) $d$ = 80\,cm, and for $\theta$(2$\pi$/3, $\pi$) $d$ = 60\,cm. Additionally, because of the proximity of hot end--caps (see Fig.~\ref{fig:BackStudy}), these regions were removed by a cone--like cut in the top and bottom of the detector as presented in Fig.~\ref{fig:fvshape}. The corresponding volume is changing in time and has a mean value of (141.83 $\pm$ 0.55)\,ton, almost twice larger than the one used for the $^7$Be--$\nu$  interaction rate measurement (75\,ton). Figure~\ref{fig:fvshape} shows an example of the $\rho$--$z$ projection of this FV in comparison with the 75\,ton one.

The temperature-dependent density $\rm{\rho_{PC}(T)}$ of pure pseudocumene, expressed in g/cm$^3$, is given by~\cite{bib:density}:
$\rm{\rho_{PC}(T) = (0.89179 \pm 0.00003) 
                                         - (8.015 \pm 0.009)  10^{-4} \cdot T}$,
where T is the temperature in degree Celsius.
For a PC + PPO mixture the density $\rm{\rho_{mix}(T)[g/cm^3]}$ is $\rm{\rho_{mix}= \rho_{PC} (T)\cdot (1+ (0.316 \pm 0.001) \eta_{PPO})}$, where
$\eta_{PPO} $ is the concentration of dissolved PPO in g/cm$^3$. Using the average values of the temperature of the scintillator of Borexino of T = (15.0 $\pm$ 0.5)\,$^\circ$C and the concentration of the dissolved PPO of $\rm{\eta_{PPO} }$= 1.45 $\pm$ 0.05 g/cm$^3$, we obtain a scintillator density of 0.8802 $\pm$ 0.0004 g/cm$^3$.  Taking into account the chemical composition of the scintillator (including the 1.1\% isotopic abundance of $^{13}$C), we get number of target electrons of $\rm{(3.307 \pm 0.003) \times 10^{31}}$ electrons/100\,ton.

\begin{table*}
\begin{center}
\begin{tabular}{l |c c c c c c } \hline \hline
  Analysis     &  $R_{\rm max}$     & $z_{\rm min}$    &  $z_{\rm max}$  &  Volume  & Mass &  $N_{e^-}$ \\  
                    &       [m]    & [m]                       &  [m]                       &  [$m^3$]  & [ton]     &   $\times 10^{31}$ \\ \hline
$^7$Be--$\nu$ rate       & 3.02  & $-1.67$           &   1.67           &    86.01                &   75.47          &      2.496\\ 
$pep$--$\nu$ and CNO--$\nu$  & 2.8    & $-1.8$              &    2.2           & 81.26   &    71.30&      2.358  \\
$^7$Be--$\nu$ rate day--night asymmetry & 3.3 &               -3.3     &3.3            &151.01   &   132.50  &4.382 \\
$^7$Be--$\nu$ rate annual modulation   (mean FV)&     &                        &             &  161.64   & 141.83     &     4.690\\
\hline \hline
\end{tabular}
\end{center}
\caption{Definition of the fiducial volumes used in the different solar neutrino analysis.}
\label{tab:FV} 
\end{table*}

\subsubsection{Dynamical reconstruction of the vessel shape}
\label{sub:VesShape}

In this section we describe a method to reconstruct the IV shape and position based on the events due to the vessel radioactive contaminants.

\begin{figure}[h]
\begin{center}
\vspace{-1 mm}
\centering{\includegraphics[width = 0.5\textwidth]{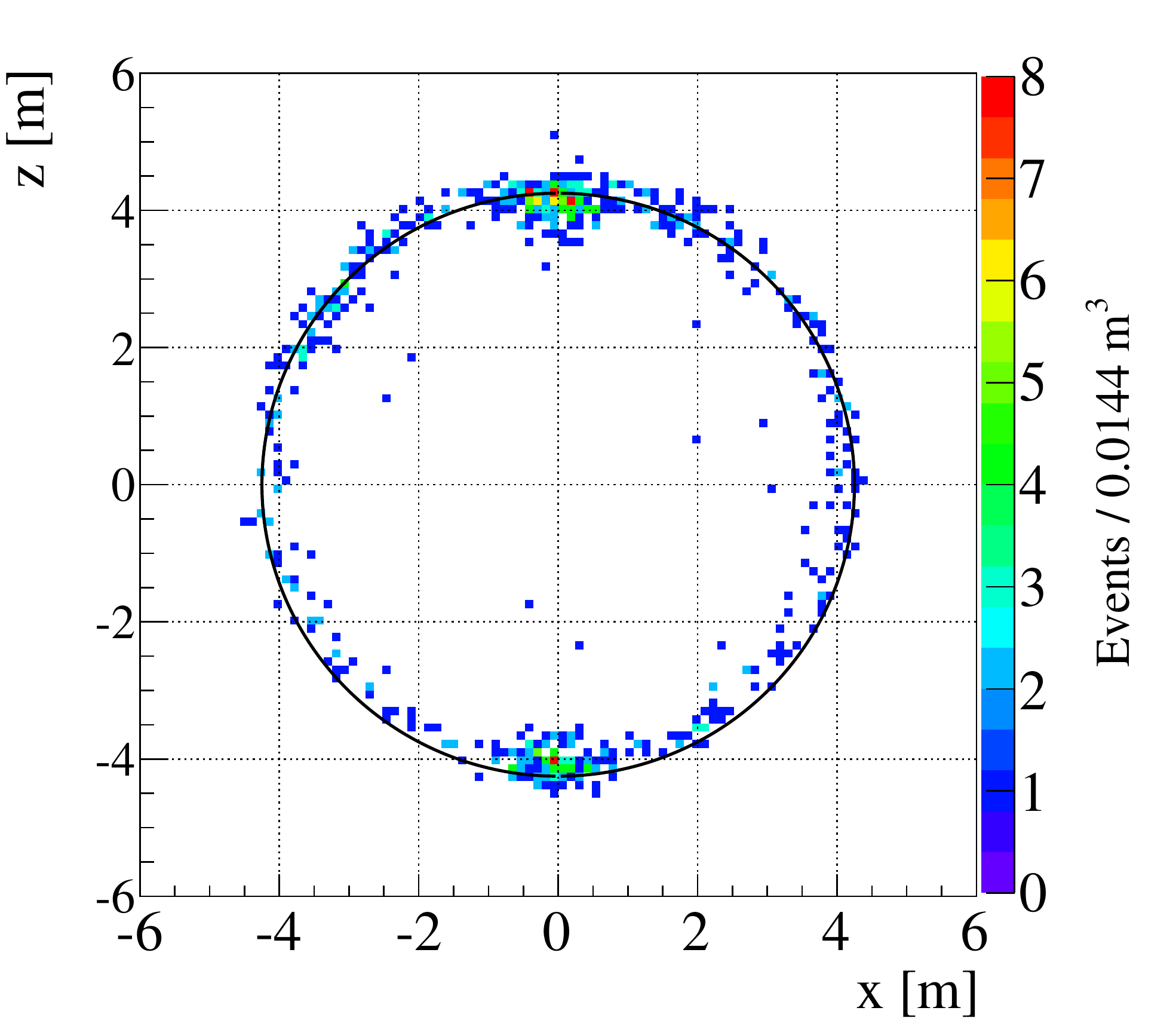}}
\end{center}
\vspace{-3 mm}
\caption{$z-x$ distribution ($|y| < 0.5$\,m) of the events in the energy region 800 - 900\,keV, which are mainly due to $^{210}$Bi contaminating the IV surface. The color axis represents the number of events per 0.0144 m$^3$ in a pixel of 0.12\,m x 1.00\,m x 0.12\,m ($x$ x $y$ x $z$). This spatial distribution reveals the IV shift and deformation with respect to its nominal spherical position shown in solid black line.}
\label{fig:Binylon}
\end{figure}

\begin{figure}[t]
\begin{center}
\centering{\includegraphics[width = 0.5\textwidth]{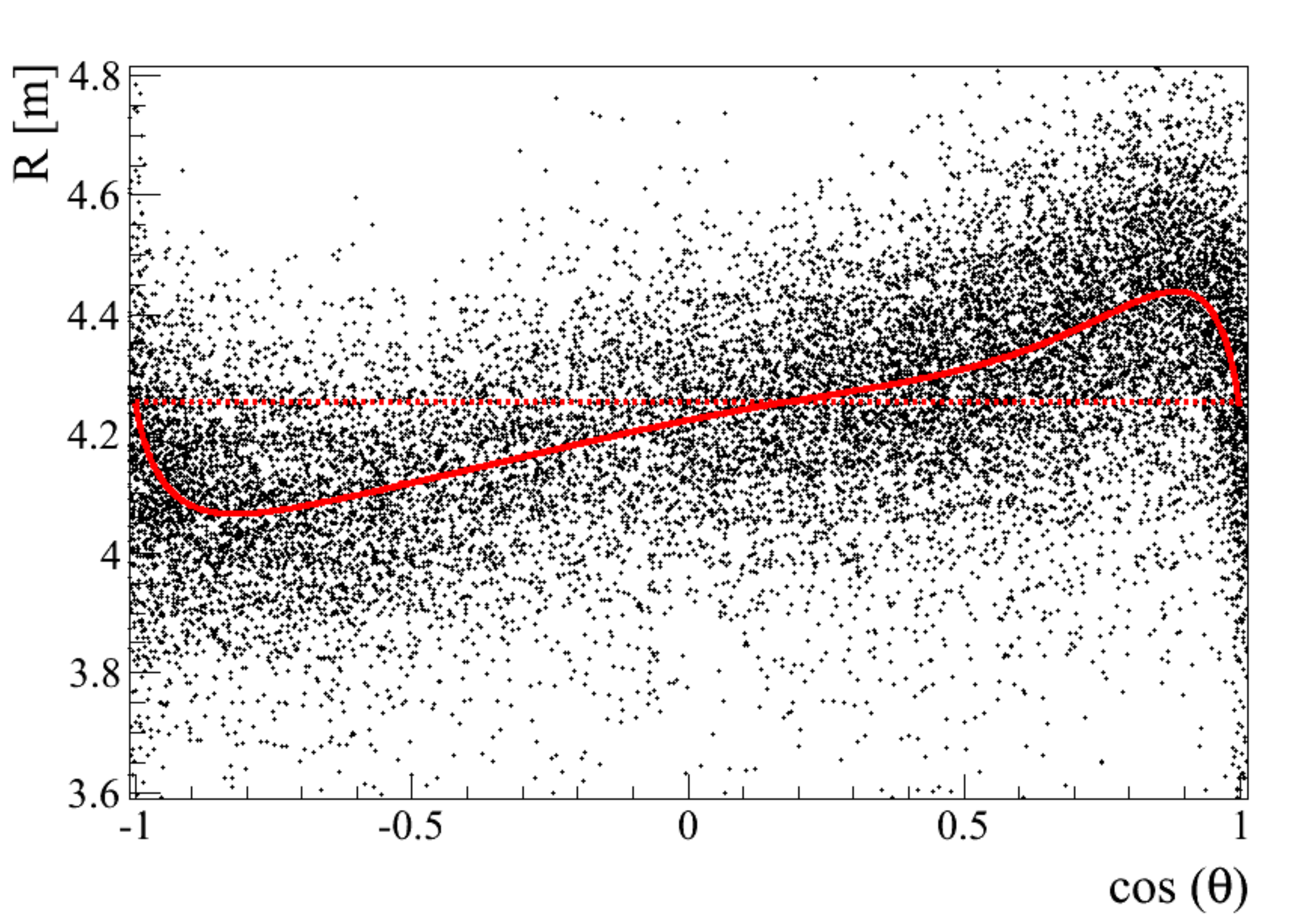}}
\caption{$R$ -- $\cos (\theta)$ distribution of the events in the energy region 800 - 900\,keV from November 2007 used for the IV shape reconstruction. The best-fit vessel shape is shown in a solid red line. The dotted red line represents the nominal spherical vessel with $R$ = 4.25\,m. } 
%\vspace{-2 mm}
\label{fig:IVshapefit}
\end{center}
\end{figure}

The IV profile is determined by using  background events reconstructed on its surface and identified as due to $^{210}$Bi decay.
Figure~\ref{fig:Binylon} shows the $z-x$ distribution of these events in the energy region 800 -- 900\,keV.
Assuming azimuthal symmetry, the dependence of the reconstructed radius $R$ on the $\theta$ angle is fitted  (see Fig.~\ref{fig:IVshapefit}) with a 2D analytical function (red line) having a Gaussian width.
The function itself is either a high--order polynomial or a Fourier series function.
The end--points are fixed in the fit at $R$ = 4.25\,m because the end--caps are hold in place by rigid supports, whereas the total length of the vessel profile is included as a penalty factor in $\chi^2$ (the vessel can deform but not expand elastically in any significant way). 
The procedure was calibrated by a method which we have used in the first year of data taking, based on inner--detector pictures taken with the CCD cameras~\cite{BxCalibPaper}. This old method requires to switch off the PMTs, so it cannot be used often. The new one does not require DAQ interrupts and allows therefore to monitor the variation of the IV shape on a weekly basis.  The precision of this method is of the order of 1\%.

\subsection{Internal background}
\label{subsec:InternalBack}

\begin{figure} [h]
\begin{center}
\centering{\includegraphics[width = 0.5 \textwidth]{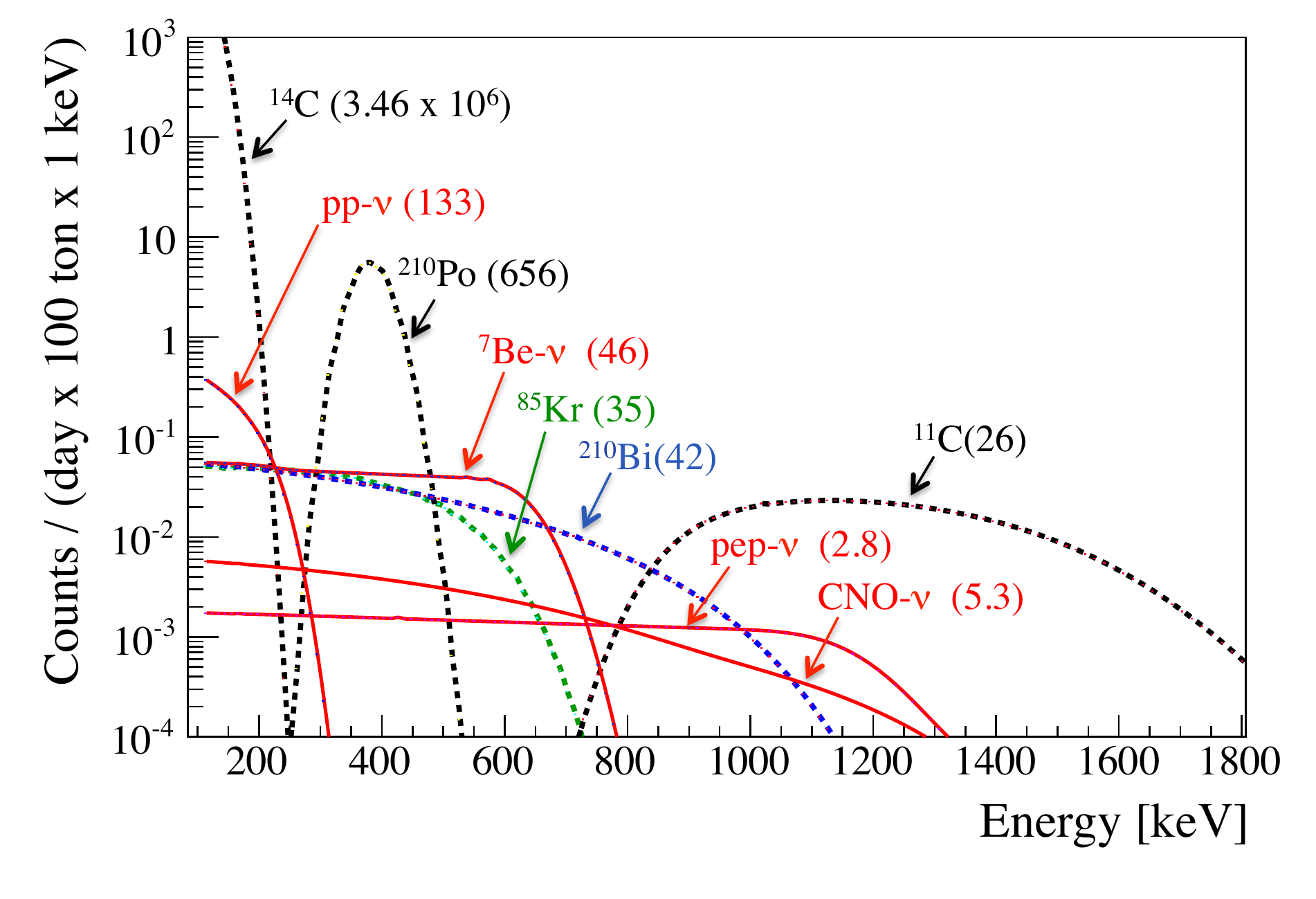}}
\caption{Calculated energy spectra due to solar neutrinos (shown as continuous red line) and of the main background components (dotted lines). The realistic Borexino energy 
resolution is included. The rates are fixed at the values shown in parenthesis and given in units of cpd/100\,ton. }
\label{fig:sb}
\end{center}
\end{figure}

We discuss here the background from radioactive isotopes contaminating the scintillator, their rates, spectral shapes, and life--times.
The contribution of the internal background producing $\gamma$ or $\beta$ decays can be  separated from the signal only through its spectral shape. Additional removal  possibilities are available when the pulse--shape discrimination procedure can be used, as for example for $\alpha$ particles. 

Figure~\ref{fig:sb} shows the expected energy spectrum in Borexino including solar neutrinos
and the relevant internal and cosmogenic background sources, taking into account the realistic energy resolution of the detector. The rates of solar neutrinos correspond to the SSM expectations while those of background components are set to values typical for Borexino Phase-I period.

Tables~\ref{tab:rad_back}, \ref{tab:u238_prod} ($^{238}$U chain), \ref{tab:th232_prod} ($^{232}$Th chain) summarize the decay characteristics of relevant isotopes that may contribute to the internal background. We underline that the values reported below refer to the background measured in the Phase-I of Borexino. Some of these backgrounds have been reduced by purification, but their values will be quoted only in future papers.
Table~\ref{tab:back_rate} summarizes the rate of the main components of the internal background.

\begin{table}[b]
\begin{center}
\begin{tabular}{l c c c} \hline \hline
Isotope          & Mean Life      & Energy  & Decay \\ 
                       &                     &  [keV] &    \\ 
\hline
$^{14}$C & $8.27 \times 10^3$ yrs & 156 & $\beta^-$\\
$^{85}$Kr & 15.4 yrs & 687 & $\beta^-$\\
$^{40}$K (89\%) & $1.85 \times 10^9$ yrs & 1310 & $\beta^-$\\
$^{40}$K (11\%) & $1.85 \times 10^9$ yrs & 1460 & EC + $\gamma$\\
$^{39}$Ar & 388 yrs & 565 & $\beta^-$\\
\hline \hline
\end{tabular}
\end{center}
\caption{ \label{tab:rad_back} List of the main internal radioactive backgrounds considered in this work, except those related to $^{238}$U and $^{232}$Th chains, which are reported in Tables \ref{tab:u238_prod} and \ref{tab:th232_prod}. }
\end{table}

\begin{table}[b]
\begin{center}
\begin{tabular}{l c c c} \hline \hline
Isotope               & Decay Rate  & Reference\\ 
                                          &  [cpd/100\,ton] &    \\ 
\hline
$^{14}$C     & $(3.46 \pm  0.09) \times10^6$    & Section~\ref{subsec:c14}\\
$^{85}$Kr                            & $(30.4 \pm 5.3 \pm 1.5)^{(a)} $ & Section~\ref{subsec:Kr85}\\ 
                    & $(31.2 \pm 1.7 \pm 4.7)^{(b)} $ & Section~\ref{sec:Be7results}\\
$^{40}$K      &  $<$0.42\, (95\% C.L.)  & Section~\ref{subsec:K40} \\
$^{39}$Ar     & $\sim 0.4$                              & Section~\ref{subsec:Ar39} \\ 
$^{238}$U   & $(0.57 \pm 0.05)$                        & Section~\ref{subsec:U238} \\
$^{222}$Rn   &  $(1.72 \pm 0.06)$                      & Section~\ref{subsec:U238}\\
$^{210}$Bi    &   $(41.0 \pm 1.5 \pm 2.3) $       & Section~\ref{subsec:Bi210}\\
$^{210}$Po  & $ 5 \times 10^2 - 8 \times 10^3$                     & Section~\ref{subsec:Po210}\\
$^{232}$Th   & $ (0.13 \pm 0.03)$                 & Section~\ref{subsec:Th232} \\
\hline \hline
\end{tabular}
\caption{Decay rate of the main internal radioactive backgrounds considered in this work; if the rate has changed in time the minimum and maximum values are quoted. For $^{85}$Kr contamination, both the results of delayed coincidence method $(a)$ and spectral fit analysis $(b)$ are reported.} 
\label{tab:back_rate}
\end{center}
\end{table}

\subsubsection{$^{14}$C}
\label{subsec:c14}

The $^{14}$C isotope   ($\beta$--emitter with 156\,keV end point and 8270~years mean--life, see Table~\ref{tab:rad_back})
unavoidably accompanies the $^{12}$C with relative abundances that may span several orders of magnitude. It is produced in the upper atmospheric layers through the interaction of cosmogenic neutrons with nitrogen. Even though  $^{14}$C has a geologically short mean--life, it is constantly being replenished by the cosmic--ray flux.
$^{14}$C is chemically identical to $^{12}$C and thus it cannot be removed from the organic scintillator
through purification. In order to reduce the levels of contamination, the Borexino scintillator is derived from petroleum from deep underground where the
levels of $^{14}$C  are reduced by roughly a factor of a million compared with the usual values in organic materials. Since the petroleum has been underground for millions of years, the remaining amount of $^{14}$C  are possibly due to underground neutron production.

The extremely low $^{14}$C/$^{12}$C ratio of $10^{-18}$ g/g  has been measured in CTF~\cite{c14CTF}. This result was a key milestone in the low--background research and development of Borexino. 
Even with this large reduction in contamination, $^{14}$C is by far the largest Borexino background and it determines the detector low--energy threshold.  The $^{14}$C rate is (3.46 $\pm$ 0.09\,) 10$^6$ cpd/100\,ton,  about $\simeq$$10^5$ times higher than the expected $^7$Be--$\nu$ signal rate. A hardware trigger threshold at $\simeq$50\,keV reduces the trigger rate to about 30 Hz. The 156\,keV $^{14}$C end--point is low enough (even after the smearing effects of the detector energy resolution) that we can safely fit the energy spectrum beyond it and keep high sensitivity to the $^7$Be--$\nu$'s.

\subsubsection{ $^{85}$Kr}
\label{subsec:Kr85}

The isotope $^{85}$Kr is a $\beta$--emitter with 687\,keV end--point (99.57$\%$ branching ratio, see below) and 15.4~years mean--life (see Table \ref{tab:rad_back}). Its spectral shape is very similar to the electron recoil spectrum due to $^7$Be--$\nu$ and it is one of the most important backgrounds in the $^7$Be--$\nu$ analysis. It is present in the air mostly because of nuclear explosions in atmosphere at the average concentration of $\sim$1\,Bq/m$^3$, thus even extremely small air exposures during the detector--filling operations would yield significant contamination. As mentioned in previous sections, the development and use of N$_2$ with low Kr content during the scintillator manipulations has been of fundamental importance.

With a small branching ratio of $0.43$\%, $^{85}$Kr decays into the meta--stable $^{85 \rm{m}}$Rb emitting a $\beta$ particle with maximum kinetic energy of 173\,keV. The $^{85 \rm{m}}$Rb then decays to the ground state $^{85}$Rb by emitting a 514\,keV $\gamma$ ray with 2.06\,$\mu$s mean--life. This fast $\beta$ -- $\gamma$ sequence is the signature used to obtain a measure  of the $^{85}$Kr concentration independent from the one resulting from the spectral fit.

Candidate $^{85}$Kr($\beta$) -- $^{85 \rm {m}}$Rb($\gamma$) sequences of events are selected looking at the triggers with two clusters (see Section~\ref{sec:electronics}) that are not identified as muons. The two candidates must be reconstructed with a distance smaller than 1.5\,m and with a time delay between 300\,ns and  5840\,ns (four times the life--time of this decay). The 300\,ns value ensures that the efficiency of the clustering algorithm can be assumed to be 100$\%$ at the energies of interest. The spatial distance cut has been tuned to maximize the selection efficiency over background and by taking into account the worsening of spatial reconstruction performances at the $^{85}$Kr--$\beta$ low energies.
The energy window of the $^{85}$Kr $\beta$ is chosen between the detector threshold (typically $\sim$50\,keV) and 260\,keV  and the one of the $^{85 \rm{m}}$Rb $\gamma$ lies in the interval 280 560 \,keV.
Some background for $^{85}$Kr($\beta$ -- $\gamma$) search originates from  $^{212}$Bi -- $^{212}$Po coincidences due to thoron emanated from the IV (see Section~\ref{subsec:Th232}). This becomes negligible by requiring that the $^{85 \rm{m}}$Rb candidate is reconstructed within a sphere of 3.5\,m radius (156.2\,ton).  

Accidental $^{14}$C -- $^{210}$Po coincidences are the most important background for the $^{85}$Kr($\beta$ -- $\gamma$) measurement. They are partially suppressed by requiring that the  $G_{\beta}$ variable (see Section~\ref{sec:shape}) of the $^{85 \rm{m}}$Rb candidates falls within the interval (-0.07, 0.02). This background has been quantified  by looking for events satisfying all the selection criteria but with a time delay in a displaced interval between 2 and 8\,ms:  3.1 fake events are expected in the 33--events data sample. 
After the accidental background subtraction,  the 29.9 surviving events correspond to a $^{85}$Kr contamination of (30.4~$\pm$~5.3~(stat) $\pm$~1.5 (sys))~cpd/100\,ton where the systematic uncertainty is mostly coming from the FV definition and from the efficiency of the $^{85}$Kr--$\beta$ energy cut. 

The cut efficiencies have been evaluated with the help of the Monte Carlo code and of the source calibration data.  The $^{14}$C source emulates the $^{85}$Kr decay while the $^{85}$Sr source, like $^{85}$Rb, produces a 514\,keV $\gamma$. A  run--by--run analysis was necessary to evaluate the detector trigger efficiency, since no low--energy cut was applied to the $^{85}$Kr candidate. The overall combined efficiency of all the cuts is 19.5$\%$.

\subsubsection{ $^{40}$K}
\label{subsec:K40}

$^{40}$K is a primordial nuclide with a mean--life of 1.85\,billion years (see Table~\ref{tab:rad_back}) and a natural abundance of 0.012\%. In addition to the domiant pure $\beta$--decay (89\% BR and 1310\,keV end--point), there is a 10.7\% BR for electron capture to an excited state of $^{40}$Ar. This results in the emission of a mono--energetic 1460\,keV $\gamma$ ray, which helps to distinguish the $^{40}$K energy spectrum from the other $\beta$ spectra, though it does also mean that $^{40}$K decays at the vessel end--caps or in components on the SSS may deposit energy within the FV.

This isotope can enter into the scintillator primarily in two ways. The first way is through micron or sub-micron dust particulates. The fraction of natural potassium in the Earth crust is about 2.5\% by weight, corresponding to roughly 800\,Bq/kg from $^{40}$K. Secondly, it was found that commercially available PPO, the wavelength shifter added to the scintillator, had a potassium contamination at the level of parts per million. Given the PPO concentration of 1.5 g/l of scintillator, this equates to $10^{-9}$\,g-K/g-scintillator or roughly to 2.7$\times 10^5$\,cpd/100\,ton, nearly 6000 times the expected $^7$Be--$\nu$ rate.

The maximum concentration of potassium that was considered acceptable during the Borexino design is $\sim$$10^{-14}$ g-K/g-scintillator. The $^{40}$K contamination was reduced through distillation, filtration, and water extraction of the PC--PPO solution~\cite{Benziger}. Unfortunately, the efficiency of these methods at removing $^{40}$K is unknown and so we cannot a priori calculate the expected rate in the scintillator. For this reason, we have included the $^{40}$K spectrum as a free parameter in all spectral fits. The upper limit of 0.42\,cpd/100\,ton (95\% C.L.) results from the $pep$--$\nu$ analysis, see Section~\ref{sec:pepCNOResults}.

\subsubsection{$^{39}$Ar}
\label{subsec:Ar39}

The isotope $^{39}$Ar is produced primarily through cosmic ray activity in the atmosphere. It is a pure $\beta$--emitter with a Q--value of 565\,keV, see Table~\ref{tab:rad_back}. With an end--point fairly close to the 665\,keV $^7$Be--$\nu$ shoulder and no accompanying $\gamma$ rays or delayed coincidence, it would be hard to disentangle the $^{39}$Ar spectrum from that of $^7$Be--$\nu$. Therefore, great care was taken in ensuring that the $^{39}$Ar contamination was as low as possible. The argon levels in the specially prepared low Ar/Kr nitrogen used for the stripping of the scintillator was around 0.005\,ppm (by volume). When mixed in equal volumes of gaseous nitrogen and pseudocumene, argon will partition itself in the ratio 4.1 : 1, respectively~\cite{Zuzel}. Given an activity of 1.4\,Bq/m$^3$ in atmospheric argon, this translates to an expected rate of less than 0.02\,cpd/100\,ton in the scintillator. However, as it was observed by the high $^{85}$Kr rate, there appears to have been a small air leak during the vessel filling. The activity of $^{39}$Ar in air (13\,mBq/m$^3$~\cite{Zuzel}) is roughly 75 times lower than that of $^{85}$Kr (1\,Bq/m$^{3}$). Assuming that all the $^{85}$Kr contamination ($\sim$30\,cpd/100\,ton) in the scintillator came from the air leak, and that the ratio of $^{39}$Ar to $^{85}$Kr was the same as in the atmosphere, the expected $^{39}$Ar contamination is ~0.4\,cpd/100\,ton. At this level, the contribution of $^{39}$Ar to the spectrum is negligible and we have not included it in the spectral fits. The hypothesis that all the $^{85}$Kr contamination
is due to the air leak is supported by the results of the recent purification
campaign of the scintillator which results very effective in reducing the 
 $^{85}$Kr contamination. The resulting value is consistent with 0\,cpd/100\,ton and it appears stable in time.

\subsubsection{$^{238}$U chain and $^{222}$Rn}
\label{subsec:U238}

\begin{table}[t]
\begin{center}
\begin{tabular}{l c c c } \hline \hline
Isotope          & Mean Life      & Energy  & Decay \\ 
                       &                     &  [keV] &    \\ 
\hline
$^{238}$U & $6.45 \times 10^9$ yrs & 4200 & $\alpha$\\
$^{234}$Th & $34.8$ days & 199 & $\beta^-$\\
$^{234m}$Pa & $1.70$ min & 2290 & $\beta^-$\\
$^{234}$U & $3.53 \times 10^5$ yrs & 4770 & $\alpha$\\
$^{230}$Th & $1.15 \times 10^5$ yrs & 4690 & $\alpha$\\
$^{226}$Ra & $2.30 \times 10^3$ yrs & 4790 & $\alpha$\\
$^{222}$Rn & $5.51$ days & 5490 & $\alpha$\\
$^{218}$Po & $4.40$ min & 6000 & $\alpha$\\
$^{214}$Pb & $38.7$ min & 1020 & $\beta^-\gamma$\\
$^{214}$Bi & $28.4$ min & 3270 & $\beta^-\gamma$\\
$^{214}$Po & $236~\mu$s & 7690 & $\alpha$\\
$^{210}$Pb & $32.2$ yrs & 63 & $\beta^-\gamma$\\
$^{210}$Bi & $7.23$ days & 1160 & $\beta^-$\\
$^{210}$Po & $200$ days & 5410 & $\alpha$\\
$^{206}$Pb & stable & -- & -- \\
\hline \hline
\end{tabular}
\end{center}
\caption{  The $^{238}$U decay chain showing isotopes, life--times, maximum released energies and type of decay. A large number of $\gamma$ line accompanying many of the decays are not reported in the table.  }
\label{tab:u238_prod} 
\end{table}

$^{238}$U is a primordial radioactive isotope with a mean--life of 6.45~billion years. It is the most common isotope of uranium, with a natural abundance of 99.3$\%$. Table \ref{tab:u238_prod} reports the relevant information about the $^{238}$U decay chain containing eight $\alpha$ and six $\beta$ decays and ending with the stable $^{206}$Pb.

The concentration of contaminants of the $^{238}$U chain in secular equilibrium can be measured by identifying the fast decay sequence $^{214}$Bi -- $^{214}$Po that offers a delayed coincidence tag: 
\begin{equation}
{^{214}}\rm{Bi} \rightarrow {^{214}}\rm{Po} +  e^-  +  \bar{\nu} _e 
\end{equation}
\begin{equation}
{ ^{214}}\rm{Po} \rightarrow {^{210}}\rm{Pb} +  \alpha , 
\end{equation}
with $\tau$~=~238\,$\mu$s (the $^{214}$Po life--time), $Q$ of the $^{214}$B--decay equal to 3272\,keV, and $\alpha$ energy of 7686\,keV. However, these two isotopes are $^{222}$Rn daughters and the hypothesis of secular equilibrium is often invalid due to radon diffusion through surfaces or a possible contamination of the scintillator with radon coming from air.

The $^{214}$Bi~--~$^{214}$Po candidates are searched for by analyzing consecutive events which are not tagged as muons; their time delay is requested to be between 20 and 944\,$\mu$s (4~life--times) and their reconstructed spatial distance less than 1\,m. The energy  must be in the range (180 - 3600)\,keV
for the $^{214}$Bi candidate and (400 - 1000)\,keV (electron equivalent)
for the $^{214}$Po candidate. The overall efficiency of these cuts has been evaluated as 90\% through a Monte Carlo simulation.

\begin{figure}[t]
\begin{center}
\centering{\includegraphics[width = 0.50\textwidth]{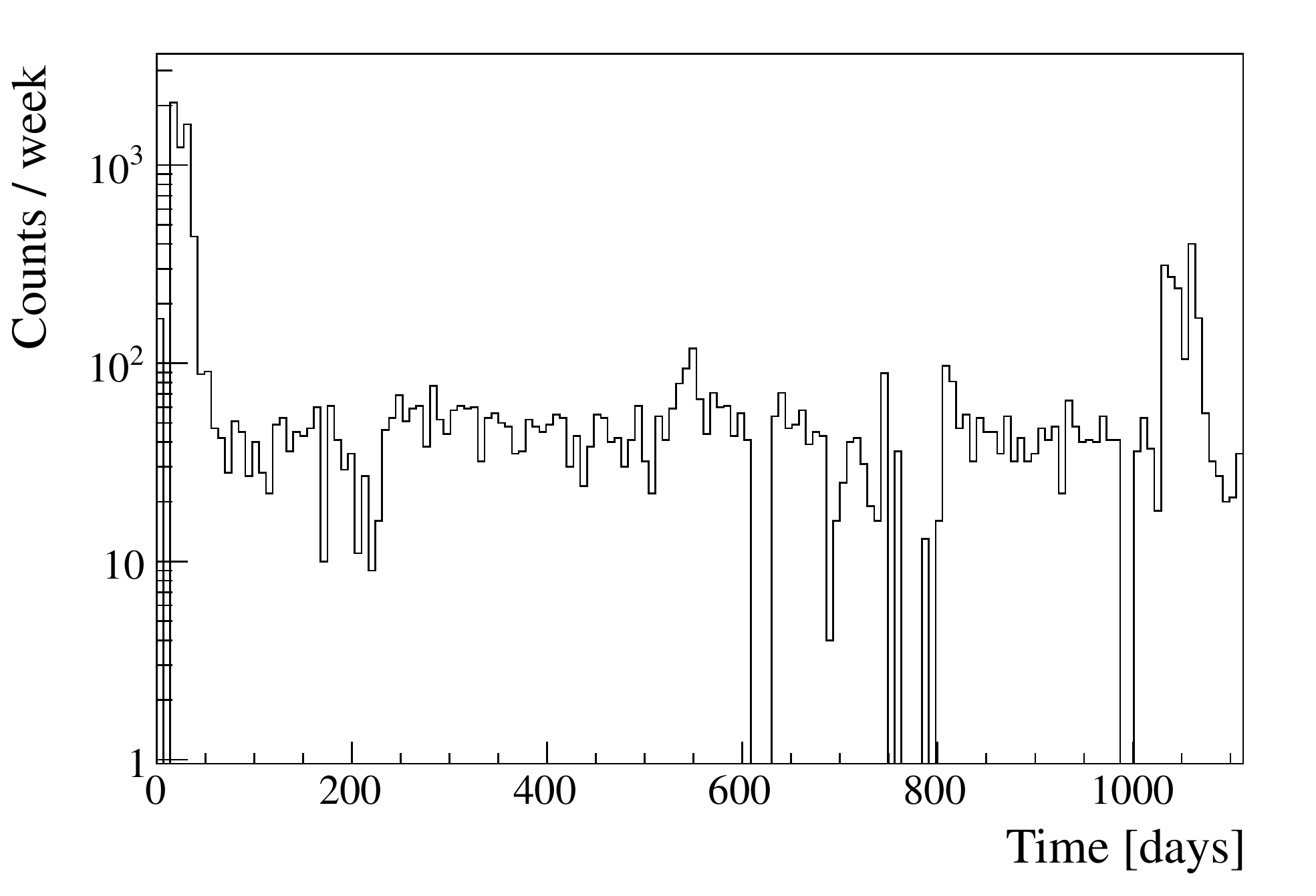}}
\caption{Number (counts/week) of $^{214}$Bi - $^{214}$Po coincidences detected in the whole scintillator volume as a function of time starting from May 16$^{\rm{th}}$, 2007 (day 0). The spikes are due to the filling operations.}
\label{fig:rnVsTime}
\end{center}
\end{figure}

\begin{figure}[t]
\begin{center}
\centering{\includegraphics[width = 0.50\textwidth]{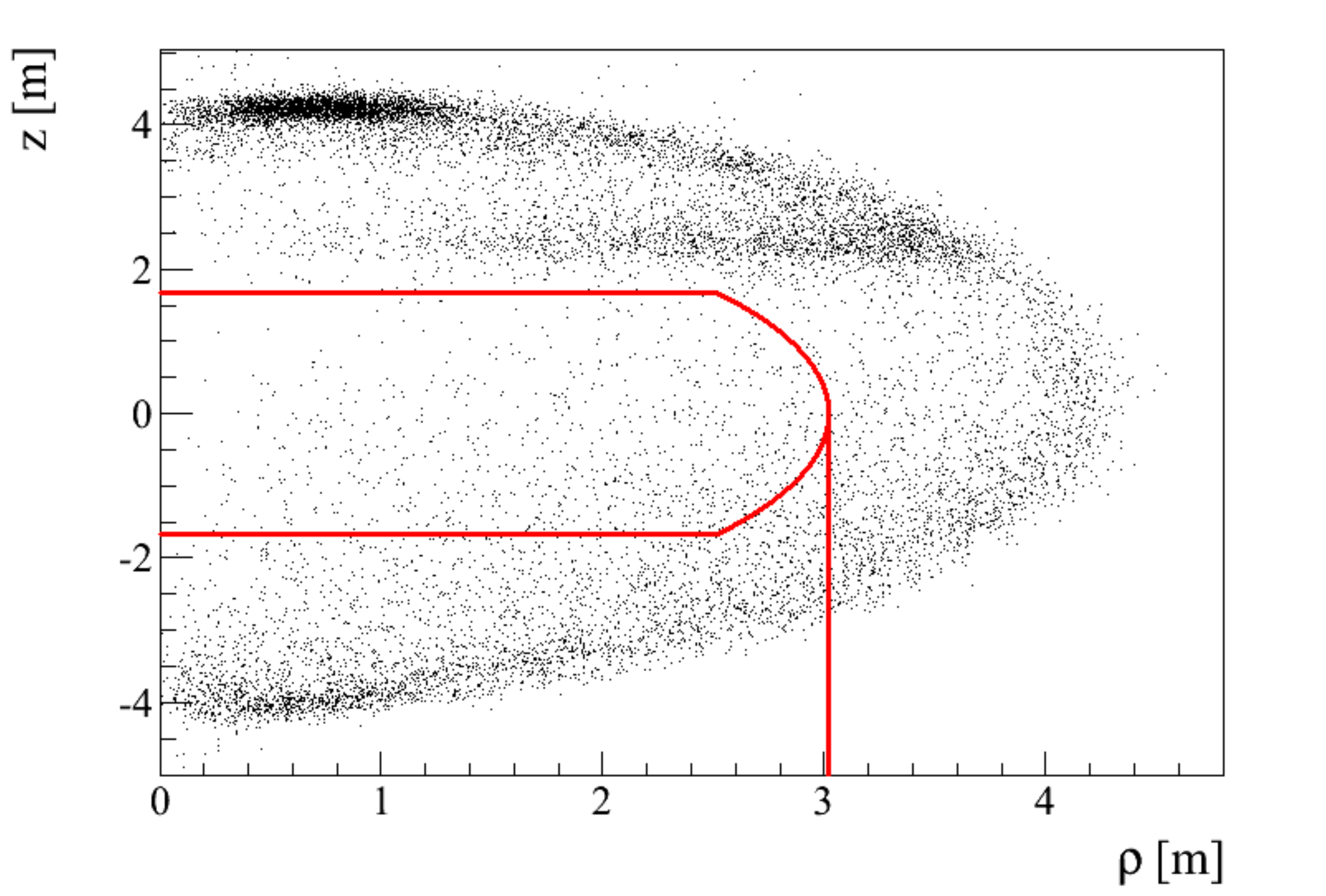}}
\caption{Spatial distribution of the $^{214}$Po events (from $^{214}$Bi - $^{214}$Po coincidences) in the $z$ - $\rho$ plane from May 2007 to May 2010. The solid red line indicates the FV used in the $^7$Be--$\nu$ analysis. The layer of events in the upper hemisphere is outside this volume and corresponds to the increased rate of radon at the beginning of data taking (see Fig.~\ref{fig:rnVsTime}).}
\label{fig:rnSpatial}
\end{center}
\end{figure}

The number of $^{214}$Bi~--~$^{214}$Po coincidences has been monitored continuously during the data taking. A sharp increase followed by the decay with the 5.5\,days mean--life of $^{222}$Rn has been observed in correlation with operations performed on the detector (like IV refilling, insertion of the calibration sources). No persistent contamination was introduced by any of such operations suggesting that the observed $^{214}$Bi~--~$^{214}$Po activity was due to the emanation of $^{222}$Rn from the pipes or from the inserted materials. 
Figure~\ref{fig:rnVsTime} shows the $^{214}$Bi -- $^{214}$Po rate versus time in the whole scintillator volume.
Figure~\ref{fig:rnSpatial} shows the $z$  -- $\rho$ ($\rho =\sqrt{x^2+y^2}$) distribution of the $^{214}$Po events in a period of no operations. This spatial distribution suggests a higher surface contamination than the bulk one.
The mean rate of $^{214}$Bi~--~$^{214}$Po coincidences in the $^7$Be--FV in the period from May 2007 to May 2010 is (1.72~$\pm$~0.06)~cpd/100\,ton.

The $^{238}$U concentration in the scintillator has been inferred from the asymptotic $^{214}$Bi~--~$^{214}$Po rate in the $^7$Be--FV in absence of operations which is (0.57~$\pm$~0.05)~cpd/100\,ton. Assuming secular equilibrium, the resulting $^{238}$U contamination is (5.3~$\pm$~0.5)~$\times$~$10^{-18}$~g/g. This is 20~times lower than the target design of Borexino. 

 From the measured number of $^{214}$Bi -- $^{214}$Po coincidences, we deduced the number of $^{214}$Pb events to be included in the spectral fit. For the $pep$ and CNO neutrino analysis, the only isotope in the chain before the $^{222}$Rn that could yield a measurable contribution in the count rate is $^{234 \rm{m}}$Pa with $Q$ value of 2290\,keV.

The natural abundance of $^{235}$U, the parent isotope of the actinium chain, is only 0.7$\%$, thus its activity is fully negligible. Searching for the double--$\alpha$ time coincidence of $^{219}$Rn and $^{215}$Po ($\tau$ = 1.78\,ms) from this chain
results in a rate of (0.05~$\pm$~0.04)\,cpd/100\,ton, consistent with the natural abundance.

\subsubsection{$^{210}$Pb}
\label{subsec:Pb210}

$^{210}$Pb is a $\beta$--emitter in the ${^{238}}$U--decay chain. Due to its long mean--life (32~years) and its tendency to adsorb on to surfaces, it is often found out of secular equilibrium with the ${^{222}}$Rn section of the chain. While $^{210}$Pb itself is not a problem, since its end--point (Q--value = 63.5\,keV) is well below the energy region of interest for solar neutrinos, its daughters, $^{210}$Bi and $^{210}$Po, are a major source of background in Borexino.

\subsubsection{ $^{210}$Bi}
\label{subsec:Bi210}

$^{210}$Bi is a $\beta$--emitting daughter of  $^{210}$Pb with 7.23~days mean--life and Q--value of 1160\,keV. Its spectrum spanning through the energy range of interest for $^7$Be, $pep$, and CNO--$\nu$'s does not exhibit any specific signature, except its spectral shape, that would help its identification. Therefore, $^{210}$Bi contamination can be measured only through the spectral fit. 

\begin{figure}[t]
\begin{center}
\vspace{-3mm}
\centering{\includegraphics[width = 0.51\textwidth]{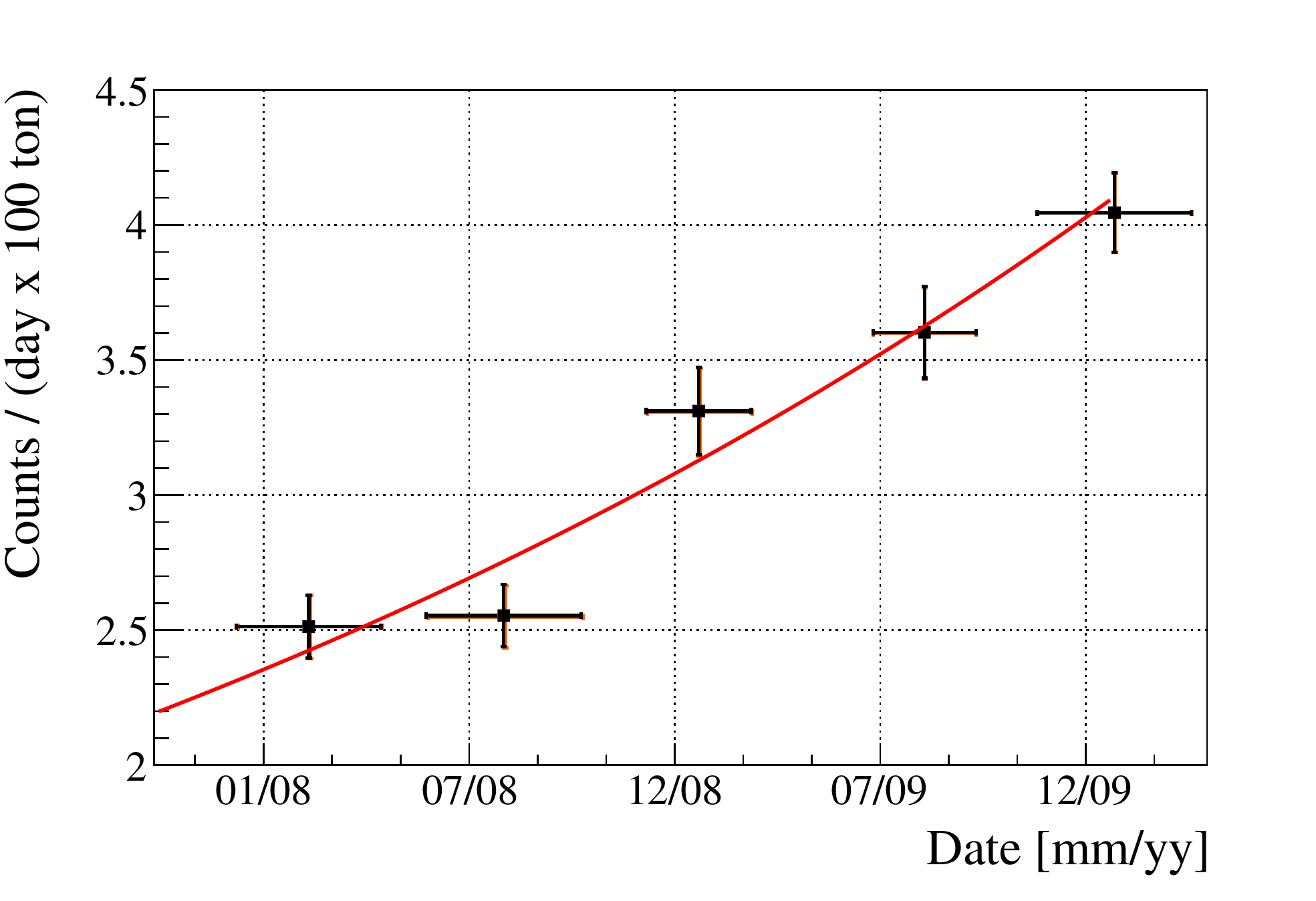}}
\caption{Count rate $R(t)$ in the energy region from 390 $<N_{pe}^d<$ 450  in the FV used for the $^7$Be--$\nu$ annual modulation analysis from the beginning of the year 2008 until the middle of year 2010. This region is dominated by the $^{210}$Bi contribution. The red line is a fit according to Eq.~\ref{eq:Bi}.}
\label{fig:Birate}
\end{center}
\end{figure}

At the start of data taking, following the initial filling of the detector, the $^{210}$Bi rate was measured as $(10 \pm 6)$~cpd/100\,ton. However, over time, the  $^{210}$Bi  contamination has been steadily increasing and at the start of May 2010 the rate was $\simeq$75\,cpd/100\,ton. The reason for this increase is currently not fully understood but it seems correlated with operations performed on the detector.
Figure~\ref{fig:Birate} shows the count rate stability in the FV used for the $^7$Be--$\nu$ annual modulation analysis (Table~\ref{tab:FV}) and in the energy region 385 $<N_{pe}^d<$ 450 which is dominated by the $^{210}$Bi. The time behavior of this count rate $R(t)$ is reasonably described by the sum of a constant background term R$_0$ and and an exponentially increasing term:
\begin{equation}
R(t)  = R_0 + R_{\rm Bi} ~ e^{\Lambda_{\rm Bi}t}
\label{eq:Bi}
\end{equation}
This variable background is a major concern for the annual modulation analysis and it will be thoroughly discussed in Section~\ref{sec:annmod}.
On the contrary, the $^7$Be and $pep$--CNO neutrino interaction rate analyses are only sensible to the mean value of the $^{210}$Bi rate and not to its relative time variations.

\subsubsection{ $^{210}$Po}
\label{subsec:Po210}

$^{210}$Po is after $^{14}$C the most abundant component of the detected spectrum. It is a  mono--energetic 5300\,keV $\alpha$--emitter (200\,days mean--life) but the strong ionization quenching of the scintillator (see Section~\ref{sec:scintillator})  brings its spectrum within the $^7$Be--$\nu$ energy region. Even though it is a direct daughter of $^{210}$Bi, the rate of $^{210}$Po was about 800 times higher than that of $^{210}$Bi at the start of data--taking. This high rate (out of equilibrium with the rest of the $^{238}$U--decay chain) may be due to $^{210}$Po washing off the surfaces of the scintillator storage tanks and pipes. The identification of $^{210}$Po in the liquid scintillator was one of the major CTF breakthroughs~\cite{bib:CTFlist1}. The pulse--shape discrimination is very effective in reducing this background component, as we will show in Section~\ref{sec:ab}.
The $^{210}$Po rate is easily measured since its high rate originates a peak clearly identified in the energy spectrum around $N_p \simeq N_h \simeq 190$ or $N_{pe} \simeq  N_{pe}^d \simeq 210$ and fitted with a Gaussian or a Gamma function (see Section~\ref{sec:analytical}).

\begin{figure}[t]
\begin{center}
\centering{\includegraphics[width = 0.5\textwidth]{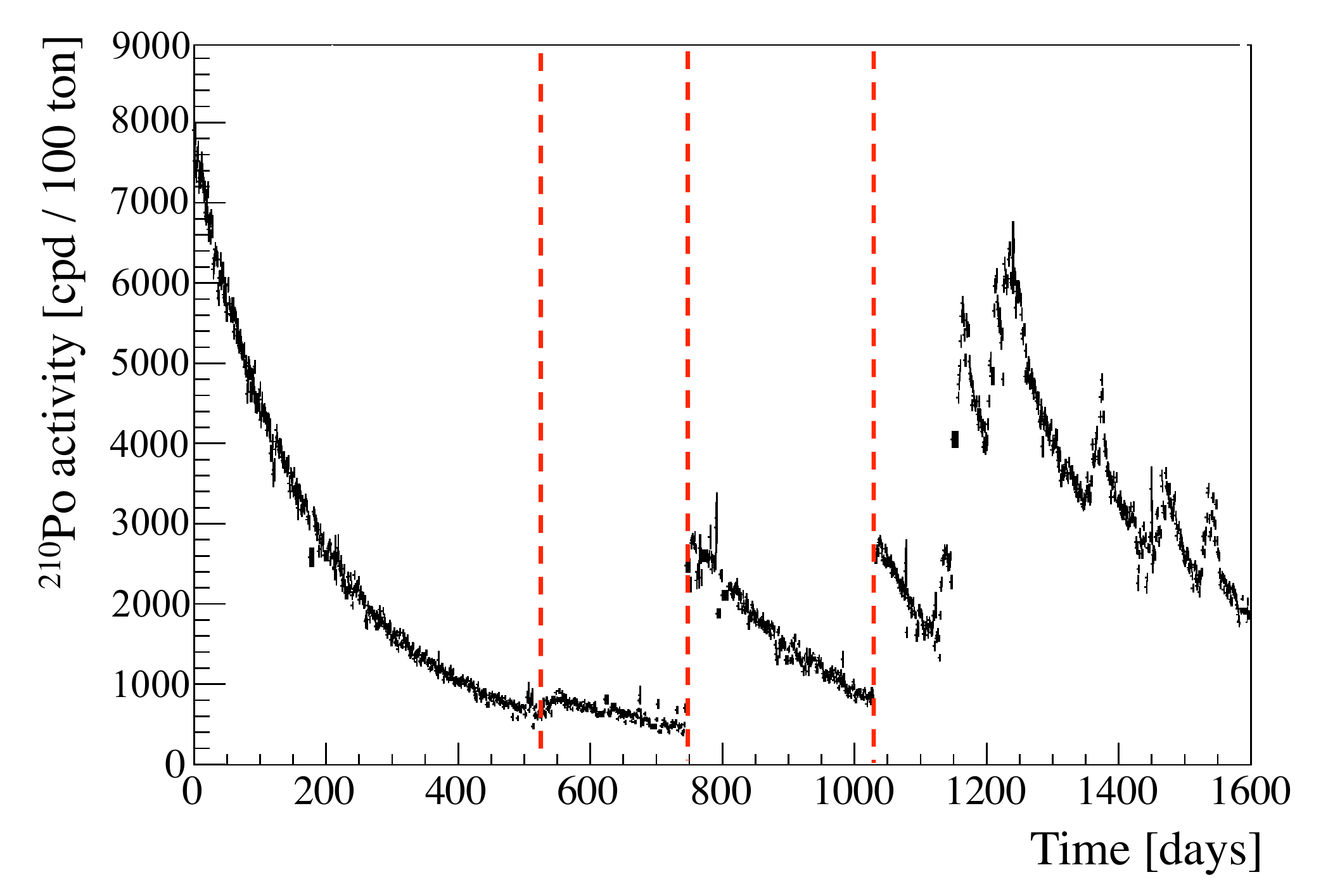}}
\caption{$^{210}$Po count rate in the FV used in the $^7$Be--$\nu$ rate analysis as a function of time. The various increases are due to the IV-filling operations shown by the three vertical lines (Section~\ref{subsec:leak}) and due to the tests of the purification methods (details in text).   }
\label{fig:PoVsTime}
\end{center}
\end{figure}

\begin{figure}
\begin{center}
\centering{\includegraphics[width = 0.43\textwidth]{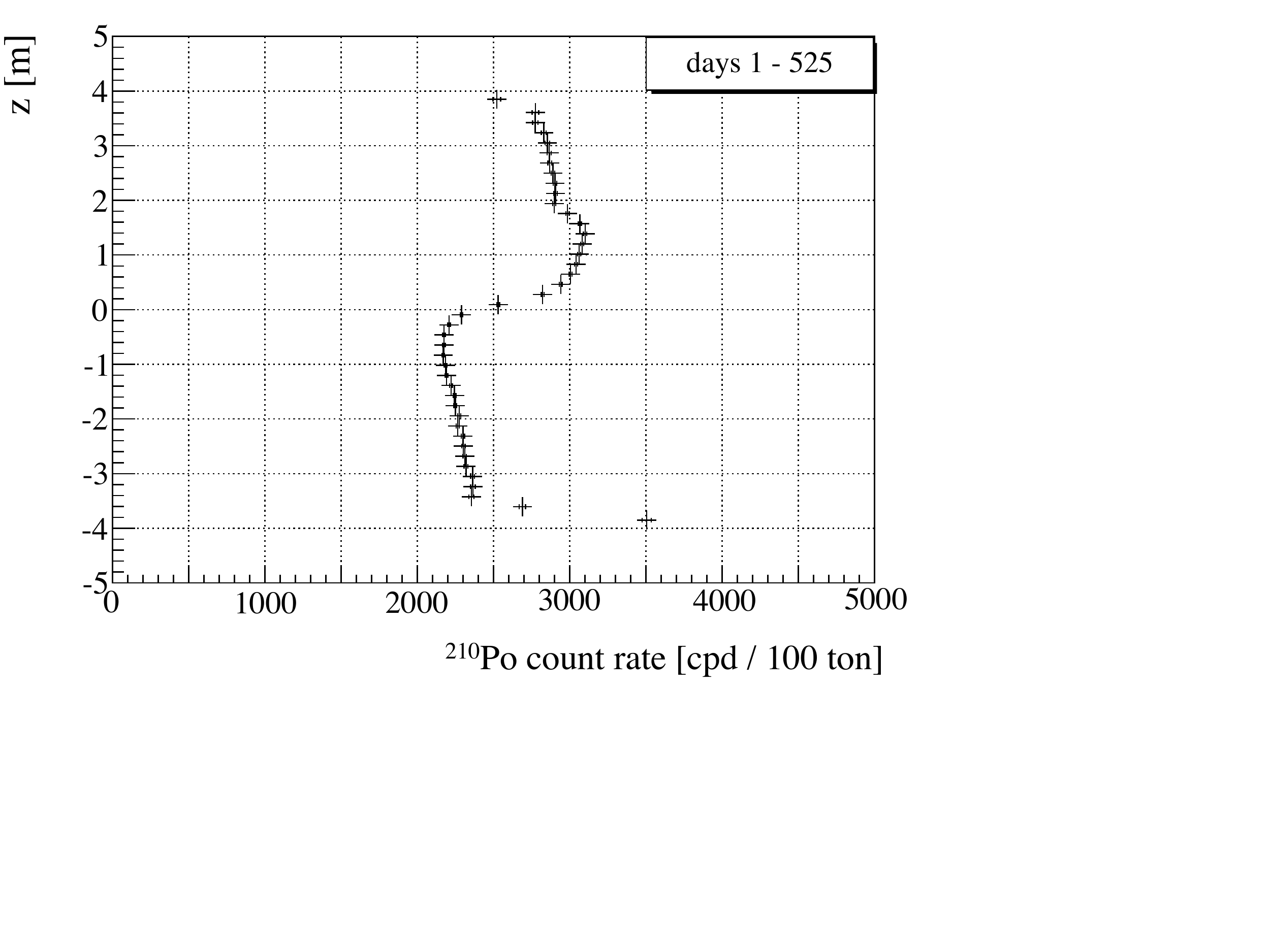}}
\centering{\includegraphics[width = 0.43\textwidth]{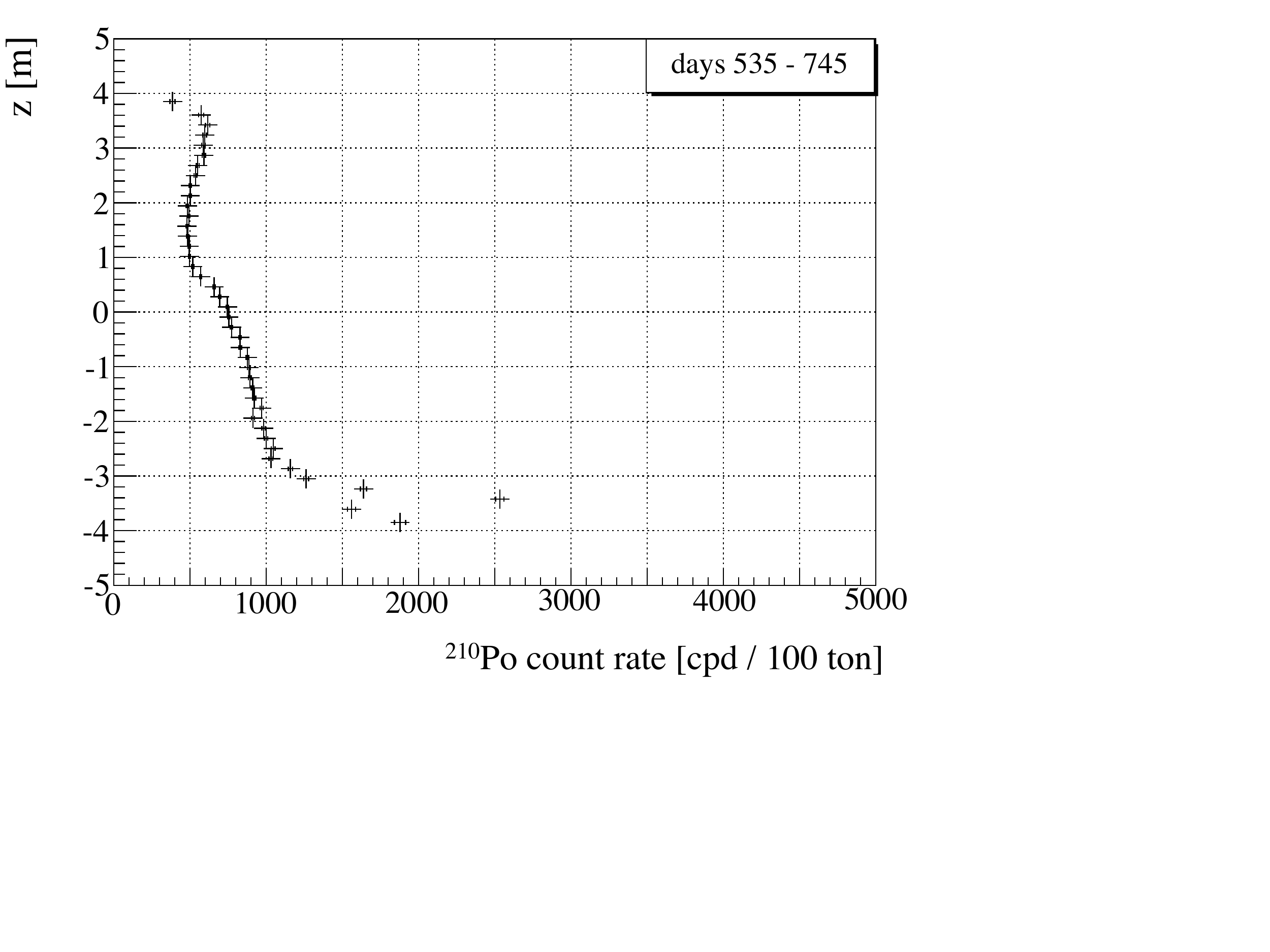}}
\centering{\includegraphics[width = 0.43\textwidth]{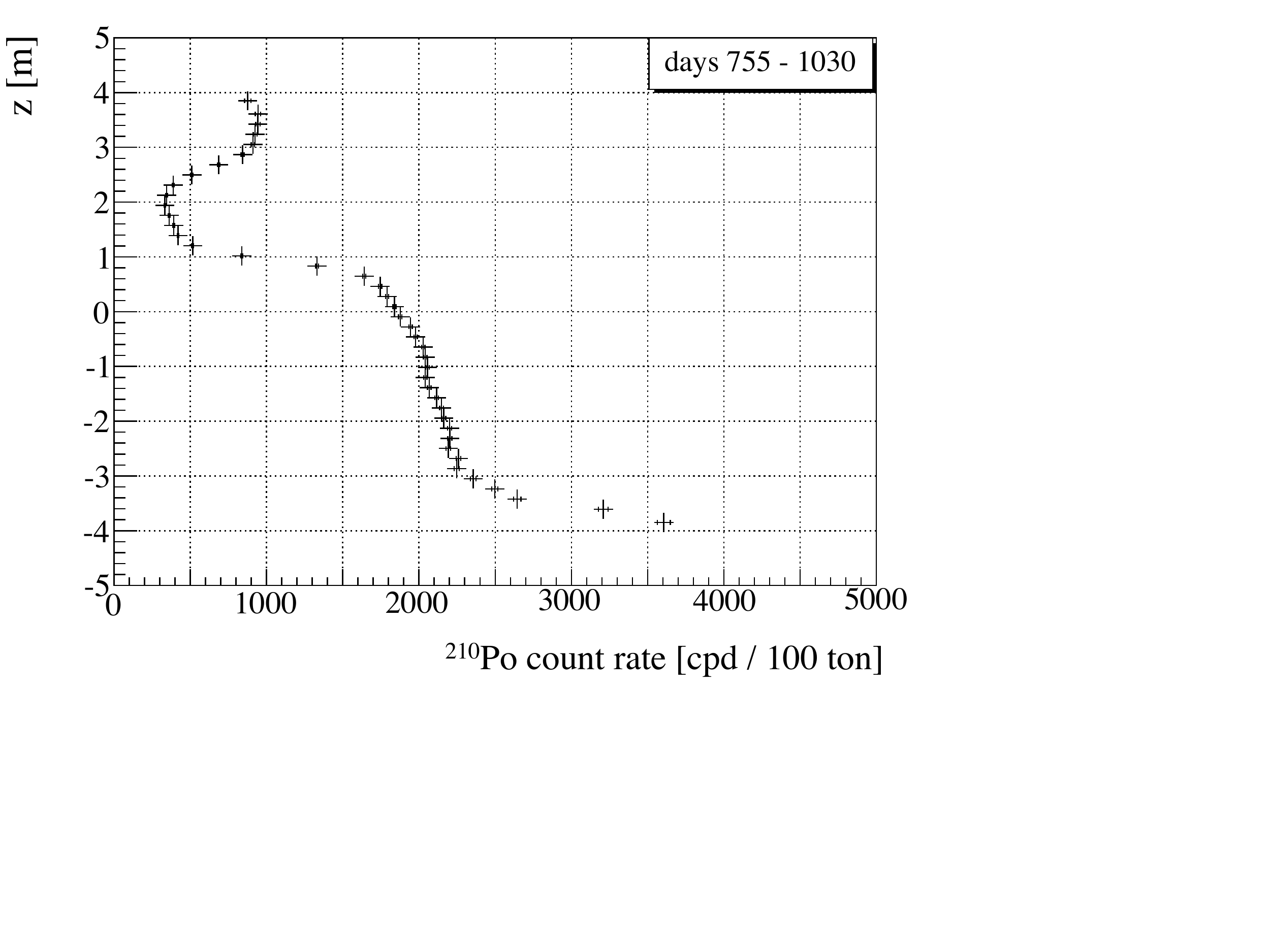}} 
\centering{\includegraphics[width = 0.43\textwidth]{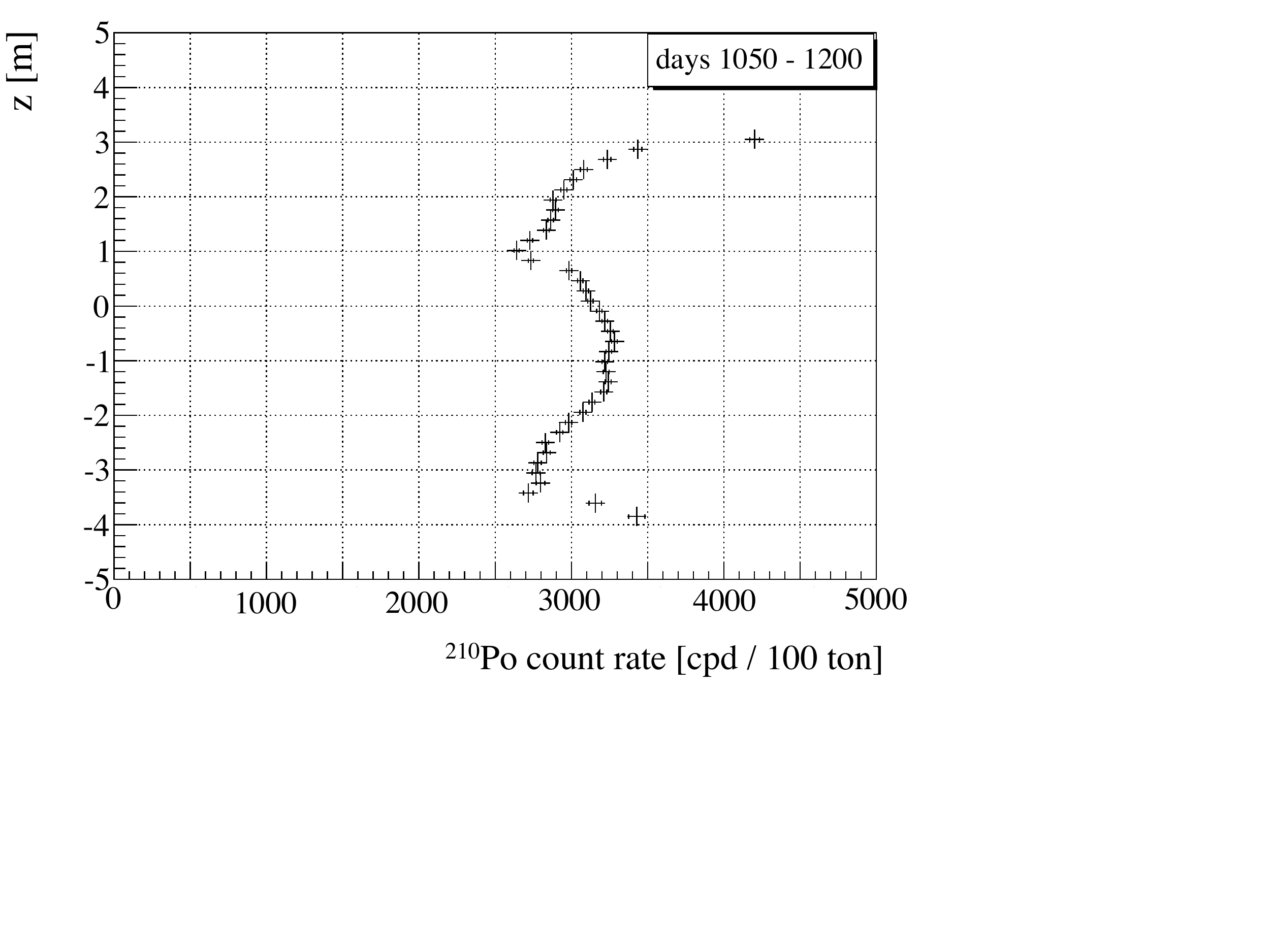}}
\caption{$^{210}$Po count rate as a function of the vertical position $z$. The four plots represent the four periods separated by the IV-filling campaigns shown by the vertical lines in Fig.~\ref{fig:PoVsTime}. }
\label{fig:PoVsz}
\end{center}
\end{figure}

Figure~\ref{fig:PoVsTime} shows the $^{210}$Po count rate in the $^7$Be--FV as a function of time. The various sudden increases of the count rate are related to the IV refilling operations described in Section~\ref{subsec:leak} or due to the tests of the purification procedures which were applied on the whole scintillator volume after the completion of the physics program described in this paper (Section~\ref{sec:concl}). 
%%The decay time, measured in the periods from xx to xx is xx days. Comment about the decay time.
%
The spatial distribution of the events in the $^{210}$Po energy region shows a significant non--uniformity, further perturbed by the detector operations and mixing. Figure~\ref{fig:PoVsz} shows the $z$--distribution of the $^{210}$Po events in four different periods separated by short periods of IV refilling, indicated by the vertical lines in Fig.~\ref{fig:PoVsTime}. 
This instability and non-uniformity does not have any significant effect on the solar neutrino rate analysis, while in the $^7$Be--$\nu$ annual modulation study (Section~\ref{sec:annmod}) pulse--shape discrimination techniques are used to fight this problem.

\subsubsection{ $^{232}$Th chain }
\label{subsec:Th232}

The primordial isotope $^{232}$Th has a mean--life of 20.3~billion years and 100$\%$ abundance in natural Th. The main decay branches of the $^{232}$Th include six $\alpha$ and four $\beta$ decays. The fast decay sequence of $^{212}$Bi -- $^{212}$Po:
\begin{equation}
{^{212}}\rm{Bi} \rightarrow {^{212}}\rm{Po} + e^- + \bar{\nu} _e
\end{equation}
\begin{equation}
{^{212}}\rm{Po} \rightarrow {^{208}}\rm{Pb}+\alpha,  
\label{eq:Po212}
\end{equation}
with $\tau$ = 433\,ns (Eq.~\ref{eq:Po212}) allows to estimate the $^{220}$Rn content of the scintillator and to infer the $^{232}$Th contamination.

The $^{212}$Bi is a $\beta$--emitter with $Q$ = 2252\,keV, while the $\alpha$ of $^{212}$Po decay has 8955\,keV energy. The thoron rate in the $^7$Be-FV is not constant in time and it changes as a consequence of the operations on the detector. Again, no time persistent contamination is introduced since we observe that the $^{212}$Bi -- $^{212}$Po rate recovers the initial value within a few days (the longest living isotope among thoron daughters is $^{212}$Pb with $\tau$ = 15.4\,hours). The intrinsic contamination of $^{232}$Th in secular equilibrium has been measured from the $^{212}$Bi -- $^{212}$Po asymptotic rate.

The $^{212}$Bi -- $^{212}$Po events are selected within gates having two clusters surviving the muon cut, reconstructed with a distance of 1\,m, having a time delay between 400\,ns and 1732\,ns (four times the life--time of the decay). The 400\,ns value ensures that the efficiency of the clustering algorithm may be safely assumed to be 100$\%$ at the energies of interest.

\begin{table}[t]
\begin{center}
\begin{tabular}{l c c c } \hline \hline
Isotope          & Mean Life      & Energy  & Decay \\ 
                       &                     &  [keV] &    \\ 
\hline
$^{232}$Th & $2.03 \times 10^{10}$ yrs & 4010 & $\alpha$\\
$^{228}$Ra & $8.31$ yrs & 46 & $\beta^-\gamma$\\
$^{228}$Ac & $8.84$ hrs & 2140 & $\beta^-\gamma$\\
$^{228}$Th & $2.76$ yrs & 5520 & $\alpha$\\
$^{224}$Ra & $5.28$ days & 5690 & $\alpha$\\
$^{220}$Rn & $80.2$ s & 6290 & $\alpha$\\
$^{216}$Po & $209$ ms & 6780 & $\alpha$\\
$^{212}$Pb & $15.3$ hrs & 573 & $\beta^-\gamma$\\
$^{212}$Bi($64\%$) & $87.4$ min & 2250 & $\beta^-$\\
$^{212}$Bi($36\%$) & $87.4$ min & 6050 & $\alpha$\\
$^{212}$Po & 431 ns & 8780 & $\alpha$\\
$^{208}$Tl & $4.40$ min & 4990 & $\beta^-\gamma$\\
\hline \hline
\end{tabular}
\end{center}
\caption{ \label{tab:th232_prod} The $^{232}$Th decay chain showing isotopes, life--times, maximum released energies and type of decay. A large number of $\gamma$ line accompanying many of the decays are not reported in the table.}
\end{table}

The energy region of the first candidate is selected to be $<$2000\,keV
 and that of the second one is required to lie in the interval 900 - 1300\,keV.
 The cut efficiency is 34$\%$. The mean counting rate of the events reconstructed within a sphere of 3.3\,m radius during periods far from any detector operations  (611\,days of life--time) is (0.13~$\pm$~0.03)\,cpd/100\,ton which corresponds to a scintillator $^{232}$Th contamination of (3.8 $\pm$ 0.8)$\times$ $10^{-18}$\,g/g at equilibrium, 20 times lower than the target design.

\begin{figure}[t]
\begin{center}
\centering{\includegraphics[width = 0.5\textwidth]{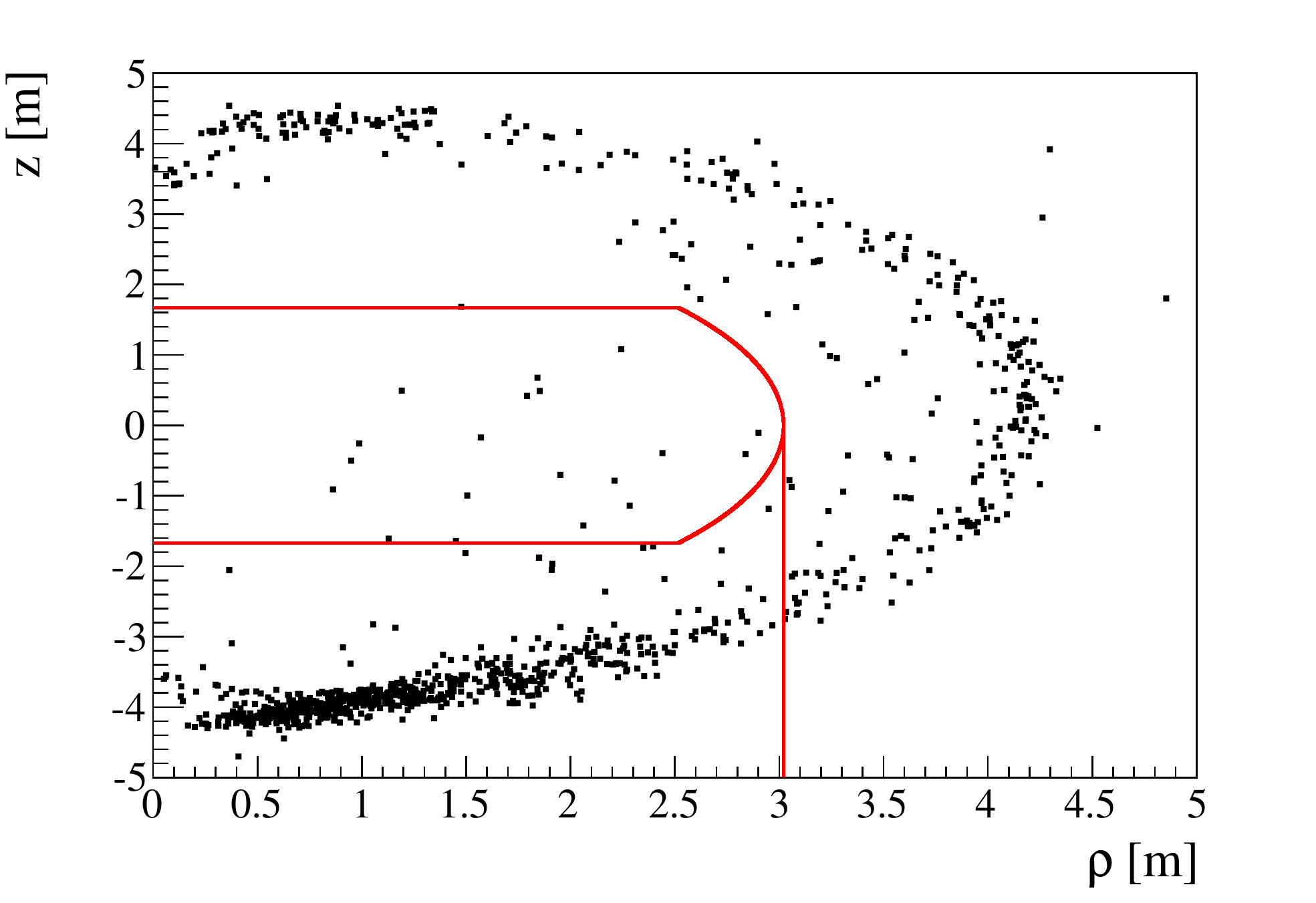}}
\caption{Distribution of the $^{212}$Po events (from $^{212}$Bi - $^{212}$Po coincidences) in the $z$ - $\rho$ plane from May 2007 to May 2010. The solid red line indicates the FV used in the $^7$Be--$\nu$ analysis. }
\label{fig:ThSpatial}
\end{center}
\end{figure}

The spatial distribution of the events reported in Fig.~\ref{fig:ThSpatial} indicates the presence of a contamination located close to the IV and higher than the bulk one. The higher rate at the bottom of the detector could suggest particulate deposition. By decreasing the low--energy cut on the $^{212}$Bi charge to $N_{pe}$=200\, it is possible to study the thoron emanation from the vessel: the typical counting rate is $\sim$5\,cpd. Given the short thoron life--time ($\tau$ = 80.6\,s), practically it does not penetrate deeply inside the scintillator, so the $^{212}$Bi -- $^{212}$Po spatial distribution reproduces very closely the vessel shape.

\subsection{Cosmic muons and cosmogenic background}
\label{subsec:CosmogenicBack}

The dominant muon--induced cosmogenic background in Borexino, $^{11}$C, represents the
biggest challenge for the measurement of $pep$ and CNO neutrinos. About 95$\%$ of this nuclide is produced by  muons through a reaction resulting in the emission of free neutrons~\cite{C11Prod}:
\begin{equation}
\mu  + {^{12}C}   \rightarrow \mu  + {^{11}C} + n \cdot
\label{eq:c11prod}
\end{equation}
$^{11}$C decays with a mean--life $\tau$ = 29.4\,min via positron emission:
\begin{equation}
 ^{11}C  \rightarrow  {^{11}B} + e^+ + \nu_e \cdot
\label{eq:c11decay}
\end{equation}
The total energy released in the detector is between 1020 and 1980\,keV ($\beta^+$ with Q--value of 960\,keV plus $2\times 511$\,keV $\gamma$--rays from $e^+$ annihilation) and lies in the energy region of interest for the detection of electron recoils from $pep$ and CNO neutrinos.
In the Borexino scintillator, the neutrons produced in association with $^{11}$C (see Eq.~\ref{eq:c11prod}) are captured with a mean life--time of (254.5~$\pm$~1.8)\,$\mu$s~\cite{BxMuons}) on hydrogen, emitting characteristic 2230\,keV $\gamma$ rays.  

The  muon flux in Borexino is reduced by about a factor of $10^6$ compared to the sea level thanks to its location deep underground in the Gran Sasso laboratory. It amounts to 1.2 muons $m^{-2} hour^{-1}$~\cite{BxMuons}. The interaction of these residual muons with $^{12}$C is expected to produce few tens of $^{11}$C nuclei per day in the FV.  The continuous cosmogenic production and the short $^{11}$C mean--life originate an equilibrium concentration of $^{11}$C that cannot be reduced by any purification procedure. On the other hand, $^{11}$C tagging through its spatial and time coincidence with muons and captured neutrons, together with pulse--shape discrimination are powerful methods to reduce its contribution, as described in Section~\ref{sec:c11}. 

\begin{table*}
\begin{center}
\begin{tabular}{l c c c c} 
\hline \hline
Isotope          & Mean Life      & Energy  & Decay  & Residual rate\\ 
                       &                    &  [keV] & [cpd/100\,ton]   \\ 
\hline
n & 255\,$\mu$s & 2230 & Capture $\gamma$ on $^1$H    & $<0.005$   \\
$^{12}$N & 15.9 ms & 17300 & $\beta^+$                & $<5 \times 10^{-5}$\\
$^{13}$B & 25.0 ms & 13400 & $\beta^-\gamma$          & $<5 \times 10^{-5}$\\
$^{12}$B & 29.1 ms & 13400 & $\beta^-$                & $(7.1\pm0.2) \times 10^{-5}$ \\
$^{8}$He & 171.7 ms & 10700 &$\beta^-\gamma n$        & $0.004\pm0.002$  \\
$^{9}$C & 182.5 ms & 16500 & $\beta^+$                & $0.020\pm0.006$ \\
$^{9}$Li & 257.2 ms & 13600 & $\beta^-\gamma n$       & $0.022\pm0.002$\\
$^{8}$B & 1.11 s &18000 & $\beta^+\alpha$             & $0.21\pm0.05$\\
$^{6}$He & 1.16 s & 3510 & $\beta^-$                 & $0.31\pm0.04$\\
$^{8}$Li & 1.21 s & 16000 & $\beta^-\alpha$           & $0.31\pm0.05$\\
$^{11}$Be & 19.9 s & 11500 & $\beta^-$                &$0.034\pm0.006$\\
$^{10}$C & 27.8 s &3650 & $\beta^+\gamma$            &$0.54\pm0.04$\\
$^{7}$Be & 76.9 days& 478 & EC $\gamma$            &$0.36\pm0.05$\\
\hline \hline
\end{tabular}
\end{center}
\caption{ \label{tab:cosmogenics_prod} Cosmogenic isotopes in Borexino. The last column shows the expected residual rate after 
the  300\,ms time veto after each muon passing through ID is applied (see Section~\ref{sec:cuts}). The total rates have been evaluated following \cite{bxB8} or extrapolating
simulations reported  in \cite{KamlandCosmogenics}.}
\end{table*}

Table~\ref{tab:cosmogenics_prod} shows 
other isotopes produced cosmogenically within the detector. Their importance is suppressed
since we veto all the detector for 300\,ms after each muon as described in Section~\ref{sec:cuts}; the last column shows the residual rate after this cut.
These background sources are relevant only for
the $pep$ and CNO neutrino analysis.  The differential rates in this table show
that only $^6$He has a value that is greater than 5$\%$ of the $pep$
signal rate and, therefore, it has been included in the fit (see
Section~\ref{sec:pepCNOResults}). A special treatment is required for
$^{10}$C: even though its starting point at 1740\,keV is past the $pep$ neutrinos
energy, its count rate of (0.54 $\pm$ 0.04)\,cpd/100\,ton is relatively
high and a large fraction of its spectrum falls within the fit region
($<$3200\,keV). The 480\,keV $\gamma$--line from $^7$Be EC decay, on
the other hand, is negligible with respect to the $^{85}$Kr, $^{210}$Bi, and $^7$Be solar neutrino
recoil spectra ($\sim$0.150\,cpd/100\,ton/keV) and can safely be excluded from the fit, even
if its total count rate is comparable to that of $^{10}$C and $^6$He.

Cosmogenic backgrounds originating from outside the detector and from
untagged muons are expected to be smaller than those presented in
Table~\ref{tab:cosmogenics_prod}. In fact, untagged muons that pass all the
standard and FV--volume cuts and enter into the final spectrum used for the 
measurement of the interaction rate of the $^7$Be neutrinos  have an expected rate $<$$3\times10^{-4}$\,cpd/100\,ton
(considering the OD absolute efficiency of 99.2$\%$ and independence
between ID and OD muon flags~\cite{BxMuons}). Except for neutrons,
none of the other cosmogenic isotopes can travel very far away from
the muon shower and, therefore, all of their decays in the FV will be
preceded by the muon shower in the scintillator, for which the tagging
efficiency is highest ($>99.992$\%~\cite{BxMuons}). 
There is a possibility though, that neutrons produced outside the IV will deposit
energy within the FV.
Those that are captured in the IV may be vetoed
by the fast coincidence condition, as proton recoils from the
thermalization will be visible near the capture
position. 
Another possible background is from proton recoils due to fast neutrons from
untagged muons that do not cross the IV, where the neutron is not
captured in the scintillator, and the reconstructed position of the
recoils is within a reduced FV. 
We have studied carefully the background induced by fast neutrons \cite{geonu} for the geo--neutrino (antineutrinos of geo--physical origin) analysis. The limit of $<1.4 \cdot 10^{-4}$\,cpd/100\,ton set there allows to conclude that the expected background due to proton recoils is negligible for all analysis presented in this paper.
%  in We do not expect such background to be higher than the limit of $<1.4 \cdot 10^{-4}$\,cpd/100\,ton already determined in \cite{geonu}, the limit on fast neutrons created in the surrounding rock and in the WT tank that mimic $\bar{\nu}$ signals, and, therefore, of relevance for our analysis.
%
Furthermore, any surviving proton--recoil signal would be subtracted from the final spectrum due to their
$\alpha$--like pulse--shape. 

\section{Shape variables and event quality estimators}
\label{sec:shape}

The time distribution of the  photons emitted by the scintillator depends on the details of the energy loss, which in turn depend on the particle type, as was described in Section~\ref{sec:scintillator}.
It is therefore possible to define shape variables that can either efficiently distinguish noise events from point--like scintillation events or disentangle different particle types. 
In this section we define such variables which are then applied in the
event selection procedure described in Section~\ref{sec:cuts}.

\begin {itemize} [leftmargin=*]

\item The ``{\em Gatti}'' parameter ($G$)

The Gatti filter~\cite{Gatti,gatti1,gatti2} allows to separate two
classes of events with different but known time distributions of the
detected light. It is used to perform $\alpha/\beta$ discrimination
and also to separate $\beta^+ / \beta^-$ events. For the two classes
of events under examination, normalized reference shapes $P_1(t)$ and
$P_2(t)$ are created by averaging the time distributions of large sample of events selected without any use of pulse--shape variables. For example, in case of $\alpha/\beta$ selection, the $^{214}$Bi -- $^{214}$Po coincidences are used to select clean samples of $\alpha$ and $\beta$ events. The functions $P_1(t)$ and $P_2(t)$ represent the probability that a
photoelectron is detected at the time between $t$ and $(t+dt)$ for events of classes 1 or 2, respectively. The
reference shape is binned for an easy comparison with the data,
obtaining $r_1(t_n)$ and $r_2(t_n)$
\begin{equation}
r_{12}(t_n) = \int_{t_0+ n \Delta t}^{t_0 + (n+1) \Delta t} P_{12}(t) dt,
\label{eq:refGatti}
\end{equation}
where $n$ is the bin number, $t_0$ is a reference point of the time distribution (either the beginning of the cluster or the position of the maximum), and $\Delta t$ is the bin width.

If we call $e(t_n)$ the distribution of the measured binned time
distribution for a generic event, then the Gatti parameter $G$ is defined as
\begin{equation}
G = \sum_n e(t_n) w(t_n),
\label{eq:GattiDef}
\end{equation}
where $w(t_n)$ are weights given by
\begin{equation}
w(t_n)  = \frac{r_1(t_n)- r_2(t_n)}{r_1(t_n)+r_2(t_n)} \cdot
\label{Gattiw}
\end{equation}
The $G$ parameter follows a probability distribution with the mean
value $\overline G_i$ that depends on particle type:
\begin{equation}
\overline G_i= \sum_n r_i(t_n) w(t_n). 
\end{equation}

\begin{figure}[t]
\begin{center}
\centering{\includegraphics[width = 0.51\textwidth]{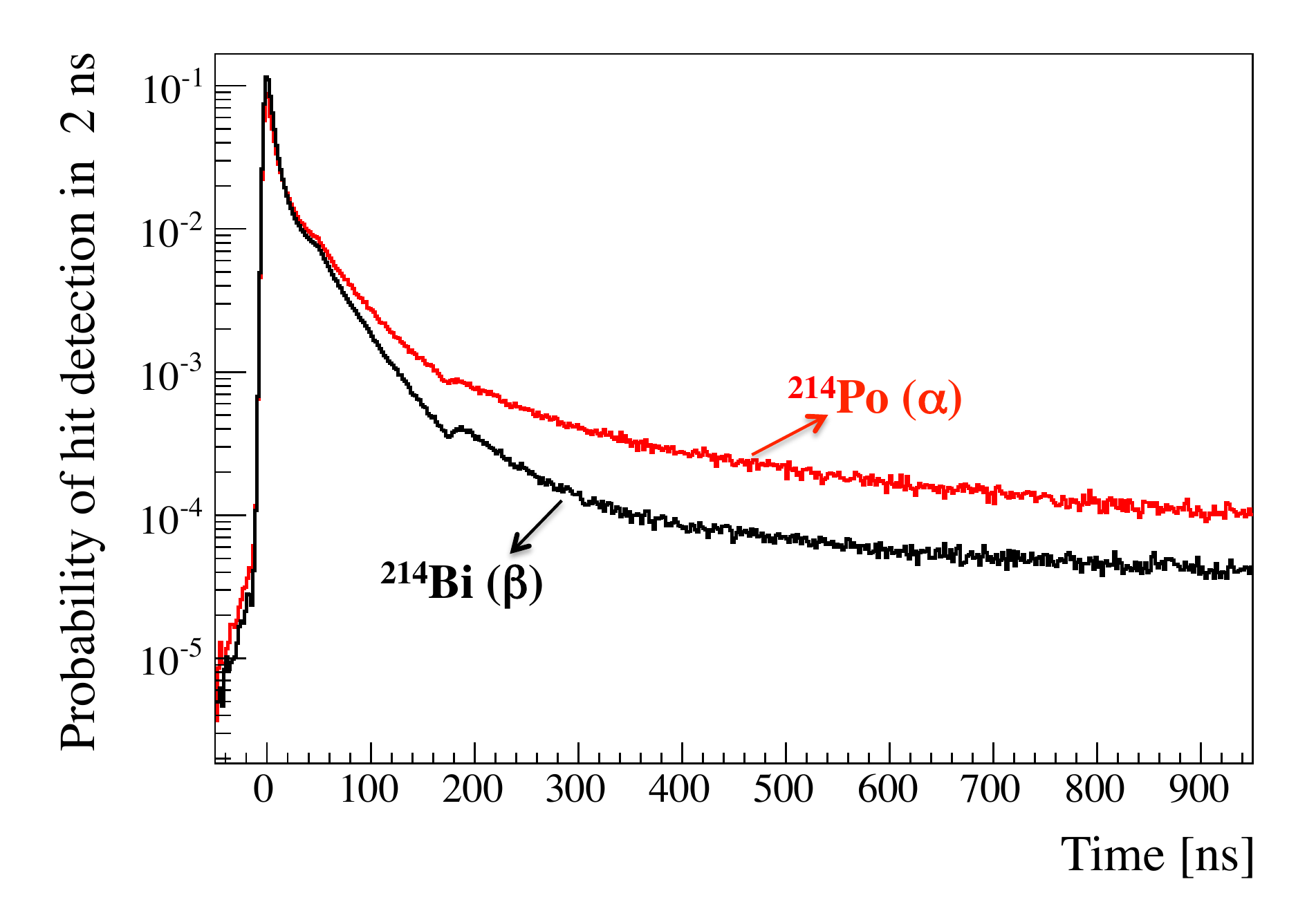}}
\caption{The reference $r_{\alpha}(t_n)$ (red) and $ r_{\beta}(t_n)$ (black) pulse shapes obtained by tagging the radon-correlated $^{214}$Bi -- $^{214}$Po coincidences. The dip at 180\,ns is due to the dead time on every individual electronic channel applied after each detected hit (see Section~\ref{sec:electronics}). The small shoulder around 60\,ns is due to the light reflected on the SSS surface and on the PMTs' cathodes.} 
\label{fig:abPulse}
\end{center}
\end{figure}

\begin{figure} [t]
\begin{center}
\centering{\includegraphics[width = 0.5\textwidth]{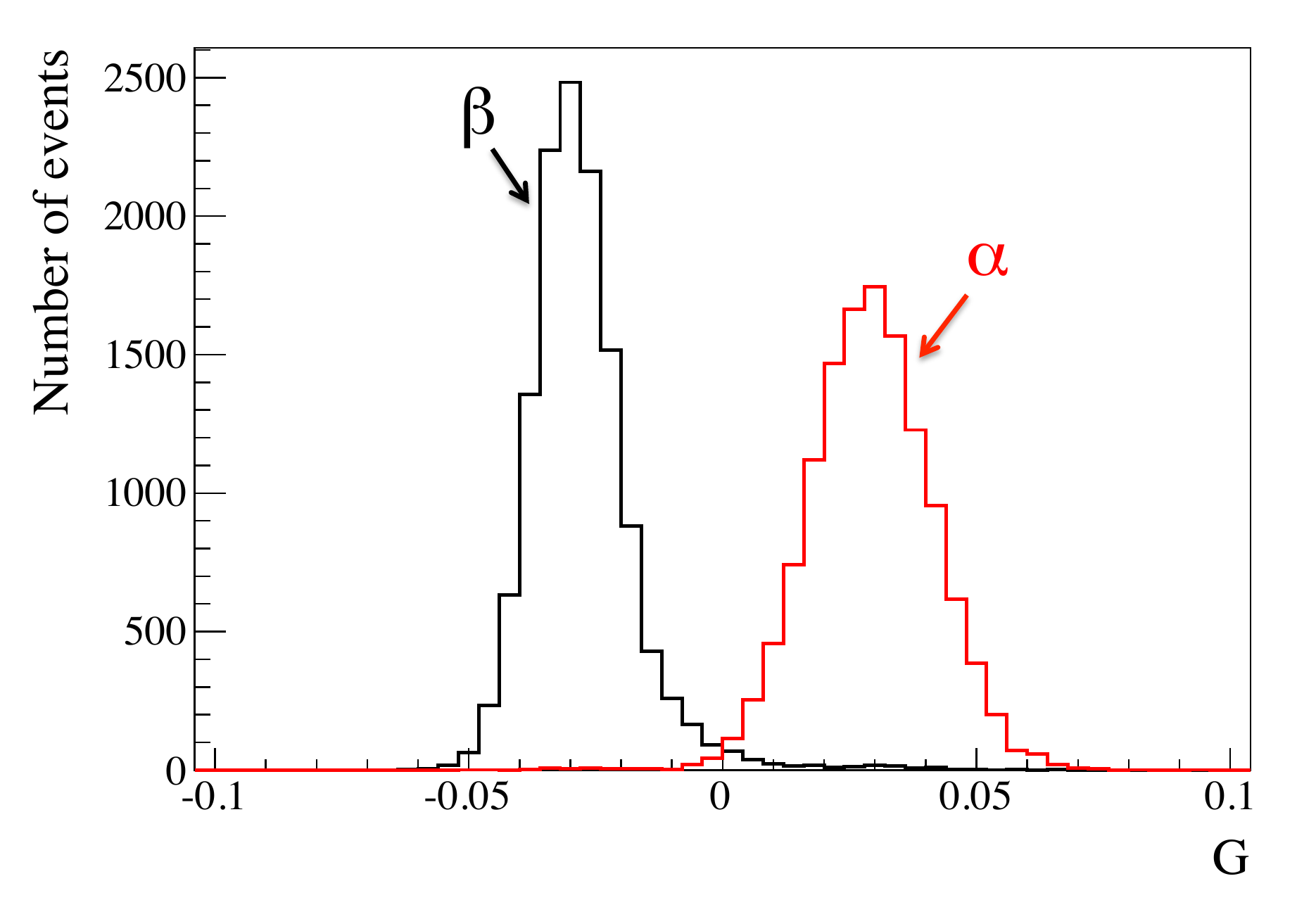}}
\caption{The distribution of $G_\alpha$ (red) and $G_\beta$ (black) (see Eq.~\ref{eq:GattiDef}) for events obtained by tagging the radon correlated $^{214}$Bi--$^{214}$Po coincidences.}
\label{fig:Gattiab}
\end{center}
\end{figure}

In the scintillator used by Borexino, $\alpha$ pulses are slower and have therefore a longer tail with respect to $\beta / \gamma$ pulses. 
The reference shapes $r_{\alpha} (t_n)$ and $r_{\beta}(t_n)$ (obtained from $^{214}$Bi($\beta$) -- $^{214}$Po($\alpha$) coincidences), are shown
in Fig.~\ref{fig:abPulse}, while the distributions of the corresponding $G$ parameters ($G_{\alpha}$ and $G_{\beta}$) 
 are shown in Fig.~\ref{fig:Gattiab}. The large separation between the $G_{\alpha}$ and $G_{\beta}$ distributions is due to different weight of the delayed scintillation light for $\alpha$ and $\beta$ particles that is summarized in Table~\ref{table:prop}. To enhance the sensitivity to this delayed light, the time duration of the event has been fixed to 1.5\,$\mu$s starting from the time of the first hit generating the trigger.
The variance of the distributions of $G_{\alpha}$ and $G_{\beta}$ depends on the energy and it sligthly increases as the energy decreases thus reducing the
discrimination power. However this fact is  important only for energy deposit lower than that considered in the analysis reported here. We have in any accounted for this effect
 in the  $\alpha$ $\beta$
statistical subtraction (discussed in Section ~\ref{sec:ab}) procedure by considering the variance of the $G_{\alpha}$ and $G_{\beta}$ distributions as free fit parameters when using  the analytical approach. The Monte Carlo simulation reproduces the effect.

Similarly, as will be discussed in Section~\ref{subsec:psd}, we have built a $G$ parameter to discriminate between $\beta^+$ and
$\beta^-$ (called $G_{\beta^+}$ and $G_{\beta^-}$) events using as $\beta^-$ reference the time distribution of $^{214}$Bi and for $\beta^+$ a sample of $^{11}$C events tagged with the TFC method (described in Section~\ref{subsec:tfc}).
The $G_{\beta^+}$ and $G_{\beta^-}$ distributions are shown in Fig.~\ref{fig:GattibetaC11}.
The separation between the $G_{\beta^+}$ and $G_{\beta^-}$ distributions is small, since it is mostly due to the delay in the scintillation introduced in case of $\beta^+$ because of the formation of positronium and its survival  time in the scintillator before annihilation. This time, as it will be discussed in detail in Section~\ref{subsec:tfc}, is of the order of only few ns.

\begin{figure} [h]
\begin{center}
\centering{\includegraphics[width = 0.5\textwidth]{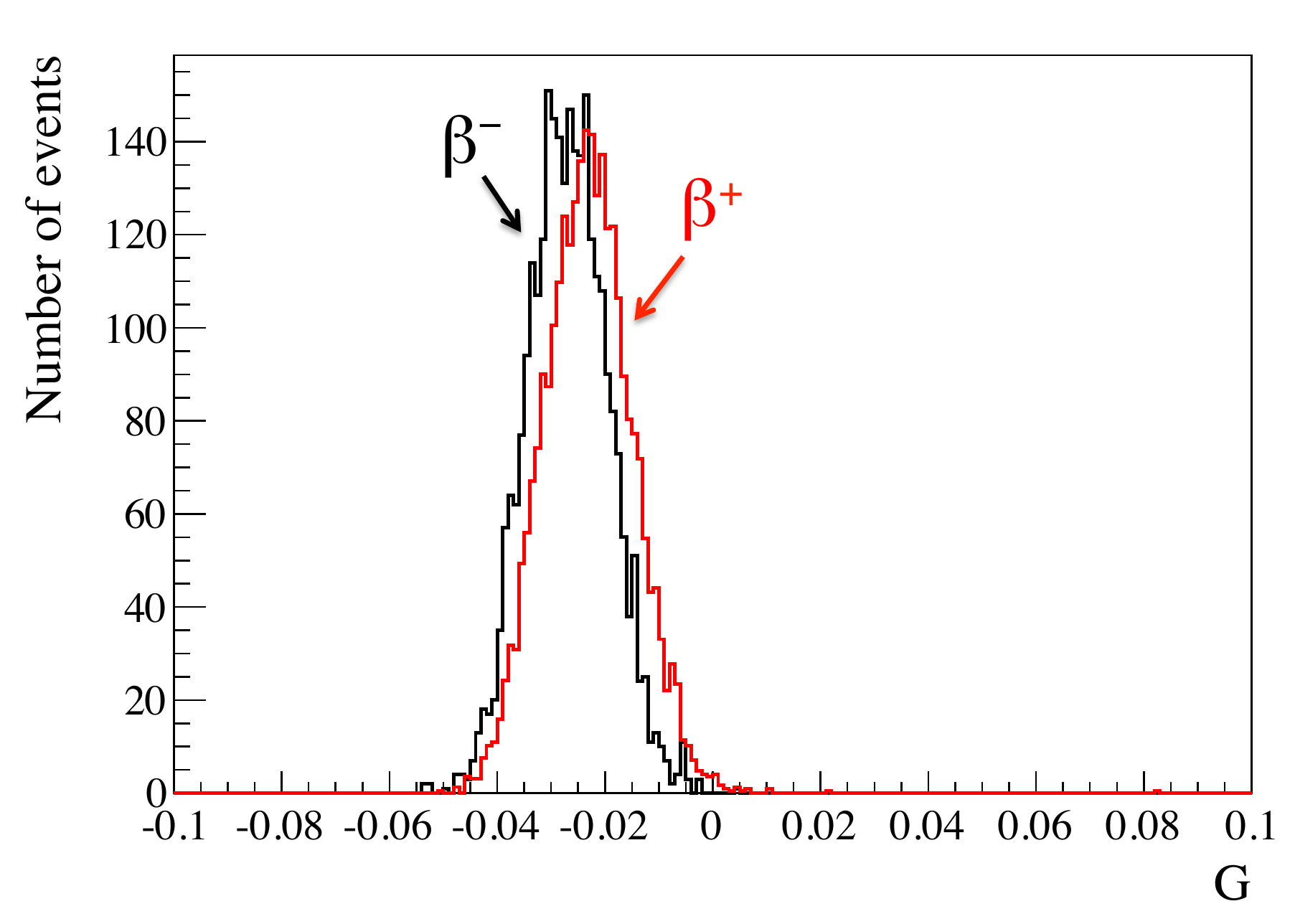}}
\vspace{-5 mm}
\caption{The distribution of the $G_\beta^-$ parameter (black) obtained from $^{214}$Bi\,($\beta^{-}$) events and of the $G_\beta^+$ parameter (red) obtained from $^{11}$C\,($\beta^{+}$)  events.}
\label{fig:GattibetaC11}
\end{center}
\end{figure}

\item Anisotropy variables $\beta_l$ and $S_p$

Noise events with anisotropic hit distributions are rejected
by characterizing the distribution of observed hits with respect to
the reconstructed position.
For localized energy deposits, such as neutrino--induced scattered electrons or
$\beta$--decays, the scintillation light is emitted isotropically from
the interaction point, while for noise events the detected hit--time distribution is likely to be anisotropic.
Two different variables describing the event
isotropy, $\beta_l$ and $S_p$, are defined.

$\beta_l$: first, the number of photoelectrons detected on each PMT is estimated by rounding the detected charge, normalized by the corresponding single photoelectron mean, to the nearest integer. Then, for every pair of photoelectrons $i$ and $j$, the angle $\theta_{ij}$, between the corresponding PMTs is
calculated with respect to the reconstructed position of the event. We sum up the
Legendre polynomials $P_l(\cos(\theta_{ij}))$ for each of the pairs,
to obtain the anisotropy parameter $\beta_l$:
\begin{align}
\beta_l \equiv \frac{2}{N(N +  1)}\sum_{i=0}^{N}\sum_{j=i+1}^{N}  P_l(\cos(\theta_{ij})),
%\beta_l \equiv \frac{2}{N_{pe}(N_{pe} + 1)}\sum_{i=0}^{N_{pe}}\sum_{j=i}^{N_{pe}}P_l(\cos(\theta_{ij})),
\label{eq:beta_l}
\end{align}   
where $N$ is the total number of photoelectrons and the sum runs over each pair of detected photoelectrons (estimated as described above).

$S_p$: the $\cos(\theta)$ and $\phi$ angular distribution of the
detected light is computed with respect to the reconstructed position
and developed in a series of ``spherical harmonic'':
\begin{equation}
Y_{l}^{m}(\theta,\phi) = \frac {2 \sqrt{\pi}}{N_{h}}  e^{im\phi} P_{1}^{m}(\cos\theta),
\label{eq:SH}
\end{equation}
with $m$ = -1, 0, 1 and $P_{1}^{m}$ the associated Legendre
polynomials, and  $N_{h}$ is the total number of detected hits.
Three complex coefficients $S_{m}$ are calculated as:
\begin{equation}
S_{m} = \sum_{i =1}^{N_h} Y^{m}_{1}(\theta_{i},\phi_{i}),
\label{eq:Sm}
\end{equation}
where the index $i$ runs on the hits in the cluster while $\theta_{ i}$
and $\phi_{i}$ are the spherical coordinates of the hit PMT in a
reference frame centered in the reconstructed vertex.
We define the $S_p$ variable as:
\begin{equation}
S_p = |S_{0}|+|S_{1}|+|S_{-1}|. 
\label{eq:SHpower}
\end{equation}

%***************************************************
\item {\em $N_{peak}$} variable

The scintillation pulse shape  due to an interaction of a single particle features only a single maximum.
A dedicated algorithm was developed for identifying the number of peaks within the time distribution of  detected hits.
Fig.~\ref{fig:Npeaks} shows an example of  an event for which $N_{peak}$ variable is 2, most probably due to pile--up of two distinct interactions occurring in a time window that was too short to be recognized as two separate clusters.

\begin{figure} [h]
\begin{center}
\centering{\includegraphics[width = 0.49\textwidth]{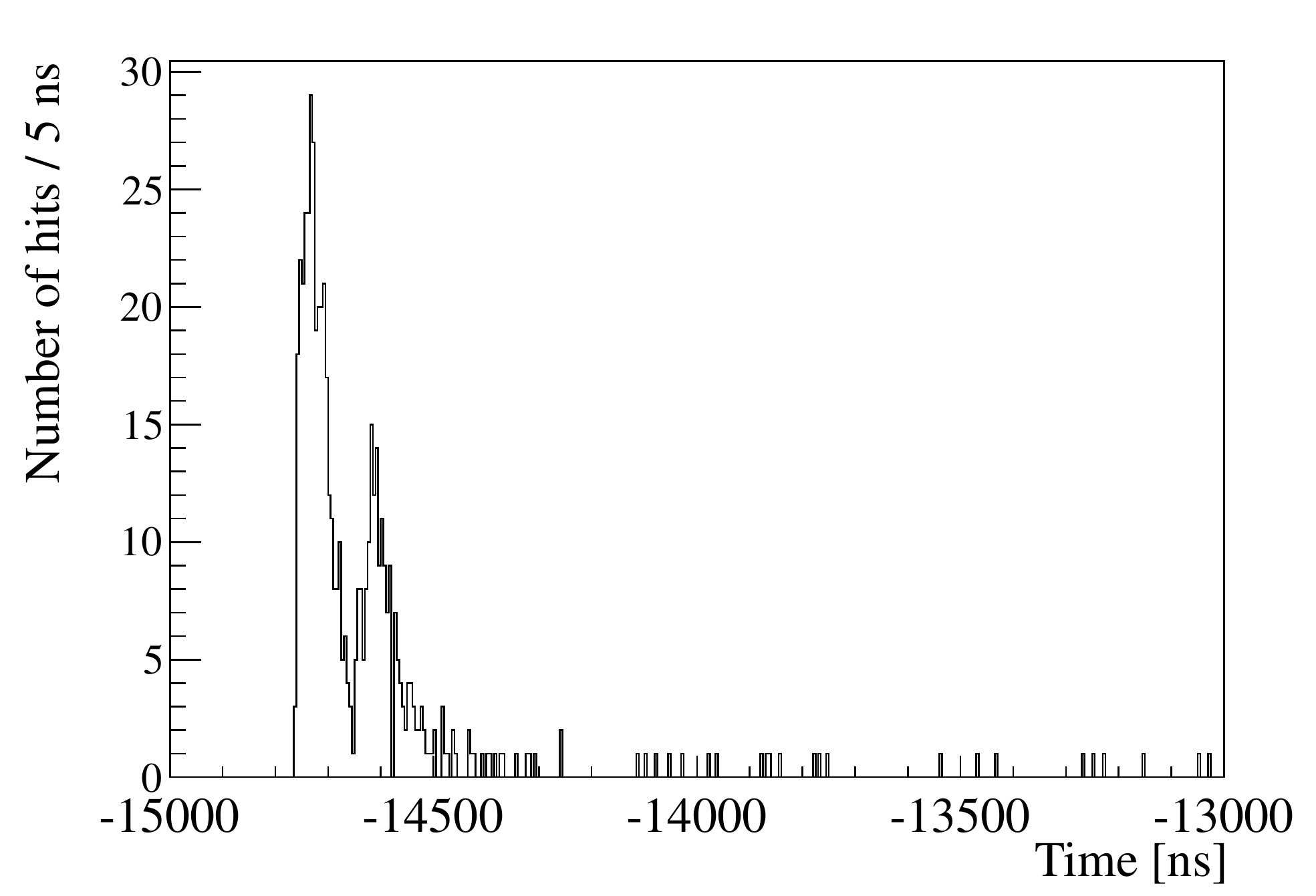}}
\caption{Example of an event with $N_{peak}$ = 2. The horizontal axis shows the time difference of each hit with respect to trigger signal, so the absolute values have no physical meaning.}
\label{fig:Npeaks}
\end{center}
\end{figure}

%***************************************************

\item $R_{pe}$ variable

The different energy estimators introduced in Section~\ref{sec:estimators} are correlated.
Once $N_p$ is known then  the expected charge variable $N_{pe}^{exp}$ can be expressed as:
\begin{equation}
N_{pe}^{exp} = \frac{-N_{tot} \cdot \ln  \left (1 - \frac{N_p}{N_{tot}} \right )}{\left (1 + g_C \ln \left (1 - \frac{N_p}{N_{tot}} \right ) \right)},
\label{eq:qexp}
\end{equation}
where $g_C$ is a geometrical correction factor defined in Eq.~\ref{NpVol}. Its value depends on the choice of the
fiducial volume and it is typically about 0.11.
The $R_{pe}$ variable is defined as a ratio of the measured and expected charge:
\begin{equation}
\begin{split}
  R_{pe} = \frac{N_{pe}}{ N_{pe}^{exp}}. 
\end{split}
\label{eq:qrec}
\end{equation}

%***************************************************
\item {\em $R_q$} variable

In order to identify events that have an abnormally large number of hits with invalid charge, we define the variable $R_q$:
\begin{equation}
\begin{split}
  R_{q} = \frac{N_{pe}}{ N_{pe-avg}} \cdot
\end{split}
\label{eq:Rq}
\end{equation}
We expect $R_q$ to be approximately equal to 1 for normal scintillation events, 

%***************************************************
\item {\em $f_{rack}$} variable

One of the common sources of noise events during data taking is electronic noise from a single electronics rack (corresponding to 160 channels)~\cite{BxNim}. If several PMTs detect some noise signal then the main trigger may fire. In order to discriminate against this type of triggers, for each event we keep track of the total fraction of hits that are recorded on the most active crate, $f_{rack}$.     

%***************************************************

\end {itemize}

\section{The event selection and cut efficiency}
\label{sec:cuts}

This section is devoted to the selection of the $^7$Be, $pep$ and CNO solar neutrino events.
Solar neutrino events cannot be distinguished from background events. However, a series 
of cuts applied on an event--by--event basis has been developed with the aim to remove taggable backgrounds and 
non--physical events.
A set of these cuts is described in Section~\ref{subsec:cuts},
while the FV cut has been already discussed in Section~\ref{sec:FV}.
Many of these cuts are correlated. We report therefore in
Section~\ref{subsec:efficiency}  the overall efficiency of
the whole chain of cuts. 

In order to obtain the spectra used for solar neutrino analysis, first, the set
of selection cuts as described in Section~\ref{subsec:cuts} is applied.
Afterwords, two different ways of exploiting the pulse--shape
capability of the scintillator, based on the use of $G_{\alpha
\beta}$ parameter, are applied.
In the first approach, the discrimination is applied on an event--by--event basis
and is described in Section~\ref{subsec:softAB}.
This energy dependent cut is used to eliminate a small fraction of non--physical events with $\alpha$--like character and therefore it unavoidably removes also some part of real $\alpha$ particles.
The second approach, where the $\alpha$ and $\beta$ contributions are separated
statistically and not on an event--by--event basis, is
described in a separate Section~\ref{sec:ab}.
The cut described in Section~\ref{subsec:softAB} is not applied in this statistical approach in order not to deform the $G_{\alpha
\beta}$ distributions.
A special use of the $G_{\beta^-}$ and $G_{\beta^+}$ parameters in the process of $^{11}$C subtraction in the $pep$ and CNO neutrino analysis is described in Section~\ref{sec:c11}.

\subsection{Event selection}
\label{subsec:cuts}

\begin{itemize}[leftmargin=*]

\item {\em Muon and muon--daughter removal}

There are about $\sim $4300 muons/day crossing the ID. The events  due to muons must be identified and removed in particular when, due to geometrical effects, their energy deposit is so small that they may fall into the energy region interesting for the solar neutrino detection; moreover the  identification of muons is  important for tagging possible  cosmogenic radioisotopes produced by their interaction in (or around) the scintillator. 
Here we give a brief overview of muon tagging methods, while a full
description is given in~\cite{BxMuons}.

Muons passing the water shield in WT produce Cherenkov light causing
the OD to trigger (Muon Trigger Flag, MTF).
The Cherenkov light is identified as a cluster of hits within the data
taken from the OD PMTs (Muon Cluster Flag, MCR).
Muons passing through ID, produce a distinct pulse
shape different from point--like scintillation events and can be
therefore identified by pulse--shape analysis (Inner Detector Flag,
IDF).
Events identified as muons by either of these three flags are excluded
from the analysis.
Muons passing the scintillator or buffer region (ID--$\mu$'s) can be
detected by any flag but a presence of a cluster of hits within the
data taken from the ID PMTs is required.
Since ID--$\mu$'s produce a large amount of light often saturating the
electronics, the whole detector is vetoed for 300\,ms after each of
them.
This time is sufficient to suppress cosmogenic neutrons (captured with
a mean life--time of (254.5~$\pm$~1.8)\,$\mu$s~\cite{BxMuons}) and other
relevant spallation products as well.
Muons passing through the OD only (OD--$\mu$'s) are identified by either
MTB or MCR flags, they do not produce a cluster of hits within the ID
data, and they do not saturate the electronics.
In this case, only cosmogenic neutrons can penetrate through the SSS
from the OD to the ID volume and a 2\,ms veto is sufficient.
The total live-time reduction introduced by all these cuts is $1.6\%$.

\item {\em Single energy deposit requirement}

Accepted events are required to correspond to a single
energy deposit within the DAQ gate (16\,$\mu$s  long).
If the clustering algorithm (Section~\ref{sec:electronics}) does not
identify any cluster or identifies multiple clusters,  the event is
rejected.
The clustering algorithm can recognize two physical events as two separate clusters if their time separation is more than $\sim$230\,ns.
The pile--up of events occurring within shorter time intervals can be identified by $N_{peak}$
variable (Section~\ref{sec:shape}) greater than one.
Therefore,  $N_{peak}$=1  is requested, which is also a powerful tool for removal of irregular noise signals often featuring several peaks in the hit--time distributions.

\item {\em Removal of coincident events}

All events reconstructed with mutual distance smaller than 1.5\,m and occurring in a
2\,ms time window are rejected.
This cut removes a small part of uncorrelated events
(Section~\ref{subsec:efficiency}), while it removes sequences of
noise events and possible correlated events of unknown origin.
This cut removes radon correlated $^{214}$Bi -- $^{214}$Po delayed
coincidences as well.

During the normal data acquisition, several calibration events (pulser,
laser, and random triggers, see Section~\ref{sec:electronics}) are
regularly generated by interrupts based on a 200\,Hz clock~\cite{BxNim}.
If a physical event occurs in a coincidence with such an interrupt,
the event is rejected, since it might be read incompletely or might
contain calibration hits as well.

\item {\em Quality control of the $N_{pe}$ variable}

The quality of the charge variable $N_{pe}$
(Section~\ref{sec:estimators}) of individual events is checked in two
independent ways:

{\em i)}   The $R_{pe}$ variable (Eq.~\ref{eq:qrec}) has to be within the interval from 0.6 to 1.6.

{\em ii)} The $R_{q}$ variable (Eq.~\ref{eq:Rq}) has to be more than 0.5.

Both these conditions are also a powerful tool for noise suppression.

\item {\em Isotropy control}

Additional noise events are further rejected by requiring that the
detected scintillation light is isotropically emitted around the
interaction point.
This is guaranteed by these two independent conditions:

{\em i)}  The variable $\beta_l$~(Eq.~\ref{eq:beta_l}) has to satisfy: 
\begin{equation}
\begin{split}
\beta_1 <  0.027 &+  \exp(-1.306 - 0.017  N_{pe}) \\
         &+ \exp(-3.199 - 0.002  N_{pe}).
\end{split}
\label{eq:NpeCheck}
\end{equation}

{\em ii)} The $S_p$ variable (Eq.~\ref{eq:SHpower}) has to be
below an energy--dependent threshold:
\begin{equation}
\begin{split}
S_p < 0.119 & + \exp(12.357 - 0.305 N_p) \\
                                 & + \exp(-0.612 - 0.011 N_p).
\end{split}
\label{eq:SHpowerCut}
\end{equation}
Figure~\ref{fig:Sp} demonstrates application if this cut on a sample of events which are not tagged as muons and which are reconstructed in the $^7$Be--FV. 

\begin{figure}
\centerline{\includegraphics[width = 0.5\textwidth]{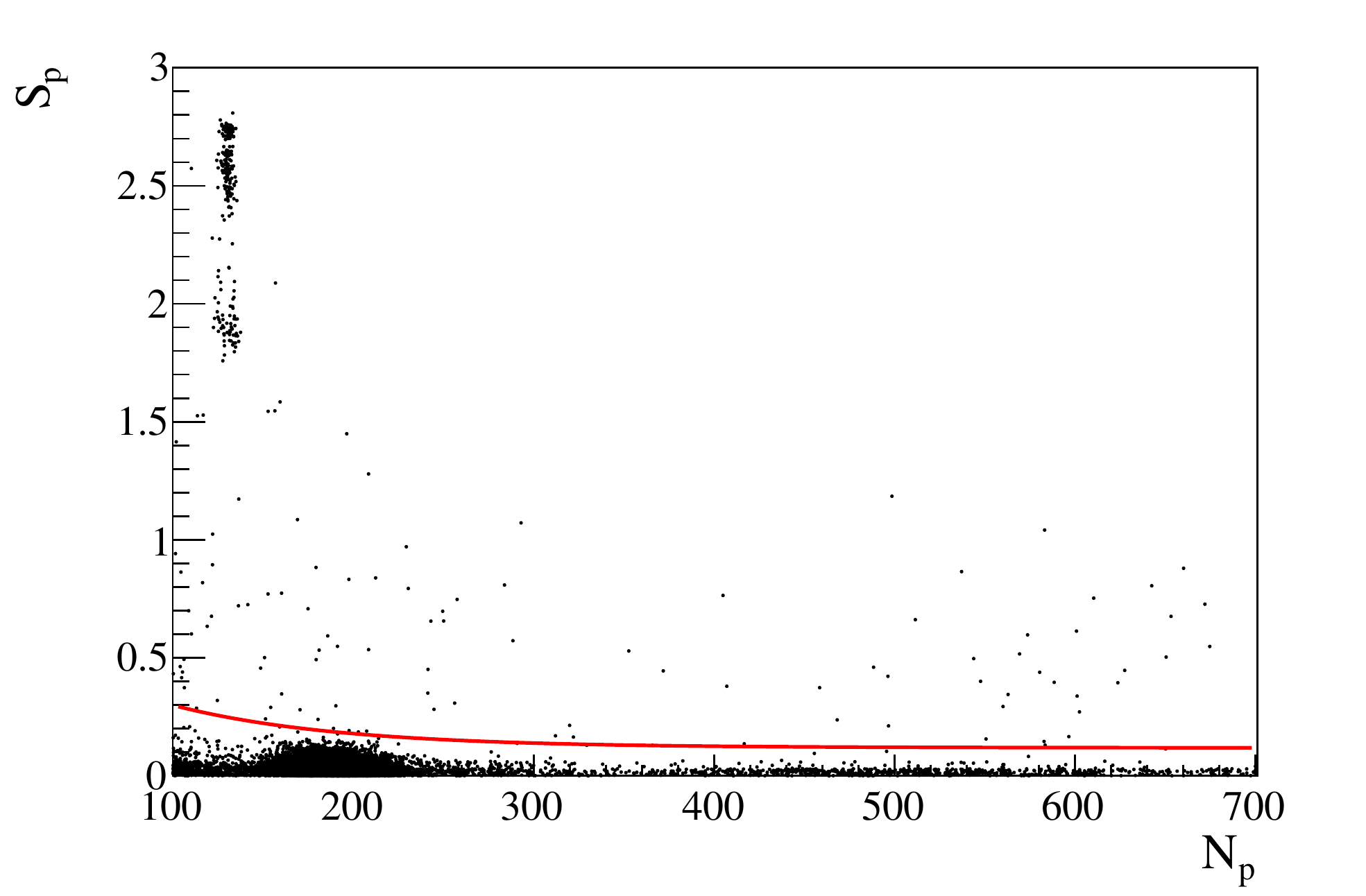}}
\caption{An example of the application of the $S_p$ energy--dependent cut (see Eq.~\ref{eq:SHpowerCut}) indiciated by the solid red line on a sample of events which are not tagged as muons and which are reconstructed in the $^7$Be--FV. The events having the $S_p$ variable above the value indicated by the line are excluded from the data set. Clearly, the cluster at energies of $N_p \sim$130 with $S_p > 1.7$ is due to anisotropic noise events.}
\label{fig:Sp}
\end{figure}

\item {\em Additional noise removal}
 
The cluster which caused the trigger generation has a
well defined position within the DAQ gate.
The $rms$ of the distribution of the cluster start time is $\sim$55\,ns
and features some tails.
An event is accepted only if its cluster starts within a conservative
1.7\,$\mu$s wide time window which has a fixed position in the DAQ
gate.

Additionally, all events for which $f_{rack} > 0.75$
(Section~\ref{sec:shape}) are rejected.

\end{itemize}

\begin{figure}[t]
\begin{center}
\vspace{-3mm}
\centering{\includegraphics[width = 0.5 \textwidth]{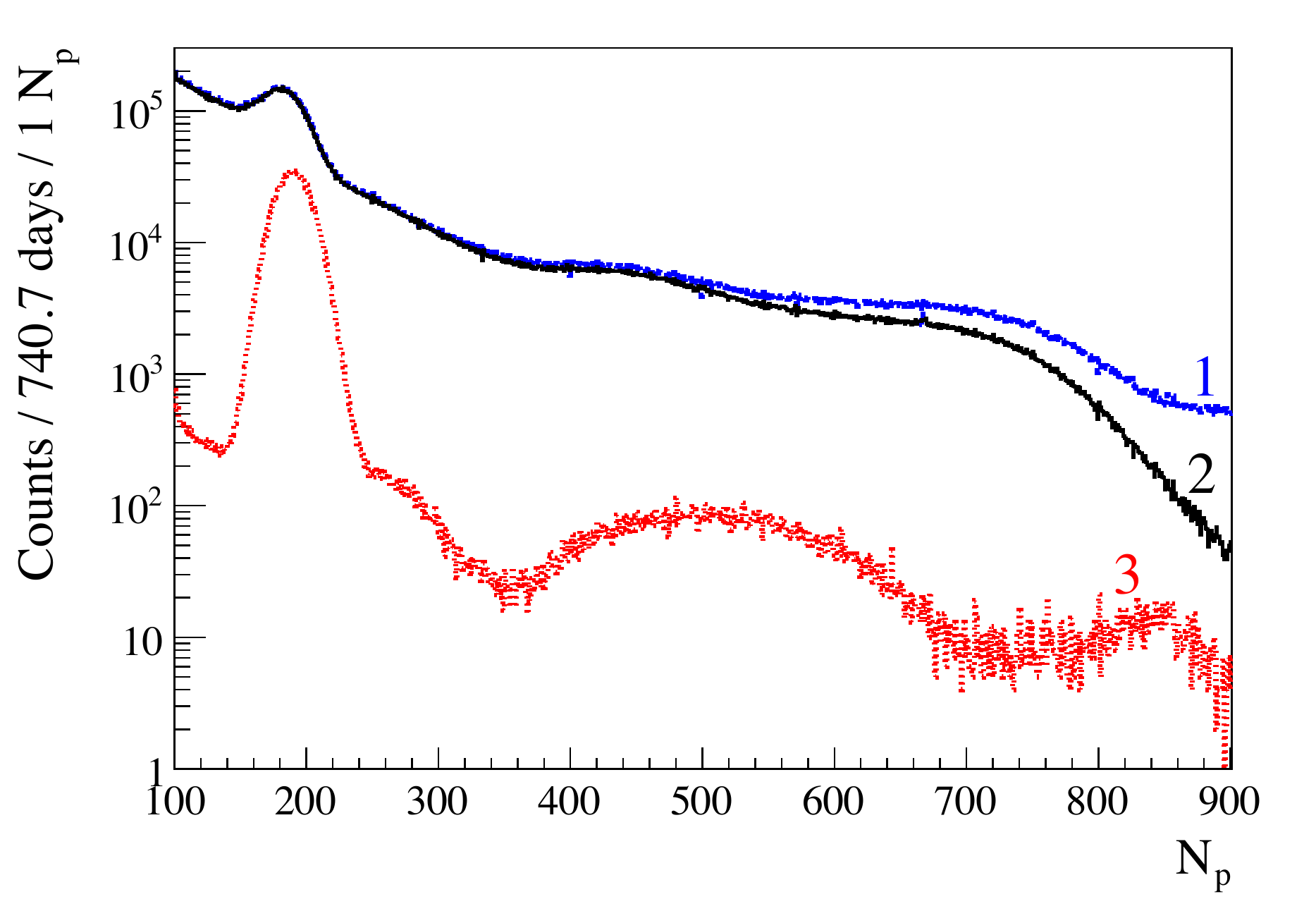}}
\caption{Effect of selection cuts on the raw spectrum in the $N_p$ variable (marked 1 and shown in blue). The black spectrum (marked 2)
shows the effect of the muon and muon daughter cut. The shape
of the final $N_p$ spectrum (marked 3 and shown in red) is dominated by the effect
of the $^7$Be--$\nu$ FV cut. The effect of other cuts described in
Section~\ref{subsec:cuts} is important at the level of fit, but
cannot be appreciated visually.}
\label{fig:spectrumAndCut}
\end{center}
\end{figure}

\begin{figure}[thb]
\begin{center}
\centering{\includegraphics[width = 0.5 \textwidth]{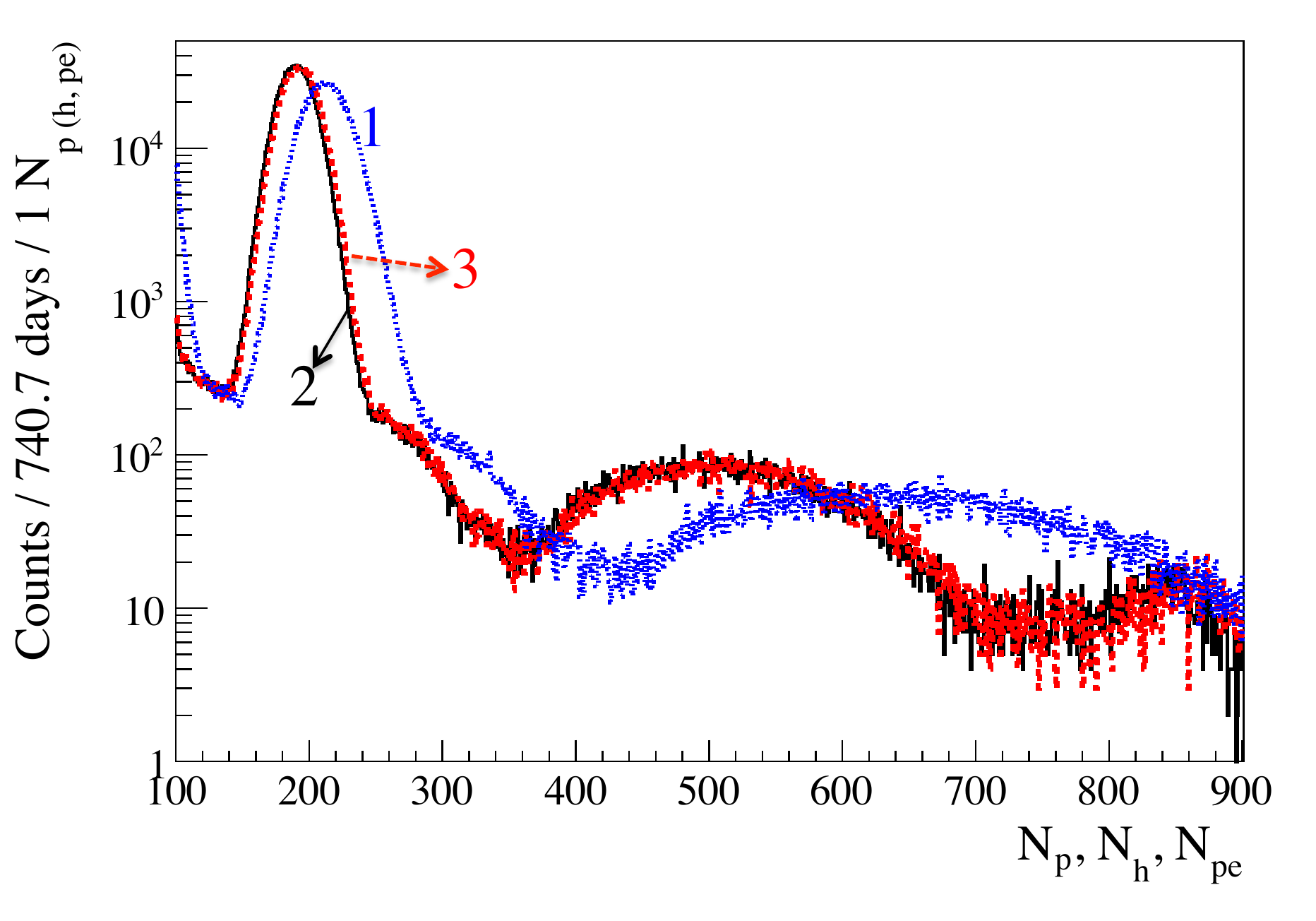}}
\caption{Final spectra for all events in the $^7$Be--$\nu$ FV passing all selection cuts, shown
in three different energy estimators: $N_{pe}$ (1-dotted blue), $N_p$ (2-solid black), and $N_h$ (3-dashed red).}
\label{fig:StandardSpectra}
\end{center}
\end{figure}

The effect of these selection cuts is shown in
Fig.~\ref{fig:spectrumAndCut} with the choice of the $^7$Be--FV.
The final spectra of events passing all the selection cuts are shown
for three energy estimators $N_p$, $N_h$, and $N_{pe}$
(Section~\ref{sec:estimators}) in Fig.~\ref{fig:StandardSpectra}.
As it was anticipated in Section~\ref{sec:estimators}, the $N_p$ and $N_h$ spectra are very similar
(but identical) in the energy region of interest since the probability of multiple hits on the same PMT is small.
The comparison of Fig.~\ref{fig:StandardSpectra} and Fig.~\ref{fig:sb} allows to identify the main contribution in the spectrum:  the $^{210}$Po peak is well visible around 190 $N_p$ and $^{11}$C is easily identified in the region between $\sim$380 and 700 $N_p$. The shoulder of the electron--recoil spectrum due to $^7$Be-$\nu$'s is visible in the region $N_p \simeq$ 250 -- 350.  In the low-energy region ($N_p \le $ 100) the count rate is dominated by the $^{14}$C. 

\begin{figure}[b]
\begin{center}
\centering{\includegraphics[width = 0.5 \textwidth]{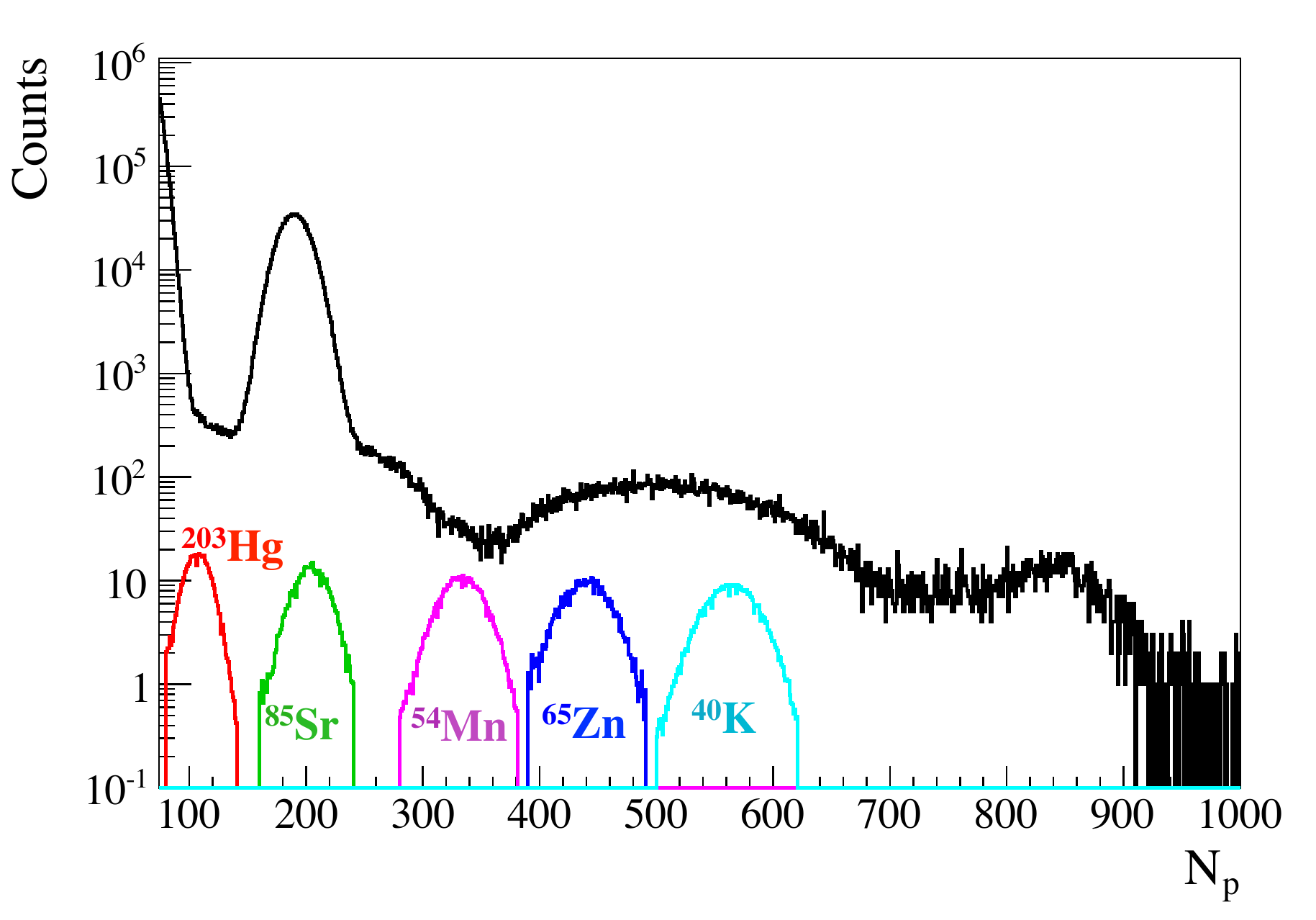}}
\caption{Final spectrum for all events passing all selection cuts is
shown in $N_p$ variable (black) with respect to the $\gamma$--sources
calibration spectra: $^{203}$Hg (red) , $^{85}$Sr (green), $^{54}$Mn
(magenta), $^{65}$Zn (blue), and $^{40}$K (cyan). The calibration
peaks are all normalized to an area of 500.}
\label{fig:SourceSpectra}
\end{center}
\end{figure}

\subsection{Cut efficiency}
\label{subsec:efficiency}

\begin{table}[t]
\begin{center}
\begin{tabular} {l c c c} \hline \hline
Isotope          & ($x$,$y$,$z$)      & Energy   &   Fraction of events \\ 
                      &  [m]                         & [keV]       &   removed \\ \hline
%$^{57}$Co        & (0,0,0)   &  122+14 (89$\%$)-136 (11$\%$)& \\
%$^{139}$Ce       & (0,0,0)   &  165    &         \\ 
$^{203}$Hg       & (0,0,0)  &  279     &  $(4.03 \pm 0.76)    \times 10^{-4}$   \\  
$^{203}$Hg       & (0,0,3)  &  279     &  $(5.84 \pm 1.76)    \times 10^{-4}$    \\ 
$^{203}$Hg       & (0,0,-3)  &  279    &  $(5.25 \pm 1.75)    \times 10^{-4}$    \\  \hline
%**************
$^{85}$Sr        & (0,0,0)  &  514     &  $(2.94 \pm 1.11)    \times 10^{-4}$     \\ 
$^{85}$Sr        & (0,0,3)  &  514     &  $(1.08 \pm 0.36)    \times 10^{-3}$    \\ 
$^{85}$Sr        & (0,0,-3)  &  514    &  $(1.12 \pm 0.37)    \times 10^{-3}$      \\  \hline
%**************
$^{54}$Mn        & (0,0,0)  &  834     &  $(1.65 \pm 0.40)    \times 10^{-4}$      \\ 
$^{54}$Mn        & (0,0,3)  &  834     &  $(1.01 \pm 0.71)    \times 10^{-3}$     \\ 
$^{54}$Mn        & (0,0,-3)  &  834    &  $(3.91 \pm 1.05)   \times 10^{-4}$      \\  \hline
%**************
$^{65}$Zn        & (0,0,0)   &  1115   &  $(3.79 \pm 1.69)    \times 10^{-4}$  \\  
$^{65}$Zn        & (0,0,3)   &  1115   &  $(2.11 \pm 2.11)    \times 10^{-4}$     \\
$^{65}$Zn        & (0,0,-3)   &  1115  &  $(1.87 \pm 0.62)    \times 10^{-3}$      \\ \hline
%**************
$^{40}$K         &  (0,0,0)   &  1460  &  $(1.89 \pm 0.50)    \times 10^{-4}$   \\ 
$^{40}$K         &  (0,0,3)   &  1460  &  $(3.82 \pm 1.27)    \times 10^{-4}$   \\ 
$^{40}$K         &  (0,0,-3)   &  1460 &  $(1.54 \pm 0.24)    \times 10^{-3}$  \\ 
 \hline \hline
\end{tabular}
\end{center}
\caption{ Fraction of $\gamma$ events from radioactive sources removed by the set of selection cuts described in Section~\ref{subsec:cuts}.
The typical number of source events used for this analysis is of the order of  $10^5$ events. The typical  count rate due to the source is of few Bq. }
\label{tab:efficiency}
\end{table}

The overall efficiency of the chain of cuts has been studied
both with Monte Carlo simulations and with the radioactive--source calibration data.

Figure~\ref{fig:SourceSpectra} compares the energy spectra of
$^{203}$Hg , $^{85}$Sr, $^{54}$Mn, $^{65}$Zn, and $^{40}$K $\gamma$
sources with respect to the energy spectrum of events passing all
selection cuts as described above. Table~\ref{tab:efficiency}
summarizes the fraction of events due to $\gamma$~rays from these
radioactive sources  which
are rejected by such selection cuts. We give are the results for source
positions in the center of the detector and outside the FV along the
vertical axis at $z$ = 3\,m and $z$ = $-3$\,m. In this test we
excluded from the selection cuts the FV cut (we tested all available
source positions), coincidence cut and multiple--cluster cut due to
increased source activity. The dominant rejection of events is due to
the IDF muon flag.

The spectra of the neutrino--induced events and events from radioactive background sources
have been simulated with the Monte Carlo following the procedure whose
details will be reported in Section~\ref{sec:MC}. The efficiency of
the cuts has been evaluated for each spectral component as a fraction of events surviving the event--selection cuts from all events reconstructed in the FV.
The muon cut in the
Monte Carlo data only includes the IDF flag. Removal of coincident
events and the abnormal delay of the cluster start time are not
simulated. Table~\ref{tab:MCefftable} reports the fraction of events
with energy higher than $N_h$ = 100 removed from each spectrum.
We conclude that the inefficiency of the cuts is negligible.

\begin{table}
\begin{center}
\begin{tabular}{l c } \hline \hline
Spectrum      & Fraction of removed events \\ \hline
                      
$^7$Be       &  $ (2.3 \pm 0.1) \times 10^{-4}$    \\ 
$^{85}$Kr    &  $ (2.7 \pm 0.2) \times 10^{-4}$    \\ 
$^{210}$Bi   &  $ (2.1 \pm 0.2) \times 10^{-4}$    \\ 
$pep$        &  $ (1.0 \pm 0.2) \times 10^{-4}$    \\
$^{210}$Po   &  $ (3.5 \pm 0.5) \times 10^{-4}$    \\ 
\hline \hline
\end{tabular}
\end{center}
\caption{The fraction of Monte Carlo events reconstructed in the
  $^7$Be--FV and thrown away by the set of selection cuts described in
  Section~\ref{subsec:cuts}.}
\label{tab:MCefftable}
\end{table}

\subsection{Event--by--event based $\alpha$ -- $\beta$ cut}
\label{subsec:softAB}

\begin{figure}[t]
\vspace{-3 mm}
\begin{center}
\centering{\includegraphics[width = 0.5\textwidth]{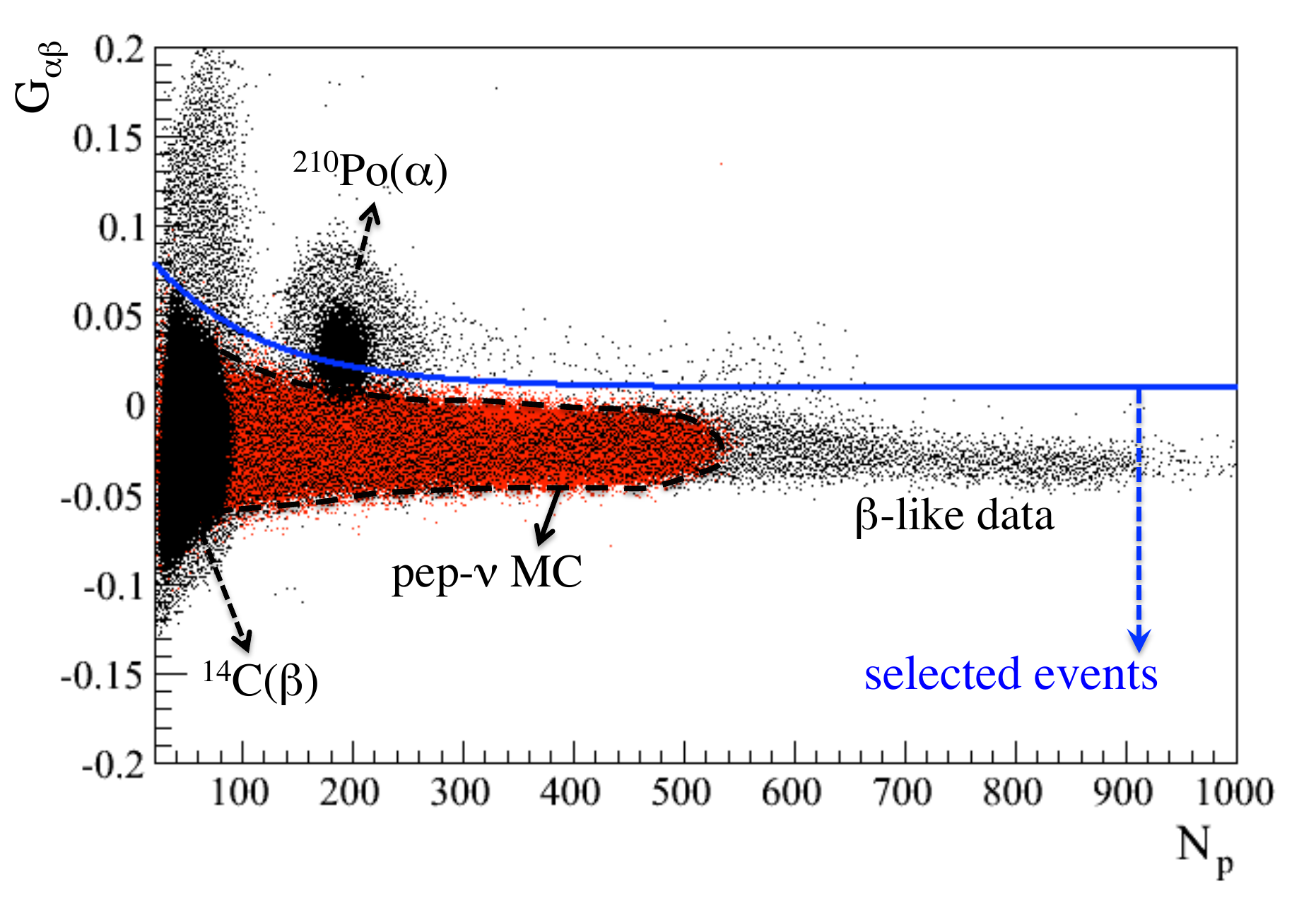}}
\caption{The distribution of the Gatti $G_{\alpha \beta}$ parameter as
  a function of energy ($N_p$ variable). 
  The continuous (blue) line shows the energy dependent cut; it removes all the events 
  with $G_{\alpha \beta}$ above the line. The (red) points enclosed within the dashed line
  show the $G_{\alpha \beta}$ distribution for MC-simulated $pep$--$\nu$'s; these events are an example of true $\beta$ events 
  with a spectrum extending over a sufficiently large energy range. The $\beta$ events are not affected by this cut, while, on the contrary, this cut removes
  a large fraction of the $\alpha$ events, as those from the $^{210}$Po decay. }
\label{fig:softAB}
\end{center}
\end{figure}

Figure~\ref{fig:softAB} shows the distribution of the $G_{\alpha \beta}$ parameter 
as a function of energy. It compares the data passing all selection
cuts as described in Section~\ref{subsec:cuts} and  the Monte
Carlo simulated $pep$ neutrinos.
These neutrinos were chosen instead of $^{7}$Be neutrinos since they
span up to higher energies.
The main structure is a band of $\beta$-- and $\gamma$--like events with
$G_{\alpha \beta}$ typically negative, while in the energy region
dominated by $^{14}$C ($N_p$ below 100) the $\alpha$ -- $\beta$
discrimination is not effective.
In the data, there are events with positive $G_{\alpha \beta}$ not
compatible with the Monte Carlo expectation for neutrino interactions,
dominated by the $\alpha$ events of $^{210}$Po at $N_p \sim$200.
However, events with such positive $G_{\alpha \beta}$ are present also
outside the energy range of the $^{210}$Po peak which can be only
partially explained by real $\alpha$'s.
Some events do not have the $G_{\alpha \beta}$ variable compatible
neither with $\beta$ or $\gamma$ nor with $\alpha$ and are explained
as remaining noise events.
Therefore, we have applied an additional energy--dependent cut based on
the $G_{\alpha \beta}$ variable (see the solid blue curve in Figure~\ref{fig:softAB}) which was
tuned both on the Monte Carlo and on the radioactive source
calibration data in order to minimize the fraction of $\beta$ events
thrown away (see Table~\ref{tab:softAB}).
The spectra obtained in this way are then fit as described below.

\begin{table}[t]
\begin{center}
\begin{tabular}{l c c c } \hline  \hline
Isotope          & ($x$,$y$,$z$)      & Energy   &   Probability \\ 
                      &  [m]                         & [keV]       & (limits at 90\% C.L.)   \\ \hline
%$^{57}$Co        & (0,0,0)   &  122+14 (89$\%$)-136 (11$\%$)& \\
%$^{139}$Ce       & (0,0,0)   &  165    &         \\ 
$^{203}$Hg       & (0,0,0)  &  279     &  $(1.45 \pm 0.15)    \times 10^{-3}$   \\  
$^{203}$Hg       & (0,0,3)  &  279     &  $(1.06 \pm 0.75)    \times 10^{-4}$    \\ 
$^{203}$Hg       & (0,0,-3)  &  279    &  $(4.67 \pm 1.65)    \times 10^{-4}$    \\  \hline
%**************
$^{85}$Sr        & (0,0,0)  &  514     &  $(1.56 \pm 0.26)    \times 10^{-3}$     \\ 
$^{85}$Sr        & (0,0,3)  &  514     &  $(3.59 \pm 2.07)    \times 10^{-4}$    \\ 
$^{85}$Sr        & (0,0,-3)  &  514    &  $(9.98 \pm 3.53)    \times 10^{-4}$      \\  \hline
%**************
$^{54}$Mn        & (0,0,0)  &  834     &  $( 3.00 \pm 0.54)    \times 10^{-4}$      \\ 
$^{54}$Mn        & (0,0,3)  &  834     &  $(2.43 \pm 0.86)    \times 10^{-4}$     \\ 
$^{54}$Mn        & (0,0,-3)  & 834    &  $( 2.80 \pm 2.80)    \times 10^{-5}$      \\  \hline
%**************
$^{65}$Zn        & (0,0,0)   &  1115   &  $(3.03 \pm 1.52)    \times 10^{-4}$  \\  
$^{65}$Zn        & (0,0,3)   &  1115   &   $< 4.9 \times 10^{-4}$ \\
%$^{65}$Zn        & (0,0,3)   &  1115   &   $<$XXX $ 10^{-}$   0 from 4735     \\
$^{65}$Zn        & (0,0,-3)   &  1115  &  $(2.08 \pm 2.08)    \times 10^{-4}$      \\ \hline
%**************
$^{40}$K         &  (0,0,0)   &  1460  &  $ 1.35 \pm 1.35 \times 10^{-5}$   \\ 
$^{40}$K         &  (0,0,3)   &  1460  &   $< 9.8 \times 10^{-5}$ \\
$^{40}$K         &  (0,0,-3)   &  1460 &   $< 8.9 \times 10^{-5}$ \\
%$^{40}$K         &  (0,0,3)   &  1460  &  $<$XXX $ 10^{-}$   0 from 23533 \\
%$^{40}$K         &  (0,0,-3)   &  1460 &  $<$XXX $ 10^{-}$   0 from 25880  \\ 
 \hline  \hline
\end{tabular}
\end{center}
\caption{The fraction of $\gamma$--source events passing the selection
  cuts described in Section~\ref{subsec:cuts} which are then thrown
  away by the energy dependent $G_{\alpha \beta}$ cut.}
\label{tab:softAB}
\end{table}

\section{$\alpha$ -- $\beta$ statistical subtraction}
\label{sec:ab}

After the application of the
cuts described in Subsection~\ref{subsec:cuts}, the $^{210}$Po
$\alpha$ peak, which falls entirely within the $^7$Be--$\nu$ energy window,
remains two to three orders of magnitude above the rest of the
spectrum at these energies, as Fig.~\ref{fig:StandardSpectra} shows.  The peak tails, if not correctly modeled
in the fit procedure, might influence the results about the $^7$Be
neutrino interaction rate.

As it can be seen in
Fig.~\ref{fig:softAB} and Fig.~\ref{fig:Gattiab}, the $G_{\alpha}$ variable of $^{210}$Po $\alpha$'s
extends to negative values and is not fully separated from the $G_{\beta}$
variable of $\beta$--like events. Therefore, an event--by--event cut
based on the $G_{\alpha \beta}$ value throwing away $\alpha$'s with
high efficiency while keeping all of the $\beta$'s is  not
possible, in particular when the number of $\alpha$ events largely exceeds that of the $\beta$.
We have then implemented a statistical separation of the
$\alpha$-- and $\beta$--induced signals. For each bin in the energy
spectrum, the $G_{\alpha \beta}$ distribution of the data is fitted to
two curves which represent the distribution of the $G_{\alpha}$ and $G_{\beta}$ variables.
The fit amplitudes are then the relative population of each species in
the energy bin. This procedure has been included in both the
analytical and Monte Carlo fit methods. 

\begin{figure}[t]
\begin{center}
\centering{\includegraphics[width = 0.5\textwidth]{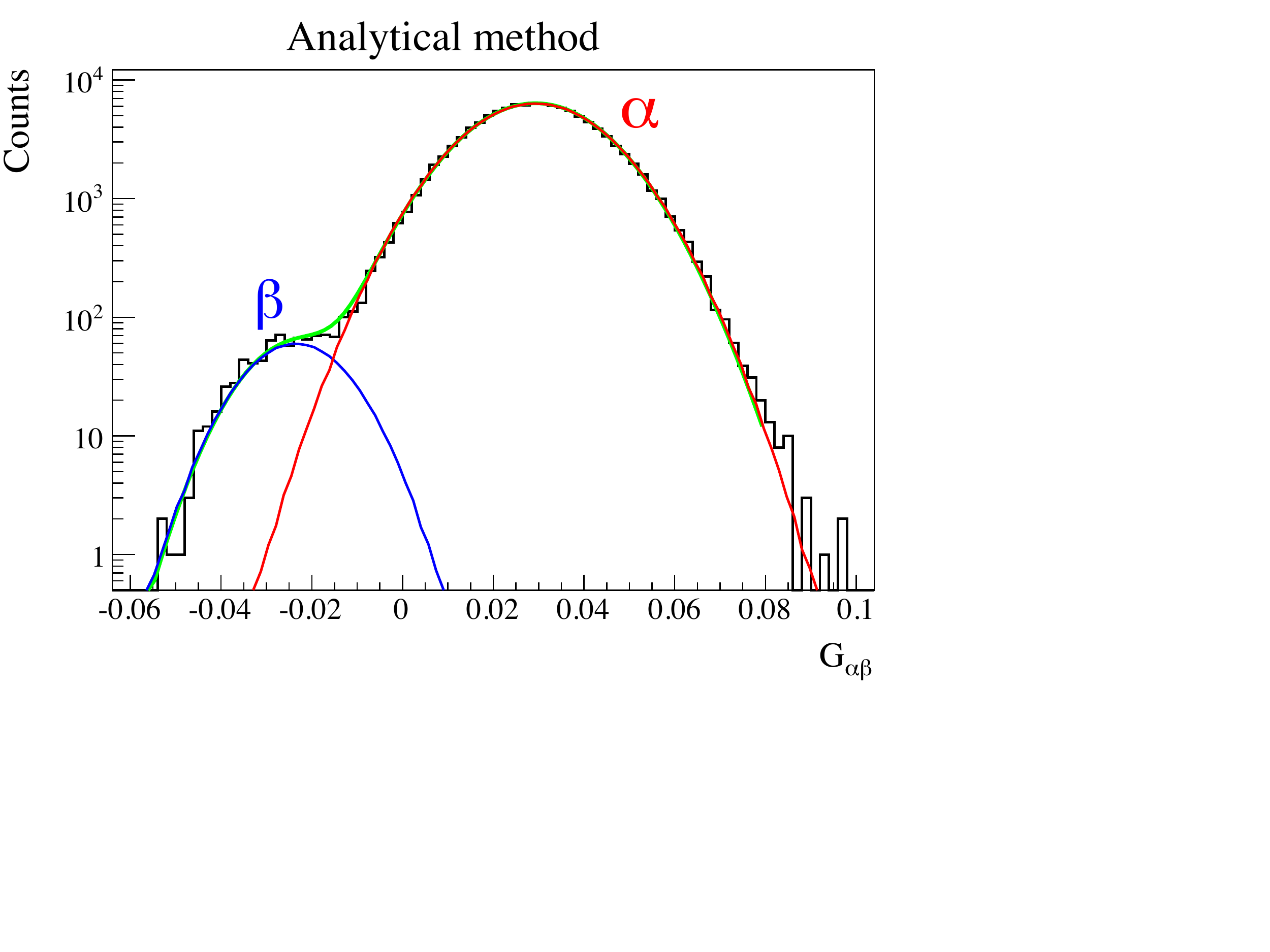}}
\caption{Example of $\alpha$ -- $\beta$ statistical subtraction with
  the analytical method for events in the energy range $200 < N^d_{pe}
  < 205$. The blue and red lines show the individual Gaussian fits to
  the Gatti parameter distributions for the $\beta$ and $\alpha$
  components, respectively, while the green line is the total fit.}
\label{fig:abAnalytical}
\end{center}
\end{figure}

In the analytical method the $G_{\alpha}$ and $G_{\beta}$ distributions
are assumed to be Gaussian, the fit is done iteratively and the
population estimates is continuously refined. In bins where one
species greatly outnumbers the other, for example in the energy region
of the $^{210}$Po peak, the means of the Gaussians are fixed to their
predicted values. Figure~\ref{fig:abAnalytical} shows an example of
the $G_{\alpha \beta}$ parameter of the data in the energy range 200 $<N_{pe}^{d}<$ 205 
 and its fit with the analytical
method. In order to estimate possible bias in the fit results, we simulated and fitted events with
known $G_\alpha$ and $G_\beta$ parameter in relative proportions as in the data. The fit results are then compared to the true
$\alpha$ and $\beta$ proportions used in the simulation.

\begin{figure}[t]
\begin{center}
\centering{\includegraphics[width =  0.5\textwidth]{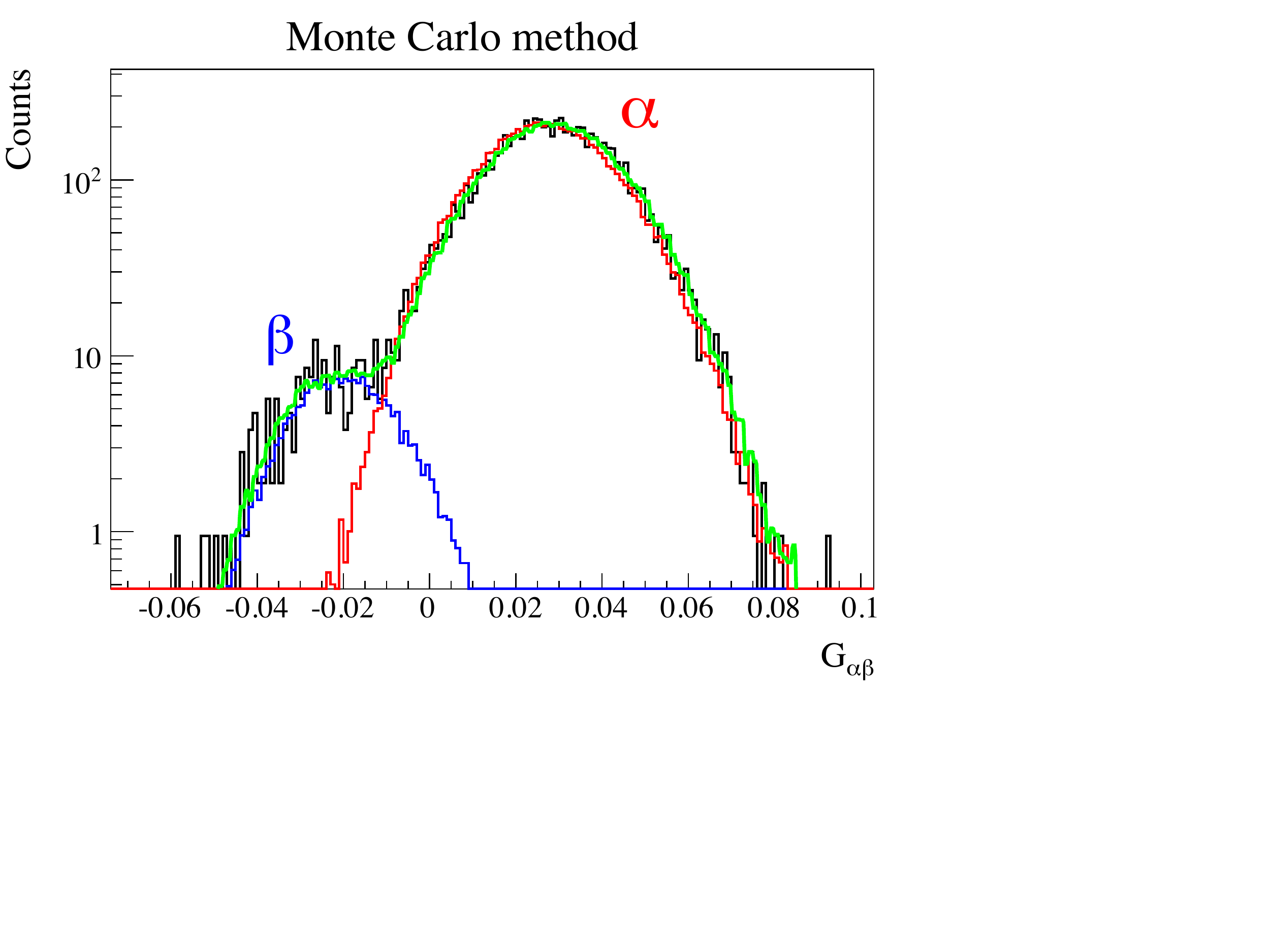}}
\caption{Example of $\alpha$ -- $\beta$ statistical subtraction with
  the Monte Carlo method for events in the energy range $168 < N_h <
  170$. The plot shows the Gatti parameter of the data (black) and the
  fit (green) aiming to separate the $\alpha$ (red) and $\beta$ (blue)
  contributions. In comparison to analytical method demonstrated in
  Fig.~\ref{fig:abAnalytical}, the blue $\beta$ and $\alpha$ shapes
  are Monte Carlo shapes which are not Gaussian. Details in text.}
\label{fig:abMC}
\end{center}
\end{figure}

In the Monte Carlo method the $G_{\alpha}$ and $G_{\beta}$ functions
are obtained simulating a large number of $\alpha$ and $\beta$ events
with the energy of interest, uniformly distributed in the IV and then
reconstructed within the FV. These curves are used as fit functions
and the free fit parameters are their amplitudes (that is the number
of $\alpha$ and $\beta$ events in the energy bin under examination)
and the shift resulting from the discussion in Section~\ref{sec:MC}.
Figure~\ref{fig:abMC} shows an example of the $G_{\alpha \beta}$
parameter for the data with $N_h$ from 168 to 170 and the fit with the
Monte Carlo procedure. 
A shift is observed in the MC distribution, which is very small:
its value is 0.002 at maximum (corresponding to two bins). The red and
blue lines in Fig.~\ref{fig:abMC} are the Monte Carlo $G_{\alpha}$ and
$G_{\beta}$ functions without the shift, while the best fit (green curve)
takes the small shift into account. The shapes of the $G_{\alpha}$ and
$G_{\beta}$ curves obtained with the Monte Carlo show some tails that
slightly deviate from a Gaussian curve. The effect is small and in fact
the results about the statistical subtraction obtained with the
analytical and Monte Carlo method are in a good agreement.

The statistical subtraction can be carried out over the entire energy
spectrum removing $\alpha$ decays of isotopes such as $^{210}$Po,
$^{222}$Rn, and $^{218}$Po or it can be applied in a restricted energy
region. The effect of the choice of the energy region where the statistical
$\alpha$ -- $\beta$ subtraction is applied on the resulting $^7$Be--$\nu$ interaction rate is
accounted in the systematic uncertainty as discussed in Section~\ref{sec:Be7results}.

\section{ $^{11}$C suppression}
\label{sec:c11}

 The $^{11}$C interaction rate in Borexino is determined via a fit of the energy spectrum  (see Section ~\ref{sec:Be7results})
 and is measured as $(28.5\pm0.2 {\rm (stat)} \pm0.7 {\rm (sys)})$\,cpd/100\,ton~\cite{be7-2011}. While $^{11}$C is not problematic for the determination of the $^7$Be--$\nu$ interaction rate, it is a relevant source of background for the measurement of the interaction rate of $pep$ and CNO neutrinos; in fact, its  rate is about ten times greater than the one from $pep$ neutrinos and the higher energy portion of of the signals induced by $pep$ and CNO neutrinos largely superimposes with its spectrum.
  Only the development of robust procedures able to subtract its contribution has allowed the $pep$ and CNO studies. Most of the events due to $^{11}$C decays has been rejected via a threefold coincidence (TFC) between the $^{11}$C positron decay, the parent muon, and the signal from capture of the free neutron (described in Subsection~\ref{subsec:tfc}) as  implemented in~\cite{pep-ctf}. The residual amount of $^{11}$C is determined using a novel pulse--shape discrimination technique (described in Subsections~\ref{subsec:psd})  applied in a Boosted Decision Tree (BDT) approach discussed in Subsection~\ref{subsec:BDT}.

\subsection{Three--fold coincidence veto}
\label{subsec:tfc}

As described in Section~\ref{subsec:CosmogenicBack}, $^{11}$C is mostly originated  by the interaction of muons in the scintillator and its production is accompanied by prompt neutrons and by the delayed 2230\,keV $\gamma$ ray resulting from the subsequent neutron capture in hydrogen. Neutron capture by $^{12}$C produces a $\gamma$ of higher energy (4950\,keV) but the probability is small compared to the hydrogen capture.
The reconstruction of the interaction positions of these $\gamma$ rays and of the tracks of parent muons are crucial for the success of the TFC technique. Muon tracking algorithms have been developed and are described in detail in~\cite{BxMuons}.

\begin{figure}[t]
\begin{center}
\centering{\includegraphics[width=0.5\textwidth]{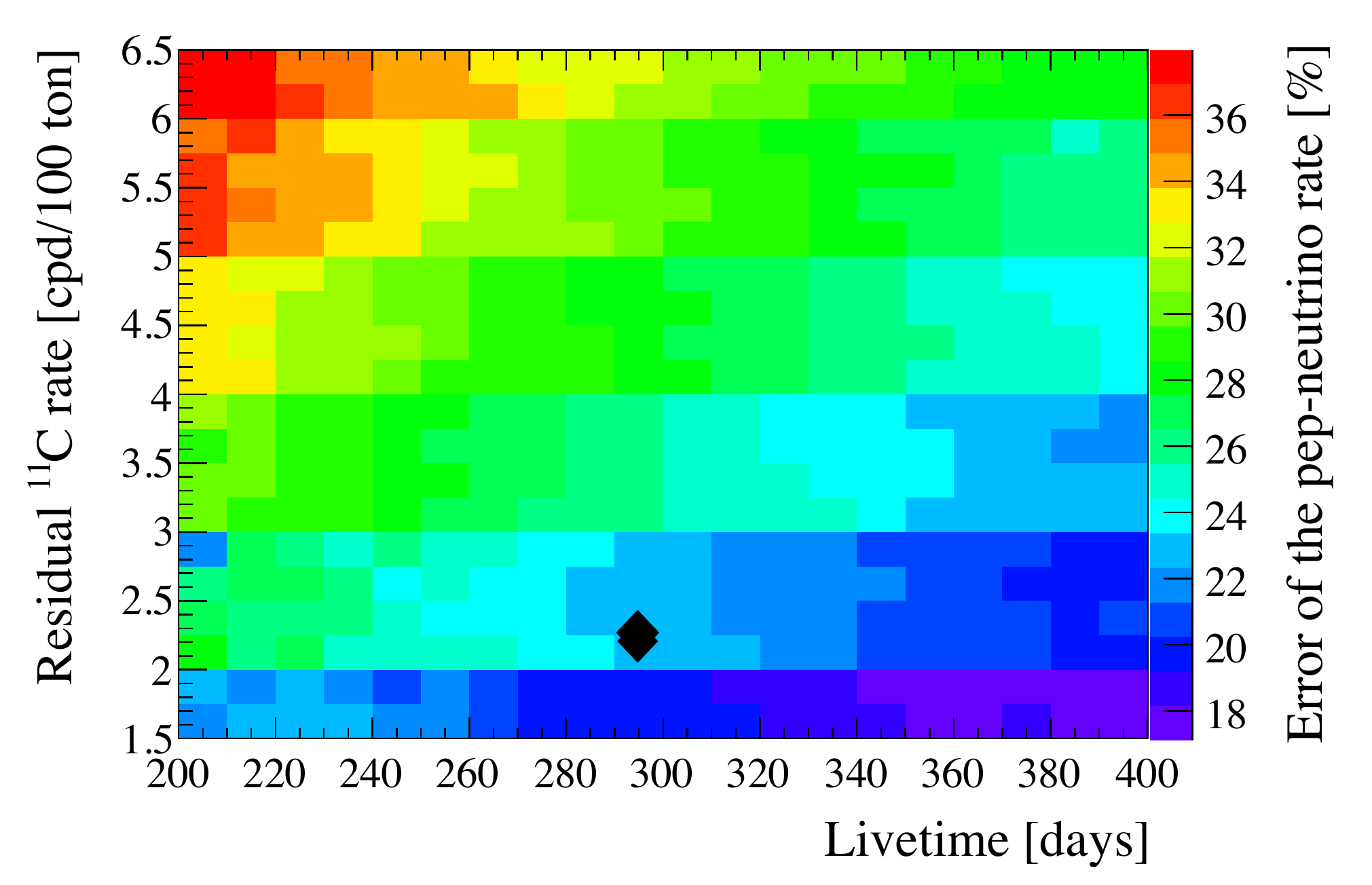}} 
\caption{Prediction of the sensitivity to the $pep$--$\nu$ interaction rate measurement ($z$-axis) as a function of the residual 
$^{11}$C rate ($y$-axis) and the effective live-time ($x$-axis) obtained by fitting the MC-simulated data. The color $z$--axis is expressed as the uncertainty (\%) on the $pep$--$\nu$ interaction rate returned by the fit. We recall that the $^{11}$C rate measured without applying the TFC subtraction is ($28.5\pm0.2 {\rm (stat)} \pm0.7 {\rm (sys)})$ cpd/100\,ton. The black diamond corresponds to the TFC-subtracted spectrum used in the analysis (see Fig.~\ref{fig:tfc_npe}). }
\label{fig:tfc_errors_coverage}
\end{center}
\end{figure}

The TFC algorithm vetoes space--time regions of the detector after muon plus neutron coincidences in order to exclude the subsequent $^{11}$C($\beta^{+}$) decays. The guiding principle for the determination of the most appropriate parameters is the search for the optimal compromise between $^{11}$C rejection and preservation of the residual exposure after the veto cuts. Figure~\ref{fig:tfc_errors_coverage} shows the Monte Carlo predictions on the sensitivity for the $pep$--$\nu$ interaction rate measurement as a function of the residual $^{11}$C rate and the effective residual exposure. This study shows that 
no significant bias in the fitted $pep$--$\nu$ interaction rate is expected from the loss of exposure.

The evaluation of the effective exposure after all veto cuts, many of which overlap in time and space, has been performed through the so-called counting method; firstly, through a simulation feeding uniformly distributed events to the veto cuts and by computing the fraction of the simulated events that survive these vetoes. The result has been compared with that obtained counting the number of $^{210}$Po events before and after the application of the TFC algorithm. The two methods agree to much better than 1$\%$.

\begin{figure}[h]
\begin{center}
\includegraphics[width=0.48\textwidth]{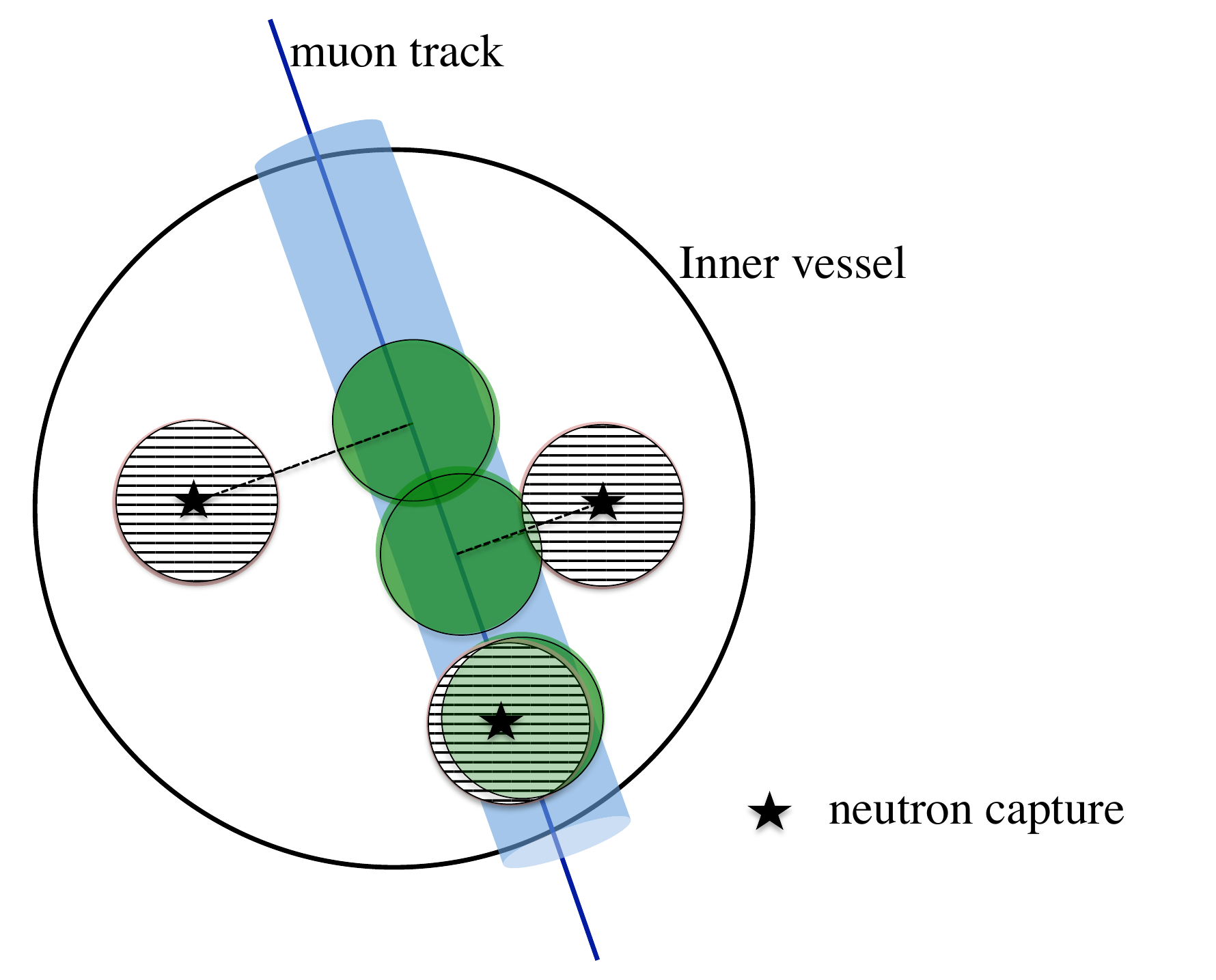}
\caption{The spatial regions vetoed in the TFC method: a cylinder around the muon track (blue) and some examples of spheres centered around the point where the $\gamma$ following the neutron capture is reconstructed (areas with horizontal lines around the stars) and their projections along the muon track (green areas).}
\label{fig:CylAndSpheres}
\end{center}
\end{figure}

\begin{figure}[!t]
\begin{center}
\centering{\includegraphics[width=0.5\textwidth]{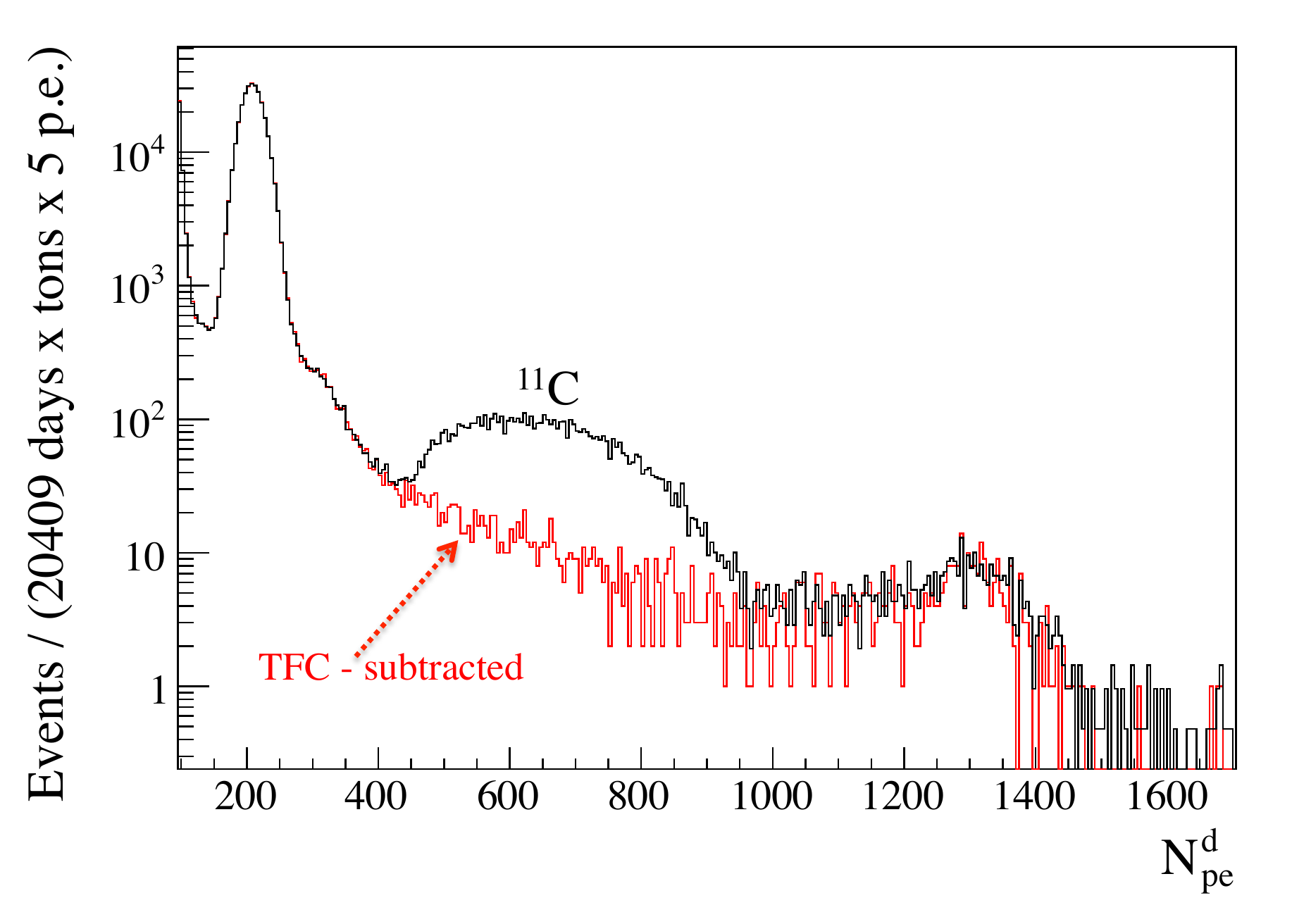}}
\vspace{-5mm}
\caption{Energy spectra ($N_{pe}^d$ energy estimator) before (black)  and after (red)  the application of the TFC technique for $^{11}$C  removal.  Both spectra are normalized to the same exposure.}
\label{fig:tfc_npe}
\end{center}
\end{figure}

The TFC procedure can be summarized as follows:
\begin{itemize} [leftmargin=*]
\item A suitable time veto has been applied at the beginning of each run, since muon plus neutron coincidences can be lost in the interval between runs. The veto time, in minutes, is obtained by $10 + 60\,\left(1 - \exp{\left(-3\Delta t/\tau\right)}\right)$, where $\Delta t$ is the dead--time interval between subsequent runs (in minutes) and $\tau$ is the neutron--capture time. $\Delta t$ is typically of the order of a minute but sometimes can reach a significant fraction of an hour.
\item A veto of 2 hours is applied after muons with high neutron multiplicity. 
\item When the reconstructed neutron position is not reliable, due to the electronics saturation effects and/or due to a large fraction of noise hits, a cylindrical veto along the parent muon track with a radius of 80\,cm for a time span of 2\,hours is applied.
\item If neutron clusters are found in a muon gate but more than 2\,$\mu$s after the muon, then the cylindrical veto along the muon track described above is applied.
\item A cylindrical veto is also applied around those OD--$\mu$'s (see Section~\ref{subsec:cuts}) tracks after which the Analogue Sum DAQ (Section~\ref{sec:electronics}) finds at least one neutron.
\item If a neutron is found and its position is considered reliable, we veto a sphere of 1\,m radius centered in this reconstructed position for 2\,hours. 
Moreover, another 1\,m spherical veto is applied around the point on the muon track that is closest from the neutron capture position.
\end{itemize}
Figure~\ref{fig:CylAndSpheres} schematically shows the vetoed regions. 
The application of this TFC algorithm results in $>$89.4\% $^{11}$C  rejection with a residual exposure of 48.5\%. Figure~\ref{fig:tfc_npe} shows the effect of the TFC veto and compares the spectra before  and after its application:  the $^{11}$C rate decreases from $\sim$28\,cpd/100\,ton to $\sim$2.5\,cpd/100\,ton with a 51.5\% loss of exposure. Only events passing the selection criteria described in Subsection~\ref{subsec:cuts} and the $pep$--FV cut described in Section~\ref{sec:FV} contribute in these spectra. The resulting exposure of the TFC--subtracted spectrum is 20409\,days $\times$\,ton, while for the spectrum of the TFC--tagged events it is 23522 days $\times$\,ton. 

\vspace{-2mm}

\vspace{-9mm}

\vspace {1 cm}

\subsection{$\beta^{+}$/$\beta^{-}$ pulse--shape discrimination}
\label{subsec:psd}

We have observed that the profile of the reconstructed emission times for scintillation photons produced by positron 
 is different than those from electrons. 
%After the positron deposits its kinetic energy, it annihilates with electrons in the scintillator emitting two back-to-back $\gamma$-rays. 
%The time lag in the physical process of positron annihilation, as well as the fact that the energy is deposited by multiple particles, leads to differences in the reconstructed emission time of the detected photons.
Prior to annihilation in two back--to--back $\gamma$ rays, the positron emitted in $^{11}$C decays may form a bound state with an electron in the scintillator, the positronium. The ground state of positronium has two possible configurations depending on the relative orientation of the spins of the electron and the positron: the spin singlet state (para--positronium), with a very short mean life--time of 125\,ps in vacuum, and the spin triplet state, called ortho--positronium, with a mean life--time in vacuum equal to 140\,ns. In liquid scintillator, however, the life--time of ortho--positronium is reduced because of interactions with the surrounding medium: processes like spin--flip, or pick--off annihilation on collision with an anti--parallel spin bulk electron, lead to the two--body decay within few ns. Laboratory measurements lead to $\sim$3\,ns mean--life and $\sim$50\% ortho--positronium formation probability~\cite{PsScint} in scintillators. This delay of the annihilation introduced by the ortho--positronium formation is comparable in size to the fast scintillation time constant $\tau_1$ (see Table~\ref{table:prop}), and therefore is expected to introduce a measurable distortion in the time distribution of hit PMTs with respect to a pure $\beta^-$ event of the same reconstructed energy. Additional distortions are expected from the diffuse geometry of events resulting from the positronium decay, due to the non--null mean free path of the ensuing $\gamma$ rays.
The direct annihilation of the positron in flight is expected to occur $<$5\% of the time following $^{11}$C decay \cite{bib:Heitler}. Considering the time-resolution of the scintillator, this process is indistinguishable from annihilation following para--positronium formation, and only contributes to a small fraction of the events assigned to that population.

\begin{figure}[b]
\centering{\includegraphics[width=0.49 \textwidth]{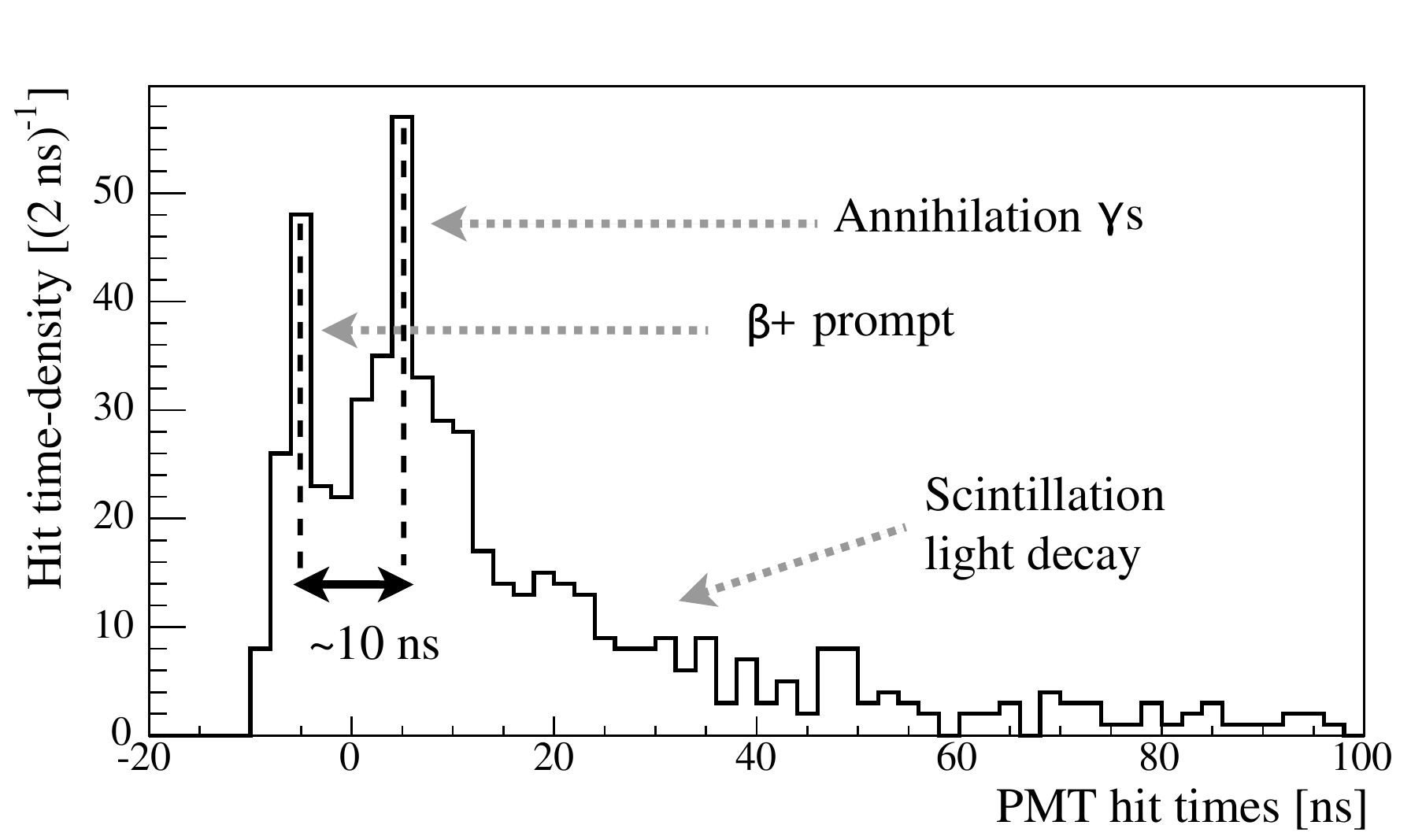}}
\caption{Hit-emission time profile of a single event due to $\beta^+$ decay, where the positron deposits its kinetic energy (first peak) and then forms ortho-positronium. The ortho-positronium exists for $\sim$10\,ns before the positron annihilates with a bulk electron to produce $\gamma$-rays (second peak). The PS--BDT value (Fig~\ref{fig:sb_bdt}) of this cluster is -0.44.} 
\label{fig:ops-example}
\end{figure}

\begin{figure} 
\centering{\includegraphics[width=0.5\textwidth]{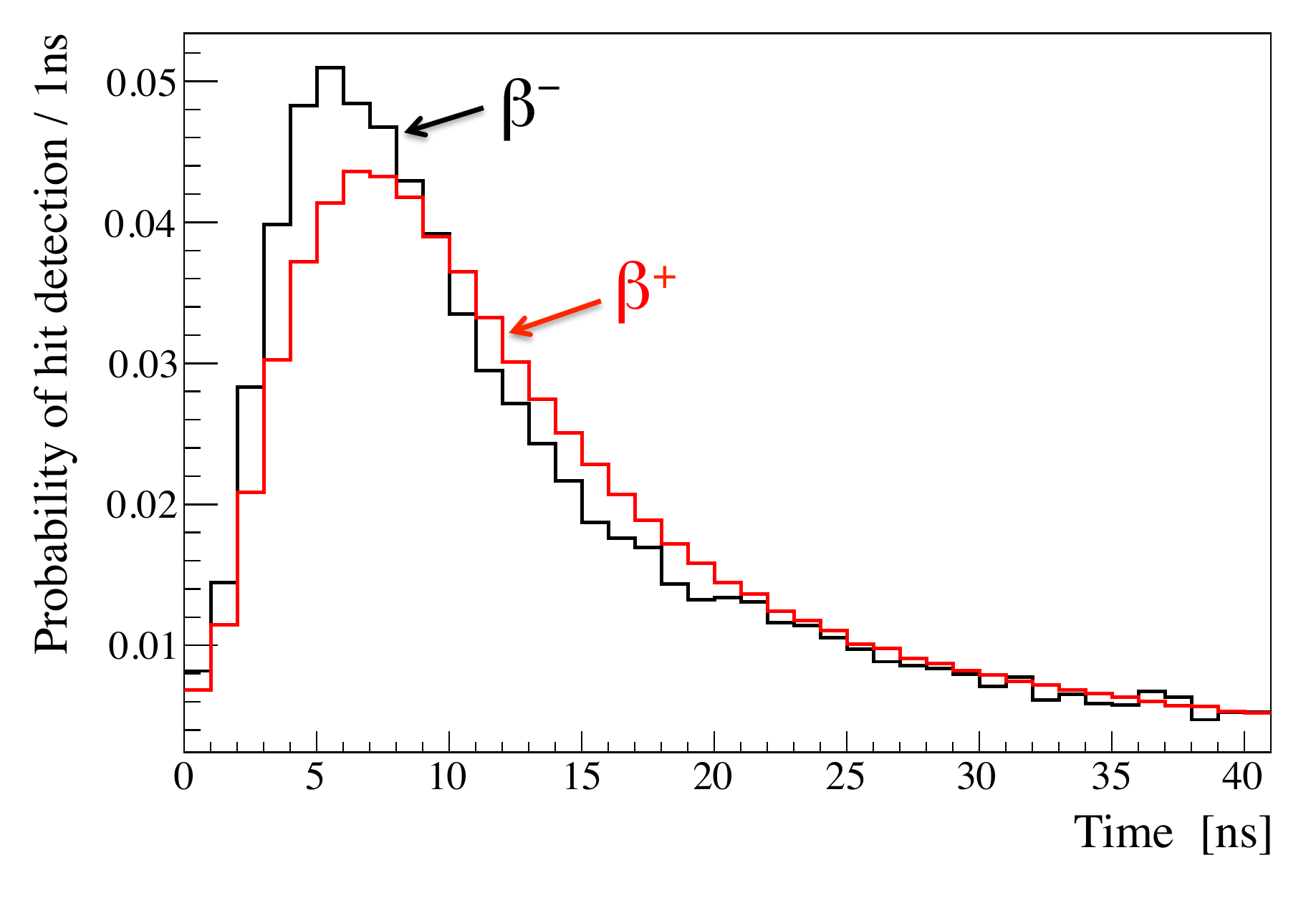}}
\caption{Reconstructed photon emission times relative to the start time of the cluster: $^{214}$Bi~($\beta^-$) events with 425~$<$~$N_h$~$<$~475 identified by a $^{214}$Bi -- $^{214}$Po fast coincidence tag (black) and $^{11}$C~($\beta^+$) events tagged by the TFC (red).}
\label{fig:bi214-c11_start}
\end{figure}

Figure~\ref{fig:ops-example} shows an event where there is a clear time-separation between the energy deposit by the positron and the sub-sequent energy deposition from the annihilation $\gamma$-rays, after the formation of ortho--positronium. Given its half--life, only $\sim$1\% of events that form ortho-positronium have a time separation that is at least this long. Generally, the separation is small enough that the two peaks are indistinguishable and only a broadening of the time distribution is observed.  

Figure~\ref{fig:bi214-c11_start} shows the distribution, averaged over many events, of the photon--emission times (hit times, once subtracted the time--of--flight from the reconstructed position) for $\beta^-$ events ($^{214}$Bi from the $^{214}$Bi -- $^{214}$Po coincidence tag) and for $^{11}$C~($\beta^+$) TFC tagged events. The delay and broadening of the peak in the average time distribution due to ortho--positronium formation is evident. %Differences may be observed in both the peak time (see Fig.~\ref{fig:bi214-c11_start} where the distributions are shown relative to the reconstructed time of the event) and in the rise and decay times of the distributions (see Fig.~\ref{fig:bi214-c11_peak} in which the two peaks are aligned on the time axis). These differences are mostly due to the delay induced by the ortho--positronium formation.

%Fig.~\ref{fig:bi214-c11_start} and~\ref{fig:bi214-c11_peak} show the distributions of the photon emission times (hit times, once subtracted the time--of--flight from the reconstructed position) for $\beta^-$ events ($^{214}$Bi from the $^{214}$Bi -- $^{214}$Po coincidence tag) and for $^{11}$C~($\beta^+$) TFC tagged events. Differences may be observed in both the peak time (see Fig.~\ref{fig:bi214-c11_start} where the distributions are shown relative to the reconstructed time of the event) and in the rise and decay times of the distributions (see Fig.~\ref{fig:bi214-c11_peak} in which the two peaks are aligned on the time axis). These differences are mostly due to the delay induced by the ortho--positronium formation.

The relative weight of the delayed annihilation energy $2 m_e c^2$ ($m_e c^2$ is the electron plus positron rest energy) with respect to the total energy deposited by the $\beta^+$ (that is $2 m_e c^2$ plus the initial $\beta^+$ kinetic energy $T$) decreases with $\beta^+$ energy increasing.  Therefore, the difference between $\beta^+$ and $\beta^-$ reconstructed emission times is energy dependent and the discrimination power of any pulse--shape based method decreases as the energy of the $\beta^+$ event increases.

\begin{figure} [t]
\centering{\includegraphics[width= 0.5 \textwidth]{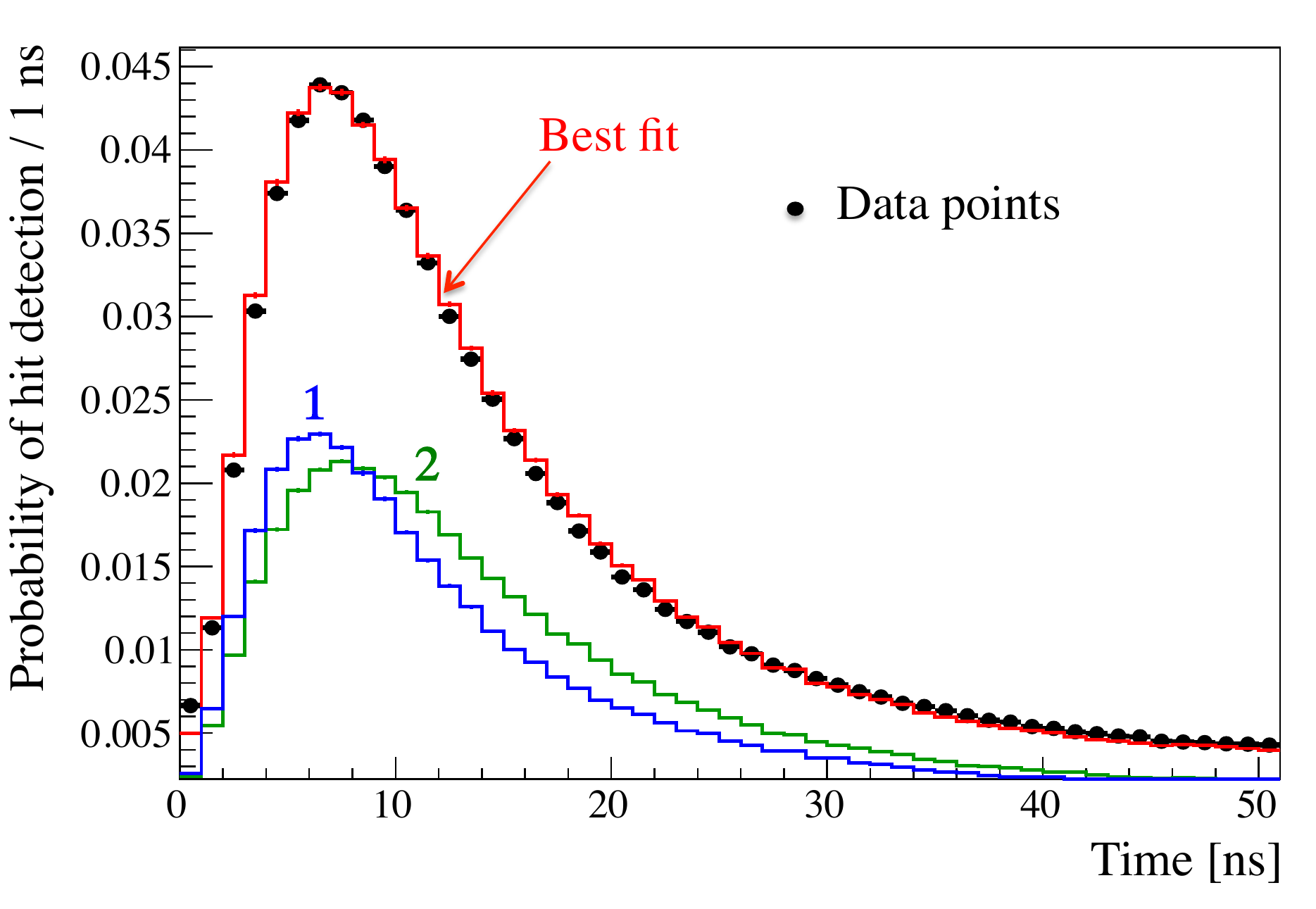}}
\vspace{-6 mm}
\caption{Reconstructed photon emission times for $^{11}$C events. The data (black points) has been fitted to the $^{11}$C Monte Carlo shapes without (1-blue) and with (2-green) ortho--positronium formation (mean--life of 3.1\,ns); the relative weights of these two shapes were left free in the fit. The best fit, the sum of the two MC-shapes, is shown red. } 
\label{fig:c11_rec_times}
\end{figure}

In order to detect and quantify this effect, as well as to develop pulse--shape variables to discriminate $\beta +$ and $\beta^-$ events, we have developed a special Monte Carlo event generator to simulate ortho--positronium formation and yield the corrected pulse shape. According to the input formation probability and life--time, the code generates  positronium decays and positron annihilations.
This process is simulated as a three--body vertex, composed by an electron, and two delayed annihilation gammas. The use of electrons instead of positrons is an approximation aimed to simplify the simulation, and motivated by the almost identical energy losses, with the exception of the annihilation process. The delay of the 511\,keV $\gamma$ rays follows an exponential law with $\tau$ set to that of the ortho--positronium mean--life. The comparison between the reconstructed emission times for simulated and measured $^{11}$C is shown in Fig.~\ref{fig:c11_rec_times}. 
The fitted ortho--positronium formation probability of 53\% is compatible with other laboratory measurements~\cite{PsScint}.

\subsection{Boosted Decision Tree}
\label{subsec:BDT}

\begin{figure}[t]
\centering{\includegraphics[width=0.5\textwidth]{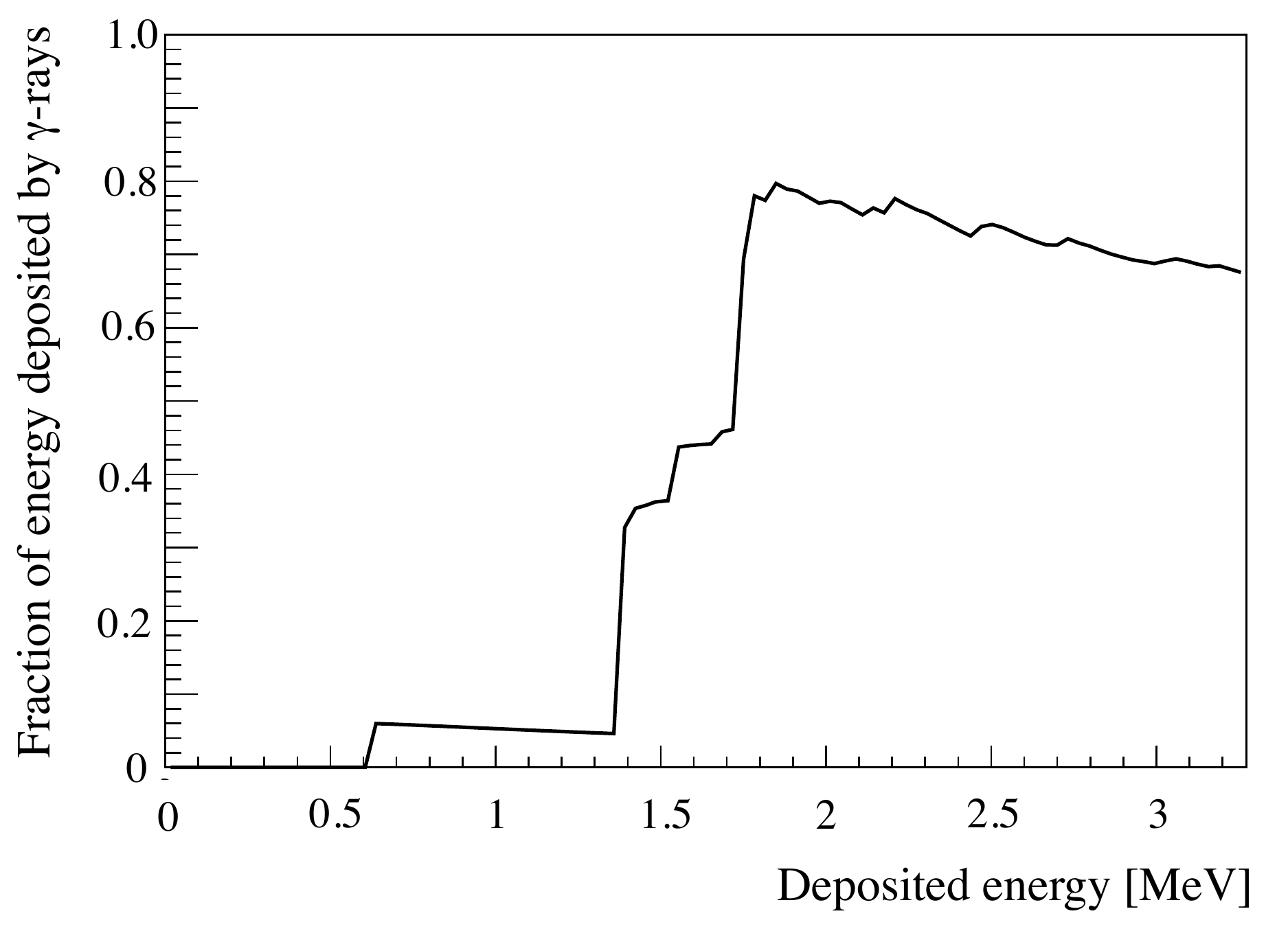}}
\caption{Average fraction of the energy deposited by $^{214}$Bi decays in the form of $\gamma$-rays. For deposited energy below 1400\,keV, only less than 5$\%$  of the energy is due to $\gamma$-rays. Therefore, low-energy $^{214}$Bi decays are a sample of mostly pure $\beta^-$ decays.}
\label{fig:214BiGammaEnergy}
\end{figure}

\begin{figure}[t]
\centering{\includegraphics[width=0.5\textwidth]{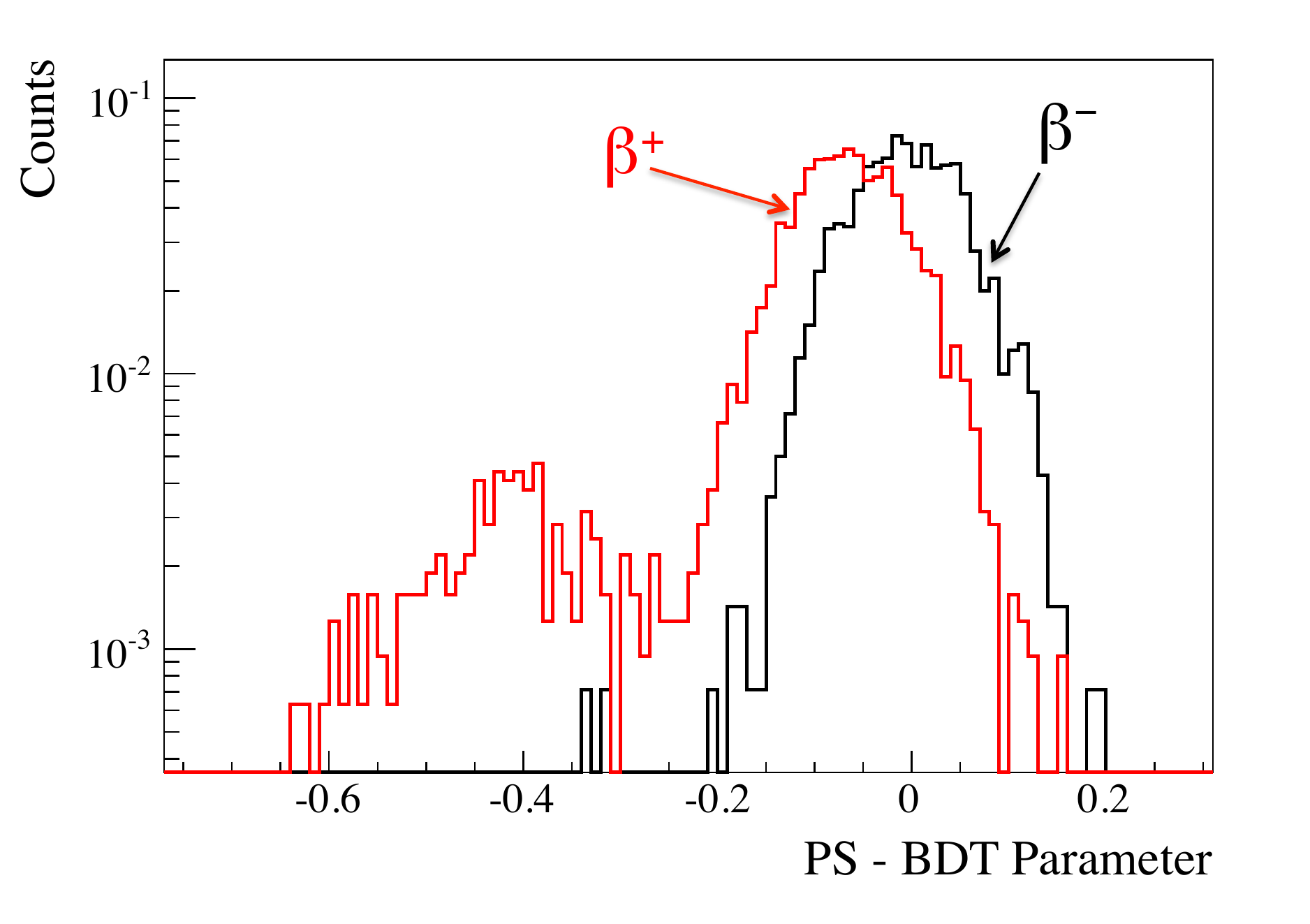}}
\caption{Distributions of the PS--BDT parameter for the test samples of $\beta^-$ (black) and $\beta^+$ (red) events as described in the text.}
\label{fig:sb_bdt}
\end{figure}

Several variables having some discrimination power between $\beta^-$ and $\beta^+$ have been used in a boosted--decision--tree algorithm
(BDT). This procedure is a powerful method to classify events and, after its training with a sample of $\beta^-$  and with another sample of $\beta^+$ events, it allows to assign a parameter PS--BDT to each event. The train samples are used to define the probability distribution function of this parameter. We selected as $\beta^-$ sample the low-energy $^{214}$Bi events (450 $<N_{pe}^d<$ 900) tagged by the $^{214}$Bi -- $^{214}$Po coincidence tag, 
where the fraction of the energy deposited by gamma rays is only $\sim$5\% as 
Fig.~\ref{fig:214BiGammaEnergy} shows.
The $\beta^+$ sample are events tagged with the TFC and with 450 $<N_{pe}^d<$ 900 and it is an almost pure ($>$98\%) $^{11}$C sample. Only events reconstructed within the FV used in the $pep$ analysis have been considered. The variables used in the BDT algorithm are:
\begin{itemize}[leftmargin = *]
\item The Gatti parameter (Section~\ref{sec:shape}) computed using as reference the $^{214}$Bi and $^{11}$C time profiles from real data, with reconstructed emission times relative to the peak.
\item The Gatti parameter computed using as reference the $^{214}$Bi and $^{11}$C time profiles from real data, with reconstructed emission time relative to the cluster start time.
\item The Gatti parameter computed using as reference the $^{214}$Bi data and ortho--positronium (Monte Carlo generated) time profiles, with reconstructed emission time relative to the cluster start time.
\item The Gatti parameter computed using as reference the Monte Carlo generated $^{11}$C time profiles with and without ortho--positronium formation, with reconstructed emission time relative to the cluster start time.
\item The Gatti parameter $G_{\alpha \beta}$ computed using as reference the $^{214}$Bi and $^{214}$Po time profiles from data.
\item The Kolmogorov -- Smirnov probabilities between the light--emission--time distribution of the event and the $^{214}$Bi and $^{214}$Po reference time profiles.
\item The reconstructed emission time, relative to the peak of the time distribution, of the earliest hit in the cluster.
\item The peak of the emission--time distribution relative to the reconstructed time of the event.
\item The first four moments of the emission--time distribution (i.e. mean, $rms$, skewness, and kurtosis) for hits up to 1.1\,$\mu$s after the cluster start.
\item Ten variables that are the fraction of the hits in the cluster after particular times (35, 70, 105, 140, 175, 210, 245, 280, 315, and 350\,ns) relative to the peak of the distribution.
\item The first four Legendre polynomials, averaged over all combinations, of the angle between any two hit PMTs relative to the reconstructed position of the event.
\item The uncertainties in the reconstructed position along an axis ($x$, $y$, and $z$, as returned by the fitter) divided by the mean of the other two uncertainties.
\item The ratio (for all axes) of the reconstructed position of the event obtained from the time--of--flight subtraction algorithm to the charge--weighted average of the hit PMT positions in the event.
\end{itemize}
The final output variable of the BDT algorithm, the PS--BDT parameter and the corresponding distributions for the test samples are shown in Fig.~\ref{fig:sb_bdt}.

\section{The energy response function}
\label{sec:resp}

The energy response function $P_{N_p}$ ($P_{N_h}$, $P_{N_{pe}}$,
$P_{N_{pe}^d}$) is the probability distribution function for the
measured energy estimator of an event when the energy $E$ is released
in a given position inside the detector. Each energy estimator defined
in Section~\ref{sec:estimators} has its response function. Besides the
energy $E$ this function depends in principle on many other
quantities:
\begin{itemize}[leftmargin = *]
\item the position $\vec r$ inside the IV where the interaction
generating the energy deposit takes place. Light absorption, optical
effects related to the light concentrators mounted around the PMTs and
the inhomogeneous distribution of dead PMTs 
make the number of detected photoelectrons position--dependent;
\item the particle type $p$ where $p=\alpha, \beta, \gamma$. The
scintillation mechanism is such that $\alpha,\beta,\gamma$ particles
depositing the same energy in the scintillator produce a different
amount of light and thus of hit PMTs and photoelectrons, as discussed in
Section~\ref{sec:scintillator};
\item parameters related to the scintillator: examples are the light
yield, the emission spectrum, the absorption and scattering length as
a function of the wavelength, the re--emission probability and so
on. We indicate the list of these parameters with the vector $\vec s$;
\item parameters describing the detector geometry and the properties
of the materials relevant for the light propagation. We generically
indicate the list of these parameters with the vector $\vec d$;
\item parameters (here indicated with $\vec e$) describing the
electronics response (dead time, gate length, multiple--hits handling),
the number and the characteristics of the active PMTs
(thresholds, gain, single--photon peak position and $rms$, dark noise,
after--pulse probability)
\item the absolute time $t$ as the properties of the detector may change in time.
\end{itemize}

Thus, in general we have $P_{N_p}(E,\vec r, p, \vec s, \vec d, \vec
e,t)$ (and similarly for $P_{N_h}$, $P_{N_{phe}}$,
$P_{N_{phe}}^d$). We will often write $P_x(E,\vec l)$ where $x$ is one
of the energy estimators and $\vec l$ stands for list of all
other variables $(\vec r, p, \vec s, \vec d, \vec e,t)$. Note
that the explicit analytical dependence on all these parameters may be
in general impossible to obtain and the models that we are going to
discuss often make use of a response function integrated over several
of the listed parameters.

We are adopting two complementary approaches to determine $P_x(E,\vec l)$:
 the first is based
on the use of analytical models and is described in the Section~\ref{sec:analytical} 
 while the second
uses a Monte Carlo method and is described in Section~\ref{sec:MC}. Both methods are
validated using the radioactive--source calibration data.

\section{The analytical procedure}
\label{sec:analytical}

A response function has to perform the transformation of the spectra from the original energy scale to the scale of the desired estimator, including the appropriate resolution effects.
This transformation is quantitatively defined by the $P_x(E; \vec l)$ relation introduced above. 

The shape of the response function is generally characterized by its central moments: the mean,
variance, and, in some cases, the skewness; mathematically it is modeled by an analytical function, whose central moments are
chosen to match those of the corresponding energy estimator. For signal and background
spectra that are not mono--energetic, the transformation can be easily generalized to obtain
the final spectrum in the domain of the energy estimator. 

This procedure has been fully developed for the three energy estimators $N_{pe}$, $N_p$, and $N_{pe}^d$, but not for $N_h$, since in this case the effect of multiple hits on a single PMT is analytically intractable. Among the others, the two $\beta^+$ decaying species $^{10}$C and $^{11}$C, cosmogenically produced in
the scintillator, require special treatment of the two associated 511\,keV annihilation gammas, especially for what concern their effective
quenching. Without entering into too many details, we can say that the $\gamma$--quenching in this
occurrence is either pre--determined through a simplified ad--hoc Monte Carlo, or added as
free parameter in the overall final fit procedure.

Needless to say, in the analytical approach several model simplifications are necessary
and the dependence upon the whole list $\vec l$ of parameters cannot be explicitly resolved.
However, the analytical procedure allows the values of some of the input parameters to be directly optimized during the fit to the data and to provide clearrelations linking
the energy to the measured quantities.

\subsection{The quenching factor and $kB$}
\label{sec:analkB}

Pre--requisite for any analytical modeling is the adoption of a practical expression for the
quenching factor $Q_p(kB; E)$ defined in Eq.~\ref{Qdef}, as well as the determination of the proper
value of the $kB$ parameter characterizing our scintillator, since the intrinsic link between the
initial energy deposit and the mean amount of produced photoelectrons is a key ingredient
in any analytical approach.

Specifically, the determination of $kB$ is performed through the 
 exploitation of the calibration data obtained with the $\gamma$ sources deployed at the center of the detector 
 with energy ranging from 250\,keV to 2230\,keV. 
The number of photons $N_{\rm ph}$ emitted in each event of energy $E$ can be expressed as: 
\begin{equation}
N_{\rm ph} (E) =  Y_0^{\rm ph} \cdot Q_p(kB; E) \cdot E +  N_{\rm Ch }^{\rm ph},
\label{Eq:1}
\end{equation}
in which the first term describes the contribution from the scintillation light and has the form of Eq.~\ref{Eq:Y_p_ph} (and in the particular case of $\gamma$--calibration sources can be replaced by the Eq.~\ref{eqn:Qgamma}), while the second term $N_{\rm Ch }^{\rm ph}$ describes the Cherenkov light contribution which, in principle, can be obtained by the integration of Eq.~\ref{Eq:Cherenkov}.

The number of ideally measured photoelectrons $N^{\rm ideal}_{pe} $ can be similarly expressed as:
\begin{equation}
N^{\rm ideal}_{pe}  =  Y_0^{pe}  \cdot Q_p(kB; E) \cdot E +  N_{\rm Ch }^{\rm pe},
\label{Eq:NpeIdeal}
\end{equation}
in which $Y_0^{pe}$ is the scintillation photoelectron yield expressed in p.e./MeV.
Again, in the specific case of the $\gamma$ rays, this relation becomes:
\begin{equation}
N^{\rm ideal}_{pe, \gamma}  =  Y_0^{pe}  \sum_i Q_\beta(kB; E_i) \cdot E_i  + \sum_i N_{\rm Ch,i }^{\rm pe},
\label{Eq:NpeIdealGamma}
\end{equation}
in which the sum $i$ goes over all electrons and positrons produced in the $\gamma$--ray interactions.

The value of $kB$ can be obtained from the $\gamma$--source calibration data using the Eq~\ref{Eq:NpeIdealGamma}. 
In order to find the best approximation of the $N_{pe}^{\rm ideal}$, we express it as follows: 
\begin{equation}
N^{\rm ideal}_{pe}  =   \left <\mu \right > N_{\rm tot},
\label{Eq:NpeIdealData}
\end{equation}
where $\left <\mu\right > $ is the average number of photoelectrons measured by a single channel and $N_{tot}$ = 2000 is the number of channels to which we normalize our energy estimators as was shown in Section~\ref{sec:estimators}.

Assuming a Poisson distribution of photoelectrons on each PMT, the average number of photoelectrons $\mu_i$ measured by a single channel $i$ can be expressed through a measurable probability $h_i$ that the channel $i$ detects at least 1 hit:
\begin{equation}
h_i =  1 - \exp (-\mu_i),
\label{Eq:himui}
\end{equation}
in which $h_i$ can be estimated by computing in what fraction of the clusters the channel $i$ registers at least one hit.
Then, for a single channel $i$ we obtain the value of $\mu_i$:
\begin{equation}
\mu_i =  - \ln \left [1 - h_i \right ]
\end{equation}
and by averaging over all channels we obtain $\left <\mu\right > $, and thus through Eq.~\ref{Eq:NpeIdealData} also $N_{pe}^{\rm ideal}$ for each $\gamma$--calibration source measurement.

\begin{figure}
\centerline{\includegraphics[width = 0.5\textwidth]{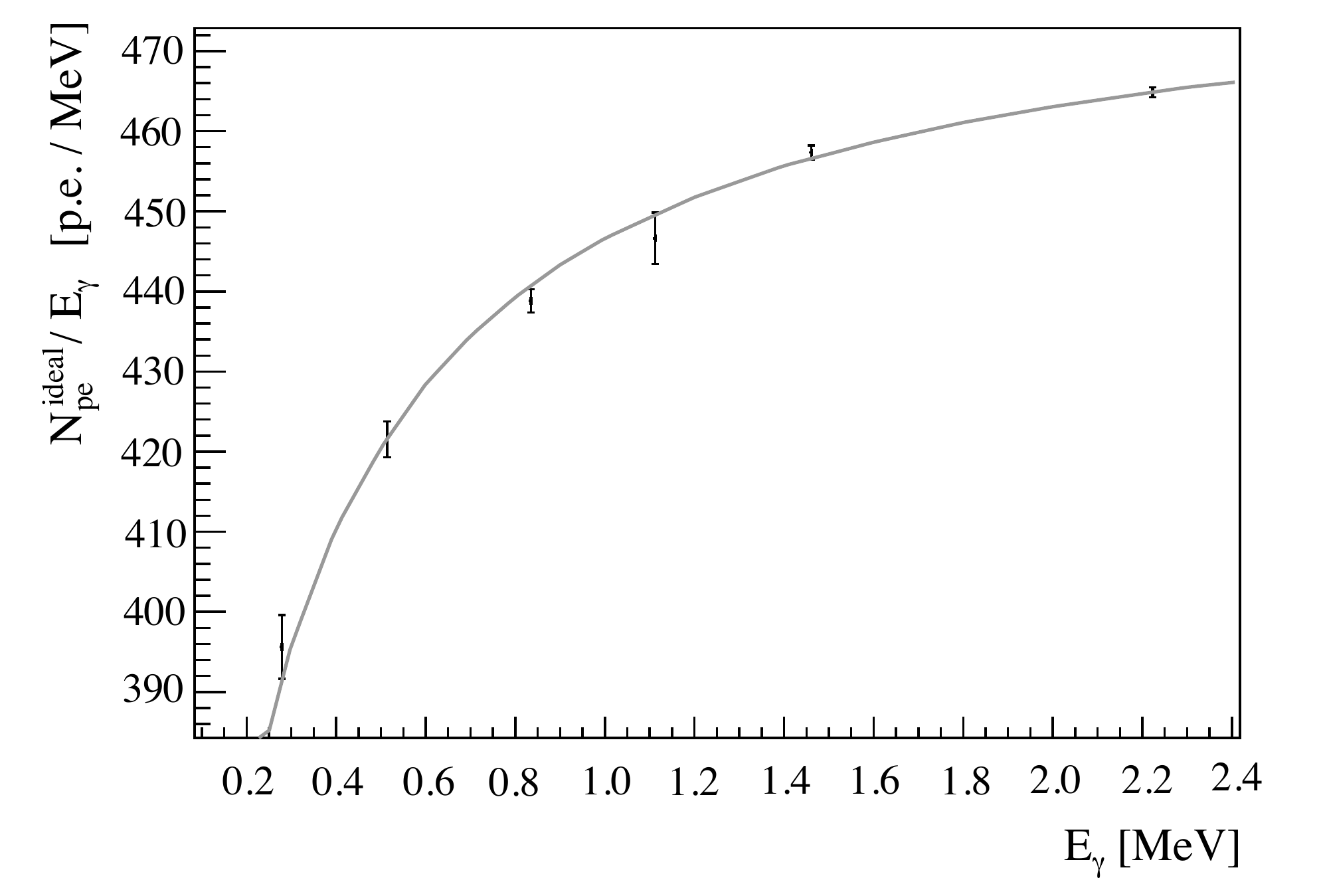}}
\caption{Data points for six different $\gamma$-ray lines: $^{203}$Hg,  $^{85}$Sr, $^{54}$Mn, $^{65}$Zn, $^{40}$K, and the 2230\,keV $\gamma$ from the neutron capture: $N_{pe}^{\rm ideal}$/MeV expressed as a function of $E_{\gamma}$. The data is fit with the function corresponding to Eq.~\ref {Eq:NpeIdealGamma} which is obtained by a dedicated ``ad hoc'' Monte Carlo using the Birk's quenching model (details in text).  The best fit values are $kB = 0.00115 \pm 0.0007$ and $ Y_0^{pe} = 489 \pm 2$ p.e./MeV. }
\label{fig:kBAnalytical}
\end{figure}

Fig.~\ref{fig:kBAnalytical} shows the data points of $N_{pe}^{\rm ideal}$ obtained as described above and shown as the function of energy of the $\gamma$ source $E_{\gamma}$.
The fit function corresponds to Eq.~\ref{Eq:NpeIdealGamma} and was obtained by a dedicated MC in the following way: 
for each of the $\gamma$--ray source energies, an event was simulated and the energy of each of the electron recoils was stored. At this stage, the quenching to each energy 
 deposit was applied ``ad hoc'' using the Birk's quenching formula of Eq.~\ref{eq:Birks}, rather than simulating fully the physical process as done in the context of 
 the Monte Carlo  evaluation described in the next Section~\ref{sec:MC}. 
This MC was then done for thousands of $\gamma$ rays and for a wide range of $kB$ values.
The Cherenkov contribution of Eq.~\ref{Eq:NpeIdealGamma} was fixed according to the full GEANT4 simulation decsribed in the next Section~\ref{sec:MC}. 
Finally, $kB$ and $Y_0^{pe}$ were left as free fit parameters and the best--fit values are
$kB = 0.00115 \pm 0.0007$ and $ Y_0^{pe} = 489 \pm 2$ p.e./MeV.

In order to compute the quenching factor $Q_\beta$ for all energies of interest and for a known $kB$ value, it is practical to use the explicit functional form taken from~\cite{kBparam}:
\begin{align}
Q_\beta(kB; E) &\equiv \left(\frac{A_1 + A_2\ln(E) + A_3\ln^2(E)
}{1+ A_4\ln(E) + A_5\ln^2(E)}\right),
\label{eqn:Qanal}
\end{align}
which has the advantage of being easy to implement in the general fitting procedure.
As explained in~\cite{kBparam}, there is a specific correspondence between the values of the $A_i$
parameters and the $kB$ value: in our case $kB=0.0115$ corresponds to the set of values
$A_i = (1.019 ,0.127 ,6.067 \times 10^{-5} ,0.117 ,0.007)$ 
with $i$ ranging from 1 to 5.

\subsection{$N_{pe}$ and $N_{pe}^d$ estimators}

Because of their similarity, we treat the two $N_{pe}$ and $N_{pe}^d$ estimators together. 
The relation between the mean number of
photoelectrons $N_{pe}$ and the energy E, is given by a generalization of the
quenching relation discussed  in Section~\ref{sec:scintillator}:
\begin{equation}
N_{pe} = N_{pe}^0+Y_0^{pe} \cdot f_R(\vec r)\cdot E\cdot Q_p(E,kB), 
\label{eqn:Charge1}
\end{equation}
where the quenching factor $Q_p(E,kB)$ is that of Eq.~\ref{eqn:Qanal} in case of $\beta$ particles (the effective quenching $Q_{\gamma}$ defined in Eq.~\ref{eqn:Qgamma} is used for $\gamma$s),  $Y_0^{pe}$ is the
scintillation photoelectron yield expressed in units of p.e./MeV, and $f_R(\vec r)$ is a function describing the dependence of the
observed signal on the event position $\vec r$ (it is convenient to set $f_R(0) = 1$ in the detector
center); $N_{pe}^0$ is a pedestal due to any kind of random noise during the duration of the cluster, mainly dark noise of PMTs. We recall the $N_{pe}^d$ estimator is obtained from the $N_{pe}$ variable through a background subtraction,
thus for the $N_{pe}^d$ estimator $N_{pe}^0 = 0$.

For data analysis purposes, we model the number of photoelectrons averaged over the FV. This
is obtained starting from Eq.~\ref{eqn:Charge1} and, in the case of uniformly distributed events, is given by:
\begin{equation}
 N_{pe} = N_{pe}^0+Y_0^{pe} \cdot  E\cdot Q_p \cdot \overline{f_R(\vec r)}.
\label{eqn:Charge2}
\end{equation}
Since $f_R(\vec r)$ is not expected to depend on energy, the scintillator photoelectron
yield and the geometrical factor can be combined into a single
parameter referred to as the FV--averaged detector scintillator photoelectron
yield $Y_{det}^{pe}$, expressed in units of p.e./MeV.  We then obtain the
final formula for $N_{pe}$:
\begin{equation}
N_{pe} = N_{pe}^0+Y_{det}^{pe}\cdot E\cdot Q_p,
\label{eqn:Charge3}
\end{equation}
where $Y_{det}^{pe} = Y_0^{pe} \cdot \overline{f_R(\vec r)}$.

The second ingredient of the model deals with the variance $ \sigma^2_{N_{pe}}$ and the third central
moment $\kappa_{N_{pe}}$.
They can be computed from the various distributions associated with the scintillation process, non--uniform light collection within the detector and the multiplication process in the PMTs,  considering also the effect of unavoidable dis--uniformities throughout the fiducial volume:  
\begin{equation}
\sigma^2_{N_{pe}} = (1+\nu_1)   c_{eq} N_{pe} + \nu_T  N^2_{pe}
\label{Equaliz1_1}
\end{equation}
\begin{equation}
\begin{split}
\kappa_{N_{pe}}  = &(1 + 3\nu_1 + \kappa_1) c^2_{eq} \cdot N_{pe} + \\
       &3(1+\nu_1) \nu_T c_{eq} N^2_{pe} + \kappa_T  N^3_{pe},
\end{split}
\label{Equaliz1_2}
\end{equation}
where $\nu_1$ and $\kappa_1$ are the relative variance and the third
central moment of the PMT single photoelectron response, respectively, and
$\nu_T$ and $\kappa_T$ are the relative variance and the third central
moment  accounting for the detector non--uniformities. 
$c_{eq}(t)$ is the equalization factor
introduced in Section~\ref{sec:estimators}, compensating for the variable number of working channels
throughout the data taking period. For a complete determination, thus, Eq.~\ref{Equaliz1_1} and Eq.~\ref{Equaliz1_2}
must be averaged over the whole data taking interval.

Furthermore, we have to consider that the variance of the ``zero''--line of the ADC (pedestal)
leads to a non--negligible contribution to the global variance: it includes digitizing error
appearing during the analog--to--digital conversion and in general any noise in the charge
measurement, that can be defined as the spread of the signal at the output of ADC with
zero input signal.
If $ N_p$ is
the number of triggered PMTs then the additional contribution to the
variance is $N_p \sigma_{ped}^2$ where $\sigma_{ped}$ is variance of
zero line for a single PMT;  at low energies $ N_p \simeq
N_{pe}$. Since pedestal noise is usually symmetric, its contribution
to the third central moment should be negligible.

Another contribution to the measured charge is due to the pick--up of
random noise (mainly from the dark rate of the PMTs). Using the data
acquired during a special random trigger, we estimated $N_{pe}^0
\simeq 1$ p.e. for cluster lengths of 1.5\,$\mu$s. Assuming a Poisson
distribution for the random noise we obtain $\sigma^2_d = \kappa_d =
1$.

Summing all contributions we finally obtain
\begin{equation}
\sigma_{N_{pe}}^{2}= \sigma_d^2 + \sigma_{ped}^2 \cdot N_p + (1+\nu_{1})\cdot c_{eq} \cdot  N_{pe}+\nu_T \cdot  N_{pe}^{2}
\label{SigQeqexp}
\end{equation}
\begin{align}
\kappa_{N_{pe}}(t) &= \kappa_d + (1+3\nu_{1}+\kappa_1) \cdot c^2_{eq}(t) \cdot  N_{pe} \notag \\
&+ 3 \cdot (1+\nu_{1}) \cdot \nu_{T} \cdot c_{eq}(t) \cdot N_{pe}^2 \notag \\
&+ \kappa_{T} \cdot  N_{pe}^3,
\label{KapQeqexp}
\end{align}
to be considered averaged over the whole period of the data taking, and obviously valid
also for the $N_{pe}^d$ estimator.
% with $\sigma^2_d = \kappa_d = 0$.
For the two estimators under consideration we used two different analytical
approximations for the response function, which are described below.

\subsection{$N_{pe}$ response function}
\label{subsubsec:npe}

For the $N_{pe}$ estimator we adopted as approximated description of the response
function the generalized gamma function proposed in~\cite{ResScintDet} for an ``ideal'' scintillation
detector. Even though Borexino is not an ideal detector, the approximation works very well
for the Borexino data, e.g. the $^{210}$Po peak. The Monte Carlo modeling shows a very good agreement with
the analytical approximation for a wide range of energies of interest.
The generalized gamma formulation is the following:
\begin{equation}
\Gamma (N_{pe};\alpha,\beta)=2\beta^{\alpha}\Gamma^{-1}(\alpha)q^{2\alpha-1}e^{-\beta q^{2}},
\label{eq:Gamma}
\end{equation}
with the parameters $\alpha$ and $\beta$ providing the match of the mean value and variance of
relation \ref{eq:Gamma} with the corresponding values of the scintillation response.
The values of the
parameters $\alpha$ and $\beta$ are defined as:
\begin{equation}
\begin{split}
& \alpha= \\
& \frac{\sigma_{N_{pe}}^2+N_{pe}^2}{\sigma_{N_{pe}}^4\left(2+\frac{3}{N_{pe}}\right)+
4N_{pe}^2 \left(\sigma_{N_{pe}}^2-2 \right)+2N_{pe}\left(6\sigma_{N_{pe}}^2-1\right)}\\
\end{split}
\end{equation}
and 
\begin{equation}
\beta=\frac{\alpha}{N_{pe}^2}.
\end{equation}
%where $\overline{N_{pe}^{2}}=\sigma_{N_{pe}}^{2}+\overline{N_{pe}}^{2}$, and the
%values of $\sigma_{N_{pe}}^2$ and $\overline{N_{pe}}$ are defined by (\ref{SigQeqexp})
%and (\ref{eqn:Charge3}) respectively 

\subsection{$N_{pe}^d$ response function}
\label{subsubsec:nped}
The analytical approximation of the response function employed for the $N_{pe}^d$ estimator is
modeled by a modified Gaussian:
\begin{equation}
P (N^d_{pe}) = \frac{1}{\sqrt{2\pi}\sqrt{a + b \cdot {N^d_{pe}}}} \exp^{- \frac {(N^d_{pe} - \lambda)^2} {2(a + b \cdot N^d_{pe})}},
\label{eq:mod_gauss_final}
\end{equation}
whose parameters $a$, $b$, and $\lambda$, are defined to match the
first three central moments of the response function (see
Section~\ref{subsubsec:npe}).
Grouping the terms contributing to the variance and the third central
moment by their dependence on the number of detected photoelectrons,
we can rewrite equations Eq.~\ref{SigQeqexp} and Eq.~\ref{KapQeqexp}
as:
\begin{align}
\sigma_{N^d_{pe}}^{2}&= g_1\cdot N^d_{pe} + g_2 \cdot  ({N_{pe}^d})^2 \\
\kappa_{N^d_{pe}}&= g_3\cdot  N^d_{pe} + 3g_1\cdot g_2\cdot  ({N_{pe}^d})^2 + g_4 \cdot  ({N_{pe}^d})^3,
\end{align}
where $g_i$ are energy--independent constants that are left free in the
fit. We note that we have ignored the small contribution of the dark
noise to the variance and the third central moment. The parameters $a$,
$b$, and $\lambda$, are then approximately related to the above energy independent
constants $g_i$ by:
\begin{align}
b &= \frac{g_2 + 3 g_1 g_3 \cdot {N^d_{pe}} + g_4 \cdot ({N^d_{pe}})^2}{3(g_1 + g_3 \cdot {N^d_{pe}})}\\
\lambda &= {N^d_{pe}} - b\\
a &= -b^2 + (g_1 - b)\cdot {N^d_{pe}} + g_3 \cdot (N^d_{pe})^2.
\end{align}
We note that also in this case the analytical expression matches the Monte Carlo simulation over a wide range of energies and input parameters.

\subsection{$N_p$ estimator}
\label{subsubsec:np}
For the $N_p$ estimator, its connection with the initial event energy $E$ is realized
through two steps: the first is the same quenching relation between energy and
photoelectrons expressed by Eq.~\ref{eqn:Charge1} (but without the volume factor, which is instead
applied later, as we will see), while the second is the more complex relationship between
$N_{pe}$ and $N_p$.

In order to determine the latter,
let us consider events with energy $E$ at the detector's center and that  all the electronics channels in the detector are equal,
i.e. for the events at the detector's center every ADC connected to a
PMT has the same probability to detect a photoelectron. If the mean total collected
number of photoelectrons is $N_{pe}$, then the number of photoelectrons
collected on average by one PMT is $\mu_{0}=\frac{ N_{pe}}{N_{tot}}$,
where $N_{tot}$ is the total number of channels defined in Section~\ref{sec:estimators}. The distribution of the detected photoelectron
number at each PMT is expected to be Poissonian. In this case the
probability $p_0$ of absence of signal is:
\begin{equation}
p_0=e^{-\mu_{0}}
\label{Formula:p0}
\end{equation}
and the probability $p_1$ to detect at least one hit on a channel is:  
\begin{equation}
p_1= 1-p_0=1-e^{-\mu_{0}}
\label{Formula:p1}
\end{equation}
In order to define the total number of the channels hit, one
can consider the number $N_{tot}$ of independent probes using
 $p_1$ defined by Eq.~\ref{Formula:p1}. The
distribution of the number of the triggered channels $N$ for events in the center obeys the binomial
distribution: 
\begin{equation}
P(N)=\left(\begin{array}{c}
N_{tot}\\
N\end{array}\right)p_{1}^{N}(1-p_{1})^{(N_{tot}-N)} 
\label{Formula:Binomial}
\end{equation}

From Eq.~\ref{Formula:p1} the mean number of the PMTs detecting a
non--zero signal is:
\begin{equation} 
N_p =N_{tot} \, p_1=N_{tot}(1-e^{-\mu_{0}}).
\label{Formula:MeanN}
\end{equation}
Taking into account from this last relation that
$p_1=N_p/N_{tot}$, being equal to $p_1$ expressed in the form of Eq.~\ref{Formula:p1}, and considering the definition of $\mu_0$ given above, we get:
\begin{equation}
N_p = N_{tot} \left( 1- e^{-{\frac {N_{pe}}{N_{tot}}}} \right ),
\label{eq:NpFin}
\end{equation}
which expresses the desired link between the measured number of hit PMTs ($N_{p}$) and the
number of photoelectrons ($N_{pe}$).
Such a relation, however, is strictly valid only at the center of
the spherical detector and for a set of identical PMTs.
In fact, for an event  
with coordinates $\vec r =\{x,y,z\}$, the mean number of detected photoelectrons is a
function of $\vec r$, a fact that leads to a generalization of expression (\ref{eq:NpFin}):
\begin{equation}
N_p=N_{tot}(1-e^{-\frac{N_{pe}}{N_{tot}}})\cdot \left(1-g_C(FV) \, \frac{N_{pe}}{N_{tot}}\right),
\label{NpVol}
\end{equation}
where the value of the geometric correction parameter $g_C$ depends on the FV used. This new formula shows good agreement with Monte Carlo simulations throughout the volume.
In summary, Eq.~\ref{NpVol} and the quenching relation between energy and photoelectrons
taken together, represent the first ingredient of the model for the $N_p$ variable, i.e. the link
between energy and $N_p$ itself.

As far as the second ingredient is concerned, i.e. the $N_p$ variance, again taking into account
from Eq.~\ref{Formula:MeanN}  that $p_1=N_p/N_{tot}$, and on the basis of the binomial nature of the detector
response, it can be expressed for events in the center as:
\begin{equation}
\sigma_{N_p}^{2}=N_{tot} \, p_1(1-p_1 )=N_p \left ( 1- \frac{N_p}{N_{tot}}  \right).
\label{Formula:SigN}
\end{equation}
Its modification due to the volumetric effect within the FV can be empirically
accounted for through an additional term quadratically dependent on $N_p$:
\begin{equation}
\sigma_{N_p}^{2}=N_p \left ( 1- \frac{N_p}{N_{tot}}  \right )+ \nu_T(N_p) N_p^2.
\label{Formula:SigNV}
\end{equation}

As $N_{pe}$, also the $N_p$ variable is defined taking into account the run--dependent number of
working PMTs and we should consider this fact while modeling the resolution by including
explicitly the equalization factor $f_{eq}(t)$ (see Section~\ref{sec:estimators}).
Therefore, the variance of the registered number of triggered PMTs in the equalized scale
is:
\begin{equation}
\sigma_{N_p}^2=N_p \left ( 1- \frac{N_p}{N_{tot}}  \right ) f_{eq}(t) +\nu_T(N_p) N_p^2
\label{NEqualiz1}
\end{equation}
to be properly averaged over the time of data taking.

The spatial non--uniformity for the $N_p$ variable is sizable, and this is why the $\nu_T$ factor is
energy dependent. It was found through Monte Carlo modeling that in the energy range of
interest this dependence is linear with respect to $N_p$, i.e. $\nu_T(Np) =\nu_T^0N_p$.
By including for completeness also the effect of random noise ($\sigma_d$ variance), the final variance
expression is:
\begin{equation}
\sigma_{N_p}^2=N_p \left ( 1- \frac{N_p}{N_{tot}}\right) \cdot f_{eq}(t) \rangle
+\nu_T^0 \cdot N_p^3+\sigma_d^2.
\label{SigNeqexp}
\end{equation}

Finally, as the response function for the $N_p$ variable, the same generalized gamma function introduced in
Section~\ref{subsubsec:npe} in Eq.~\ref{eq:Gamma} is adopted replacing the $N_{pe}$ variable with $N_p$.

\section{The Monte Carlo procedure} 
\label{sec:MC}

This method of evaluation of the detector response function is based on a Monte Carlo that models and predicts the expected shapes of the signal and background. The Borexino Monte Carlo code is an ab--initio simulation of all the processes influencing the energy deposit of each type of particle in the scintillator and in the materials building the detector. It is important to model the scintillation and Cherenkov light emission, light propagation processes including the scattering, absorption--reemission and reflection, light detection, and the  electronics response. All the $\vec l$ parameters introduced in Section~\ref{sec:resp} are used as input values of the Monte Carlo code.
The simulation of the energy deposit uses the standard GEANT4 package~\cite{Geant4} describing the energy loss of the various particle types in different materials. The photons of the produced light are tracked one--by--one until they reach a PMT and are possibly detected or until they are absorbed elsewhere. 
A detailed model of the response of the electronics is also included.
Some of the $\vec l$ parameters correlated with the light generation and propagation, as well as with the electronics response, were measured with dedicated laboratory set--ups. These include the $\tau_i$ and $w_i$ values (Table~\ref{table:prop}), the PPO and PC emission spectra as functions of the wavelength $\lambda$, the PC, PPO, and DMP molar extinction coefficients as
functions of  $\lambda$ \cite{BxNim}. The (PC + PPO) refraction index was measured for $\lambda$'s from 245.5\,nm to 600\,nm, while for smaller ultraviolet wavelengths we use the values extrapolated by comparison with the results for PC with benzene. The knowledge of the dispersion relation of the refraction index is an important input in the Monte Carlo because it allows to correctly consider the group velocity for individual photons which is important for the light tracking as well as for the simulation of the Cherenkov light emission.

The various PMTs do not have identical probability to produce a signal when a photon hits the photocathode. The PMT quantum efficiencies  as  $\lambda$--functions have been provided by the manufacturer as well as the distribution of the peak quantum efficiency at $\lambda$ = 420\,nm, having a mean value of 24.7$\%$ and $rms$ of 1.9$\%$. It results from the Borexino data (mono--energetic calibration sources located in the detector center and $^{14}$C events reconstructed within a sphere of 50\,cm radius around the detector center) that the $N_p$ mean value distribution has the $rms$  about 1.5 times larger than that resulting from the pure quantum--efficiency curves. In the Monte Carlo we introduce this effect by rescaling the peak value of the quantum--efficiency curve according to the measured efficiency of each PMT.

The simulation reproduces  the real distribution of active PMTs, the measured dark noise, and the real gain and the shape of the single--photoelectron response of each PMT following the run--by--run changes. It includes the simulation of the after--pulses and of the measured transit--time spread. The Monte Carlo code finally produces a set of raw data with the same format as that of the measured one, allowing an identical data processing.

It is required that both the energy estimators and the hit--times distributions, which are naturally highly correlated, are fully reproduced by the simulation. The Monte Carlo optimization has been performed iteratively. Several input parameters have been varied until the differences between the measured versus the simulated distributions were minimized. An effort has been made to correctly model all physical phenomena and to minimize the number of "effective" parameters.
 
\begin{figure}[top]
\begin{center}
\centering{\includegraphics[width = 0.48\textwidth]{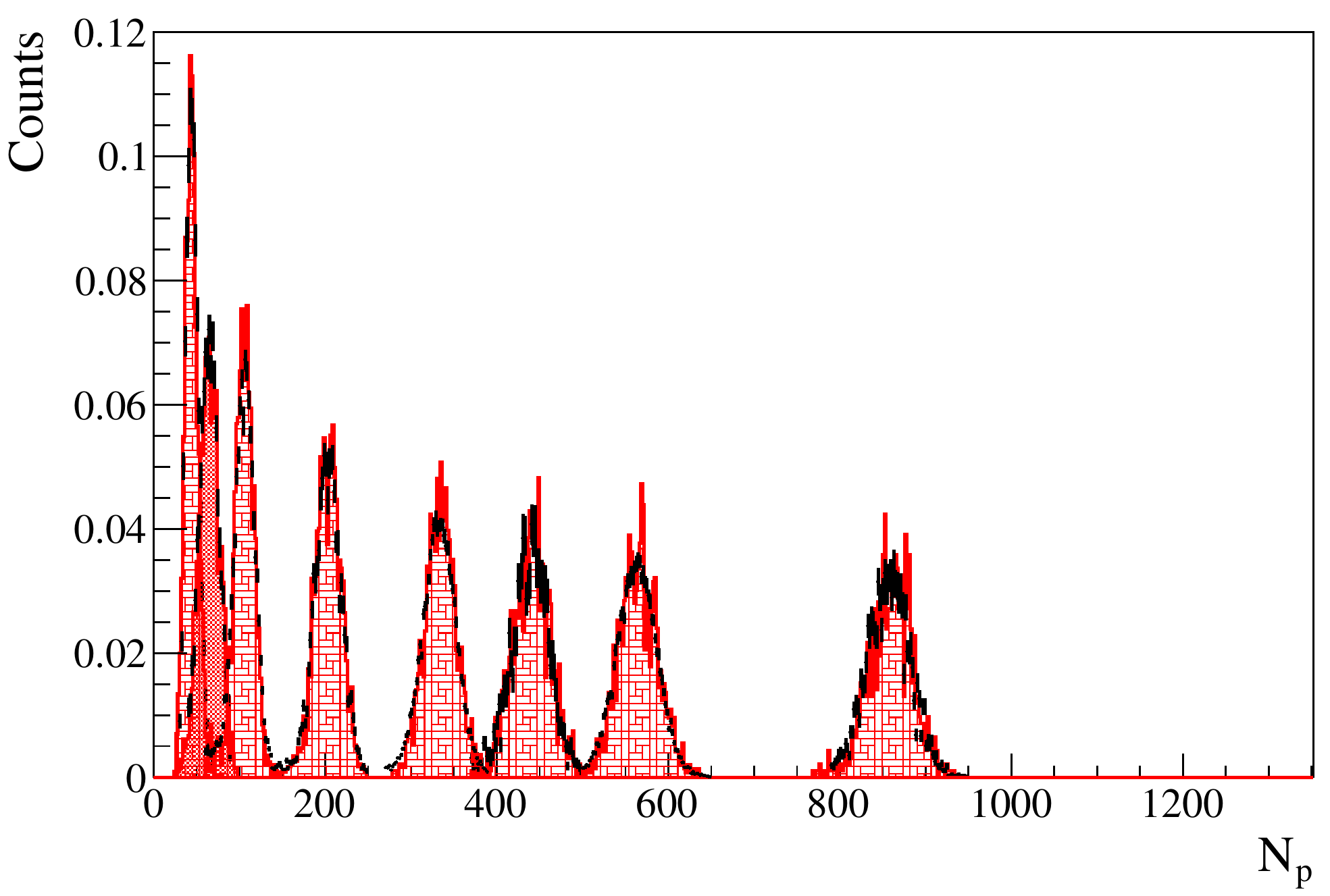}}\\
\centering{\includegraphics[width = 0.48\textwidth]{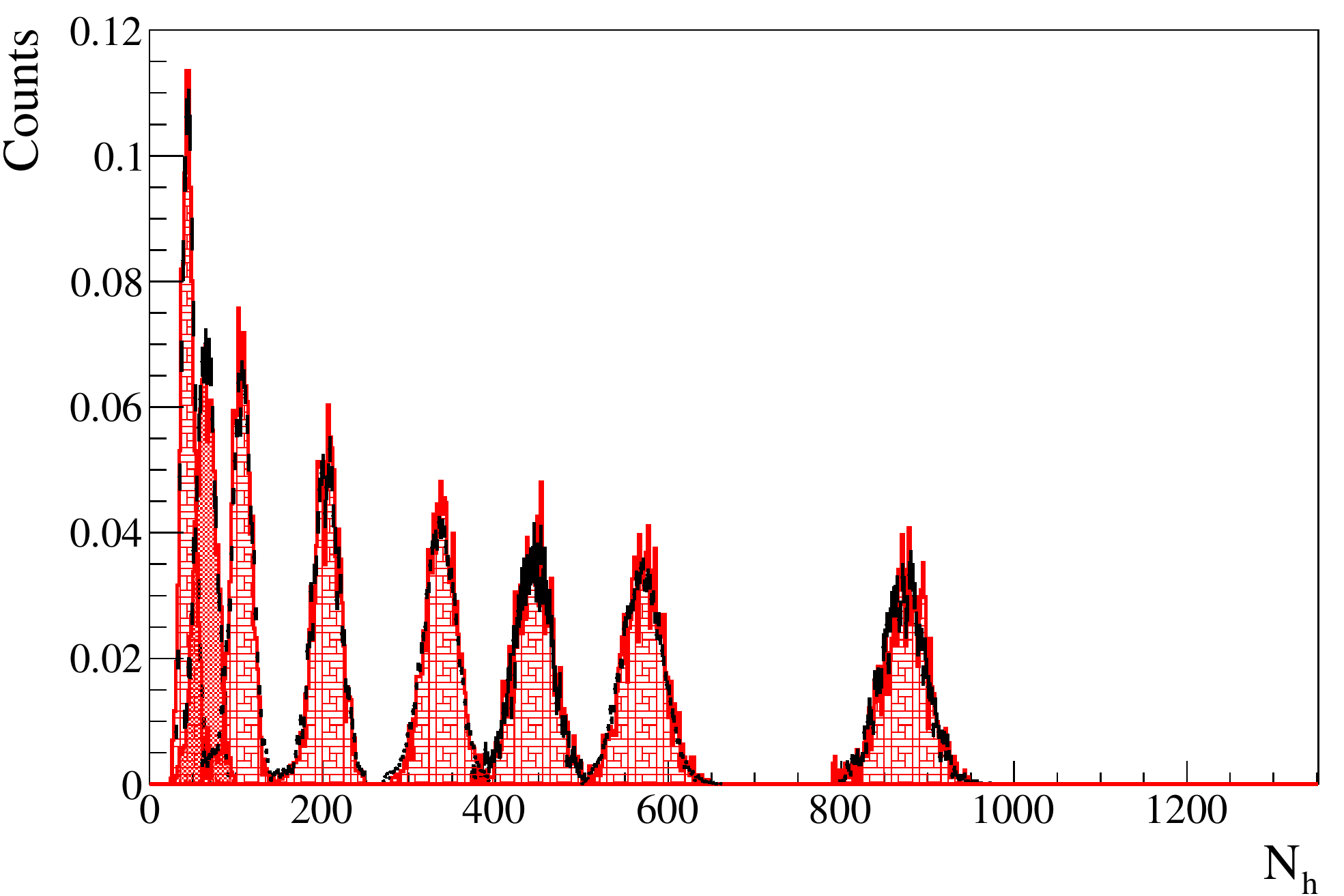}}\\
\centering{\includegraphics[width = 0.48\textwidth]{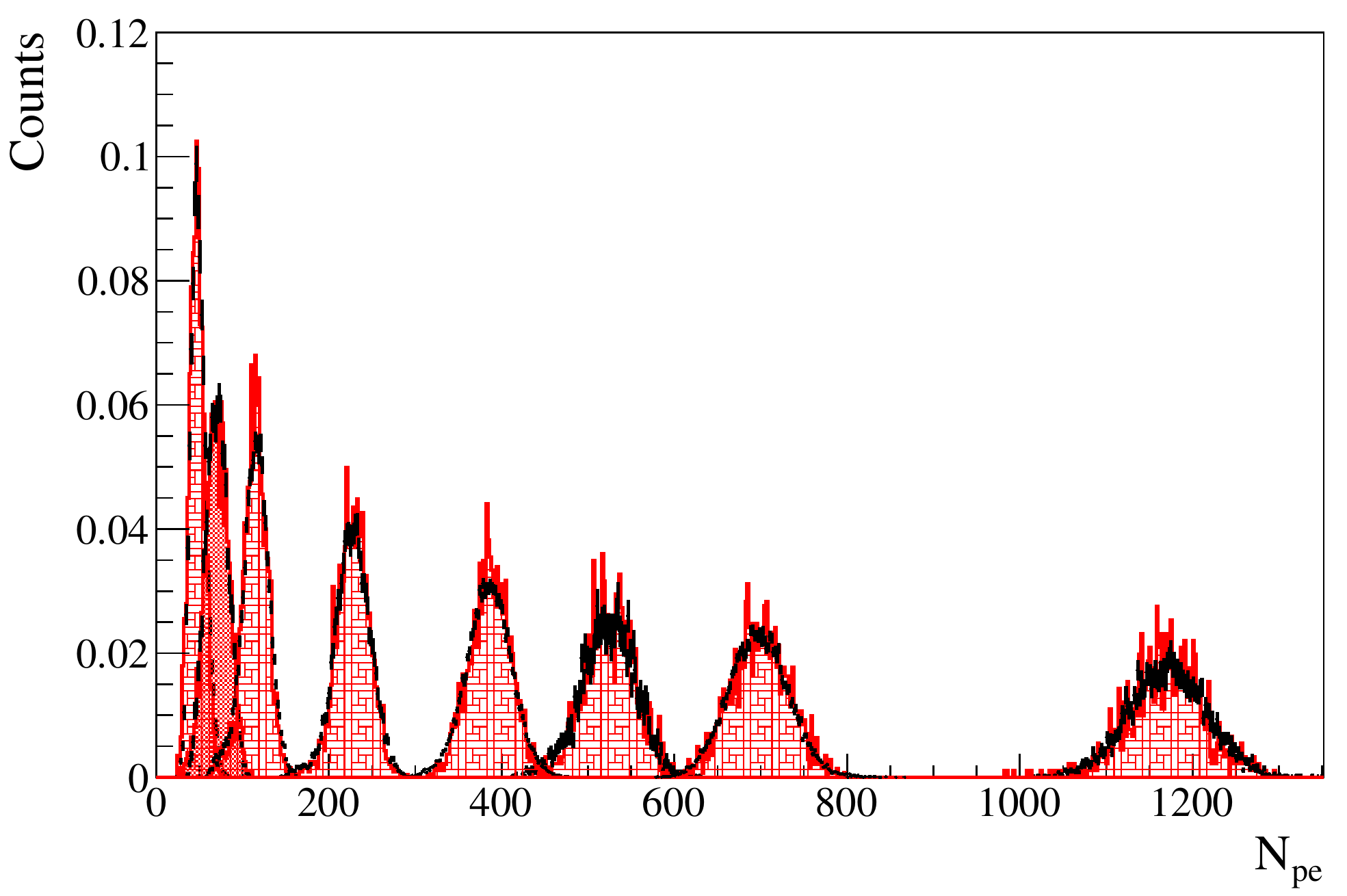}}\\
\end{center}
\caption{ Energy spectra ($N_p$, $N_h$, and $N_{pe}^d$ variables) of the calibration sources placed in the detector's center: measured data (black lines) versus the Monte Carlo simulation (areas dashed with red lines). The peaks represent (from the left to the right) the total $\gamma$ decay energy of $^{57} $Co, $^{139}$Ce, $^{203}$Hg, $^{85}$Sr, $^{54}$Mn, $^{65}$Zn, $^{40}$K, and $^{60}$Co.}
\label{fig:GammaPeak}
\end{figure}

In particular, the $\gamma$ sources placed in the detector center have been used to determine the light yield $Y_0^{\rm ph}$ and the electron quenching parameter $kB$, both introduced in Section~\ref{sec:scintillator}. The geometry of the $\gamma$--source vial has been fully included in the simulation. 
The $\gamma$--sources events have been simulated scanning the values of $Y_0^{\rm ph}$ and $kB$. The resulting distributions of all the energy estimators have been compared with the measured ones, calculating the $\chi^2$ as a function of $Y_0^{\rm ph}$ and $kB$.
The value of $kB$ corresponding to the minimum of the $\chi^2$ is (0.0109~$\pm$~0.0006)\,cm/MeV, compatible with the value obtained analytically as described in Section~\ref{sec:analkB}. 
Figure~\ref{fig:GammaPeak} shows the comparison between the measured energy distributions of the $\gamma$ sources and the simulation 
obtained with the best value of $Y_0^{\rm ph}$ and $kB$. The agreement between the data and the simulation is very good for all the three energy estimators. 
Table~\ref{tab:dataVSMC} (obtained using the data of the previous plots) give the measured and the simulated peak positions and the resolutions for the $N_p$, $N_h$,  and $N_{pe}$ energy estimators, respectively.
The peak position and the resolution of the $\gamma$ source in the detector center are reproduced by the Monte Carlo with an accuracy better than 
1$\%$.

\begin{table*}
\begin{center}
\begin{tabular}{lccccr}
\hline
\hline
Source & $N_p$ peak (data)   & $N_p$ peak (MC) & Sigma (data) & Sigma (MC) \\ %& $\chi^2$/NDF \\
\hline
$^{57}$Co  &  45.4 $\pm$ 0.2 &  44.6 $\pm$ 0.3 &  8.5 $\pm$ 0.4 &  7.2 $\pm$ 0.7 \\ %& 69/31 \\
$^{139}$Ce &  65.4 $\pm$ 0.2 &  66.0 $\pm$ 0.4 & 11.3 $\pm$ 0.3 & 11.0 $\pm$ 0.9 \\ %& 41.1/45 \\
$^{203}$Hg & 106.4 $\pm$ 0.1 & 105.7 $\pm$ 0.3 & 11.3 $\pm$ 0.3 & 10.1 $\pm$ 0.7 \\ %& 41.8/42 \\
$^{85}$Sr  & 204.3 $\pm$ 0.2 & 205.9 $\pm$ 0.3 & 15.0 $\pm$ 0.5 & 15.4 $\pm$ 0.6 \\ %& 47.3/54 \\
$^{54}$Mn  & 333.9 $\pm$ 0.1 & 336.0 $\pm$ 0.1 & 18.4 $\pm$ 0.4 & 18.2 $\pm$ 0.3 \\ %& 48.2/59 \\
$^{65}$Zn  & 440.1 $\pm$ 0.4 & 440.7 $\pm$ 0.9 & 21.6 $\pm$ 0.3 & 21.9 $\pm$ 0.5 \\ %& 41.2/40 \\
$^{40}$K   & 564.5 $\pm$ 0.2 & 565.7 $\pm$ 0.7 & 23.6 $\pm$ 0.8 & 23.8 $\pm$ 1.0 \\ %& 60.3/58 \\
$^{60}$Co  & 858.0 $\pm$ 0.3 & 859.8 $\pm$ 0.7 & 24.2 $\pm$ 0.9 & 24.2 $\pm$ 0.7 \\ %& 42.0/39 \\
\hline
\hline
\end{tabular}
\end{center}
\begin{center}
\begin{tabular}{lccccr}
\hline
\hline
Source & $N_h$ peak (data)  & $N_h $peak (MC) & Sigma (data) & Sigma (MC) \\ %& $\chi^2$/NDF \\
\hline
$^{57}$Co  &  45.6 $\pm$ 0.2 &  44.9 $\pm$ 0.3 &  8.6 $\pm$ 0.3 &  7.5 $\pm$ 0.7 \\ %& 74/28 \\
$^{139}$Ce &  66.0 $\pm$ 0.2 &  66.3 $\pm$ 0.4 & 11.9 $\pm$ 0.5 & 11.3 $\pm$ 0.9 \\ %& 45.7/45 \\
$^{203}$Hg & 107.3 $\pm$ 0.1 & 106.8 $\pm$ 0.3 & 11.1 $\pm$ 0.4 & 10.1 $\pm$ 0.8 \\ %& 41.8/41 \\
$^{85}$Sr  & 205.8 $\pm$ 0.2 & 205.9 $\pm$ 0.3 & 14.4 $\pm$ 0.5 & 14.9 $\pm$ 0.6 \\ %& 53.0/54 \\
$^{54}$Mn  & 336.9 $\pm$ 0.2 & 336.0 $\pm$ 0.1 & 18.4 $\pm$ 0.4 & 18.4 $\pm$ 0.3 \\ %& 55.3/60 \\
$^{65}$Zn  & 443.8 $\pm$ 0.4 & 445.0 $\pm$ 0.7 & 21.5 $\pm$ 0.3 & 21.7 $\pm$ 0.5 \\ %& 32.2/38 \\
$^{40}$K   & 571.1 $\pm$ 0.2 & 571.8 $\pm$ 0.6 & 24.2 $\pm$ 0.5 & 24.6 $\pm$ 0.5 \\ %& 60.4/58 \\
$^{60}$Co  & 872.4 $\pm$ 0.4 & 874.6 $\pm$ 0.8 & 26.0 $\pm$ 0.9 & 25.8 $\pm$ 0.6 \\ %& 32.2/38 \\
\hline
\hline
\end{tabular}
\end{center}
\begin{center}
\begin{tabular}{lccccr}
\hline
\hline
Source & $N_{pe}$ peak (data)  & $N_{pe}$ peak (MC) & Sigma (data) & Sigma (MC) \\ %& $\chi^2$/NDF \\
\hline
$^{57}$Co  &   48.8 $\pm$ 0.2 &   47.8 $\pm$ 0.3 &  8.5 $\pm$ 0.4 &  7.7 $\pm$ 0.6 \\ %& 111/36 \\
$^{139}$Ce &   71.0 $\pm$ 0.2 &   71.6 $\pm$ 0.4 & 14.3 $\pm$ 0.7 & 14.0 $\pm$ 1.3 \\ %& 54/58 \\
$^{203}$Hg &  116.4 $\pm$ 0.1 &  115.9 $\pm$ 0.4 & 13.3 $\pm$ 0.4 & 13.9 $\pm$ 1.6 \\ %& 73/48 \\
$^{85}$Sr  &  228.5 $\pm$ 0.2 &  229.4 $\pm$ 0.4 & 18.6 $\pm$ 0.7 & 18.9 $\pm$ 0.7 \\ %& 36.4/32 \\
$^{54}$Mn  &  386.0 $\pm$ 0.2 &  384.9 $\pm$ 0.5 & 24.0 $\pm$ 0.4 & 23.4 $\pm$ 0.4 \\ %& 78.0/68 \\
$^{65}$Zn  &  525.1 $\pm$ 0.5 &  526.0 $\pm$ 1.0 & 31.3 $\pm$ 0.2 & 30.9 $\pm$ 0.7 \\ %& 28.2/34 \\
$^{40}$K   &  697.5 $\pm$ 0.2 &  699.1 $\pm$ 0.9 & 33.8 $\pm$ 0.7 & 33.6 $\pm$ 0.6 \\ %& 52.5/57 \\
$^{60}$Co  & 1171.7 $\pm$ 0.6 & 1169.1 $\pm$ 1.4 & 44.5 $\pm$ 0.4 & 41.2 $\pm$ 1.5 \\ %& 35.2/34 \\
\hline
\hline
\end{tabular}
\end{center}
\label{tab:dataVSMC}
\caption{Comparison between the measured and Monte Carlo simulated peak positions and the resolutions for the $\gamma$ calibration sources located in the detector center for the three energy estimators $N_p$, $N_h$, and $N_{pe}$. }
\end{table*}

The same energy deposits occurring in various detector positions give rise to non--equal, position--dependent values of the energy estimators, $N_h$, $N_p$, and $N_{pe}$. This is due to the light absorption, the geometrical effects as for example the presence of the light concentrators mounted on some PMTs, the different response of individual electronics channels as well as non--uniform distribution of non--working electronics chains. The broken PMTs are concentrated close to the bottom of the detector
 thus giving  a higher light loss for off--center events in the bottom hemisphere with respect to the  ones in the upper hemisphere. 

\begin{figure}[h]
\vspace{-2 mm}
\begin{center}
\centering{\includegraphics[width = 0.51\textwidth]{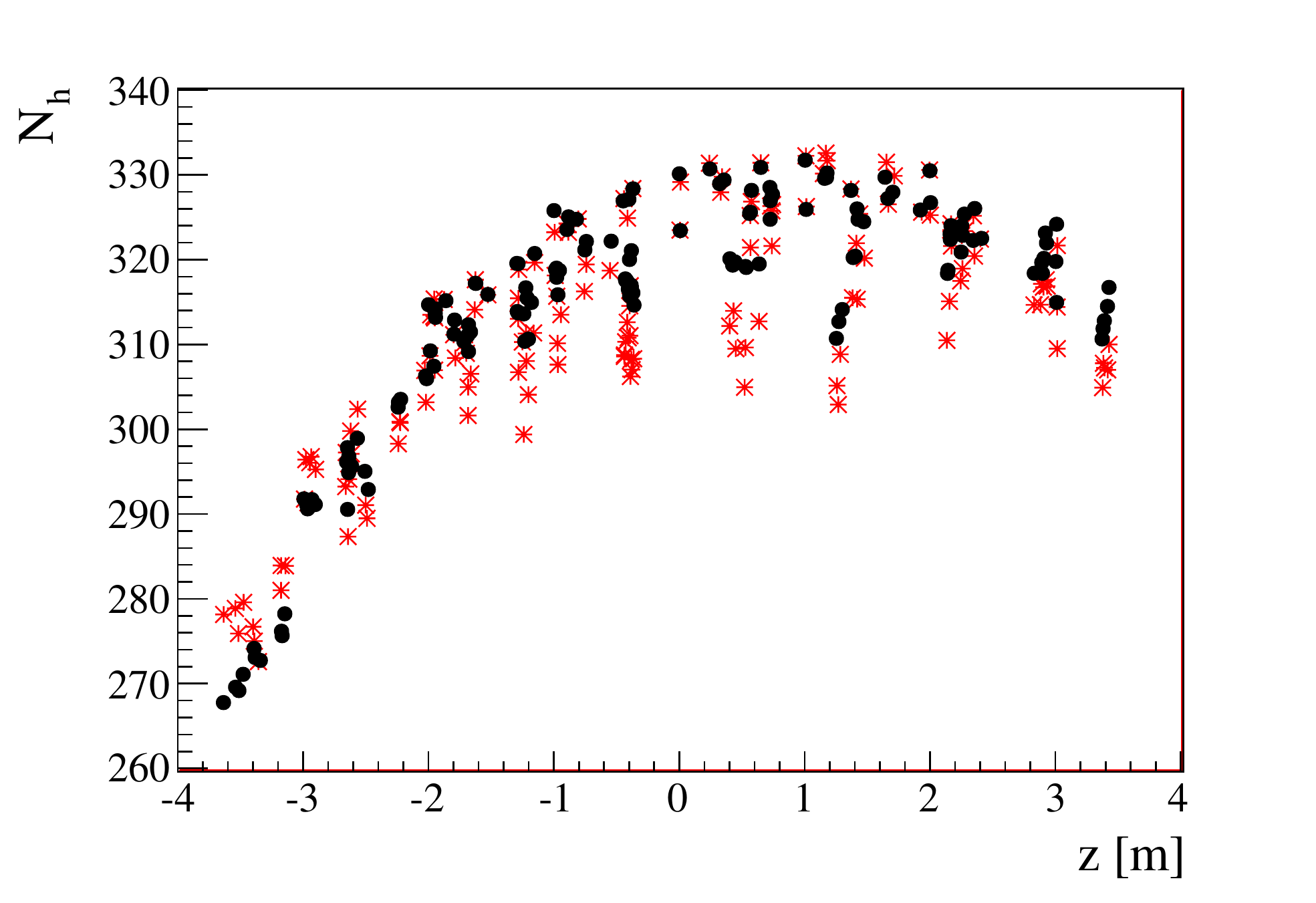}}
\caption{The $N_h$ peak position vs $z$--coordinate of the $^{214}$Po $\alpha$ peak from the radon calibration source, shown for the data (black circles) and the Monte Carlo simulation (red stars). The various points at fixed $z$ position correspond to different $x$ and $y$ coordinates. The reduction of the collected light for negative $z$ is due to the concentration of broken PMTs close to the detector's "South pole". }
\label{fig:Po214VsZ}
\end{center}
\end{figure}

\begin{figure}[h]
\begin{center}
\centering{\includegraphics[width = 0.5\textwidth]{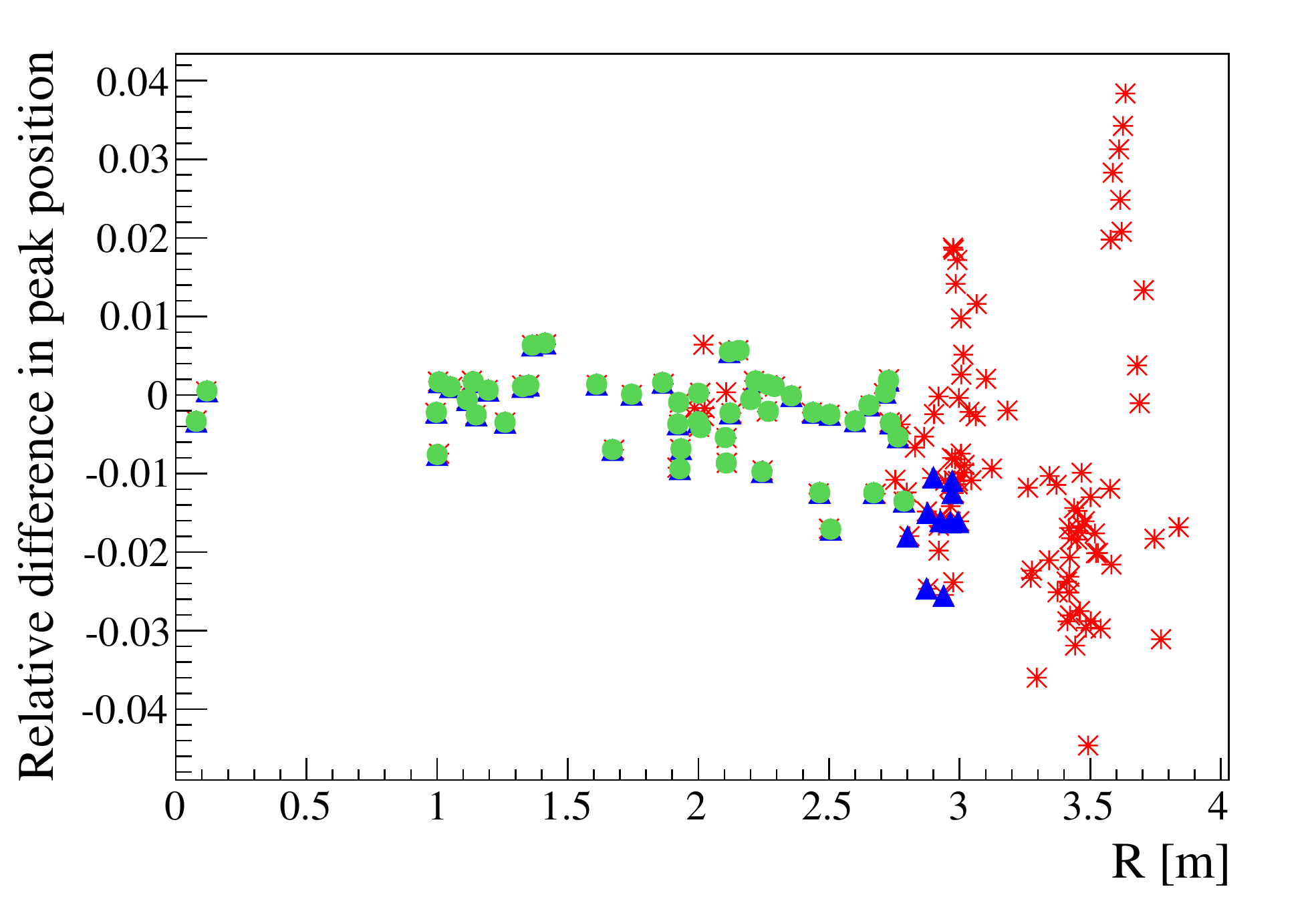}}
\caption{The relative difference $\frac {N_h (MC) - N_h (data)} {N_h (data)}$  as a function of the radial position $R$ of the $^{214}$Po $\alpha$ peak from the radon calibration source. Blue triangles: $^7$Be--$\nu$ FV,  green circles: $pep$--$\nu$ FV, red stars: outside both FVs.}
\label{fig:Po214VsR}
\end{center}
\end{figure}

The geometrical non--uniformity of the energy response has been measured with the radon source comparing the energy estimators of the $^{214}$Po $\alpha$--peak of the data and the Monte Carlo. The Monte Carlo data has been generated with the input parameters optimized to reproduce the source calibration data located in the detector center. As an example, Fig.~\ref{fig:Po214VsZ} demonstrates the $z$--dependency of the $N_h$ estimator both for the data (black circles) and for the Monte Carlo simulation (red stars). 
Figure~\ref{fig:Po214VsR} shows the percentage difference between the $N_h$ peak position of the Monte Carlo and the data normalized to the data peak. The source locations within the FVs used for the $pep$ and $^7$Be neutrino analysis and locations outside both these FVs are shown in different colors. As demonstrated in Fig.~\ref{fig:Po214VsR}, the Monte Carlo underestimates the energy  for events close to the $^7$Be--FV border by 2\% at maximum. For this reason, the events uniformly distributed in this FV are generated with the light yield $Y_0^{\rm ph}$ multiplied by a correction factor of about $1.01$. The exact value of this correction factor is optimized based on the spectrum of $^{11}$C events uniformly distributed in this FV and selected as described in the  Section~\ref{sec:c11}.
This correction factor is not included in Fig.~\ref{fig:Po214VsR}.

\begin{figure}[h]
\begin{center}
\includegraphics[width = 0.5\textwidth]{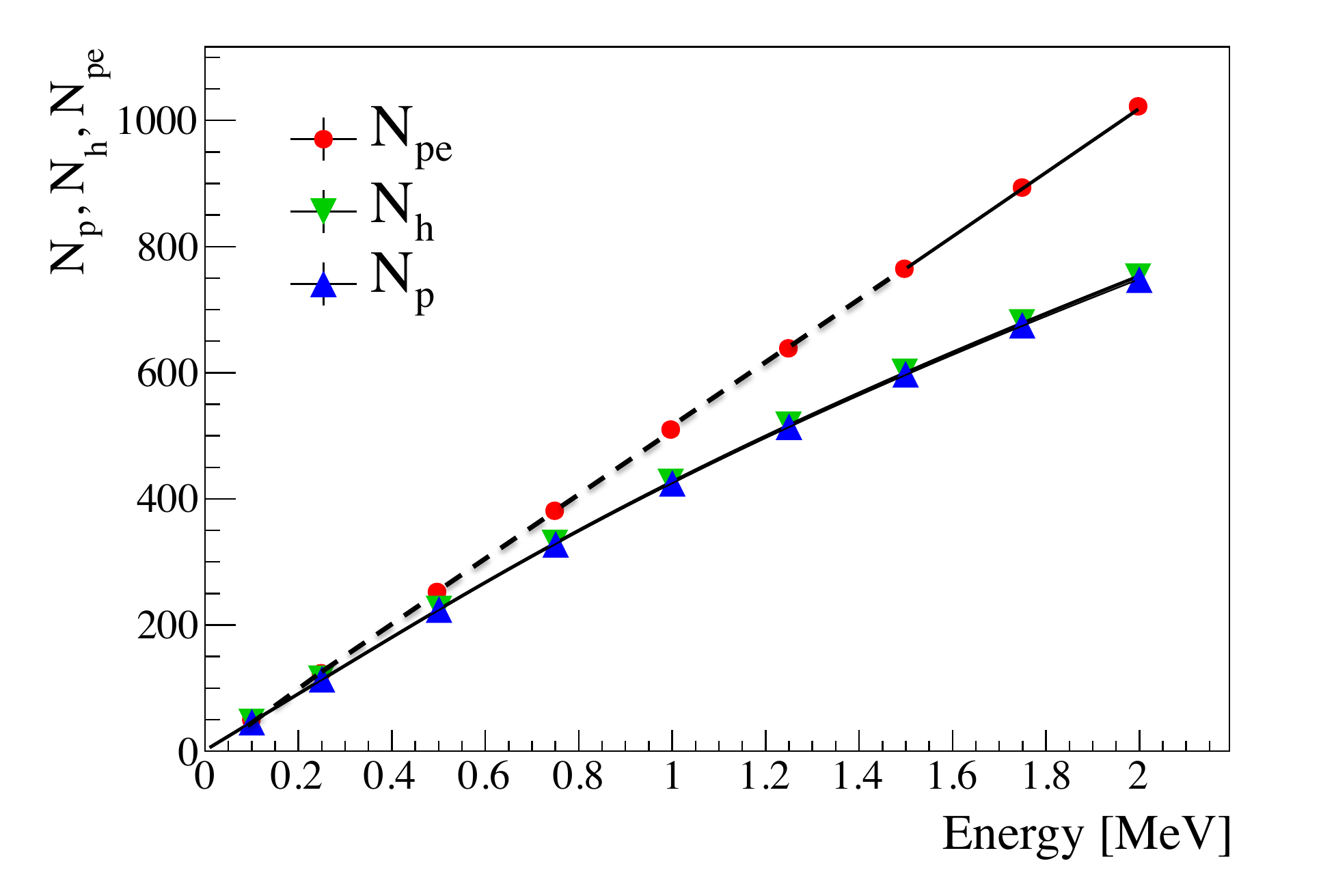}
\caption{The energy estimators $N_p$, $N_h$, and $N_{pe}$ versus energy for $\beta$ events uniformly generated in the $^7$Be--FV as obtained with the Monte Carlo simulation.}
\label{fig:Yield}
\end{center}
\end{figure}

Figure~\ref{fig:Yield} shows the relation between the energy estimators $N_p$, $N_h$, and $N_{pe}$ and the energy for $\beta$--particles with positions reconstructed within the $^7$Be--FV. Events have been generated uniformly within a sphere of 3.5\,m radius, following the run--by--run variations (described above) during the whole data taking period used in the $^7$Be --neutrino analysis. The relative contribution of the Cherenkov light is shown in the Table~\ref{tab:Ceren}.
The dark noise of the PMTs is included in the simulation according to the run--by--run measured values.
As expected, $N_h$ and $N_p$ versus energy shows significant deviation from linearity; 
the difference between them is too small to be visible in the graph. The curve is a fit with a polynomial.
 $N_{pe}$ is linear with energy at high energy: the dotted line is the extrapolation of the fit in the low--energy region where small deviation from linearity due to the quenching effect are present.
\begin{table}
\begin{center}
\begin{tabular}{c c } \hline \hline
 Energy    & $ (N_h - N_h ^{\rm{No Cer}}) / N_h$   \\  
$[$keV$]$     & $[\%]$ \\
 \hline
 250      & 1.25  \\
 500       & 3.7 \\
 1000         & 5.1 \\
 2000         & 5.6  \\
\hline
\hline
\end{tabular}
\caption{Relative differences between the $N_h$ and $N_h^{\rm {No Cer}}$ resulting from the Monte Carlo simulations of mono--energetic $\beta$'s with and without the generation of the Cherenkov light, respectively.}  
\label{tab:Ceren}
\end{center}
\end{table}

\begin{figure} [t]
\begin{center}
\centering{\includegraphics[width = 0.5\textwidth]{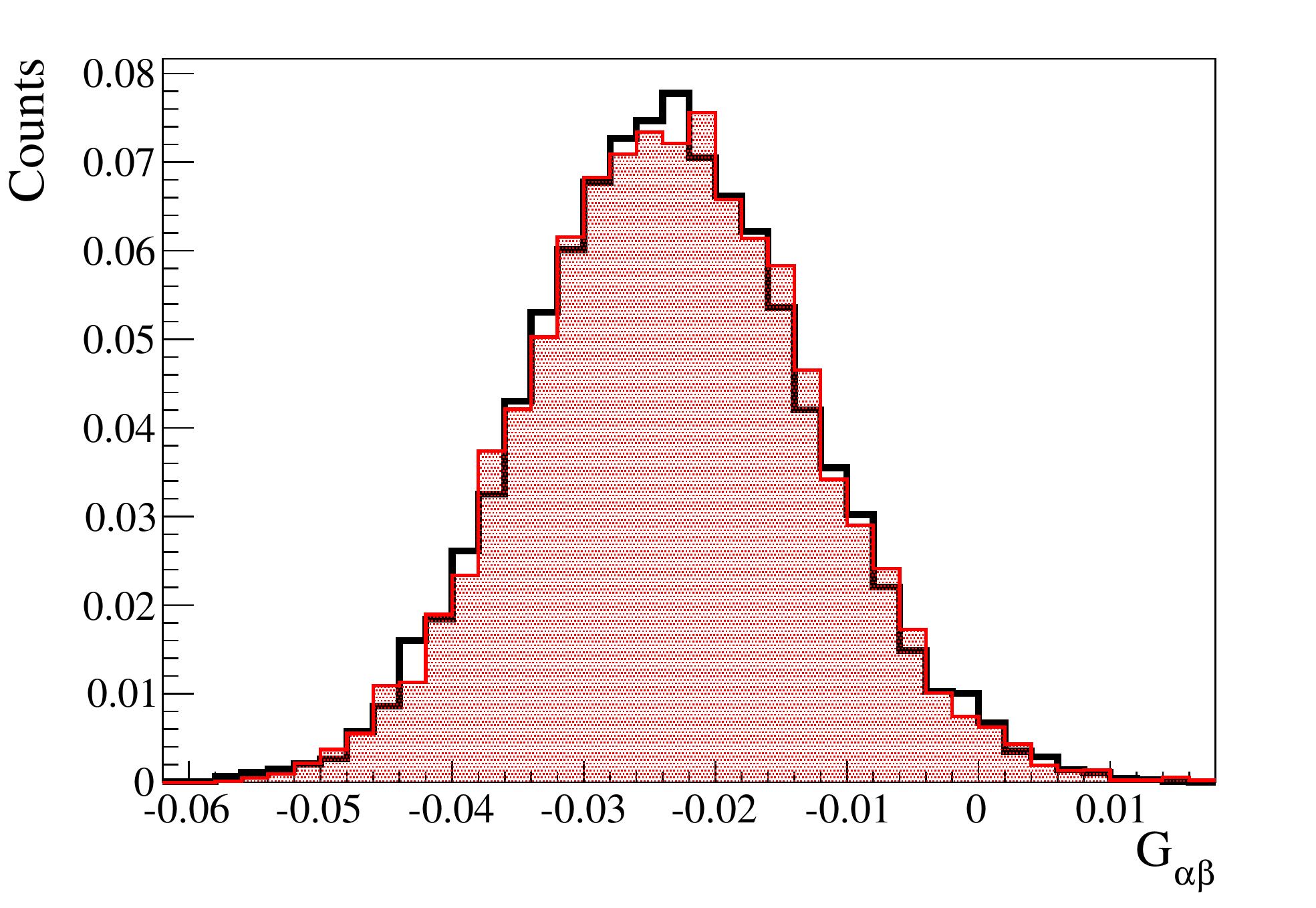}}
\caption{The $G_{\alpha\beta}$ variable for the $^{85}$Sr $\gamma$--calibration source placed at the location $(x,y,z) = (0,0,3)$\,m, for measured (black line) and Monte Carlo simulated (red filled area) events. }
\label{fig:GattiSr}
\end{center}
\end{figure}

The Monte Carlo accurately models also the hit--time distributions, making possible to reproduce the shape variables which has been described in Section~\ref{sec:shape}. As a consequence, the Monte Carlo can then be used to evaluate the efficiency of cuts as described in Section~\ref{sec:cuts}. In addition, it is possible to implement the $\alpha$ -- $\beta$ statistical subtraction described in Section~\ref{sec:ab} using the Monte Carlo distributions of the $G_{\alpha\beta}$ parameter for different energy intervals. Figure~\ref{fig:GattiSr} compares this $G_{\alpha\beta}$ variable for the $^{85}$Sr calibration source placed at the position $(x,y,z)$ = $(0,0,3)$\,m, as obtained from the data (black line) and from the Monte Carlo simulation (red filled area). Note that this $\gamma$ source produces the events measured with approximately the same number of hits/photoelectrons as the $\alpha$ events of $^{210}$Po. In general, for different energies and positions, the Monte Carlo reproduces the shapes of the $G_{\alpha\beta}$ distributions but a small shift between the measured and simulated distributions may be present. This is at maximum $\pm$0.002, corresponding to one bin in Fig.~\ref{fig:GattiSr}.

\section{The $\alpha$ energy scale}
\label{sec:alphaQ}

Though most $\alpha$ decays produce particles with energy above 4\,MeV which is well above the energy range of interest for the determination of both the $^7$Be-- and $pep$--neutrino interaction rates, the high density of ionization produced by these particles lead to a number of scintillation photons corresponding to electrons with about ten times lower energy and then falling in the energy region of interest. This strong $\alpha$ quenching was already outlined in Section~\ref{sec:scintillator}. In order to include the effect of this
background in the fit, it is important to determine the $\alpha$ energy scale.
The dominant $\alpha$ background is originated from $^{210}$Po which has an average rate of few thousands cpd/100\,tons (see Fig.~\ref{fig:PoVsTime}). The peak produced in the energy spectrum is well visible and its position is easily fitted. There are however other $\alpha$ backgrounds (see Tables~\ref{tab:u238_prod} and \ref{tab:th232_prod}) with much lower rate of decay that do not produce a visible peak in the energy spectrum; they can still contribute significantly to the count rate and thus it is important to know their position in the spectrum, so they may be fixed in the fit.

To determine $\alpha$ energy scale in the fiducial volume used for the $^7$Be and $pep$ neutrino analyses, we considered the $\alpha$ decays of $^{210}$Po, $^{212}$Po, $^{214}$Po,
$^{216}$Po, and $^{220}$Rn. With the exception of  $^{210}$Po, the $\alpha$ events have been identified by searching for time--correlated decays. We used also the higher--rate $\alpha$'s of $^{214}$Po, $^{222}$Rn, and $^{218}$Po from the $^{222}$Rn calibration source.
Table~\ref{tab:alphaQval} reports the results of this analysis. By fitting the data of this Table with a linear function, we have obtained the effective quenching factor Q$_\alpha$ for events distributed in the $^7$Be-- and $pep$--FVs, reported in the last column. We did not attempt to determine $kB$ for $\alpha$ particles. 
\begin{table*}
\begin{center}
\begin{tabular}{lccccr}
\hline
\hline
Isotope      & $E_{\alpha}$ (keV)  & Data set &  Mean ($N_{pe}^d$)                            & Q$_\alpha$\\ 
\hline
$^{210}$Po  &   5310                          &   data    &  209.5 $\pm$ 0.02       & 0.079 \\
$^{222}$Rn$^*$  &   5490                          &   source &  226.5 $\pm$ 0.2      & 0.082 \\
$^{218}$Po$^*$  &   6000                          &   source &  268.4 $\pm$ 0.2      & 0.089 \\
$^{220}$Rn  &   6290                          &   data    &  282.1 $\pm$ 3.8     & 0.089 \\
$^{216}$Po  &   6780                          &   data    &  338.6 $\pm$ 3.8     & 0.099 \\
$^{214}$Po  &   7690                          &   data    &  422.1 $\pm$ 3.8      & 0.109 \\
$^{214}$Po$^*$  &   7690                          &   source &  422.1 $\pm$ 0.2     & 0.109 \\
$^{212}$Po  &   8780                          &   data    &  548.1 $\pm$ 4.6     & 0.125 \\
\hline
\hline
\end{tabular}
\end{center}
\caption{Observed energies of $\alpha$ decays in $N_{pe}^d$ energy estimator. Source data (labeled with the asterisk) have been rescaled upwards by a factor $\simeq$1.8 such that the $^{214}$Po peak from the source and regular data match. The calibration source in fact showed an additional quenching, probably related to the source--assembly procedures. The last columns shows the quenching factor Q$_\alpha$ for events distributed in all the $^7$Be/$pep$--fiducial volume. }
\label{tab:alphaQval}
\end{table*}

\section{Fit of the energy spectra}
\label{sec:EnergyFit}

The fit of the measured energy spectra is performed using both the analytical and the Monte Carlo--based detector response functions, as described in Sections~\ref{sec:analytical} and~\ref{sec:MC}. In both approaches, the free fit parameters are the amplitudes of the solar--neutrino components and of the different backgrounds. In the Monte Carlo approach, once the simulation is correctly optimized, all detector--related parameters are intrinsically built--in. Instead, in the analytical approach, several parameters related to the energy scale are additionally left free in the fit. Thus, while the model for the detector energy response is determined analytically, the values of its parameters are determined by the data. This analytical approach has therefore more free parameters but has the advantage of accounting for the correlations among different parameters while estimating the systematic uncertainties. 

The global energy--scale parameters left free in the analytical fit are the FV--averaged detector photoelectron yield $Y_{det}^{pe}$ (see Eq.~\ref{eqn:Charge3}), and, depending on the choice of the energy estimator, some of the parameters of the response functions defined in Section~\ref{sec:analytical}, as the resolution parameters $\nu_T$ and  $\sigma_{ped}$, $\alpha$ and $\beta$ parameters of the generalized gamma function or $a$ and $b$ parameters of the modified Gaussian.
The parameters $A_i$ (Eq.~\ref{eqn:Qanal}) are fixed to allow the fitter to converge within a reasonable time. The position of the $^{210}$Po peak and the starting point of the $^{11}$C spectrum, with respect to the $\beta$--energy scale, are also left free in the fit, as the high rate and distinct spectral shapes of these components allow the fitter to determine these values more accurately directly from the normal data than from the source--calibration data. For other background components that include $\alpha$ and $\gamma$ emissions, such as $^{214}$Pb and $^{222}$Rn, their relative positions in the $\beta$ energy scale are fixed using the source--calibration data. 

The Monte Carlo--based fit approach requires the generation of energy spectra of solar neutrinos and all background components. These events have been generated with uniform spatial distribution in the IV and the same position--reconstruction algorithm as the one used for the data (see Section~\ref{sec:position}) has been used to select the events in the FV. Each run included in the analysis has been simulated individually and the number of simulated events in each run is proportional to the run duration. This ensures to weight the distribution of the working channels in the Monte Carlo in the same way as it is in the real data. The number of generated events is about 100 times higher than the typical number of events expected in the spectrum allowing to neglect the statistical fluctuations of the Monte Carlo spectra. The only exception is the $^{210}$Po.

The fit of the energy spectra was fully sufficient to extract the $^7$Be solar neutrino interaction rate (Section~\ref{sec:Be7results}). However, to extract the $pep$ and CNO solar neutrino results (Section~\ref{sec:pepCNOResults}), the multivariate fit, including apart the energy spectra also the PS--BDT and radial distributions, was developed. This fit approach is described in detail in Section~\ref{sec:Multivariate}.

\section{ Multivariate Fit}
\label{sec:Multivariate}

\begin{table*}[t]
\begin{center}
\begin{tabular}{lccll}
		\hline \hline
		Species & Rate                  & Common                & PS--BDT  & Rad. distrib. \\% & Notes  \\
		        & (free or fixed)       & to both spectra       &               &               \\
		\hline
		Solar neutrino  & & & & \\
\hline
		$pep$             & free & Yes             & $\beta^-$ & Bulk\\
		CNO             & free & Yes             &  $\beta^-$ & Bulk* \\
		$^7$Be          & free & Yes             &  $\beta^-$* & Bulk* \\
		$pp$              & fixed to 133 cpd/100\,ton  & Yes &  $\beta^-$* & Bulk* \\%& From High-Z solar model \cite{smodels-2011} \\
		$^8$B      & fixed to 0.49 cpd/100\,ton & Yes &  $\beta^-$ & Bulk \\%& From High-Z solar model \cite{smodels-2011} \\ 
		\hline
	            Background  & & & & \\
\hline
		$^{214}$Pb      & fixed to 1.95 cpd/100\,ton & Yes &  $\beta^-$ & Bulk* \\%& obtained from \bipofo\ fast coincidence search\\
		$^{210}$Bi      & free & Yes             &  $\beta^-$ & Bulk* \\
		$^{10}$C        & free & No              &  $\beta^+$ & Bulk \\
		$^{11}$C        & free & No              &  $\beta^+$ & Bulk\\
		Ext. $^{214}$Bi & free & Yes             &  $\beta^-$ & External\\
		Ext. $^{40}$K   & free & Yes             &  $\beta^-$ & External\\
		Ext. $^{208}$Tl & free & Yes             &  $\beta^-$ & External\\
		$^6$He          & free & No              &  $\beta^-$ & Bulk\\
		$^{40}$K        & free & Yes             &  $\beta^-$ & Bulk\\
		$^{85}$Kr       & free & Yes             &  $\beta^-$* & Bulk* \\
		$^{234m}$Pa     & free & Yes             &  $\beta^-$ & Bulk \\
		\hline \hline
\end{tabular}\end{center}
\caption{Background and neutrino species considered in the multivariate fit. The $pp$ and $^8$B  solar neutrino interaction rates have been fixed to the central values from the high--metallicity solar model including MSW--LMA (see Table~\ref{table:Rate}). The value for $^{214}$Pb was estimated to be $1.95\pm0.07$ cpd/100\,ton from the $^{214}$Bi -- $^{214}$Po coincidence rate. The third column refers to whether the rates for a species in both the TFC--subtracted and TFC--tagged spectra are a  parameter with the same value for both spectra (Yes) or they are left free to assume different values (No). The last two columns refer to the PS--BDT parameter and the expected radial distribution in the FV. The asterisk (*) denotes species that, due to the energy range considered for the fits in the PS--BDT or radial position dimensions, are effectively excluded from the corresponding fit.}
\label{tab:fitpep}
\end{table*}

The detection of $pep$ and CNO neutrinos is more challenging than the $^7$Be one, as their expected interaction rates are $\sim$10 times lower, only a few counts per day in a 100\,ton target. The $pep$ neutrino interaction rate and the limits of the CNO neutrino rate have been determined by extending the fitting procedure used to evaluate the $^7$Be--$\nu$ interaction rate (described in  Section~\ref{sec:EnergyFit}): the energy spectra were simultaneously fit together with the distribution of the PS--BDT parameter and with the radial distribution of events. We have used a multivariate approach based on the maximization of the {\em"total"} binned likelihood function $L_T(\vec \theta)$, which depends on a set of parameters $\vec \theta$ and is a product of four factors:

\begin{equation}
\begin{split}
& L_T(\vec \theta)  = \\
&= L_E^{{\rm TFC_{sub}}} (\vec \theta) \cdot L_E^{{\rm TFC_{tagged}}}(\vec \theta) \cdot
L_{{\rm BDT}} (\vec \theta) \cdot L_{{\rm Rad}} (\vec \theta),
\end{split}
\label{eq:liktotal}
\end{equation}
where $ L_E^{{\rm TFC_{sub}}} (\vec \theta)$ is the likelihood function of the energy spectrum obtained after applying the TFC method (see Subsection~\ref{subsec:tfc}) to reduce the $^{11}$C content; $ L_E^{{\rm TFC_{tagged}}} (\vec \theta)$ is the likelihood of the complementary energy spectrum containing events tagged by the TFC; $L_{{\rm BDT}} (\vec \theta)$ is the likelihood of the PS--BDT parameter, and finally $L_{{\rm Rad}} (\vec \theta)$ refers to the likelihood of the radial distribution.

The first two terms in the product of Eq.~\ref{eq:liktotal} are the standard Poisson likelihoods:
\begin{equation}
L_E^{{\rm TFC_{sub}}} (\vec{\theta}) = \prod^{n_e}_{i=1} \frac{\lambda_i(\vec{\theta})^{k_i} e^{-\lambda_i(\vec{\theta})}}{k_i!},
\label{eq:ll}
\end{equation}
where the product is over all the energy bins $i$, $n_e$ is the total number of energy bins, $\lambda_i(\vec{\theta})$ is the expected number of entries in the bin $i$ given the fit parameters $\vec{\theta}$, and $k_i$ is the measured number of entries in the bin $i$. A similar relation holds for $ L_E^{{\rm TFC_{tagged}}} (\vec \theta)$.

The two energy spectra (TFC--tagged and TFC--subtracted) are fit keeping the rate of the most part of the components in common. The only species whose rates are different parameters in the two energy spectra are of course $^{11}$C but also $^{10}$C and $^6$He, since their origin is cosmogenic and it may be correlated with neutron production. Table~\ref{tab:fitpep} shows the different solar--neutrino fluxes and backgrounds considered in the fit.

Two different energy estimators have been used to fit the energy spectrum to get the $pep$ and CNO neutrino results with the multivariate approach: $N_h$ and $N_{pe}^d$. The probability density function (PDF) for $N_h$ was produced with the Monte Carlo method while the one for $N_{pe}^d$ with the analytical method. 

The definition of the last two terms in $L_T(\vec \theta)$ in Eq.~\ref{eq:liktotal} considers that the PDFs of the corresponding variables are produced from the data ({\em e.g.}, the pulse--shape PDF for $\beta^-$ is taken from tagged $^{214}$Bi). The statistics are limited and there is no analytical model to produce precise multi--dimensional PDFs. Therefore, we have projected the events, integrated over an energy range larger than the energy--spectrum binning, into one--dimensional histograms of the PS--BDT and radial distribution variables and computed the corresponding likelihood. In this case, we introduce a correlation between the number of counts in different histograms, as events that are in the energy spectrum will also be entries in the projections.
To handle this issue, we normalize the PDFs of the hypothesis to the total number of entries in the projected data histograms to fit.
Consequently, we define the likelihood of the PS--BDT parameter as:
\begin{equation}
L_{{\rm BDT}}(\vec{\theta}) = \prod^m_{j=1} \frac{a\lambda_j(\vec{\theta})^{k_j}e^{-a\lambda_j(\vec{\theta})}}{k_j!} 
\label{LBDT}
\end{equation}
the scaling factor, $a$, enforces the normalization and is set such that
\begin{equation}
N = a \sum^m_{j=1} \lambda_j(\vec{\theta}),
\label{eqn:adef}
\end{equation}
where $N$ is the total number of entries in the projected histogram.
Here, $\lambda_j(\vec{\theta})$ represents the expected content of bin $j$ of the PS--BDT histogram, $k_j$ is the actual number of entries in that bin, and $m$ is the total number of bins of the PS--BDT histogram. 

$L_{{\rm Rad} }(\vec{\theta})$ is defined in a way similar to $L_{{\rm BDT}}(\vec{\theta})$.
The radial dependence is assumed uniform for all the species except the external background. The PDFs of the radial distribution of the external background and its energy dependence has been obtained with the Monte Carlo, as described in Section~\ref{sec:FV}.

We have performed Monte Carlo tests with data--like samples to show that the statistical interpretation of likelihood--ratio tests holds for our computed total likelihood.

\section{The $^7$Be--neutrino interaction rate}
\label{sec:Be7results}

The first measurement of the $^7$Be--$\nu$ interaction rate was published by Borexino after only few months of data taking~\cite{BxBe} and an update was reported in~\cite{BxBe2}. The accuracy of those measurements was
significantly  improved in 2011~\cite{be7-2011} using the results of the calibration campaign (see Section~\ref{sec:calibration}), a better understanding of the detector response, and increased statistics. The data were  collected in the period from May 16$^{\rm{th}}$, 2007 to May 8$^{\rm{th}}$, 2010 and they 
corresponds to 740.7\,live--days after cuts and to 153.6\,ton $\times$ year fiducial exposure. The resulting interaction rate of the 862\,keV $^7$Be line~\cite{be7-2011} is:
\begin{equation}
 R(^7{\rm Be}) = 46.0 \pm 1.5 ({\rm stat}) \, ^{+1.5}_{-1.6} ({\rm sys}) \, {\rm cpd/100\,ton}
\end{equation}
and its corresponding $\nu_e$--equivalent flux is (2.79 $\pm$ 0.13) $\times 10^9$\,cm$^{-2}$\,s$^{-1}$. The $\nu_e$--equivalent flux is calculated by assuming that the total observed interaction rate is due to electron flavor neutrinos only. Considering the 3-flavor neutrino oscillations, the equivalent flux is (4.43 $\pm$ 0.22) $\times 10^9$\,cm$^{-2}$\,s$^{-1}$, which can be compared with the expected SSM flux of Table~\ref{table:Rate}.

The $^7$Be--$\nu$ interaction rate has been obtained fitting only the energy spectra (Section~\ref{sec:EnergyFit}). The lower bound of the fit region was chosen to avoid pile--up between two $^{14}$C $\beta$ decays (Q$_\beta$ = 156\,keV) and it corresponds to 270\,keV. The higher bound of the fit region is 1250\,keV  in the analytical fit approach, in which the contribution of the external background ($^{208}$Tl, $^{214}$Bi) is not included. The Monte Carlo fit includes the simulated spectrum of the external background allowing to extend the fit region up to 1600\,keV.
                                         
The weights for the $^7$Be neutrino signal and the main radioactive background components ($^{85}$Kr, $^{210}$Po, $^{210}$Bi,
and $^{11}$C) were left as free parameters in the fit, while the contributions of the $pp$, $pep$, CNO, and $^8$B solar neutrinos were fixed to the GS98--SSM predicted rates assuming MSW--LMA neutrino oscillations (see Table~\ref{table:Rate}).
 %with $tan^2 \theta_{12} = 0.47 ^{+0.05}_{-0.04}$ and $\Delta m^2_{12} =   (7.6 \pm 0.2)\times 10^{-5}$~eV$^2$ as reported in Table~\ref{table:Rate}. 

The 384 and 862\,keV branches of the $^7$Be solar neutrinos (see Fig.~\ref{fig:SolarNuSpectrum}) are combined into a single spectrum. The production ratio between the two branches is 10.52 : 89.48. Accounting for the energy--dependent survival probability and interaction cross--sections, the ratio between the interaction rates is 3.9 : 96.1 . Similarly, we have combined the  $^{13}$N, $^{15}$O, and $^{17}$F recoil spectra into a single spectrum, referred to as the CNO solar neutrino spectrum. The rates of $^{222}$Rn, $^{218}$Po, and $^{214}$Pb surviving the cuts were fixed using the measured rate of $^{214}$Bi -- $^{214}$Po delayed coincidence events.

Due to the slight eccentricity $\varepsilon$ = 0.01671 of the Earth's orbit around the Sun, the flux $\Phi_E$ of solar neutrinos reaching the Earth is time dependent:
\begin{equation}
\Phi_E(t) = \frac {R_{{\rm Sun}}} {4 \pi r^2(t)} \simeq   \frac {R_{{\rm Sun}}} {4 \pi r_0^2} \left(1 + 2 \varepsilon cos \left (\frac {2 \pi t } {T} \right ) \right)
\label{eq:dist}
\end{equation}
where $R_{{\rm Sun}}$ is the neutrino production rate at the Sun, $t$ is the time in days from January 1$^{th}$, $T$ is one year, $r(t)$ is the time dependent Earth--to--Sun distance and $r_0$ is its mean value. We are interested in the neutrino flux averaged over one year, while the data acquisition periods are unevenly distributed over a few years time interval. We have calculated the expected flux for each period used in the data analysis using Eq.~\ref{eq:dist}. Thus, we have obtained the correction to be applied to convert the measured flux into the yearly averaged flux. The result is a multiplicative factor of 1.0003, a negligible correction within the accuracy of the present data set.
 %has to be applied to the measured neutrino flux during the periods used in the analysis in order to obtain the average annual flux.

\begin{figure}[t]
\begin{center}
\centering{\includegraphics[width = 0.5\textwidth]{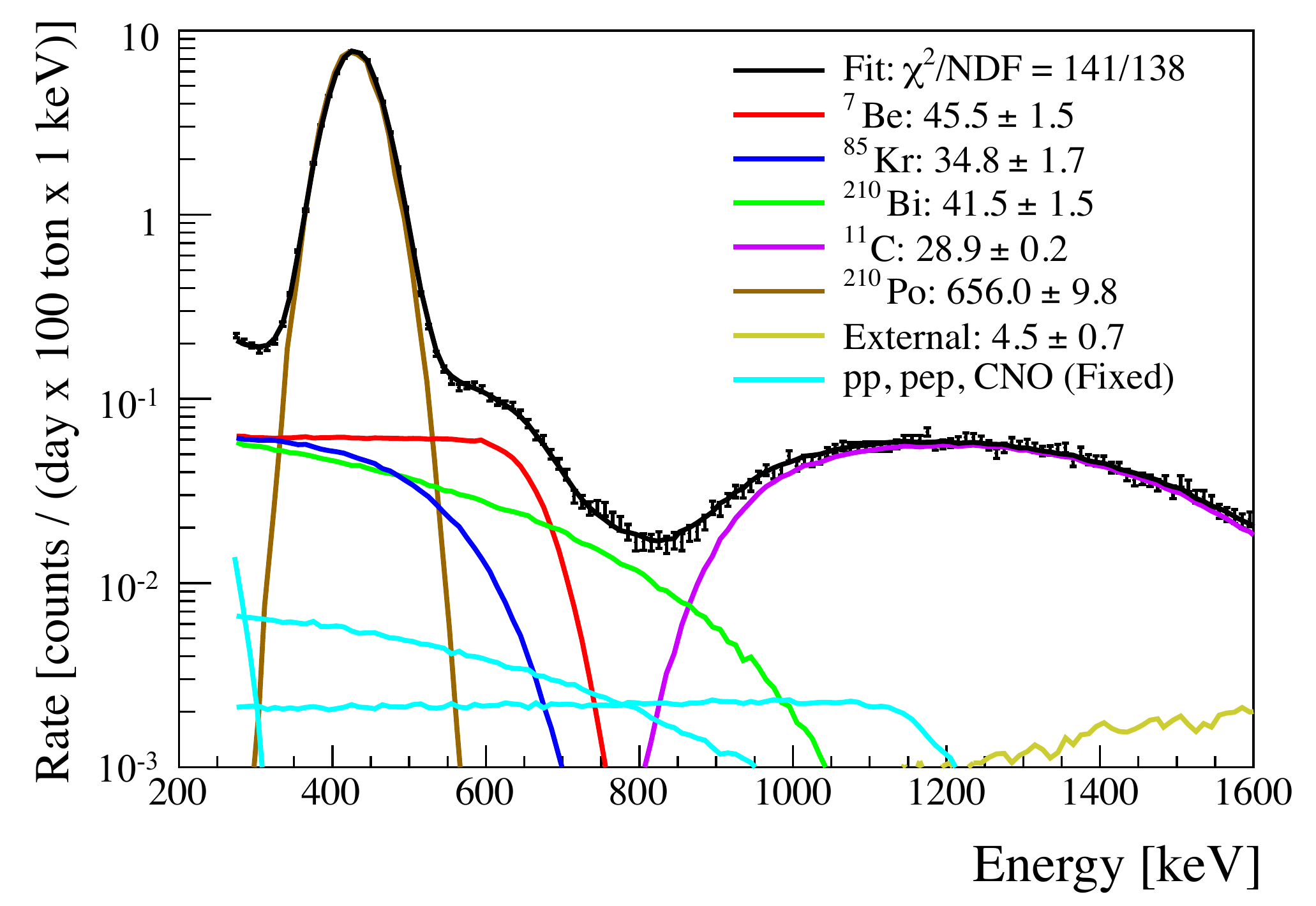}}
\end{center}
\caption{Example of fit of the energy spectrum obtained using the Monte Carlo method without $\alpha$ -- $\beta$ statistical subtraction. The fit was done using the $N_h$ energy estimator. After the fit, the horizontal axis was converted into energy scale in keV. The values of the best-fit parameters, the rates of individual species, are given in cpd/100\,ton.}
\label{fig:fitResults1}
\end{figure}

\begin{figure}[t]
\begin{center}
\centering{\includegraphics[width = 0.5\textwidth]{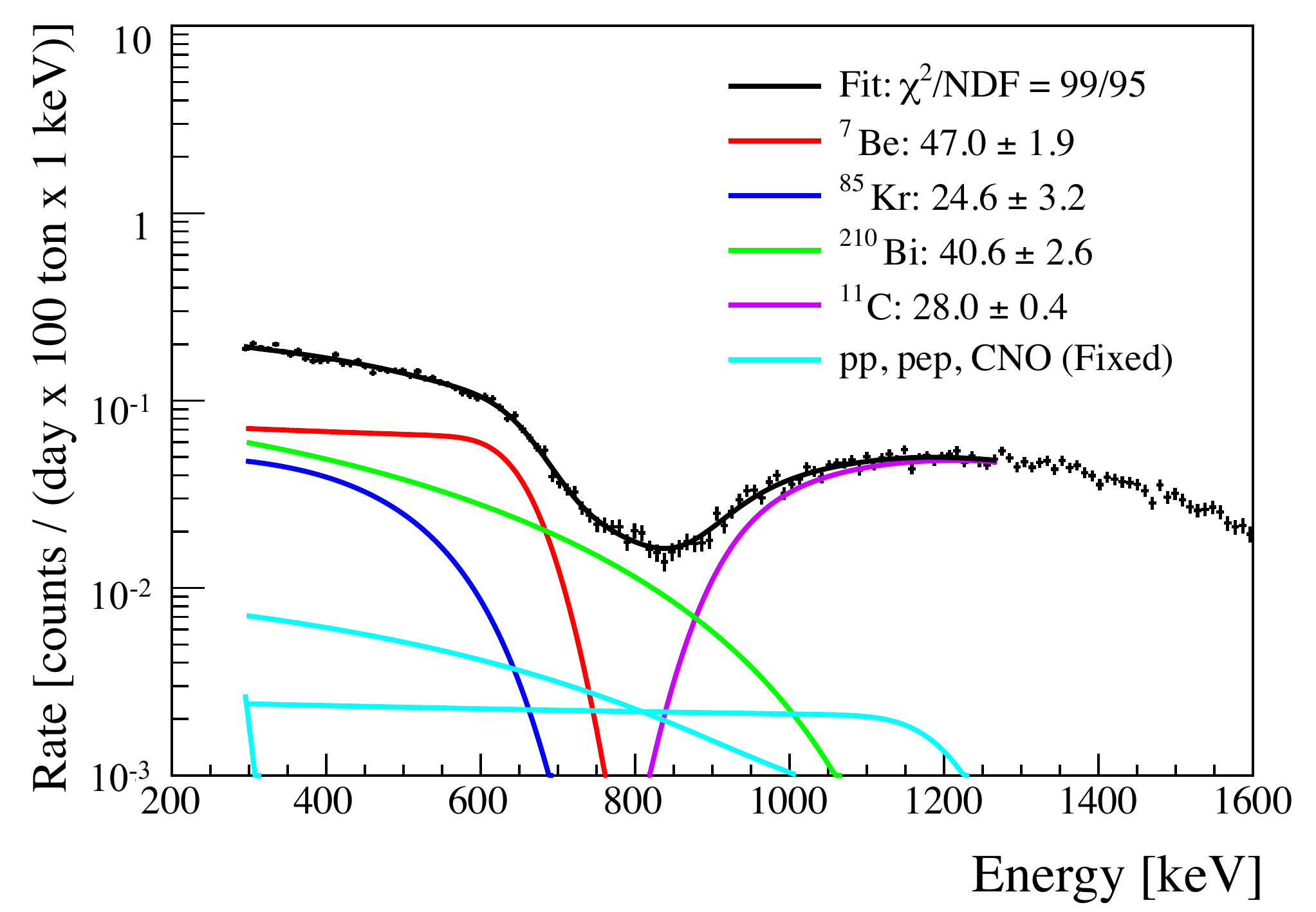}}
\end{center}
\caption{Example of fit of the energy spectrum obtained using the  analytical method with $\alpha$ -- $\beta$ statistical subtraction. The fit was done using the $N_{pe}^d$ energy estimator. After the fit, the horizontal axis was converted into energy scale in keV. The values of the best-fit parameters, the rates of individual species, are given in cpd/100\,ton.}
\label{fig:fitResults2}
\end{figure}

\begin{figure}[t]
\begin{center}
\hspace{10 mm}
\centering{\includegraphics[width = 0.5\textwidth]{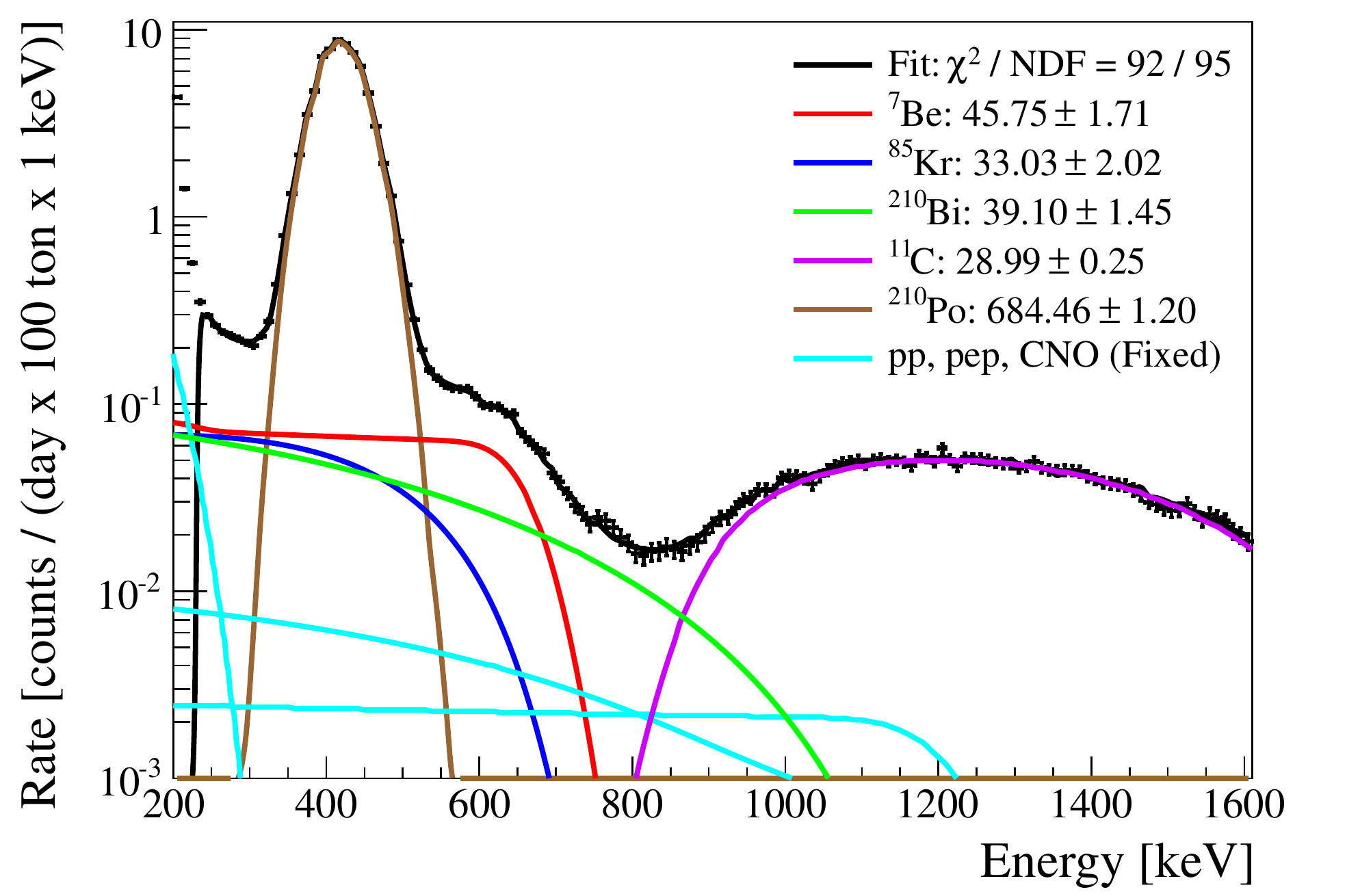}}
\end{center}
\caption{Example of fit of the energy spectrum obtained using the  analytical method without $\alpha$ -- $\beta$ statistical subtraction. The fit was done using the $N_p$ energy estimator. After the fit, the horizontal axis was converted into energy scale in keV. The values of the best-fit parameters, the rates of individual species, are given in cpd/100\,ton.}
\label{fig:fitResults3}
\end{figure}

\begin{figure}[t]
\begin{center}
\centering{\includegraphics[width = 0.5\textwidth]{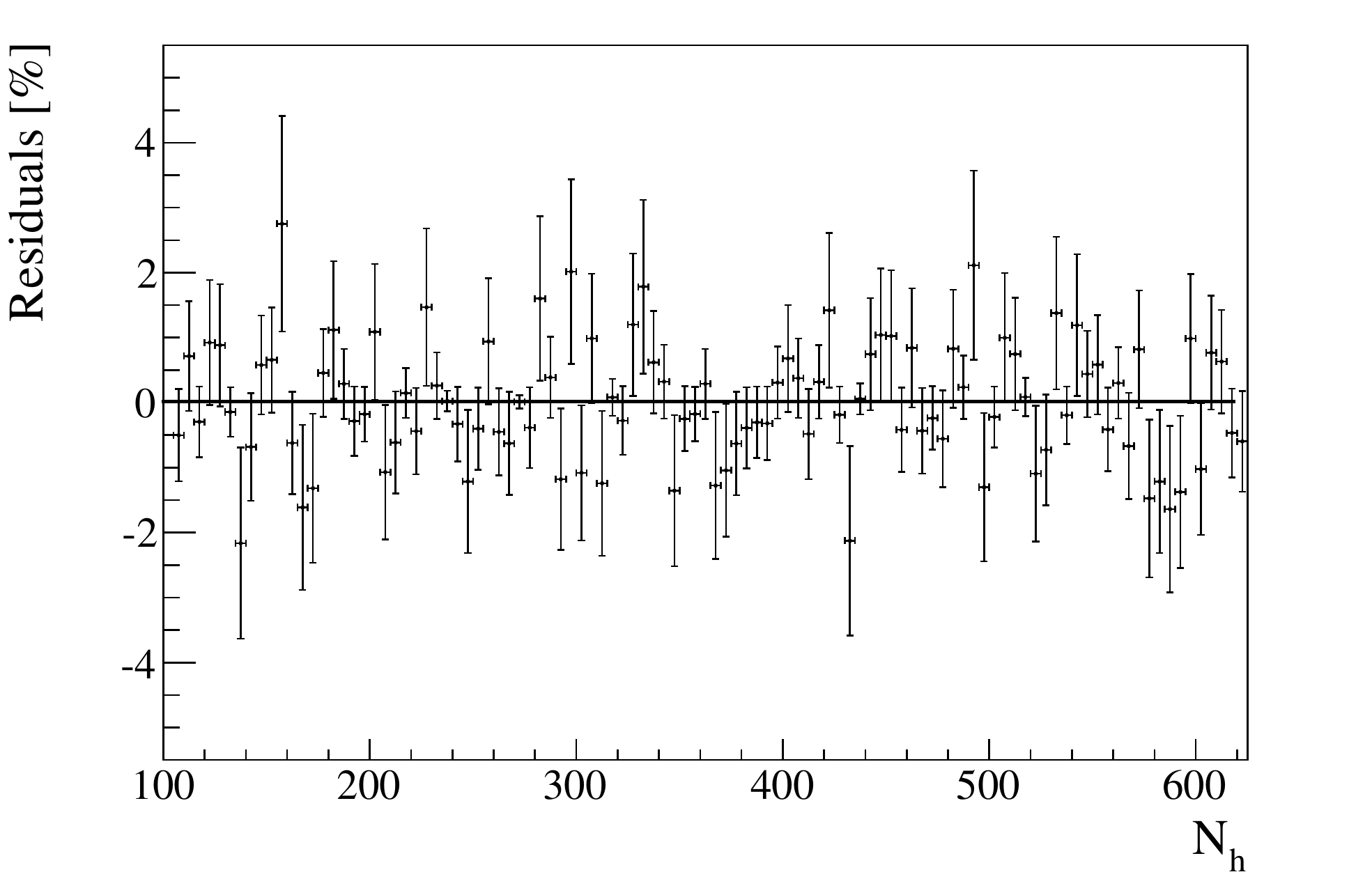}}
\caption{Typical example of the distribution of the residuals of the fit. This plot corresponds to the fit shown in Fig.~\ref{fig:fitResults3}. }
\label{fig:Be7Residual}
\end{center}
\end{figure}

All events accepted in the final energy spectra used in the fit have to pass the selection criteria discussed in Subsection~\ref{subsec:cuts}. As described in Section~\ref{sec:cuts}, the fit procedure has been implemented both with and without statistical subtraction of the $^{210}$Po--$\alpha$ peak (Section~\ref{sec:ab}). When statistical subtraction is not applied, the additional $G_{\alpha \beta}$--based energy--dependent cut described in Subsection~\ref{subsec:softAB} is used.
Figure~\ref{fig:fitResults1}, Fig.~\ref{fig:fitResults2}, and Fig.~\ref{fig:fitResults3} show some examples of fit results obtained using various procedures. Figure~\ref{fig:fitResults1} refers to the Monte Carlo fit without $\alpha$ -- $\beta$ statistical subtraction. The fit is performed by minimizing the $\chi ^{2}$  between the measured and Monte Carlo generated spectra using the $N_h$ energy estimator. Finally, after the fit procedure, the plot is transformed in the energy scale in keV. Similarly, Fig.~\ref{fig:fitResults2} shows an example of analytical fit using the energy estimator $N_{pe}^d$ with $\alpha$ -- $\beta$ statistical subtraction. The plot in Fig.~\ref{fig:fitResults3} demonstrates the fit using the $N_p$ variable based on the analytical approach without $\alpha$ -- $\beta$ statistical subtraction while Fig.~\ref{fig:Be7Residual} shows its corresponding residuals.
 
The  shape of the $^{85}$Kr energy spectrum and the one due to the electron recoil following a $^7$Be--$\nu$ interaction are very similar, as can be seen by comparing the blue and red curves in  Fig.~\ref{fig:fitResults1}, Fig.~\ref{fig:fitResults2}, and Fig.~\ref{fig:fitResults3}. These two fit components are correlated and their relative weight is influenced by details of the energy scale. The amount of $^{85}$Kr returned by the fit is also sensitive to the count rate in the low--energy portion of the spectrum at the beginning of the fit region. These effects translate in a dependence of the resulting $^{85}$Kr rate on the fit procedure (analytical or Monte Carlo) and,  particularly, on the use or not of the $\alpha$ -- $\beta$  statistical subtraction. The statistical subtraction procedure is generally the one giving the lowest krypton count rate. The similarity of the spectrum of $^7$Be--$\nu$ and $^{85}$Kr produces a systematic uncertainty in the determination of the $^7$Be--$\nu$ interaction rate. However, the absolute value of the 
uncertainty associated to the $^7$Be--$\nu$ interaction rate is smaller than the one associated to the $^{85}$Kr: the reason is that  the determination of the $^7$Be--$\nu$ interaction rate is also constrained by the energy region between 550 and 750\,keV where the weight of $^{85}$Kr is significantly reduced. The accuracy of the $^{85}$Kr direct measurement obtained with the rare delayed coincidence branch (see Subsection~\ref{subsec:Kr85} ) is not sufficient to constrain the weight of the krypton in the fit. The different fit approaches produce slightly different values for the $^{85}$Kr rate; all these values are self consistent and consistent with the direct measurement. The results displayed  in the Fig.~\ref{fig:fitResults1}, Fig.~\ref{fig:fitResults2}, and Fig.~\ref{fig:fitResults3} clearly show this effect.
Table~\ref{tab:Be7Back} summarizes the results about the background rates obtained by the fit of the energy spectra.
\begin{table}[!b]
\begin{center}
\caption{Background rates obtained fitting the energy spectra used to measure  the $^7$Be neutrinos interaction rate.}
\begin{tabular}{lc}
\hline\hline
Species                                  &rate [cpd/100 ton]\\
\hline
$^{85}$Kr			&31.2$\pm$1.7$({\rm stat})$$\pm$4.7$({\rm syst})$ \\
$^{210}$Bi			&41.0$\pm$1.5$({\rm stat})$$\pm$2.3$({\rm syst})$ \\
$^{11}$C		&28.5$\pm$0.2$({\rm stat})$$\pm$0.7$({\rm syst})$ \\
\hline\hline
\end{tabular}
\label{tab:Be7Back}
\end{center}
\end{table}

The CNO--$\nu$ and $^{210}$Bi spectra are very similar. This trend is  weakly influencing the $^7$Be--$\nu$ interaction rate measurement.

All the available fit methods have been used to study these and other systematic effects on the $^7$Be--$\nu$ rate. The evaluation has 
been performed  repeating the fit procedure many times by varying one parameter and fixing all the others, and then repeating this procedure for all the relevant parameters. These are the binning, the choice of the energy estimator, the energy range used in the fit, the use or not of the  $\alpha$ -- $\beta$ statistical subtraction and the energy region where this procedure is applied, the exact values of the fixed components of the neutrino spectra varied within the theoretical uncertainties, and the amount of residual radon correlated background rates inferred from the $^{214}$Bi -- $^{214}$Po coincidence rates.

The energy scale is a free fit parameter in the analytical method while it is fixed in the Monte Carlo method. In the Monte Carlo method the uncertainty of the fit results originated by the one of the energy scale has been studied by repeating the fit using Monte Carlo energy spectra obtained with the $\beta$ energy scale changed by $\pm 2 \%$. It results that changes of the energy scale larger than $\pm 1.5 \%$ produce fit with not acceptable $\chi^2$ and they were discarded.

We have built the distributions of all the fit results obtained by scanning the values of the above listed parameters with all the fit methods; we have discarded those fits producing  a non acceptable $\chi^2$ and then we considered as systematic uncertainty the $rms$ of the resulting distribution. The systematic effect due to the uncertainty in the energy scale when all the remaining parameters are kept fixed at their best values produces a systematic uncertainty of $2.7\%$ on the $^7$Be--$\nu$ interaction rate. The contribution of all other listed effects is included in the Table~\ref{tab:Erbe} as "Fit Methods" and it amounts to $2\%$.

The additional important source of systematic uncertainty is the knowledge of the FV since it determines the target mass and, thus, the neutrino interaction rate calculation. Details about this item and the accuracy of the FV uncertainty obtained using the calibration sources are described in~\cite{BxCalibPaper}. Here we only recall that the uncertainty about the FV definition has been evaluated by selecting the source data corresponding to source positions at the border of the $^7$Be--FV. For this data set, the distributions of $\Delta R$ and $\Delta z$, i.e. the difference between the reconstructed and the nominal value of the radius and of the vertical coordinate, were calculated. The FV systematic uncertainty results from the comparison between the nominal value (86.01\,m$^3$, see Table~\ref{tab:FV}) and the values obtained by varying $R$ and $z$ between the minimum and maximum $\Delta R$ and $\Delta z$. Based on this, the FV contribution to the total systematic uncertainty budget of the $^7$Be neutrino rate is +0.5\% and -1.3\%. The systematic shift of 4\,cm in the $z$ direction described in Section~\ref{sec:position} has a negligible impact on the selected FV, i.e. less than 0.01\%.

\begin{table}[t]
\begin{center}
\begin{tabular}{lc}
\hline\hline
Source                                  &Value [\%] \\
\hline
Trigger efficiency and stability        &$<$0.1 \\
Live--time                              &0.04 \\
Scintillator density                    &0.05 \\
Sacrifice of cuts                       &0.1 \\
Fiducial volume                         &$^{+0.5}_{-1.3}$ \\
Fit methods                             &2.0 \\
Energy response                         &2.7 \\
\hline
Total Systematic Uncertainty                  &$^{+3.4}_{-3.6}$ \\
\hline\hline
\end{tabular}
\caption{Systematic uncertainties of the $^7$Be solar neutrino rate measurement.}
\label{tab:Erbe}
\end{center}
\end{table}

The live--time for each run is calculated very precisely by taking the time difference between the first and the last valid trigger. The trigger time is obtained from a GPS clock having 100\,ns accuracy. However, there are additional sources of systematic uncertainties related to the live--time evaluation. As it results from Table ~\ref{tab:Erbe}, the overall  value of this uncertainty is small: the $0.04\%$ is dominated by the contribution associated to the 300\,ms dead time vetoing the detector after each ID muon. The uncertainty due to all the electronics and DAQ dead times amounts to only few $10^{-3} \%$ and other cuts which involve vetoing sections of the detector for varying periods of time give a contribution of the same order.

Table~\ref{tab:Erbe} summarizes all the systematic uncertainties described above.

\section{Search for a day--night asymmetry in the $^7$Be--neutrino interaction rate}
\label{sec:DN}

We have searched for a possible asymmetry between the day and night $^7$Be--solar--neutrino interaction rates.
As discussed later in Section~\ref{sec:GbAn}, this asymmetry is expected in particular regions of the oscillation parameters or it could be a signal for non--standard neutrino interaction.

The day--night asymmetry A$_{dn}$ of the $^7$Be--$\nu$ count rate is defined as:
\begin{equation}
A_{dn} = 2~\frac{R_N - R_D}{R_N + R_D} = \frac{R_{\mathrm{diff}}}{<R>},
\label{eq:Adn}
\end{equation}
where $R_N$ and $R_D$ are the night and day $^7$Be--neutrino interaction
rates, $R_{\mathrm{diff}}$ is their difference, and $<R>$ is their mean. 

With the data collected in the same period used to measure the $^7$Be--$\nu$ interaction rate, we have found a result well
consistent with absence of asymmetry~\cite{DNLetter}: 
\begin{equation}
A_{dn}=-0.001 \pm 0.012({\rm stat}) \pm 0.007({\rm sys}).
\end{equation}

\begin{figure}
\begin{center}
\centering{\includegraphics[width = 0.5\textwidth]{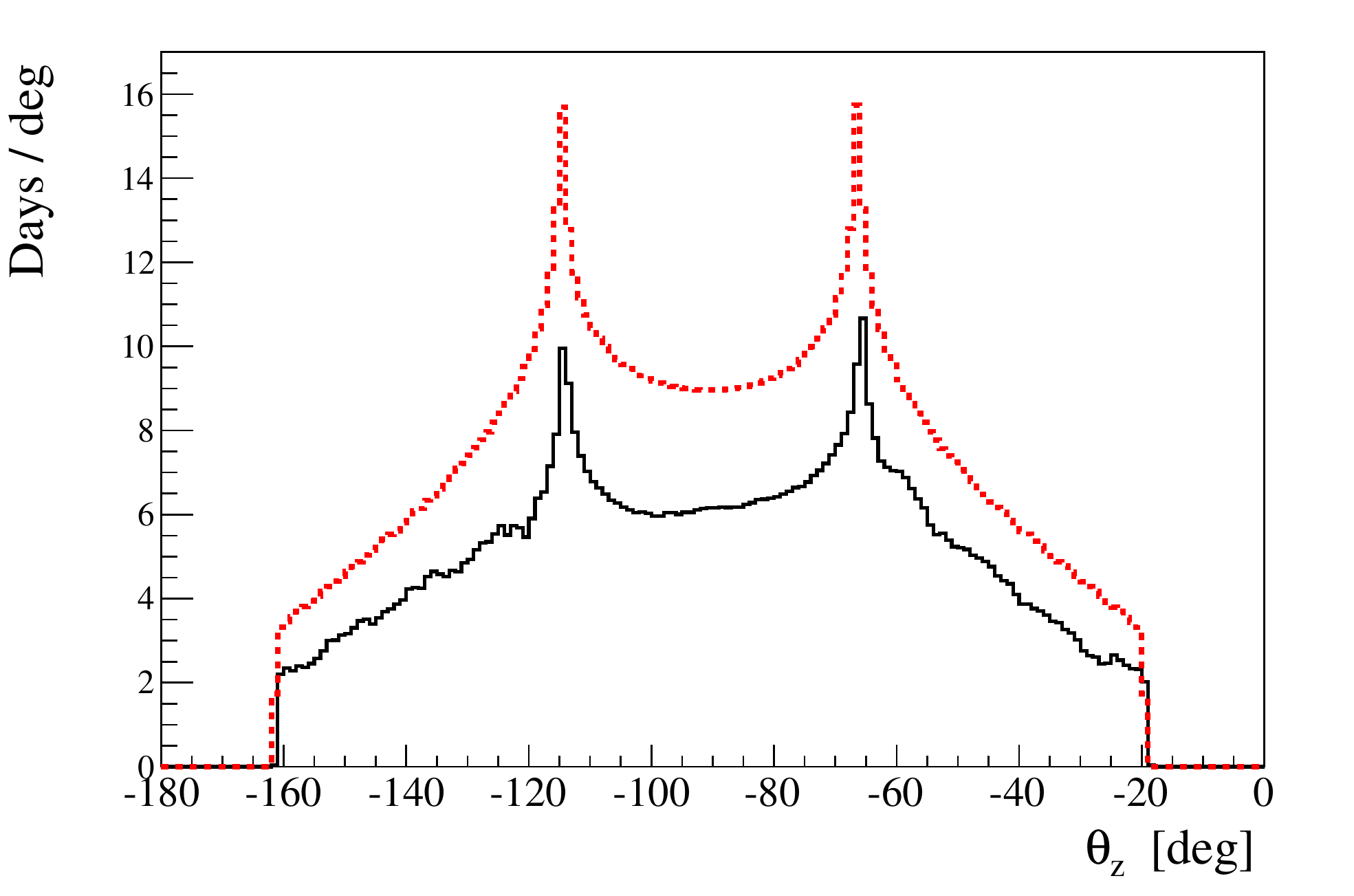}}
\caption{The experimental exposure function (black continuous line) and the ideal exposure function (red dotted line) corresponding to 3 years of data taking without interruptions at LNGS, as functions of the $\theta_z$ angle (1 deg/bin).
The interval from $-180^\circ$ to $-90^\circ$  corresponds to day (360.25 days) and the one from $-90^\circ$ to $0^\circ$ to night time
(380.63 days). We recall that at LNGS latitude the Sun is never at the zenith.}
\label{fig:exposure}
\end{center}
\end{figure}

The data have been classified as belonging to day or night according to the value of the angle $\theta_z$ between the vertical $z$--axis of the detector (positive upwards) and the vector pointing from the Sun to the detector,
 following the definition of \cite{Kamiokande}.
During the day $\theta_z$ is in the interval from $-180^\circ$ to $-90^\circ$ while during the night it is in the interval from $-90^\circ$ to $0^\circ$. Our day and night live--times are 360.25 and 380.63 days, respectively.
We have built the distribution of the $\theta_z$ corresponding to the live--time (experimental exposure function).
This is shown in black continous line in Fig.~\ref{fig:exposure} and compared with the ideal exposure  (red dotted line) corresponding to a data taking period of three years without interruptions. The experimental exposure correctly accounts for any interruption in the data taking and it is slightly asymmetric.
The shape of the distribution of the events as a function of $\theta_z$ should match exactly the experimental exposure function 
if none day--night asymmetry of neutrino rate is present. Equivalently, if there is no day--night  asymmetry, the distribution of the data as a function of $\theta_z$ normalized to the experimental exposure function should be flat.

\begin{figure}[!ht]
\begin{center}
\centering{\includegraphics[width = 0.5\textwidth]{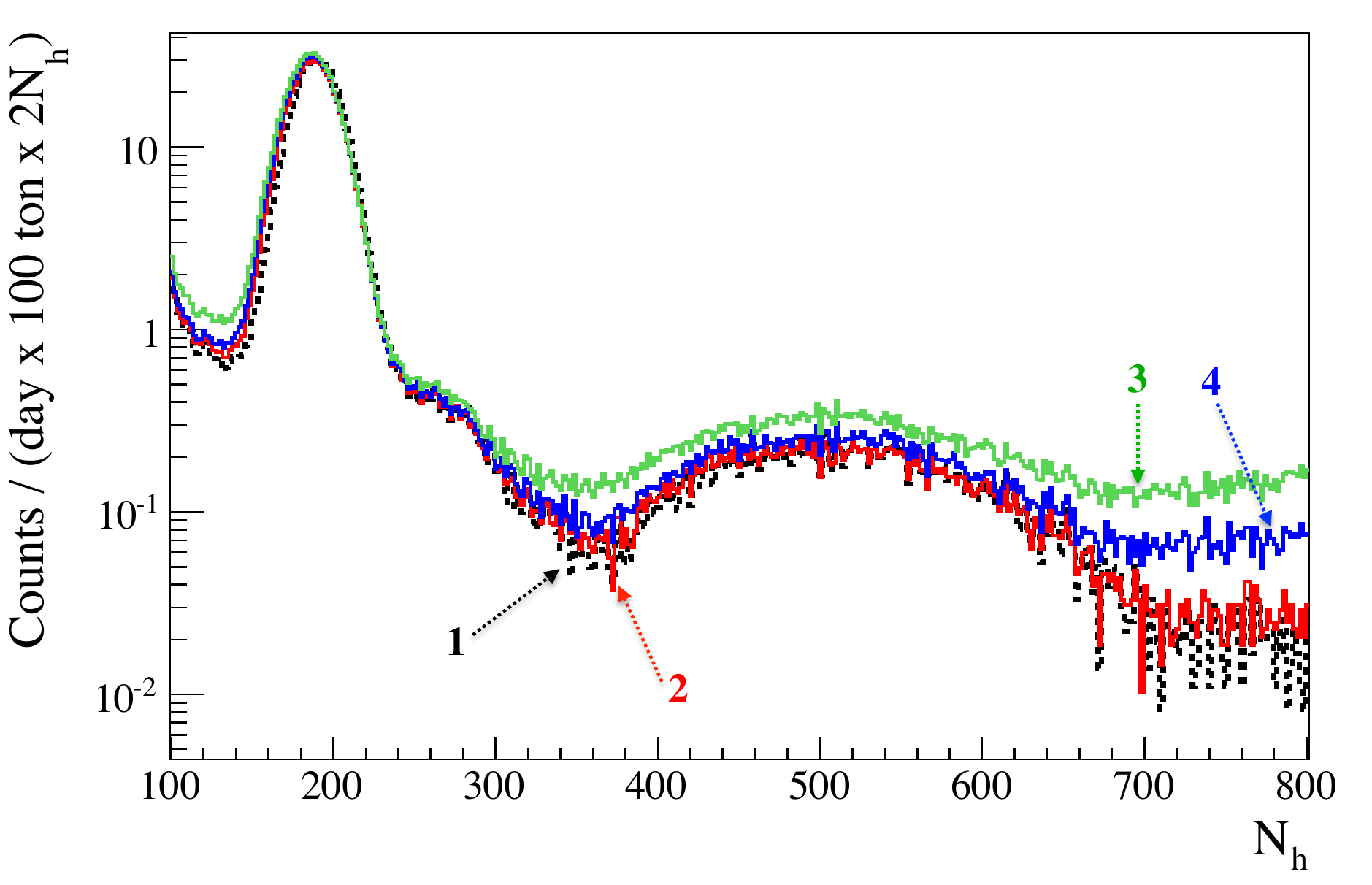}}
\caption{Energy spectra ($N_h$) during the day time in different volumes. The two curves showing the lowest count rate and almost superimposed have been obtained selecting events within the standard FV used for the $^7$Be--$\nu$ rate analysis (1 - black dotted line) and within a sphere of 3\,m radius (2 - red solid line). The curve with the highest count rate (3 - green) shows the events selected within a 3.5\,m radius sphere while the curve with the intermediate count rate (4 - blue) refers to a 3.3\,m radius sphere.}
\label{fig:CompareSpectra}
\end{center}
\end{figure}

\begin{figure}[t]
\begin{center}
\centering{\includegraphics[width=0.5\textwidth]{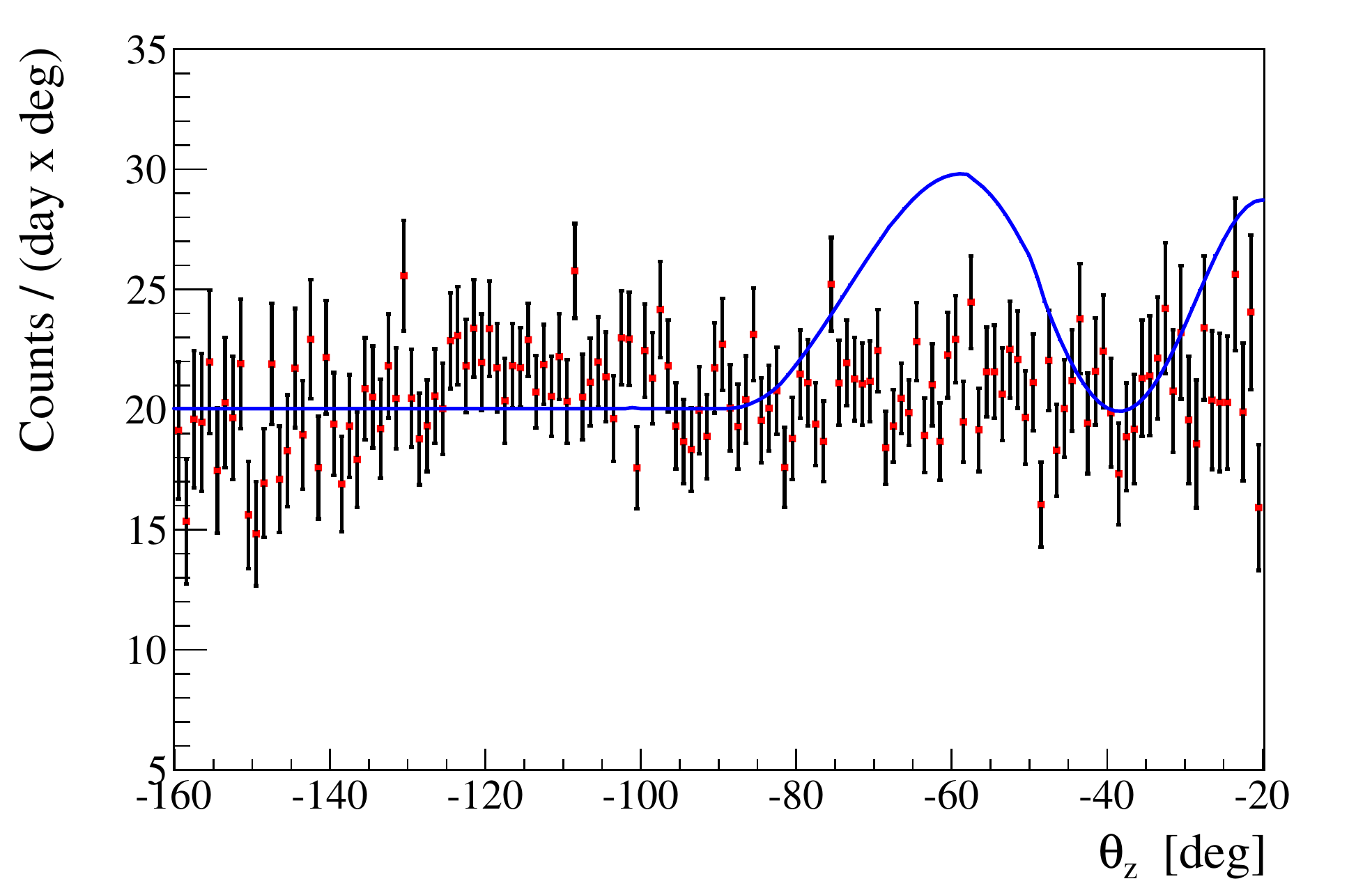}}
\end{center}
\caption{Normalized $\theta_z$--angle distribution of the events in the $^7$Be--$\nu$ energy window and reconstructed within  the enlarged FV. The effect of the Earth elliptical orbit has been removed.  The fit with a flat straight line yields $\chi^2$/NDF = 141.1/139. The blue solid line shows the expected effect with the LOW solution $\Delta m^2_{12}= 1.0 \cdot 10^{-7}$ $eV^2$  and tan\,$\theta^2_{12}= 0.955$. }
\label{fig:ZenithNeutrino}
\end{figure}

While the data--taking period is the same as for the $^7$Be--$\nu$ interaction rate analysis, the fiducial volume here used is larger:
we have selected   the events whose position is reconstructed in a spherical fiducial volume of 3.3\,m radius in order to increase the size of the data sample.
This FV corresponds to 132.50\,ton fiducial mass containing $4.382 \times 10^{31}$\,$e^-$ (a factor 1.75 larger than the $^7$Be--FV).
The additional external background that enters in the spectrum is not expected to be different during the day and night time.
Figure~\ref{fig:CompareSpectra} compares the energy  spectrum ($N_h$ energy estimator) obtained selecting events in different volumes
and, as an example, during the day time. The change of the shape of the spectrum is mostly due to the contribution of the external background
(higher at larger radius).
The energy region where the signal--to--background is maximal is the interval 550 -- 800\,keV ($N_h$ in the interval 244 -- 348).
The number of events falling in this energy window has been  plotted as a function of $\theta_z$ and then this resulting distribution has been normalized to the experimental exposure function (black solid line from Fig.~\ref{fig:exposure})  obtaining the result shown in Fig.~\ref{fig:ZenithNeutrino}.  
Note that the experimental exposure function has been corrected to take into account the change of the neutrino flux due to the annual variation of the Earth--Sun distance: in case of slightly different day and night life--times during the year this annual variation could mimic a fake day--night effect. In our conditions this effect increases the $^7$Be count rate by 0.37 $\%$ during the day and
it decreases it by 0.39 $\%$ during the night.
The fit with a straight line of the data of Fig.~\ref{fig:ZenithNeutrino} gives a $\chi^2$ probability of 0.44 demonstrating  that the two samples are statistically identical. We conclude that the rate of the events in the $550-800$\,keV energy window including both the background and the $^7$Be solar neutrino induced events is consistent with the hypothesis of no day--night effect. A similar result is obtained using the events in the smaller fiducial volume used for the $^7$Be--$\nu$ rate analysis. 

Note that a similar  plot referred to the energy region dominated by the cosmogenic $^{11}$C (800 -- 1600\,keV) when fitted with
a constant line it returns a bad $\chi^2$ ($\chi^2/$NDF = 216/141) indicating that the rate of the events as a function of $\theta_z$ does not follow the experimental exposure function of Fig.~\ref{fig:exposure}. This is not surprising since  $^{11}$C has a cosmogenic origin. The annual modulation of the muons
has been discussed in~\cite{BxMuons} and it is not related to the Earth--Sun distance.

\begin{figure}[t]
\begin{center}
\centering{\includegraphics[width=0.5\textwidth]{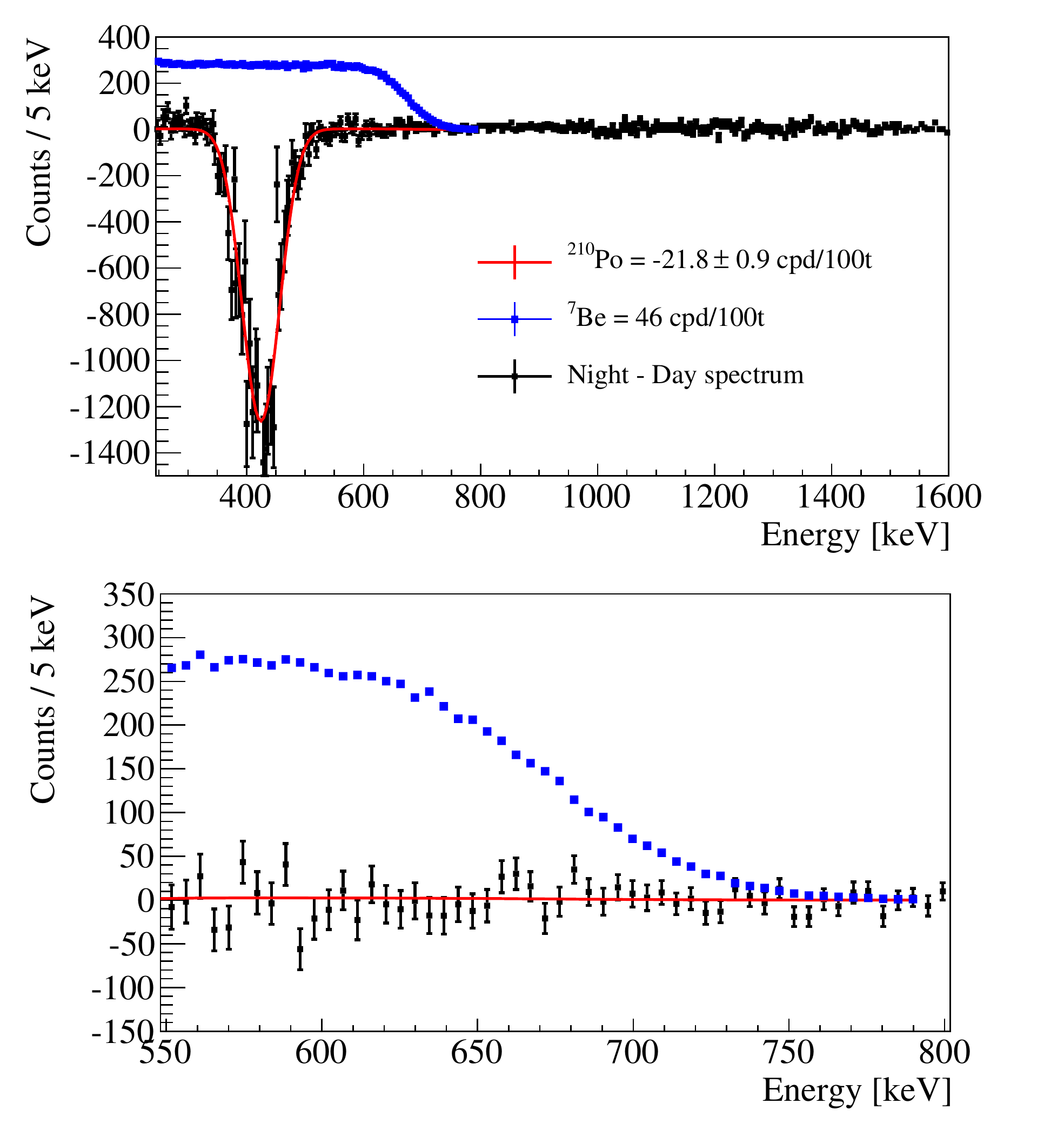}}
\caption{Difference of the night and day spectra in the enlarged FV.  The top panel shows an extended energy range including the region dominated by the $^{11}$C background while the bottom panel is a zoom  in the $^7$Be--$\nu$ energy window. The fit is performed in the $^7$Be--$\nu$ energy region (between 250 and 800\,keV) with the residual $^{210}$Po spectrum and the electron recoil  spectrum due to the $^7$Be solar neutrino interaction. The blue curve plots the $^7$Be--$\nu$ spectrum used in the fit with the amplitude $<R>$ = 46 cpd/100\,ton.
The residual $^7$Be--$\nu$ resulting from the fit is too small to be shown. } 
\label{fig:Diff_energy}
\end{center}
\end{figure}

The asymmetry of the neutrino signal alone and thus the A$_{dn}$ value is determined with limited precision by fitting the day and
night spectra separately. The most sensitive way to extract A$_{dn}$ is obtained by $i)$ assuming that the main background like $^{85}$Kr and $^{210}$Bi
are the same during the day and during the night, $ii)$ subtracting the day and night spectra properly normalized to the same
life--time and $iii)$ searching for a residual component $R_{\rm diff}$ having the shape of the electron recoil due to $^7$Be neutrinos in the resulting spectrum (following  the second term in Eq.~\ref{eq:Adn}).

The subtraction produces a flat spectrum consistent with zero except in the region of the $^{210}$Po peak as shown in Fig.~\ref{fig:Diff_energy}. 
The peak arises from the combination of the decay of the $^{210}$Po
background ($\tau_{1/2}$ = 138.38\,days) with the distribution of
the day and night live--time during the 3 years of data taking. The $^{210}$Po
count rate was highest at the time of the initial filling in May 2007, and has since
decayed. Therefore, the $^{210}$Po count rate has been overall higher during the
summers (when days are longer), leading to a noticeable effect in the
subtracted spectrum. 
The spectrum of the difference has been fitted to obtain the residual $^{210}$Po decay rate and the $R_{\rm diff}$ value for the $^7$Be--$\nu$ 
interaction rate. The fit  between 250 and 800\,keV gives $R_{\rm diff}$ = 0.04 $\pm$ 0.57 (stat)\,cpd/100\,ton.
There are only two  major sources of systematic uncertainties that contribute to the final result: the fit procedure and the variation of the $^{210}$Bi content with time.
We have repeated the analysis fitting  the spectrum of the difference of the day and night counts obtained after having applied the statistical subtraction of the   $^{210}$Po separately from the day and night spectra. In this case we only have $R_{\rm diff}$
as a single fit component. Then we also repeated the analysis using the data of different periods corresponding to different mean values
of the $^{210}$Bi decay rate. The $^{210}$Bi rate changed smoothly during the data taking time and in principle it should not 
produce a significant day--night asymmetry unless for effects due to not evenly distributed day and night live--times.

Table~\ref{tab:DNSys} reports different contributions to the systematic uncertainty.
The one associated to the fiducial volume does not enter in the determination of $R_{\rm diff}$. The energy scale in--determination
that may affect the shape of the $^7$Be--$\nu$ recoil spectrum produces negligible effects.

\begin{table}[b]
\begin{center}
\begin{tabular}{lc} \hline \hline
Source of uncertainty     &    Uncertainty on $A_{dn}$  \\ \hline
Live--time  &  $<\,$5$\cdot$10$^{-4}$ \\
Cut efficiencies  & 0.001 \\
Variation of $^{210}$Bi with time & $\pm$0.005 \\ 
Fit procedure & $\pm$ 0.005 \\ \hline
Total systematic uncertainty & 0.007 \\ \hline \hline
\end{tabular}
\end{center}
\caption{List of systematic uncertainties on $A_{dn}$.}
\label{tab:DNSys}
\end{table}

\section{Annual modulation of the $^7$Be--neutrino interaction rate}
\label{sec:annmod}

We present here novel results about the search for the annual modulation of the $^7$Be--$\nu$ interaction rate 
induced by the annual variation of the distance between the Earth and the Sun.
Similar results for $^8B$ solar neutrinos have been reported in ~\cite{SNOSeas},~\cite{SKSeas}.
The flux of neutrinos reaching the detector is expected to sinusoidally vary versus time with one year period  according to 
Eq.~\ref{eq:dist} and with a peak--to--peak amplitude of $\simeq$7$\%$.

\subsection{Analysis approach}

The spectral--fit analysis developed to measure the total average $^7$Be--$\nu$ interaction rate does not work well to search for its annual modulation. The main reason is that, with only three years of data, the statistics is not sufficient to allow independent spectral fits in sub--periods that are at the same time long enough to give meaningful fits and sparse enough to yield a good sensitivity to the modulation.

For this reason we have implemented three alternative analysis approaches, optimized for their sensitivity to the modulation. 

In all of them, the starting point is the definition of a set of time bins $t_k$ and of the corresponding normalized event rate $R(t_k)$, obtained by selecting all events falling within a given energy window and by applying a proper time normalization.

 In the first approach (fit of the rate versus time)  we fitted $R(t_k)$
   as a function of time searching for the sinusoidal signal of Eq.~\ref{eq:dist}.

The second approach consists of using  the  Lomb--Scargle method \cite{bib:Lomb}, \cite{bib:Scargle} to extract the periodical signal from $R(t_k)$.
 The Lomb--Scargle method is an extension of the Fast Fourier Transform, well suited in our conditions since it allows to account for data sample not evenly distributed in time (there are in fact  time gaps in the Borexino data taking) and it  determines the statistical significance of the identified periodicities.

The third method is the Empirical Mode Decomposition (EMD) \cite{bib:Huang98}. 
EMD decomposes a given signal into time--dependent components called \emph{intrinsic mode function} (IMF) \cite{bib:Wang2010},
which form a quasi--orthogonal and complete set.
The method provides, as in the case of the  Fast Fourier Transform, a global power spectrum by 
summing the instantaneous frequencies
of  each IMF weighted by the square average of the corresponding amplitude. 
The amplitude and the phase can be thought of as a distribution of instantaneous information contained in each IMF.

The IMFs (also called modes) are extracted from the original function through an iterative procedure (\emph{sifting} 
algorithm). 
The basic idea is to interpolate at each step the local maxima and minima of the initial signal, 
calculate the mean value of these interpolating functions, and subtract it from the initial signal. Then, we repeat the same procedure also on the residual signal (so after the relative subtraction) until suitable stopping criteria are satisfied.
These latter (slightly different in literature according to each approach, e.g.~\cite{bib:Huang98}, \cite{bib:Rilling2003}) are numerical conditions fixed to give the IMFs two general features, in common with the harmonic functions: first, the number of extrema (local maxima and minima) 
has to match the number of zero crossing points or differ from it  at most by one; second, the mean value
of each IMF must be zero.

The $i-$th IMF obtained by the $k-$th iteration is given by 
\begin{equation}
   { \rm IMF}_i(t)  = x_i(t)  - \sum_{j = 1}^k m_{ij} \, ,		   
\end{equation} 
where $x_i(t)$ is the residual signal when all "$i-1$" IMF's have been subtracted from the original signal R(t),
$x_0(t)=R(t)$, and the $m_{ij}$ are the average of the max and min 
envelopes at each $j$--th iteration.
Following the results of  a detailed study performed with   simulations, we have fixed the number of sifting iteration to 20, 
instead of around 10 as suggested in \cite{bib:WuHuang2004}. The number
20 guarantees a better symmetry of the IMF  with respect to its mean value, preserving the dyadic 
property of the method (that is each IMF has an average frequency that is the half of that of  the
previous one, see \cite{bib:WuHuang2010}).
 
In order to avoid meaningless or negative quantities, 
the instantaneous frequency, the amplitude, and the phase distributions are extracted from
the IMF's by means of the Normalized Hilbert Transform~\cite{bib:WuHuang2009-IF}.

Having a signal $R(t)$ sampled in time as in our case ($R(t_k)$), the maximum number  of IMF extracted 
(called number of modes $N_{\rm modes}$) is related to the maximum number of time bins $n_{\rm bins}$ through
\begin{equation}\label{eq:nyquist}
   N_{\rm modes} =\lfloor \log_{2}(n_{\rm bins}/2) \rfloor, 
\end{equation} 
where the $\lfloor x \rfloor$ operator represents the integer part of the real number $x$.
It is worth to point out that the first modes absorb the statistical fluctuations, while the latest ones contain
the low--frequency components of $R(t)$. In particular, the last IMF is the total trend of the data--set
and could contain relevant information about the change of the background contamination during  time. 
Since the EMD behaves as a dyadic filter, in general a given frequency $\nu$ is contained between the mode $i$ and $i$ + 1:
\begin{equation} 
N_{IMF}< -\log_2(\nu) < N_{IMF}+1 \, .
\label{eq:NIMFfreq} 
\end{equation}

The EMD approach shows two main issues: first, the method is strongly dependent on small changes of
the initial conditions; second, mode mixtures could occur for a physical component present
in the data--set especially when the ratio between signal and background is low (about 0.2, in our case).
In order to fight these problems, a white noise can be added to the
signal (\emph{dithering}) several times taking the average of all the IMFs extracted. 
Using a detailed Monte Carlo simulation we
tuned the amplitude of the white noise by minimizing the $\chi^2$ defined as the 
difference of the amplitudes and the periods extracted from the simulation and
the corresponding theoretical values. The best value
for the dithering added to the number of events in each bin is 10\% of the square root of the bin content. 
In addition, the simulation fully validated the method since it showed that the procedure  is sensitive to the phase and the frequency 
of the annual modulation when the simulated data--set has the same composition of signal and background as the Borexino real data.

\subsection{Event selection}
\label{sec:datasel}

The two main challenges of this analysis were enlarging the fiducial volume as much as possible to increase the statistical significance of the modulated data and studying the time stability of the background.

As described in Section~\ref{sec:FV}, we defined a FV (see also Fig.~\ref{fig:fvshape}) obtained including all the events whose stand--off distance from the measured, time--dependent surface of the vessel is $\ge$0.5\,m.  The corresponding volume is changing in time and has a mean value of (141.83 $\pm$ 0.55)\,ton, almost twice larger than the one used for the $^7$Be--$\nu$ interaction rate measurement (75\,ton), see Table~\ref{tab:FV}.

\begin{figure}[t]
\begin{center}
\centering{\includegraphics[width = 0.5\textwidth]{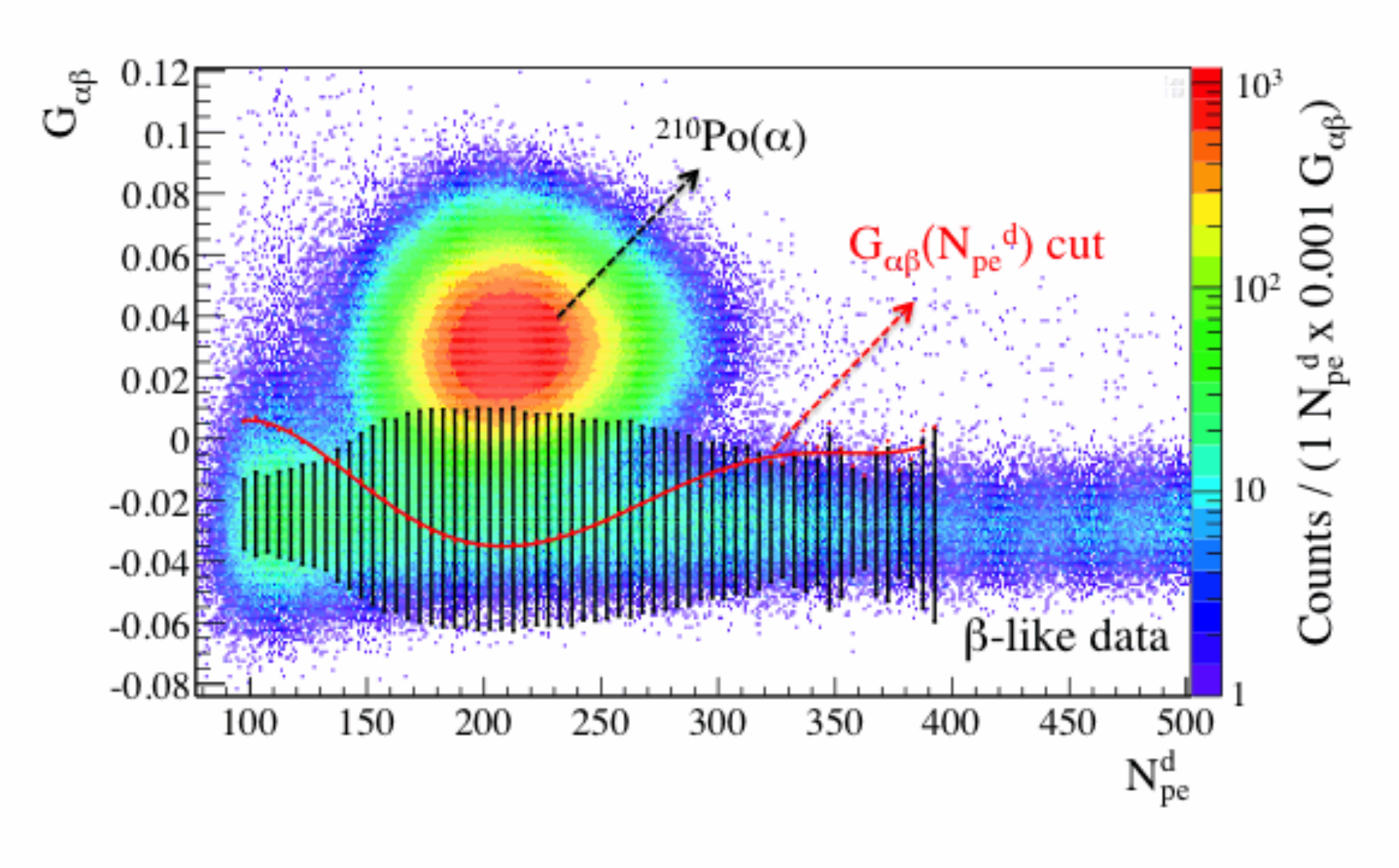}}
\caption{$G_{\alpha\beta}$ parameter in the FV used in the seasonal modulation analysis shown as a function of energy ($N_{pe}^d$ estimator). The red line indicates where the Gatti-cut was placed: in the window of $(105,380)$ $N_{pe}^d$, 66\% of $\beta$'s survive while almost 100\% of $\alpha$'s is rejected.  Black bars represent, for different energy bins, the $G_{\beta}$ interval covering 99.9\% of $\beta$-events.}
\label{fig:GattiABBe7}
\end{center}
\end{figure}

The events were selected using all cuts required for the $^7$Be--$\nu$ flux measurement  with an exception for the $\alpha$-$\beta$ cut:  this was replaced
by a new cut removing all the $\alpha$--like events at a cost of a large reduction of $\beta$ events (less than a half) in the energy window of interest. This cut allows to remove all the $^{210}$Po events whose rate is not stable in time as described in Section~\ref{subsec:InternalBack}.

The red curve in Fig.~\ref{fig:GattiABBe7} represents this new, energy--dependent $G_{\alpha\beta}$ cut, isolating the $\alpha$--contribution towards the positiive $G_{\alpha\beta}$ from the cut. The remaining $\beta$ events used in the signal $R(t)$ are those with $G_{\alpha\beta}$ towards the more negative values. This cut was later taken into account in the Monte Carlo simulations.
We selected the energy region 105 $< N_{pe}^d<$ 380 for this seasonal modulation analysis. Referring to Fig.~\ref{fig:sb},
we see that in this energy window, after the removal of the $\alpha$ events of $^{210}$Po, the only significant contribution to the
background is originated from $^{85}$Kr and $^{210}$Bi. The ratio between all the neutrino--induced signals and background (as obtained with the Monte Carlo) is $\simeq$1. The contribution of the $^7$Be neutrinos is $\simeq$70$\%$ of the whole solar neutrino--induced signals in this energy window.

\subsection{Background and detector--response stability}
\label{sec:stab}

We present four major factors that had the most significant impact on the $R(t)$ in the selected energy window.
\begin{itemize}[leftmargin = *]
\item {\it Change of the $^{210}$Bi rate.}
The observed change of the $^{210}$Bi has been already discussed in Section~\ref{subsec:InternalBack}.
We have compared the result of Fig.~\ref{fig:Birate} with the one obtained from the spectral fit in six--month long time periods using the FV used for the $^7$Be--$\nu$  rate analyis.
All the spectral components were fixed to their best known values except for $^{210}$Bi
and we have found a confirmation that an exponential function (as shown in Fig.~\ref{fig:Birate}) would reasonably well describe the increase
of the $^{210}$Bi contamination. It is important to be able to subtract this trend from the data, because the  Lomb--Scargle method misidentifies
the trend as an actual significant modulation and returns false results for the $\nu$--signal periodicity.
\item {\it Time stability of the energy scale.}
In order to verify the stability of the energy scale, we looked at the distribution of $N_{pe}^d$ obtained from the $^{210}$Po peak as a function of time in the FV used in this seasonal analysis 
It's clear from Fig.~\ref{fig:Poen} that we can trust the energy scale on a long term to within 2 $N_{pe}^d$ (that is to 1$\%$)
which is fully satisfying for our purposes.
\item {\it Time stability of the position reconstruction.}
We selected three time periods  when the $^{222}$Rn rate was temporarily high: 1) the initial detector filling in 2007;  2) the first off--axis
calibration campaign in 2009, and 3) another re--filling in 2010 (needed due to the small leak as explained in Section~\ref{subsec:leak}). Next, we plotted the absolute distance between the reconstructed  $^{214}$Bi and $^{214}$Po events and we normalized the histograms for each period to their total integrals. Results are shown in Fig.~\ref{fig:BiPoRes} where it is clear how
well all the three histograms align.  

\begin{figure}[t]
\begin{center}
\centering{\includegraphics[width = 0.5 \textwidth]{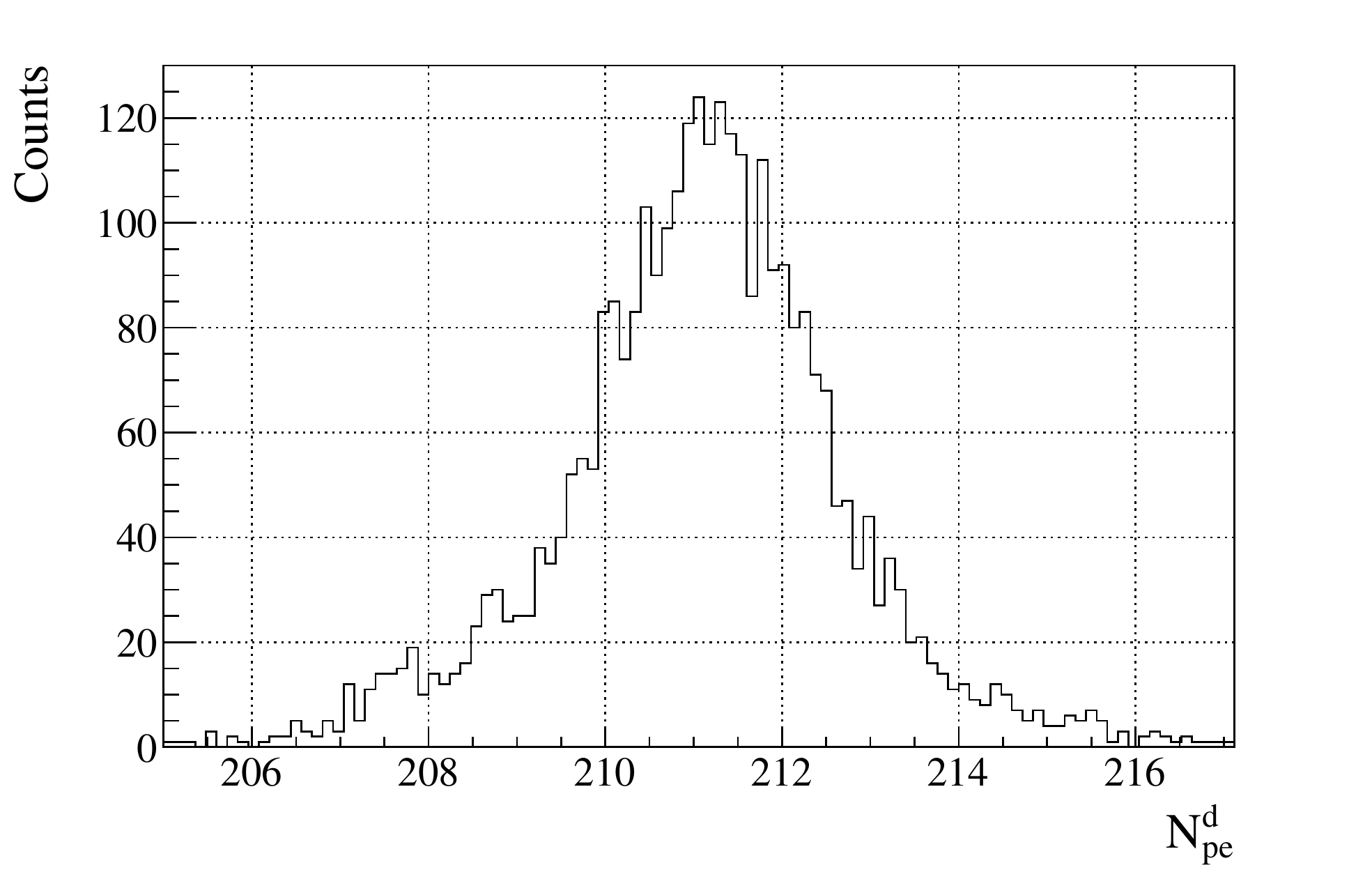}}
\caption{Energy distribution of the $^{210}$Po($\alpha$) peak (expressed in $N_{pe}^d$ estimator) in the FV used for the $^7$Be--$\nu$ annual modulation analysis.} 
\label{fig:Poen}
\end{center}
\end{figure}

\begin{figure}[h]
\begin{center}
\centering{\includegraphics[width = 0.5 \textwidth]{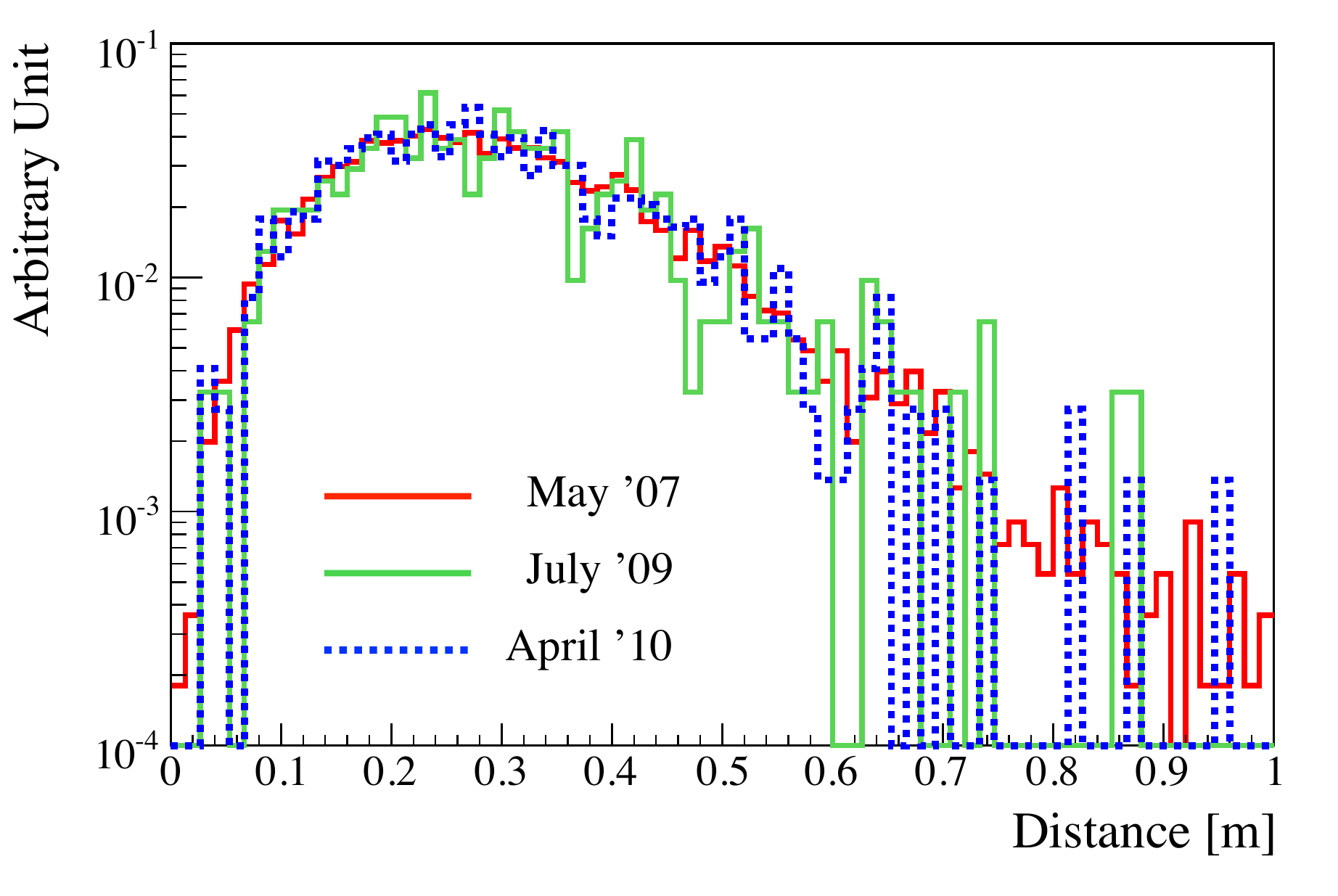}}
\caption{Absolute distance between $^{214}$Bi - $^{214}$Po fast coincidence events in three periods: May 2007 (solid red), July 2009 (solid green), and April 2010 (dotted blue).}
\label{fig:BiPoRes}
\end{center}
\end{figure}

\item {\it $^{222}$Rn contamination.}
The active volume has been frequently exposed to effects of external operations. Calibrations and refillings resulted in a temporary increased count rate of $^{222}$Rn background. Fortunately though, its short decay time did not pose any long--term danger on the overall purity of the detector.
The first six months of data taking have been excluded for this analysis due to  an increase of the number of radon events in the upper hemisphere following the detector filling. Note that the choice of the 75.47\,ton FV for the $^7$Be--$\nu$ rate analysis automatically excludes this region (see Fig.~\ref{fig:rnSpatial}) thus allowing to use in that analysis the first six months of data. Summarizing, the seasonal modulation analysis presented here refers to the period from January, 2008 to May, 2010.
\end{itemize}

\subsection{Results}

We present here the results on the annual modulation of the $^7$Be--$\nu$ interaction rate  
obtained with the three previously described methods.
The results are consistent and in agreement with the expectations.
 
 \subsection*{Fit of the rate versus time}

\begin{figure}[top]
\begin{center}
\centering{\includegraphics[width = 0.5 \textwidth]{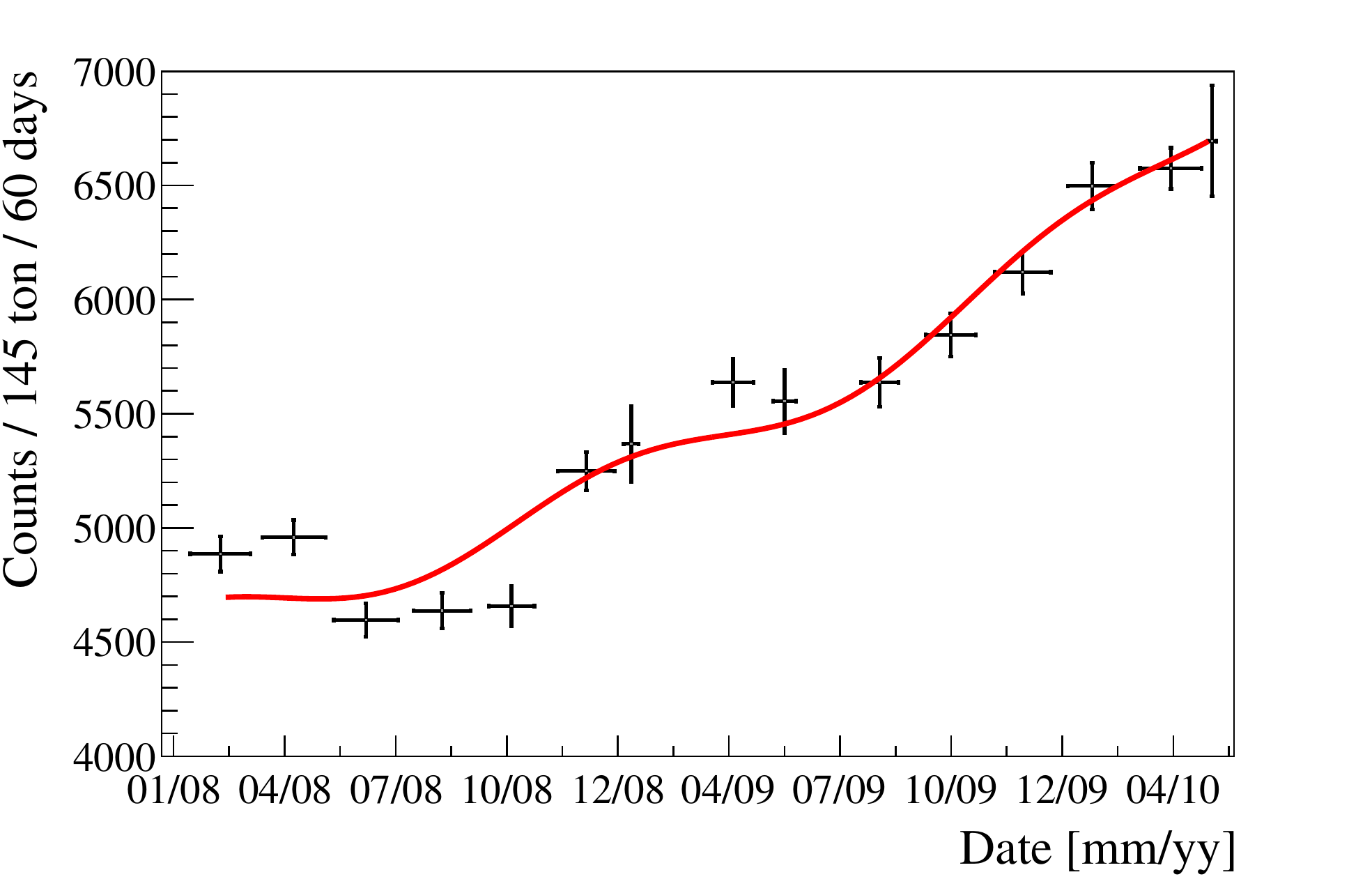}}
\caption{Results obtained with rate analysis. The continuous line is the curve of Eq.~\ref{eq:mod}. The data are grouped in bins of 60 days.}
\label{fig:Rfit_100_080_060_EC_Exp}
\end{center}
\end{figure}

The selected data are grouped in 60-days long bins and fit with:
\begin{align}
R(t)& = R_0 + R_{\rm Bi} e^{\Lambda_{\rm Bi} t} + \bar{R}\left[1 + 2\varepsilon\cos\left(\frac{2 \pi t}{T} - \phi\right)\right] \label{eq:mod}, 
\end{align}
where  $R_0$ is the background rate not depending on time~$t$ and the exponential term describes the time variation of the  $^{210}$Bi rate as discussed in 
Section~\ref{subsec:Bi210}. The third term describes the sinusoidal sesonal modulation, in which $\bar R$ is the mean neutrino interaction rate, $\varepsilon$ is the eccentricity of the Earth's orbit which defines the amplitude of the sinusoid, $T$ is the period, and $\phi$ is the phase. 
Figure~\ref{fig:Rfit_100_080_060_EC_Exp} demonstrates that the expected function (\ref{eq:mod}) is in good agreement with the data.
We have performed a fit with $\bar{R}$,  $\varepsilon$, and $T$ as free parameters,  phase $\phi$  was constrained with a penalty, and the first two terms describing the contribution of non--neutrino background were fixed, based on the study of the time variation of signal in the background--dominated region, see Fig.~\ref{fig:Birate}.
The eccentricity $\varepsilon$ and the average neutrino rates $\bar{R}$ returned by the fit are in agreement to within 2$\sigma$ with the expected ones.
The expected period $T$ of 1 year, and phase $\phi$ of 0 days, are compatible with our fit results of 1.01 $\pm$ 0.07\,year and 11.0 $\pm$ 4.0\,days,  respectively.

\begin{figure}[t]
\begin{center}
\centering{\includegraphics[width = 0.5\textwidth]{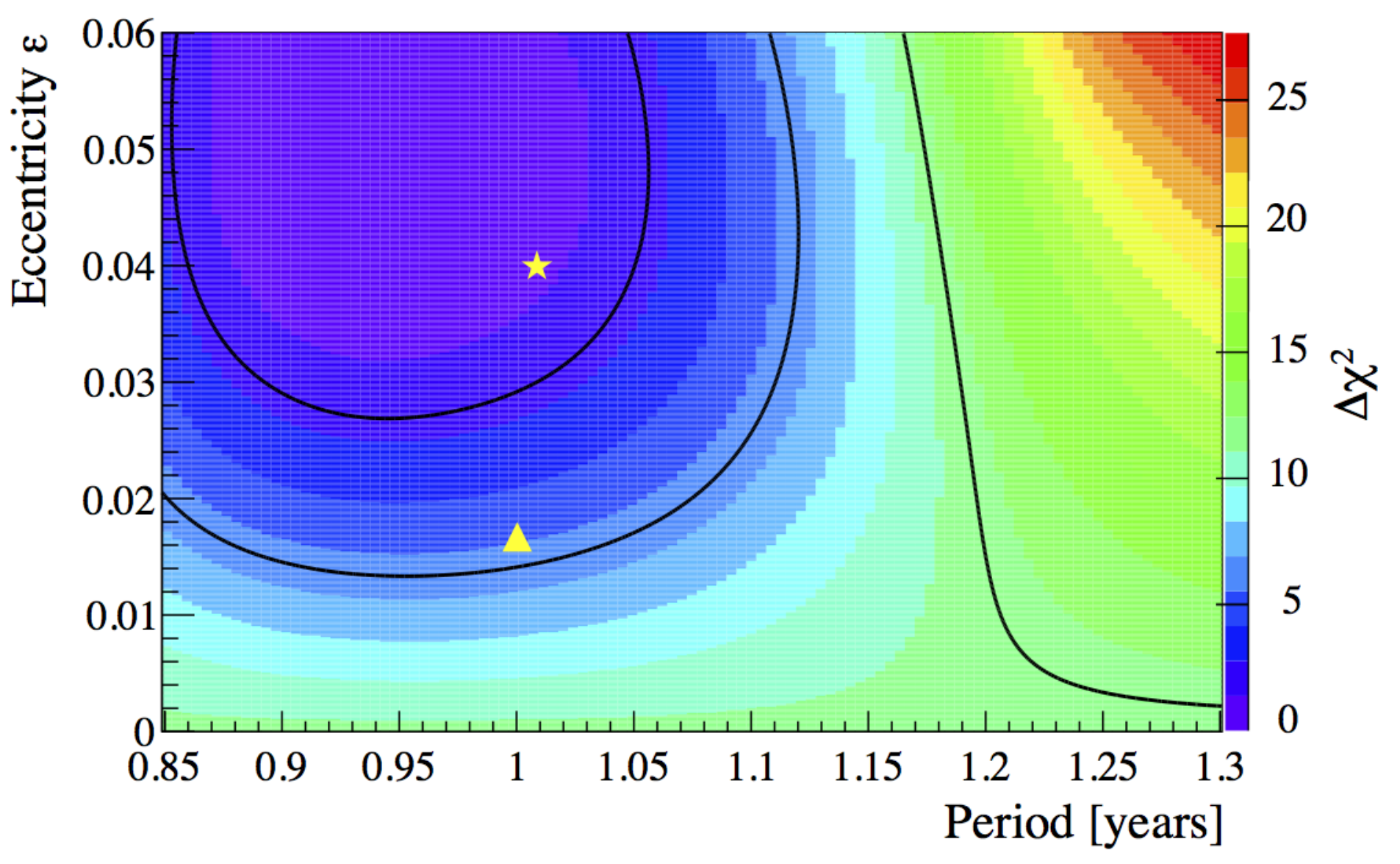}}
\caption{The results obtained via the fit of the rate-versus-time distribution shown in Fig.~\ref{fig:Rfit_100_080_060_EC_Exp}: a $\Delta \chi^2$ ($\Delta\chi^2_R$ = $\chi^2_{R_{min}}$ - $\chi^2_{R_{xy}}$) map (vertical axis) as a function of eccentricity $\varepsilon$ and period. The yellow star indicates the best--fit results, while the yellow triangle the expected values. Confidence contours of 1, 2, and 3$\sigma$ are indicated with black solid lines. }
\label{fig:Chi3DP100_080_060_EC}
\end{center}
\end{figure}

Figure~\ref{fig:Chi3DP100_080_060_EC} shows contour plots of the allowed ranges for the eccentricity $\varepsilon$
and period $T$ at 1, 2, and 3$\sigma$ C.L. Our best result (yellow star) is within the 2$\sigma$ region
of the expected values $\varepsilon$ = 0.01671 and $T$ = 1 year indicated by the yellow triangle. 

\subsection*{Results with the Lomb--Scargle method }

\begin{figure}[t]
\begin{center}
\centering{\includegraphics[width = 0.5\textwidth]{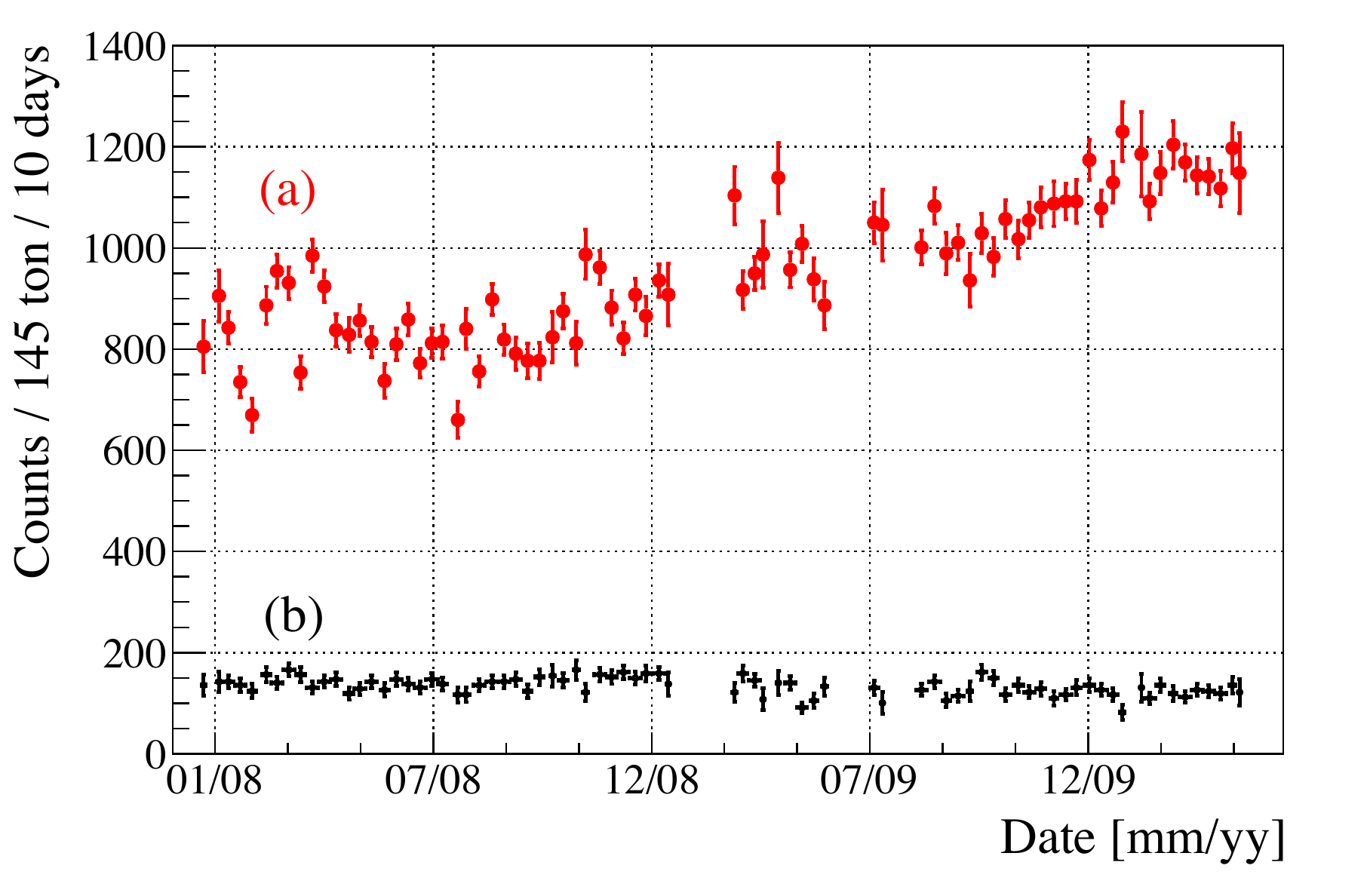}}
\caption{Rate of events in the energy region 105 $<N_{pe}^d<$ 380 and in the FV used in the seasonal modulation analysis as a function of time shown with 10-day binning: the red data points (a) were scaled by a constant factor. For comparison, the black plot (b) shows the count rate due to external $\gamma$s which is stable in time since it is not correlated with the changes of the IV shape. }
\label{fig:R050}
\end{center}
\end{figure}

The data selected after the cuts are  now grouped into 10--day bins. Such choice of binning was justified with a Monte Carlo 
simulation where we have checked that the significance of a Lomb--Scargle peak does not change drastically with a bin size varying between 1 -- 14 days.

\begin{figure}[h]
\begin{center}
\centering{\includegraphics[width = 0.5 \textwidth]{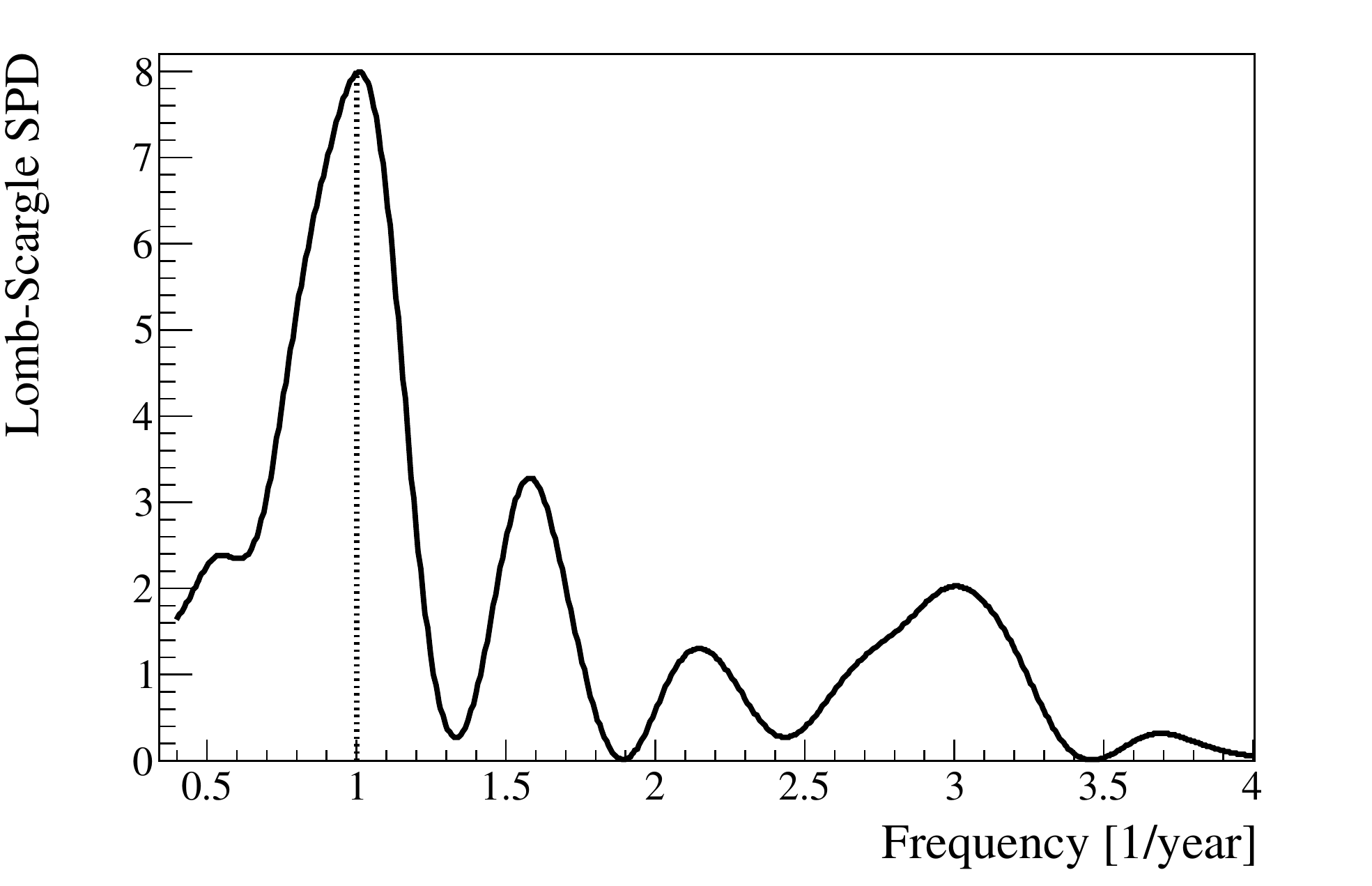}}
\caption{Lomb--Scargle periodogram for the red data points marked (a) in Fig.~\ref{fig:R050} after the subtraction of the exponentail trend due to the $^{210}$Bi contamination. The Spectral Power Density at 1--year is 7.961, as indicated by the vertical line.}
\label{fig:LS050}
\end{center}
\end{figure}

Figure~\ref{fig:R050} shows the count rate $R(t)$ in the energy region of 105 -- 380 $N_{pe}^d$ (red points (a))
 together with the background counts from external $\gamma$'s (black points (b)). Before performing the frequency analysis on the red data (a), we need to
implement a correction which consists in subtracting from these data the exponential trend due to the $^{210}$Bi contamination. This trend causes the Lomb--Scargle algorithm to misidentify the annual peak. The resulting Lomb--Scargle periodogram is shown in Fig.~\ref{fig:LS050}. Clearly, there is a peak which corresponds to 1 year period. The Spectral Power Density (SPD), that is the value of the periodogram, at the frequency that corresponds to 1 year is 7.961. 

\begin{figure}[h]
\begin{center}
\centering{\includegraphics[width = 0.5 \textwidth]{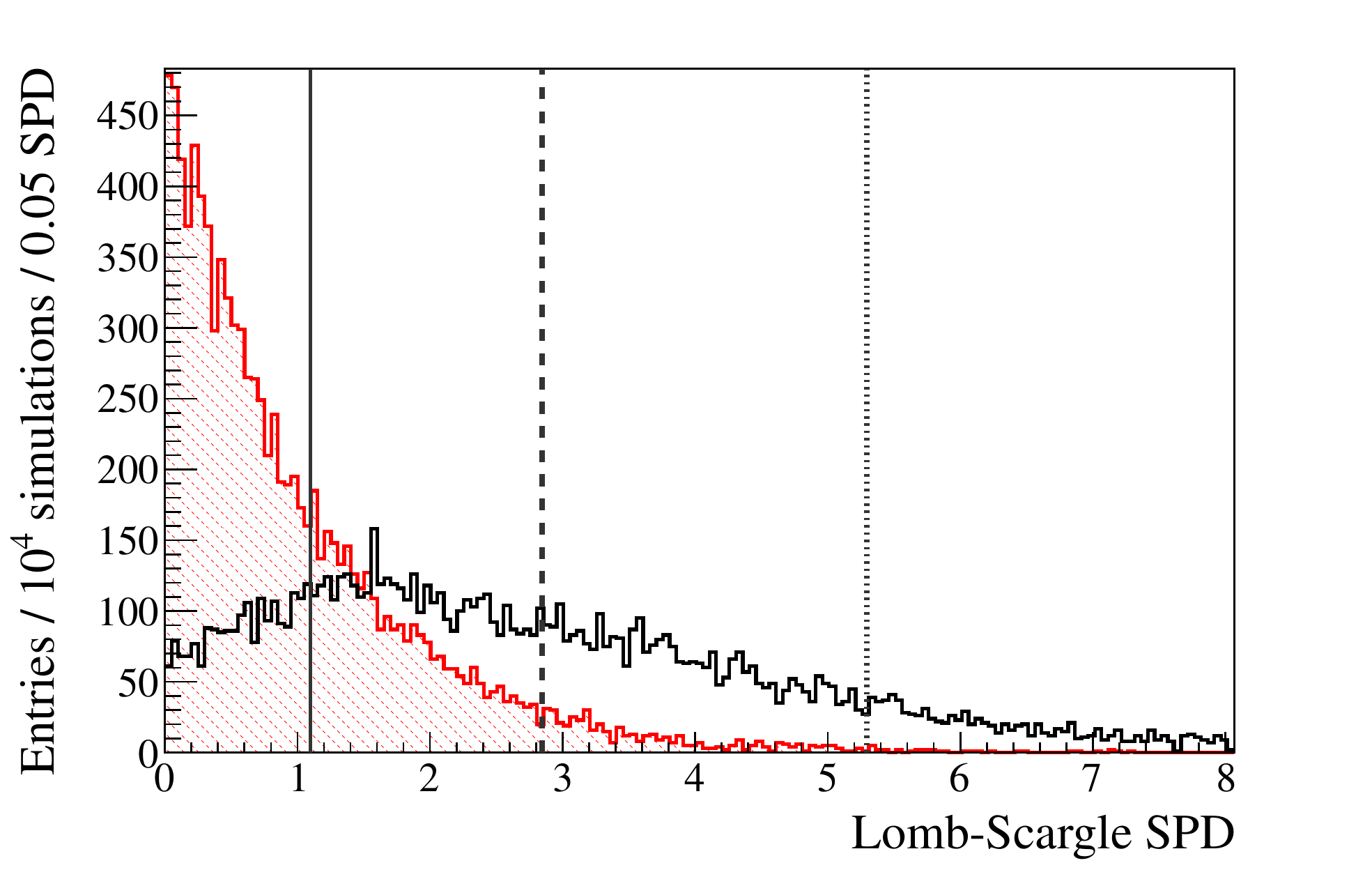}}
\caption{Distributions of the Lomb--Scargle Spectral Power Density (SPD) at frequency corresponding to 1\,year for $10^{4}$ simulations of a 7\% solar-neutrino annual flux modulation with realistic background (solid black line) and the same number of white-noise simulations of background without any seasonally-modulated signal (red area). Indicated with vertical lines are the sensitivity thresholds of 1$\sigma$ (solid), 2$\sigma$ (dashed), and 3$\sigma$ (dotted) C.L. with corresponding detection probabilities of 81.62, 43.54, and 11.68\%, respectively. The detected SPD(1\,year) = 7.961 (see Fig.~\ref{fig:LS050}) represents an evidence of the annual--modulation signal with a significance higher than 3$\sigma$; our chance of detecting the annual modulation at this level of significance is 11.68\%.}
\label{fig:LSMC050}
\end{center}
\end{figure}

The significance of this results and of the Lomb--Scargle analysis is studied with a Monte Carlo simulation with realistic signal--to--background ratio and is shown in Fig.~\ref{fig:LSMC050}. The red filled area shows the SPD(1\,year)-distribution of 10$^4$ simulations corresponding to the null hypothesis (no seasonal modulation of the neutrino signal) and the black line shows the SPD(1\,year)-distribution of another 10$^4$ simulations where the expected seasonal modulation was considered.
Indicated with vertical lines are the sensitivity thresholds of 1$\sigma$ (solid), 2$\sigma$ (dashed), and 3$\sigma$ (dotted) C.L. with corresponding detection probabilities of 81.62, 43.54, and 11.68\%. Thus, the SPD(1\,year) = 7.961 of our data (see Fig.~\ref{fig:R050}) represents an evidence of the annual--modulation signal with a significance higher than 3$\sigma$; our chance of detecting the annual modulation at this level of significance is 11.68\%.

 \subsection*{Results with the EMD method}

In order to avoid a distorted reconstruction of IMFs due to the empty bins during
the data taking, we grouped the selected data in 1--day bins and we filled these empty bins with white noise.
In contrast to the Lomb--Scargle method, the subtraction of the exponential trend due to the $^{210}$Bi contamination is not needed in application of the EMD method.
As a mean value for the white noise we used an average of the count rates from the whole data--set and as the sigma its square--root. 
We have repeated the procedure 100 times and we have built the distribution  of the 
amplitude, phase, and frequency of the IMF. The final result has been obtained by fitting these distributions.
The  simulations show that 100 extraction are enough to obtain
results not limited by the statistical fluctuations introduced by this procedure.

\begin{figure}[b]
\begin{center}
\centering{\includegraphics[width = 0.5 \textwidth]{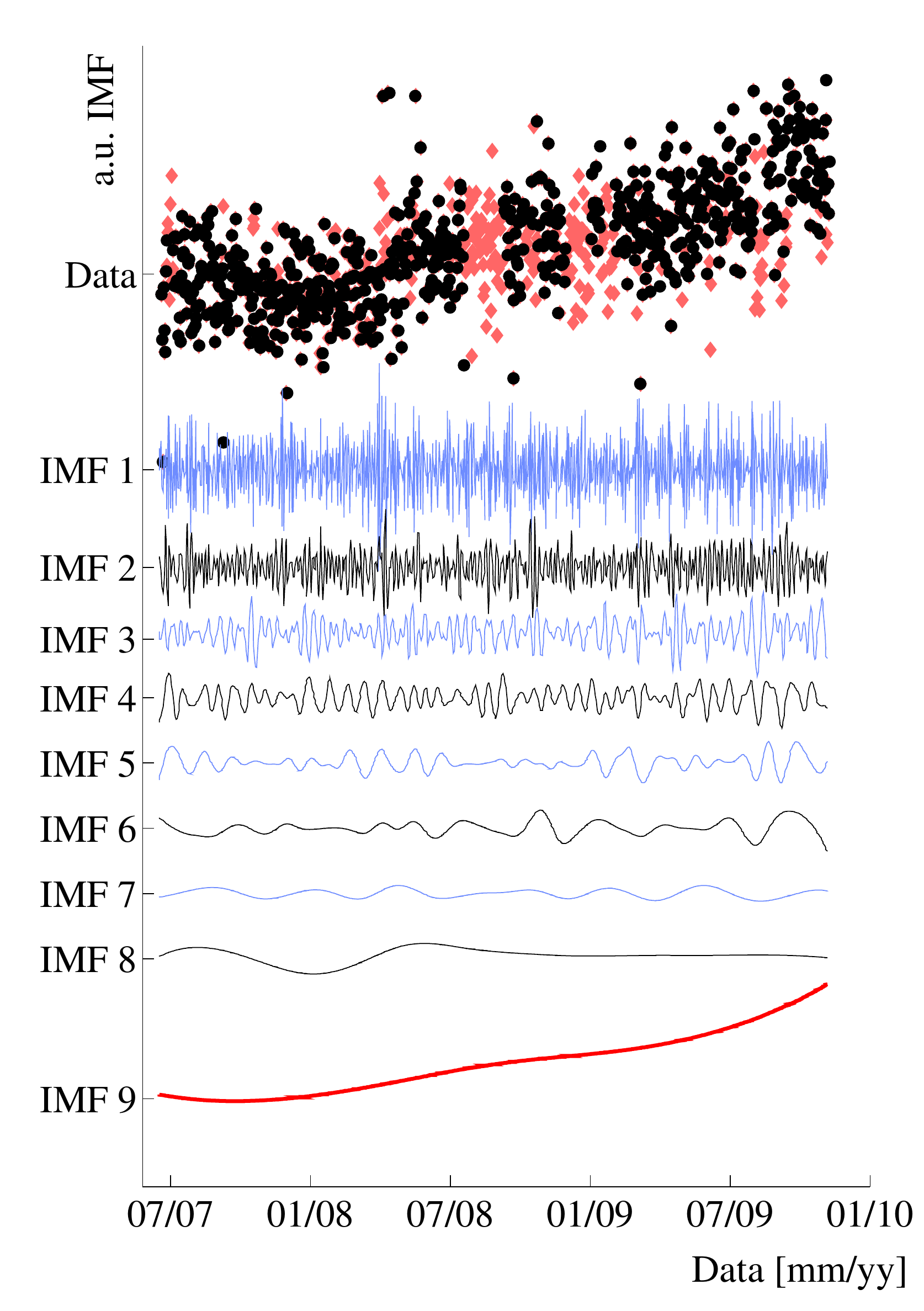}}
\caption{A sequence of the IMF extracted from the data (black circles) plus white noise (red diamonds) by means
of a sifting algorithm using 20 iterations for each one. The IMF$_{8}$ is the one used to extract the results about the seasonal modulation,
The last IMF$_{9}$, also called \emph{``trend''}, is the best representation of the background variation in time.}
\label{fig:IMFSequence}
\end{center}
\end{figure}

Figure~\ref{fig:IMFSequence} shows one example of the results of the application of the EMD method and one set of IMF extracted with the described algorithm.
The  expected annual modulation signal should  be contained in the 
mode number 8.
From the simulations done with artificial signals and from literature we found that a single signal
can be shared between closest IMFs.
Note that the statistical fluctuations can  attenuate the
signal until it may disappear. This happens in the particular example shown in Fig.~\ref{fig:IMFSequence}  in the second year of the data taking. This fact clearly explains
why we need to use the technique of the dithering before to decompose the signal 
with the EMD and why we cannot perform the decomposition just one time.

\begin{figure}[thb]   
\centering
\includegraphics[width = 0.48\textwidth]{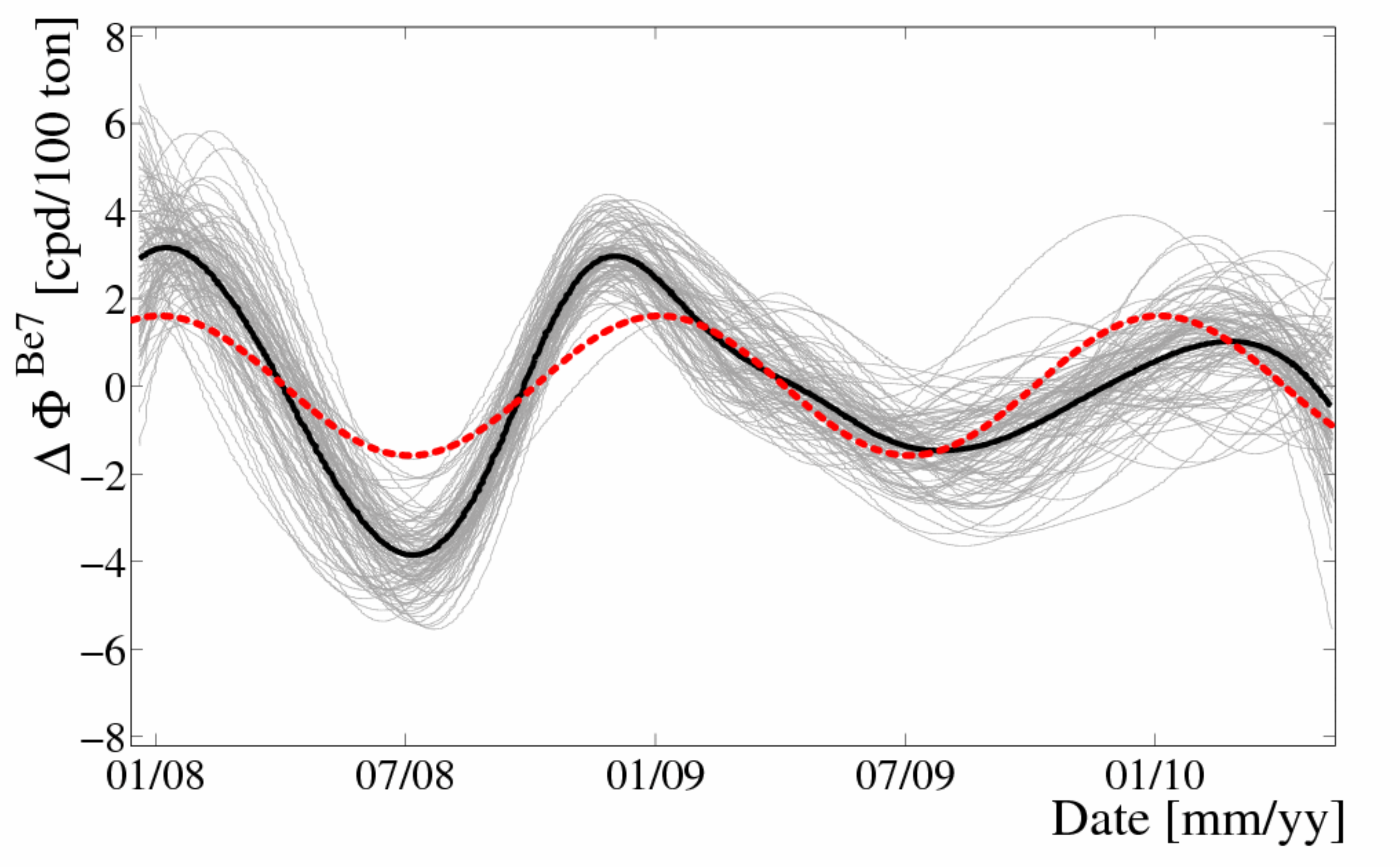}
\caption{Set of 100 intrinsic mode functions IMF$_{8}$ extracted after the addition of dithering
(grey lines). The black-solid line is their average and the dashed-red line is the annual
modulation from Eq.~\ref{eq:mod}. The number of the IMF is the expected one 
for the annual frequency of $\nu_{\rm year}=0.00274\; [d^{-1}]$.}
\label{fig:IMFSeasonal}
\end{figure}

In Figure~\ref{fig:IMFSeasonal} we show in gray all the IMF$_{8}$ ensemble obtained by applying the decomposition 100 times:
the solid--black line is the average and the dashed--red line is the expected modulation from the last term of Eq.~\ref{eq:mod}.

A good agreement for the frequency and phase is clearly visible in the trend in the picture, but we found 
a slightly larger amplitude in the first half of the data--set. 
Little changes in frequency and phase are also visible in the total trend. These are driven by
fast changes in the background behavior (e.g. due to the increasing of the $^{210}$Bi). 

The EMD method does not require an assumption on the time behavior of the IMFs: then the fact that a quasi-sinusoidal trend is clearly visible
is a proof that the annual modulation is actually detectable in our data--set.

\begin{figure}[thb]
\begin{center}
\centering{\includegraphics[width = 0.5\textwidth]{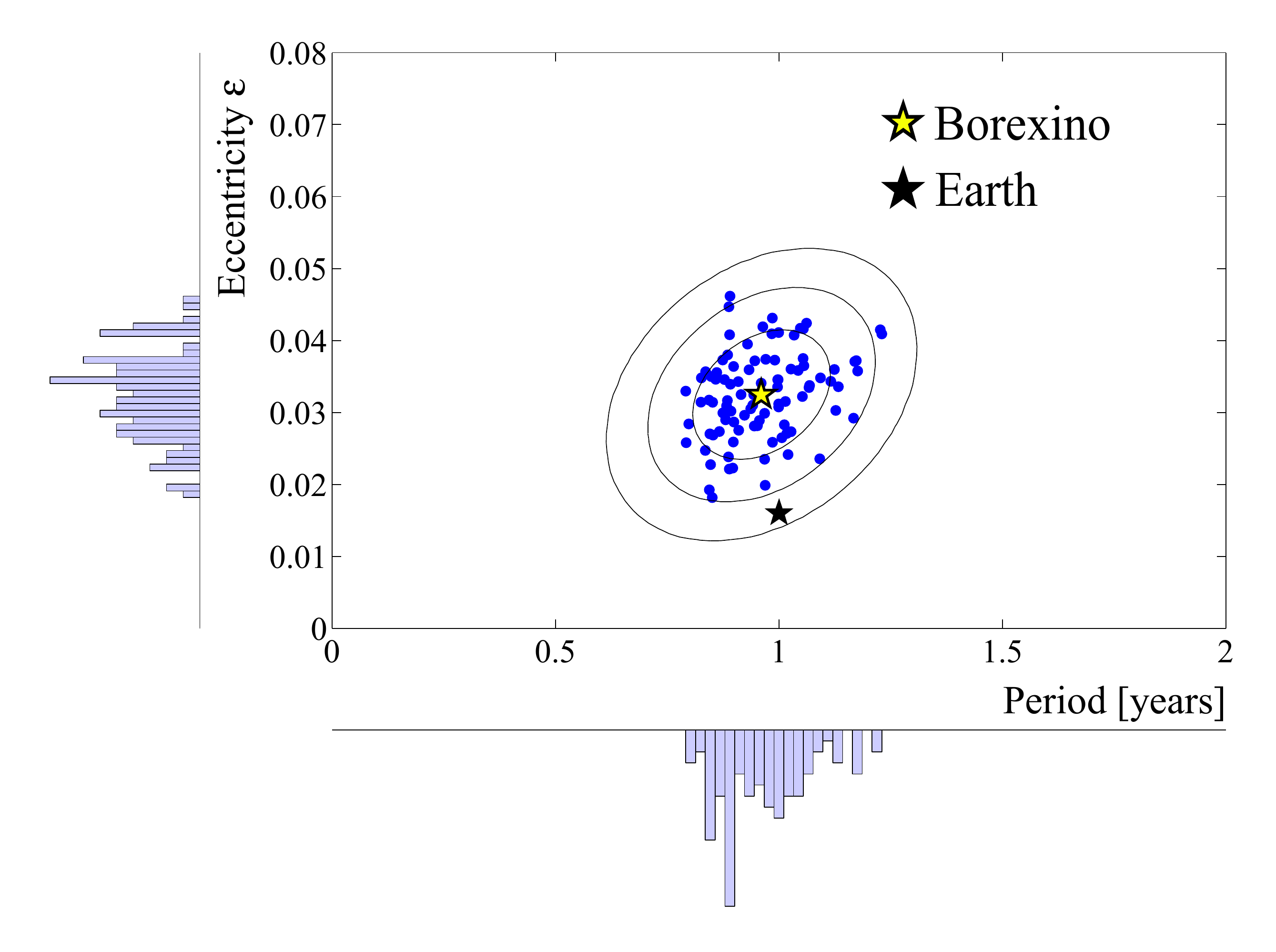}}
\caption{Blue circles show the eccentricities and periods obtained with the EMD method with 100 IMF$_8$'s (see Fig.~\ref{fig:IMFSeasonal}). The projections shown on the vertical and horizontal axis are well described by Gaussian curves. The solid--black line of Fig.~\ref{fig:IMFSeasonal} is represented by the yellow star in the middle of 1, 2, and 3$\sigma$ C.L. contours. The black star represents the expectations. The measured period is in perfect agreement with 1\,year, while the eccentricity is compatible within $3\sigma$.}
\label{fig:EccPeriodSymm}
\end{center}
\end{figure}

In Fig.~\ref{fig:EccPeriodSymm} we show the 2D distribution of the results about the eccentricity and period obtained with the 100 IMF$_8$'s as well as the corresponding projections on the $x$ and $y$ axes. We compare the mean value of the eccentricity and the period obtained from this method (yellow star) with the values expected for the terrestrial orbit (black star). The period agrees with the 1 year value within $1\sigma$ and the eccentricity is in agreement with the orbital one within $3\sigma$.

\section{The $pep$ and CNO neutrino interaction rates}
\label{sec:pepCNOResults}

\begin{figure*} [t]
\centering
\includegraphics[width=0.9\linewidth]{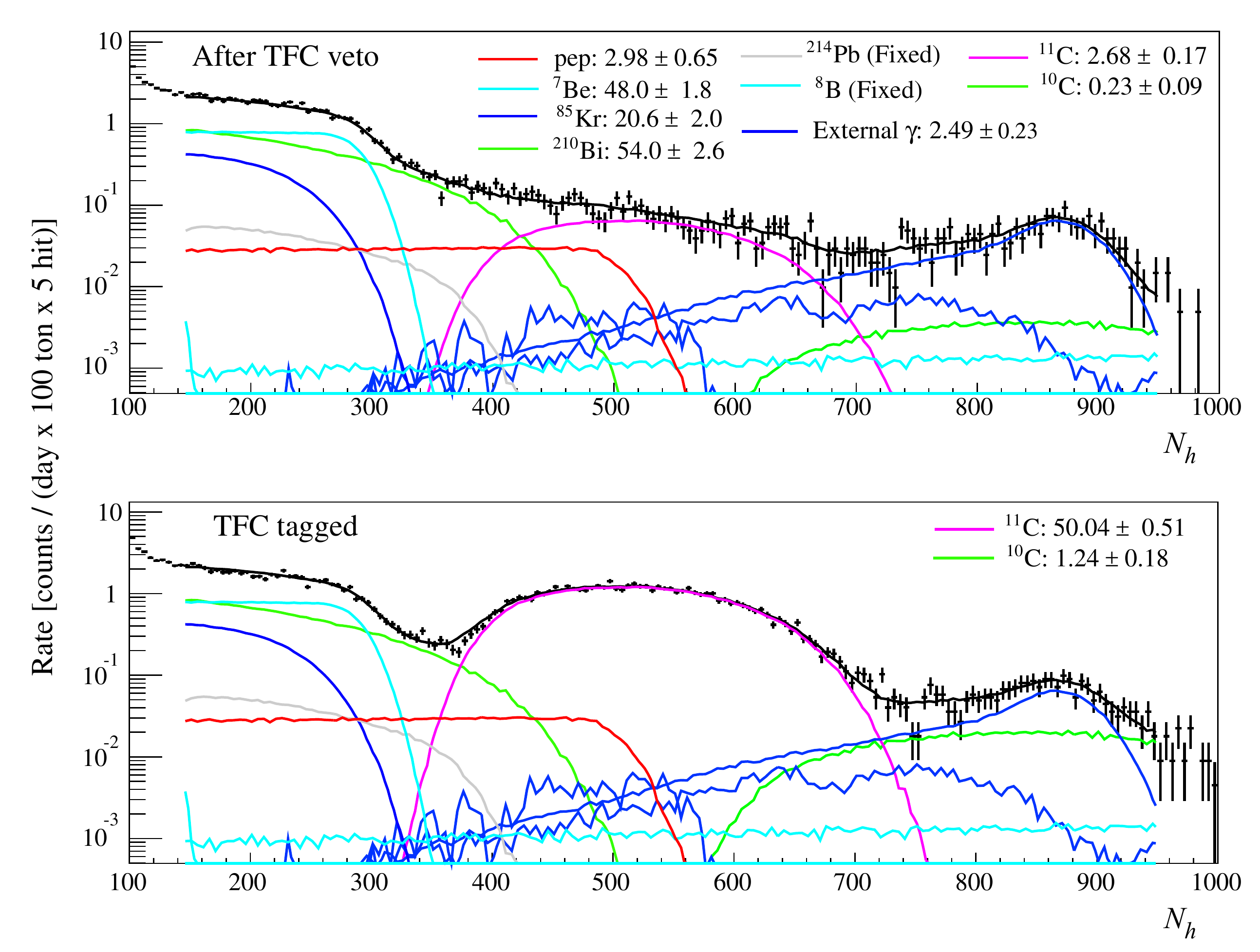}
\caption{Energy spectra and the best-fit results performed on the $N_h$ energy estimator in the $pep$ and CNO neutrino analysis. The top panel shows the best fit to the TFC--subtracted spectrum, while the bottom one presents the fit to the complementary, TFC--tagged events. The best--fit values for the rates of the species included in the fit are shown in the legend. Units are cpd/100\,ton.}
\label{fig:pep_Ene_nhits}
\end{figure*}

\begin{figure*}
\centering
\includegraphics[width=0.9\linewidth]{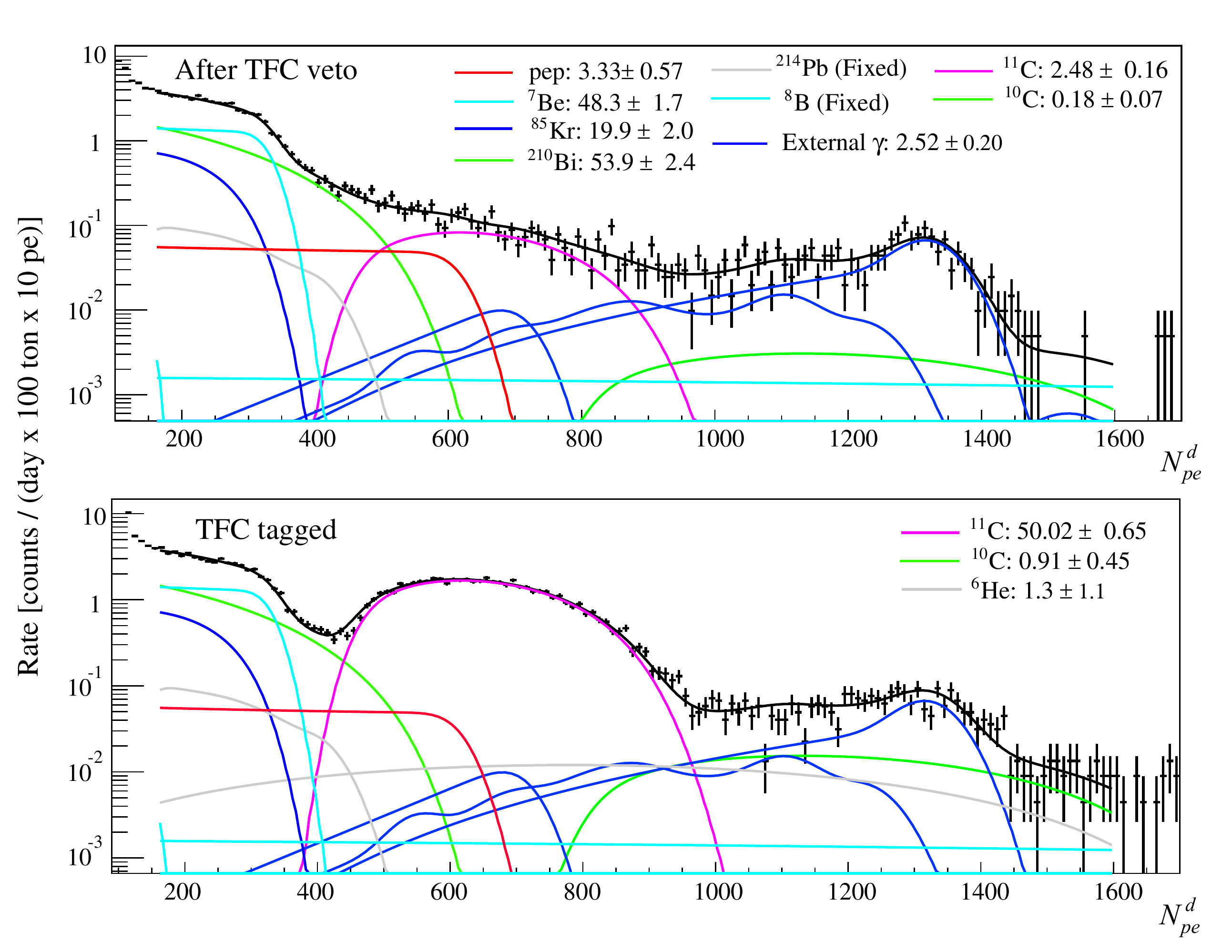}
\caption{Energy spectra and the best-fit results performed on the $N_{pe}^d$ energy estimator in the $pep$ and CNO neutrino analysis. The top panel shows the best fit to the TFC--subtracted spectrum, while the bottom one presents the fit to the complementary, TFC--tagged events.  The best--fit values for the rates of the species included in the fit are shown in the legend. Units are cpd/100\,ton. }
\label{fig:pep_Ene_npe}
\end{figure*}

\begin{figure*}
\centering
$
\begin{array}{cc}
\centering{\includegraphics[width=0.50\linewidth]{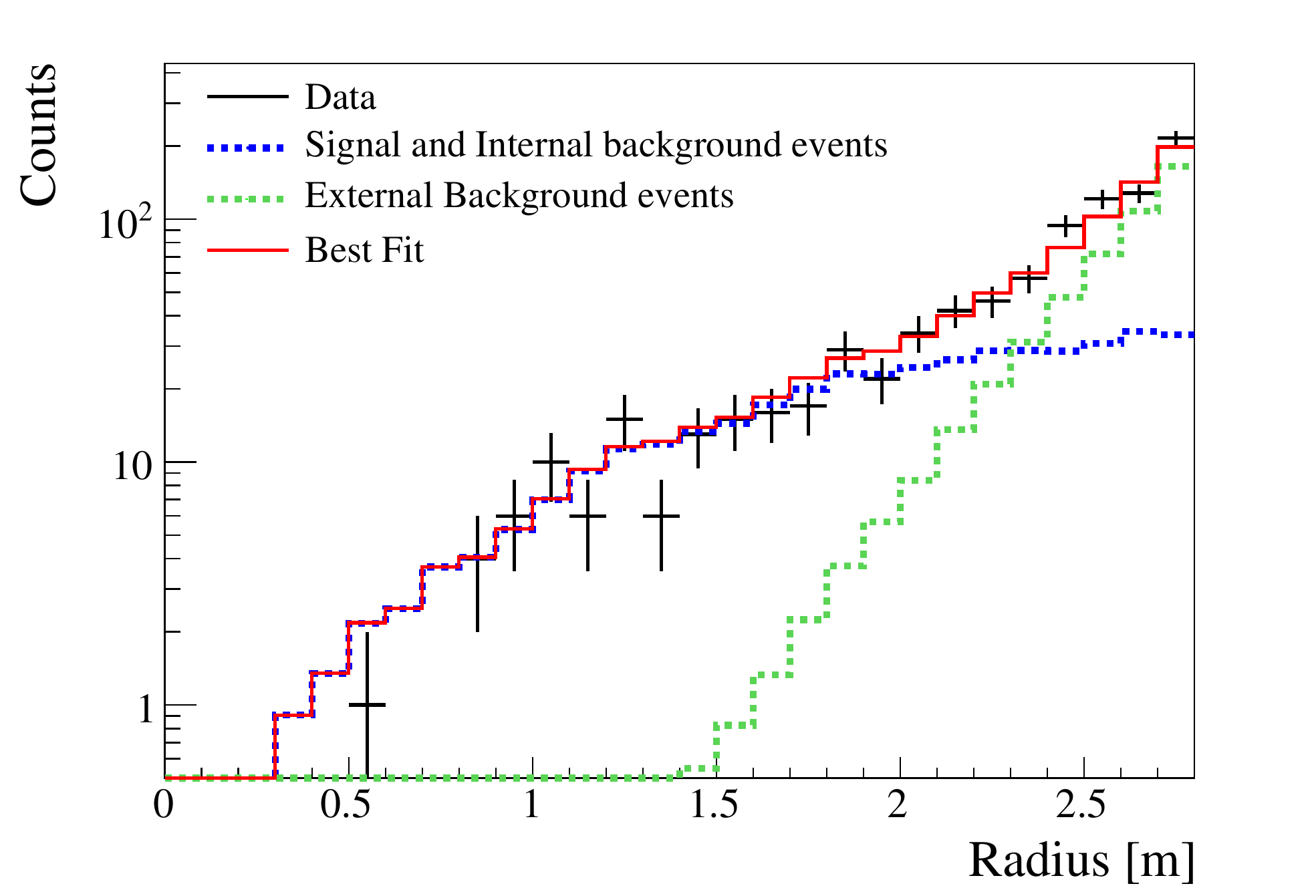}}
\centering{\includegraphics[width=0.50\linewidth]{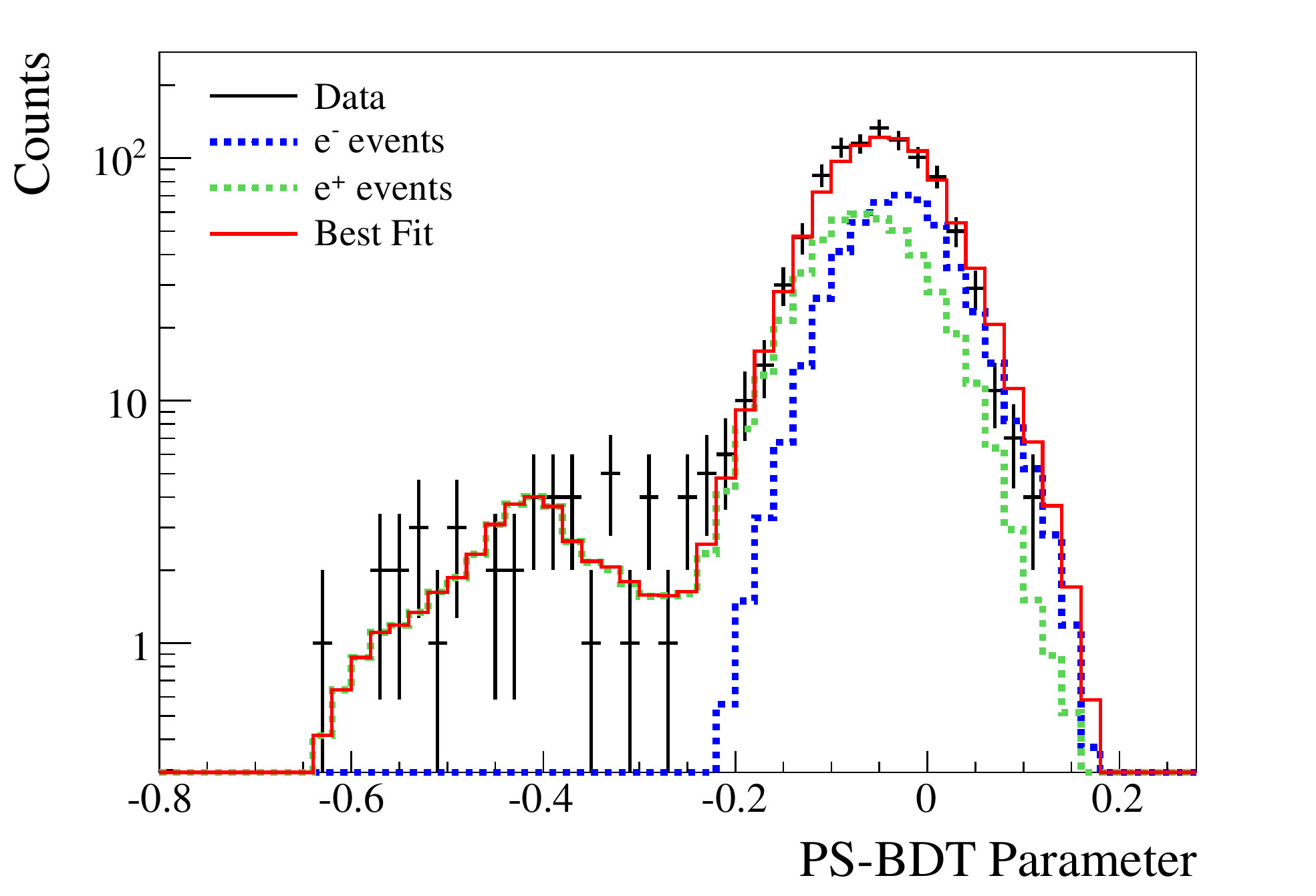}}
\end{array}
$
\caption{Left: Radial distribution of events in the energy range $N_h$ = 500 - 900 (black data points) and the best fit (solid-red line) obtained performing the multivariate fit with the $N_h$ energy estimator. The $y$-axis scale is logarithmic. A similar and statistically consistent result is obtained using the variable $N_{pe}^d$. Right: Distribution of the PS--BDT parameter for events in the energy range 450 - 900 $N_{pe}^d$ (black data points) and the best fit (solid-red line) obtained performing the multivariate fit with the variable $N_{pe}^d$. The scale is logarithmic. Similar and consistent results are obtained with $N_h$ variable.}
\label{fig:BDT_radial}
\end{figure*}

Borexino provided the first measurement of the $pep$ solar neutrino interaction rate and the strongest  limit to--date on the CNO solar neutrino  interaction rate~\cite{PepBorex}.
The measured $pep$ interaction rate is:
\begin{equation}
 R ({pep}) = 3.1 \pm 0.6_{\rm stat} \pm 0.3_{\rm syst} \,\, {\rm cpd/100\,ton},
\end{equation}
and the CNO rate is constrained to:
\begin{equation}
R({\rm CNO}) <7.9   \,\,\, {\rm cpd/100\,ton\,at\,95\%\,C.L.}
\end{equation}
Regarding the $pep$--$\nu$ interaction rate measurement, the corresponding $\nu_e$--equivalent flux is (1.00 $\pm$ 0.22) $\times 10^8$\,cm$^{-2}$\,s$^{-1}$. Considering the 3-flavor neutrino oscillations, the equivalent flux is (1.63 $\pm$ 0.35) $\times 10^8$\,cm$^{-2}$\,s$^{-1}$, which can be compared with the expected SSM flux of Table~\ref{table:Rate}.

The necessary sensitivity was achieved by adopting novel techniques for the rejection of $^{11}$C cosmogenic background (TFC--veto and BDT parameter described in Section~\ref{sec:c11}) dominating the 1 -- 2\,MeV energy region. The event selection criteria are described in Subsection~\ref{subsec:cuts} and the FV in Section~\ref{sec:FV}. A multivariate binned maximum likelihood fit procedure (Section~\ref{sec:Multivariate}) was developed. It is based on the simultaneous fit of the energy spectra, shown in Figs.~\ref{fig:pep_Ene_nhits} and \ref{fig:pep_Ene_npe}, and radial and PS--BDT distributions shown in Fig.~\ref{fig:BDT_radial}. 
The TFC --subtracted and complimentary TFC--tagged energy spectra have been fitted simultaneously.
For the simultaneous signal extraction, events with $N_{pe}^d$ 150 -- 1600 are considered in the energy spectrum. While the energy intervals for the events considered for the fits to the radial and PS--BDT distributions are restricted to 500 -- 900 and 450 -- 900, respectively. These ranges have been chosen as to include in the fit to those parameter spaces only the species whose distributions are precisely known.

\begin{figure*}
\centering
$
\begin{array}{cc}
\centering{\includegraphics[width=0.5\textwidth]{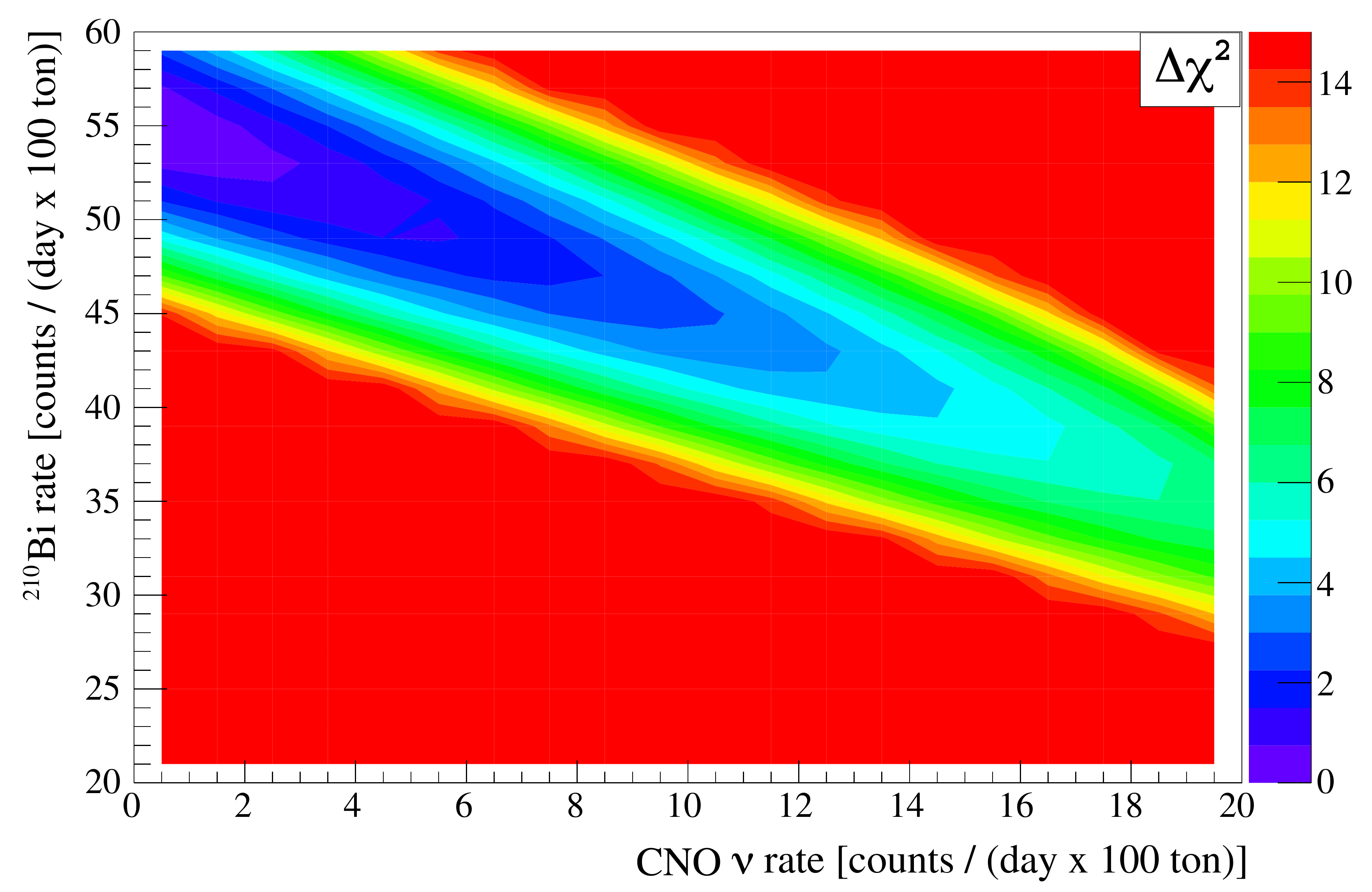}}
\centering{\includegraphics[width=0.5\textwidth]{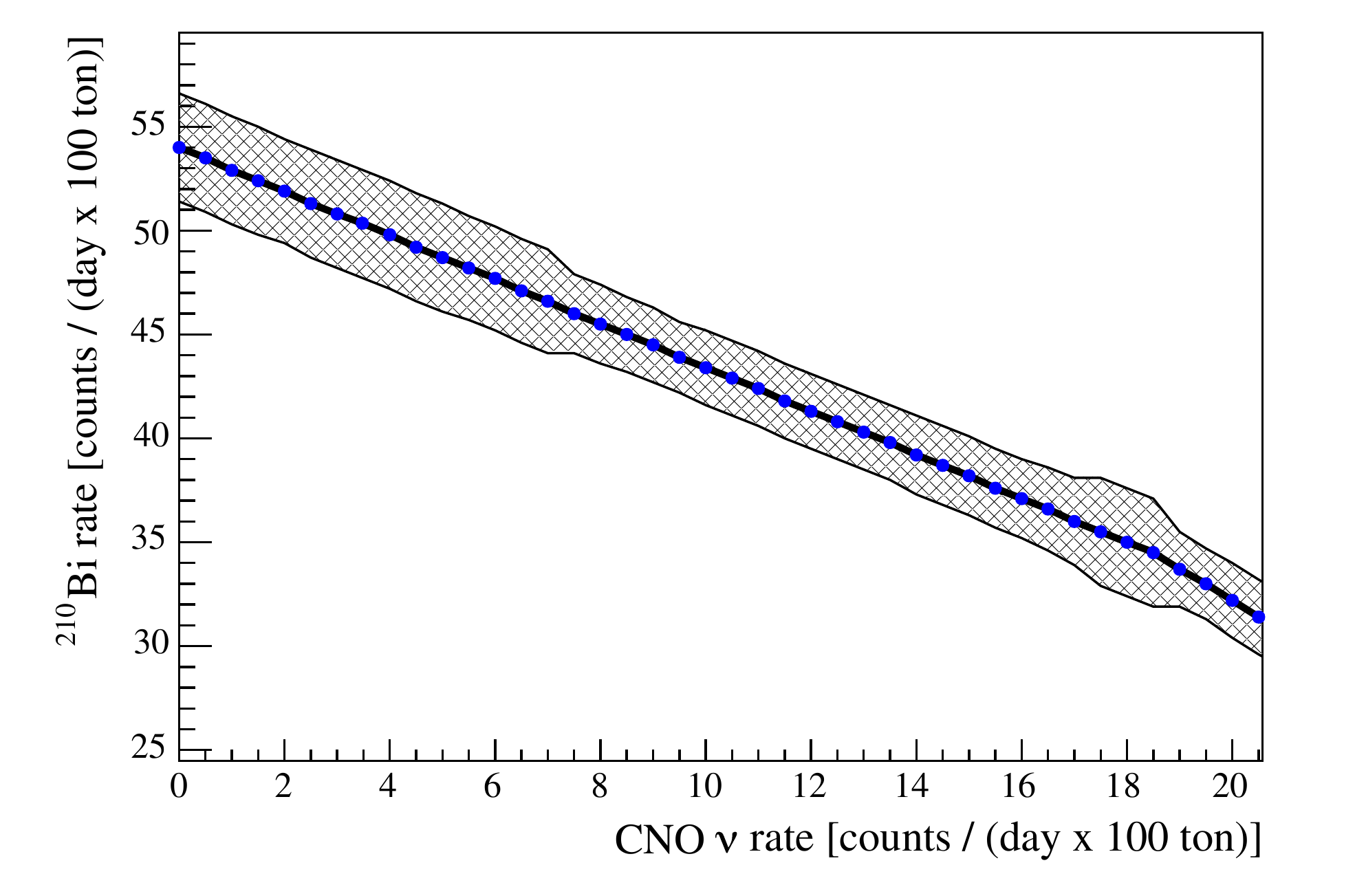}}
\end{array}
$

\caption{Correlation between CNO--$\nu$ and $^{210}$Bi rates. Left: $\Delta \chi^2$ map obtained from a likelihood--ratio test between the likelihood of the best--fit result and the maximum likelihood returned by the fit when the $^{210}$Bi and CNO--$\nu$ interaction rates are fixed to different values.
 The right column gives the colors corresponding to $\Delta \chi^2$ values; note that the same red color has been used to plot all $\Delta \chi^2 \geq 14$ values to allow to visualize color  variations in the relevant region of the plot. The plot has been obtained using  $N_h$ energy estimator. The $pep$--$\nu$ rate is fixed to the standard solar model prediction.
 Right: $^{210}$Bi interaction rate returned by the fit for different (fixed) CNO--$\nu$ interaction rates. The shaded area is the statistical uncertainty.}
\label{fig:Bi210_vs_CNO}
\end{figure*}

Table~\ref{tab:pepCNO_fit_results} summarizes the fit results, the central values with the corresponding statistical and systematic uncertainties, limits, and expected values for the rates of $pep$ and CNO neutrinos and for the background components left free in the fit.
As seen from this Table, the predictions of standard solar models are consistent with our $pep$--$\nu$ interaction rate measurement.
The fit prefers a CNO--$\nu$ interaction rate railed at zero. Therefore, we have performed a likelihood--ratio test to estimate the upper limits also reported in this Table. The ability to measure the CNO--$\nu$ interaction rate mostly depends on the presence of the $\beta$ background from $^{210}$Bi decays. The $^{210}$Bi spectral shape is very similar to that of the recoil of the electrons scattered by CNO neutrinos. The similarity of the two spectra induces a correlation between the two components in the fit.
The $^{210}$Bi count rate in the present data set  is more than 10 times higher than what expected from CNO neutrinos. Reducing  $^{210}$Bi background as much as possible is the main challenge any experiment willing to measure the CNO neutrinos in liquid scintillators need to tackle. Figure~\ref{fig:Bi210_vs_CNO} shows this effect for the present data set.

\begin{table*}[t]
\begin{center}
\begin{tabular}{lccccc}
\hline \hline
Species            & Result                                                           & Expected value    &  reference & &\\ %& Notes  \\
 		         &  [cpd/100\,ton]                                            &  [cpd/100\,ton]     &    &\\%&  \\
\hline
$pep$               &  $3.1 \pm 0.6 \pm 0.3$                     & $2.73 \pm 0.05$ $(2.79 \pm 0.06)$  & Table~\ref{table:Rate} & & \\%  &     \\
$^7$Be              &  $48.3 \pm 2.0 \pm 0.9$                   & $46.0 \pm 1.5 \pm 1.6$    &  Section~\ref{sec:Be7results}& &\\ %&       \\
$^{85}$Kr           & $19.3   \pm 2.0     \pm 1.9$           & $30.4 \pm 5.3 \pm 1.5 $    & Table~\ref{tab:back_rate} & &\\%&\\
$^{210}$Bi          & $54.5    \pm  2.4    \pm 1.4 $                 & NA                                  &   & &\\%& \cno~at 0 cpd/100\,ton\\
		
$^{11}$C            & $27.4     \pm  0.3    \pm 0.1$                 & 28.5 $\pm$ 0.2$\pm$ 0.7     & Table~\ref{tab:Be7Back} & &\\ %&  \\
$^{10}$C            & $0.62     \pm  0.2    \pm  0.1 $              & 0.54 $\pm$ 0.04               & Section~\ref{sec:FV}  & &\\ 
$^6$He              & $0.7 (0)   \pm 0.6 (0.5)   \pm 1$            & 0.31 $\pm$ 0.04               & Section~\ref{sec:FV} & &\\
Ext. $^{208}$Tl  ($N_{pe}^h$)    & $1.64  \pm     0.11 \pm 0.01 $         & NA      & see the caption & &\\ 
Ext. $^{208}$Tl  ($N_h$)    &     $1.94  \pm     0.13 \pm 0.02$             & NA      & see the caption & &\\ 
Ext. $^{214}$Bi  ($N_{pe}^h$)    & $0.67   \pm 0.12    \pm 0.01$             & NA     & see the caption  & &\\ 
Ext. $^{214}$Bi   ($N_h$)   & $0.41  \pm 0.13   \pm 0.02$             & NA     & see the caption  & &\\ 
Ext. $^{40}$K       & $0.16   \pm 0.1         \pm 0.03$                    &  NA      & see the caption  & &\\ %&  \\
Total Ext. Bkg.    &  2.49             $\pm$ 0.2               $\pm$ 0.04                    &  NA               &  see the caption & &\\ %& \\
		\hline
		\hline
		    &  68\% Limit          & 95\% Limit &  99\% Limit &  Expected value& reference\\ %&   \\
		\hline
CNO                 &  4                   & 12         & 19                                & $5.24 \pm 0.54$ $(3.74 \pm 0.37)$  & Table~\ref{table:Rate}\\ 
$^{40}$K            & 0.11                 & 0.42       &  0.69                             & NA &\\%&\\
$^{234\rm{m}}$Pa    & 0.12                 & 0.46       & 0.75                              & $1.78\pm0.06$  &Section~\ref{sec:FV} \\%&\\
		\hline \hline
\end{tabular}
\end{center}
\caption {Summary of the final $pep$,  CNO and background rate results and their corresponding statistical and systematic uncertainties. The statistical uncertainty is the one returned by the fitter. For $^{210}$Bi, the symmetric uncertainty returned by the fitter does not represent well the $\Delta \chi^2$ profile, as expected from the strong correlation with the CNO interaction rate. For species that fit to zero, the upper confidence limits are obtained from the $\Delta \chi^2$ profile. 
Differences between best--fit rates from the $N_{pe}^d$ and $N_h$ fits have been included within the systematic uncertainty.  Exceptions to this are external $^{208}$Tl and $^{214}$Bi, for which the rates and uncertainties are given separately. Note, that the rates of total external background ($^{208}$Tl  + $^{214}$Bi + $^{40}$K) obtained from the $N_{pe}^d$ and $N_h$ fits agree within 1\% and this small difference is considered in the systematic error. For the case of $^6$He, its central value is zero in the 
$N_h$ fit and the number given in parentheses for the uncertainty is the 68\% upper limit, while the corresponding 95\% upper limit is 1.5\,cpd/100\,ton. The last column shows the expected values for the different species based on other sources.  The expected rates from backgrounds from the PMTs  are taken as lower limits on the total external background. No previous valid estimates for $^{210}$Bi  and $^{40}$K are available.}
\label{tab:pepCNO_fit_results}
\end{table*}

\begin{figure*}[!ht]
\centering
$
\begin{array}{cc}
\vspace{-2 mm}
\centering{\includegraphics[width=0.5\textwidth]{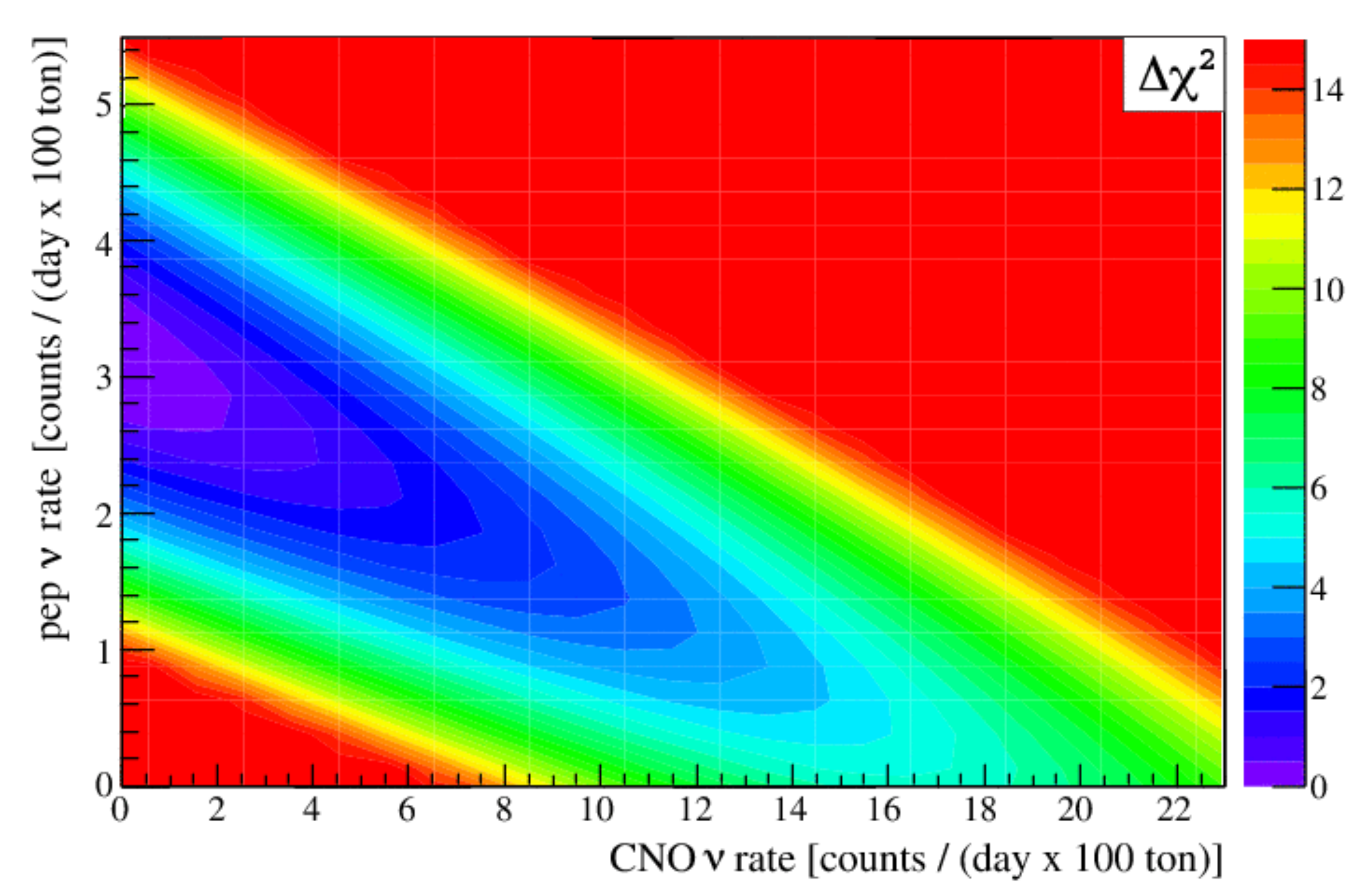}}
\centering{\includegraphics[width=0.47\textwidth]{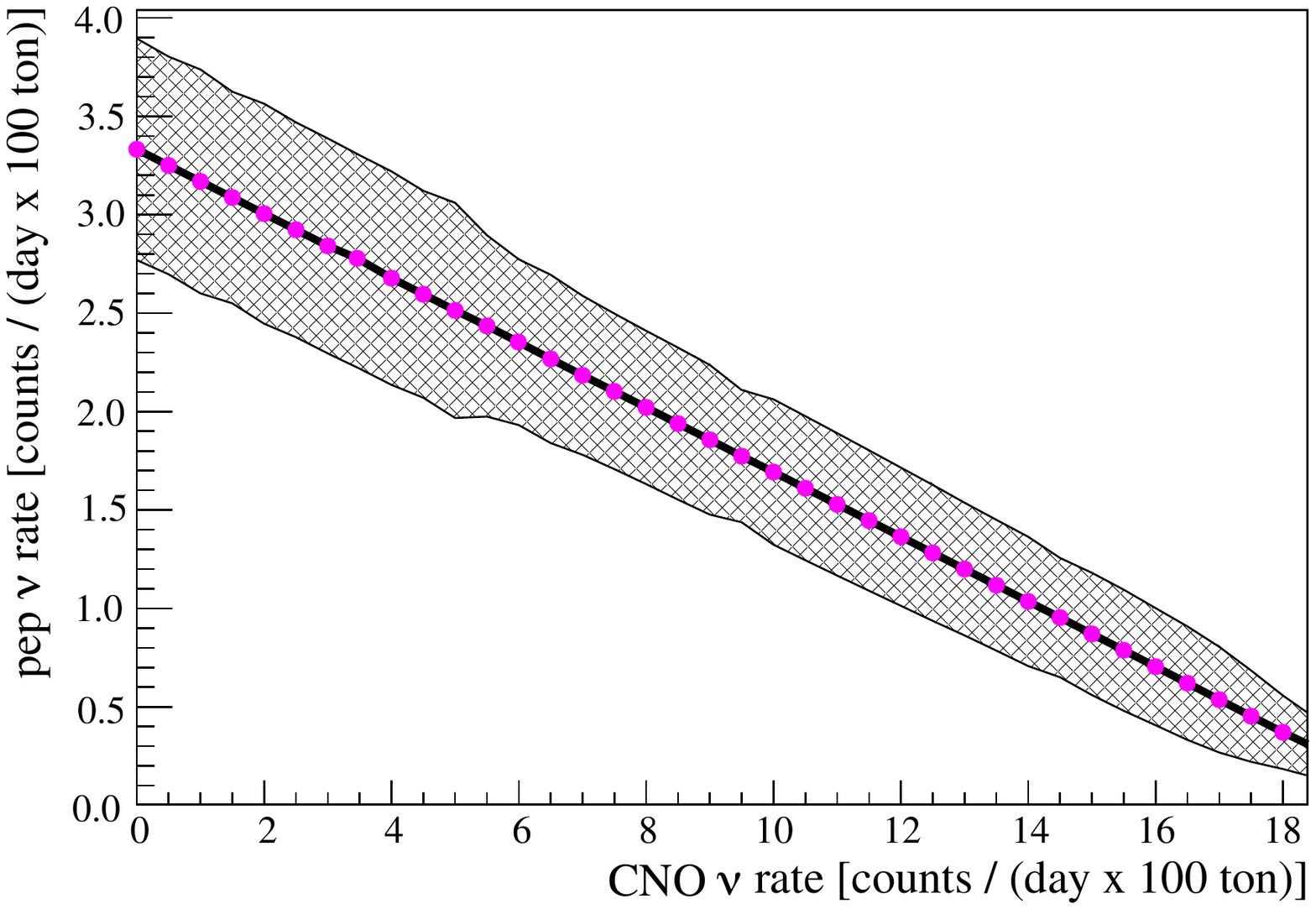}}
\end{array}
$
\caption{Correlation between the $pep$ and CNO neutrino rates. Left: $\Delta \chi^2$ map obtained from a likelihood--ratio test between the likelihood of the best--fit result and the maximum likelihood returned by the fit when $pep$ and CNO neutrino interaction rates are fixed to different values.  The right column gives the colors corresponding to $\Delta \chi^2$ values; note that the same red color has been used to plot all $\Delta \chi^2 \geq 14$ values to allow to visualize color  variations in the relevant region of the plot. The numerical values of this map are given in Table~\ref{MegaTable}. The plot has been obtained using  $N_{pe}^d$ energy estimator; we get similar and consistent numbers using the $N_h$ variable. Right: $pep$--$\nu$ interaction rate returned by the fit for different (fixed) CNO neutrino interaction rates. The shaded area is the statistical uncertainty.}
\label{fig:pep_vs_CNO}
\end{figure*}

\begin{figure}[t]
\begin{center}
\hspace{4 mm}
\includegraphics[width = 0.49\textwidth]{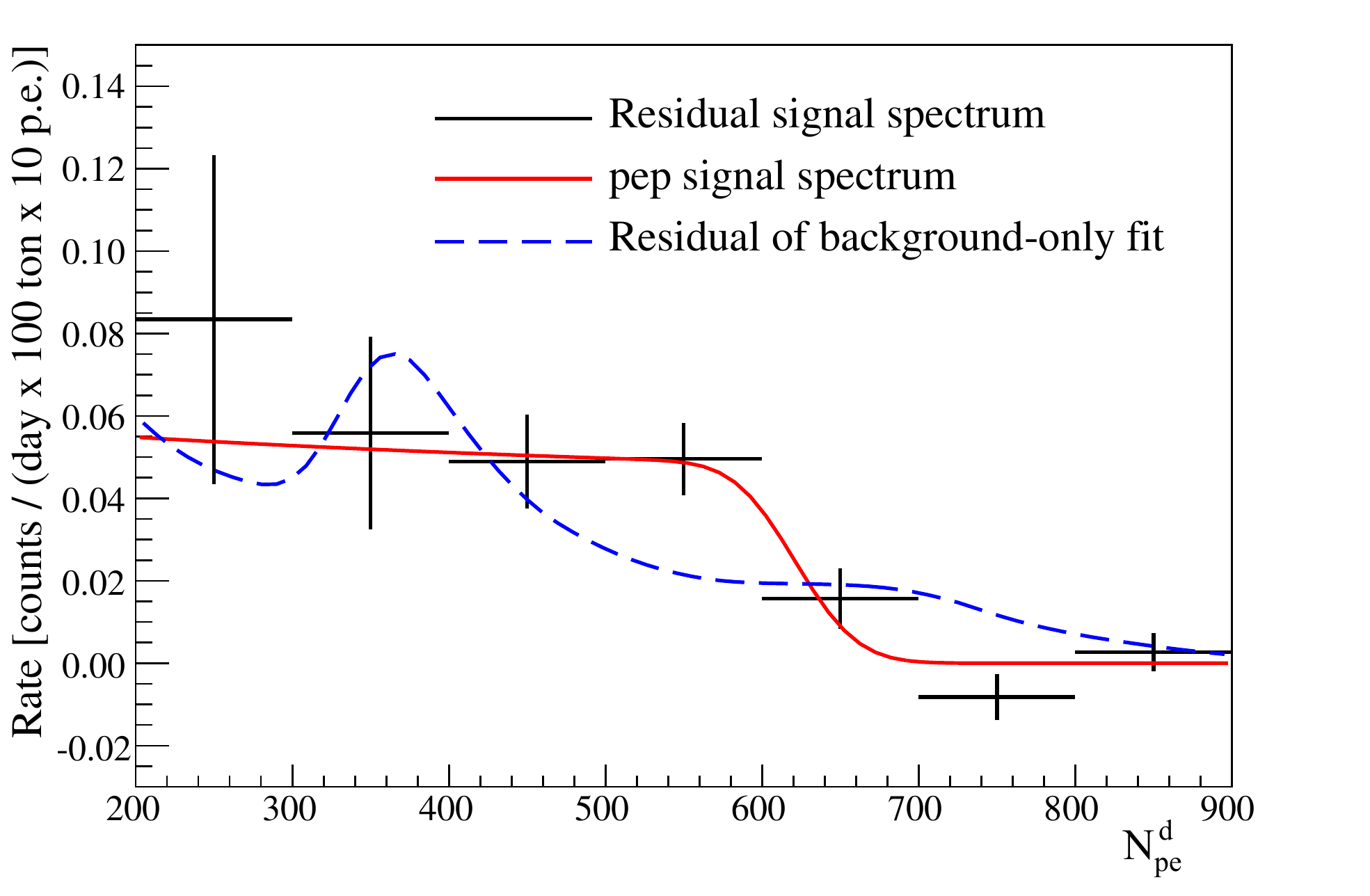}
\caption{Residuals after the best-fit values of all species except the signal from $pep$ and
CNO solar neutrinos are subtracted from the data energy spectrum (black data points). The $e^-$ recoil spectrum
due to $pep$--$\nu$ is shown in solid-red line. 
A second fit has been performed excluding the pep and CNO signals. The dashed blue line gives the resulting best-fit spectrum, after the subtraction of the best estimates for the backgrounds, as obtained from the fit used for the signal extraction.
The two isotopes showing  a significant difference in the output of the background only fit and in the one with the solar neutrinos are $^{210}$Bi and $^{234m}$Pa.
}
%The dashed-blue line gives the difference with respect to the best-fit when no signal from $pep$ or CNO $\nu$'s is included.

\label{fig:AlvaroFig9.5}
\end{center}
\end{figure}

\begin{table*}[tp]
\centering
\begin{tabular}{l|cccccccccc}
\hline\hline
CNO & 0 & 1 & 2 & 3 & 4 & 5 & 6 & 7 & 8 & 9 \\
$pep$   &    &    &     &    &     &    &     &    &    &\\
\hline
0&	33.7	&	31	&	28.5	&	26	&	23.6	&	21.4	&	19.2	&	17.2	&	15.2	&	13.5	\\
0.25&	29.2	&	26.7	&	24.3	&	22	&	19.8	&	17.7	&	15.7	&	13.8	&	12.2	&	10.7	\\
0.5&	25	&	22.6	&	20.3	&	18.2	&	16.1	&	14.2	&	12.5	&	10.9	&	9.54	&	8.37	\\
0.75&	21	&	18.8	&	16.7	&	14.7	&	12.8	&	11.2	&	9.7	&	8.43	&	7.34	&	6.44	\\
1&	17.2	&	15.2	&	13.3	&	11.5	&	9.94	&	8.56	&	7.38	&	6.38	&	5.56	&	4.93	\\
1.25&	13.8	&	11.9	&	10.2	&	8.77	&	7.48	&	6.38	&	5.47	&	4.74	&	4.19	&	3.82	\\
1.5&	10.6	&	9.04	&	7.65	&	6.46	&	5.44	&	4.62	&	3.97	&	3.5	&	3.21	&	3.09	\\
1.75&	7.9	&	6.6	&	5.49	&	4.56	&	3.82	&	3.25	&	2.87	&	2.65	&	2.61	&	2.74	\\
2&	5.6	&	4.58	&	3.74	&	3.07	&	2.59	&	2.28	&	2.15	&	2.19	&	2.39	&	2.76	\\
2.25&	3.72	&	2.96	&	2.39	&	1.98	&	1.76	&	1.7	&	1.81	&	2.09	&	2.54	&	3.15	\\
2.5&	2.25	&	1.75	&	1.43	&	1.28	&	1.3	&	1.49	&	1.84	&	2.36	&	3.04	&	3.88	\\
2.75&	1.17	&	0.927	&	0.856	&	0.953	&	1.22	&	1.65	&	2.24	&	2.99	&	3.9	&	4.96	\\
3&	0.478	&	0.483	&	0.656	&	0.994	&	1.5	&	2.16	&	2.98	&	3.96	&	5.09	&	6.37	\\
3.25&	0.161	&	0.409	&	0.82	&	1.39	&	2.13	&	3.02	&	4.06	&	5.26	&	6.61	&	8.11	\\
3.5&	0.212	&	0.696	&	1.34	&	2.14	&	3.1	&	4.22	&	5.48	&	6.9	&	8.46	&	10.2	\\
3.75&	0.621	&	1.34	&	2.21	&	3.24	&	4.42	&	5.75	&	7.23	&	8.86	&	10.6	&	12.6	\\
4&	1.38	&	2.32	&	3.42	&	4.66	&	6.06	&	7.61	&	9.3	&	11.1	&	13.1	&	15.2	\\
4.25&	2.48	&	3.64	&	4.95	&	6.42	&	8.02	&	9.78	&	11.7	&	13.7	&	15.9	&	18.2	\\
4.5&	3.91	&	5.29	&	6.82	&	8.49	&	10.3	&	12.3	&	14.4	&	16.6	&	19	&	21.5	\\
4.75&	5.67	&	7.26	&	8.99	&	10.9	&	12.9	&	15	&	17.3	&	19.8	&	22.3	&	25	\\
5&	7.75	&	9.55	&	11.5	&	13.6	&	15.8	&	18.1	&	20.6	&	23.2	&	26	&	28.9	\\
5.25&	10.1	&	12.1	&	14.3	&	16.5	&	19	&	21.5	&	24.2	&	27	&	29.9	&	33	\\
5.5&	12.8	&	15	&	17.4	&	19.8	&	22.4	&	25.2	&	28	&	31	&	34.1	&	37.4	\\
\hline
\hline
CNO & 10 & 11 & 12 & 13 & 14 & 15 & 16 & 17 & 18 & 19 \\
$pep$   &    &    &     &    &     &    &     &    &    &\\
\hline

0&	11.9	&	10.6	&	9.42	&	8.45	&	7.66	&	7.06	&	6.63	&	6.38	&	6.31	&	6.41	\\
0.25&	9.44	&	8.37	&	7.49	&	6.78	&	6.26	&	5.92	&	5.75	&	5.76	&	5.93	&	6.28	\\
0.5&	7.38	&	6.58	&	5.96	&	5.52	&	5.26	&	5.17	&	5.25	&	5.51	&	5.93	&	6.52	\\
0.75&	5.73	&	5.19	&	4.83	&	4.65	&	4.64	&	4.8	&	5.13	&	5.63	&	6.29	&	7.11	\\
1&	4.48	&	4.2	&	4.1	&	4.17	&	4.4	&	4.81	&	5.38	&	6.11	&	7	&	8.05	\\
1.25&	3.62	&	3.59	&	3.74	&	4.05	&	4.53	&	5.17	&	5.98	&	6.94	&	8.06	&	9.33	\\
1.5&	3.14	&	3.36	&	3.75	&	4.31	&	5.02	&	5.89	&	6.93	&	8.11	&	9.45	&	10.9	\\
1.75&	3.04	&	3.5	&	4.13	&	4.91	&	5.86	&	6.96	&	8.21	&	9.62	&	11.2	&	12.9	\\
2&	3.3	&	4	&	4.85	&	5.87	&	7.04	&	8.36	&	9.83	&	11.5	&	13.2	&	15.1	\\
2.25&	3.92	&	4.84	&	5.92	&	7.16	&	8.55	&	10.1	&	11.8	&	13.6	&	15.6	&	17.7	\\
2.5&	4.88	&	6.03	&	7.33	&	8.79	&	10.4	&	12.1	&	14	&	16.1	&	18.3	&	20.6	\\
2.75&	6.18	&	7.55	&	9.07	&	10.7	&	12.6	&	14.5	&	16.6	&	18.9	&	21.2	&	23.7	\\
3&	7.81	&	9.39	&	11.1	&	13	&	15	&	17.2	&	19.5	&	21.9	&	24.5	&	27.2	\\
3.25&	9.76	&	11.6	&	13.5	&	15.6	&	17.8	&	20.2	&	22.7	&	25.3	&	28.1	&	31	\\
3.5&	12	&	14	&	16.2	&	18.5	&	20.9	&	23.4	&	26.1	&	28.9	&	31.9	&	35	\\
3.75&	14.6	&	16.8	&	19.1	&	21.6	&	24.2	&	27	&	29.9	&	32.9	&	36	&	39.3	\\
4&	17.5	&	19.9	&	22.4	&	25.1	&	27.9	&	30.8	&	33.9	&	37.1	&	40.4	&	43.8	\\
4.25&	20.7	&	23.3	&	26	&	28.8	&	31.8	&	34.9	&	38.2	&	41.5	&	45	&	48.7	\\
4.5&	24.1	&	26.9	&	29.8	&	32.9	&	36	&	39.3	&	42.7	&	46.3	&	50	&	53.7	\\
\hline \hline 
\end{tabular}
\caption{$\Delta \chi^2$ values obtained from a likelihood--ratio test between the likelihood of the best--fit result and 
the maximum likelihood returned by the fit when $pep$ and CNO neutrino interaction rates are fixed to different values.
The rates are expressed in cpd/100\,ton. These values have been obtained using $N^d_{pe}$
energy estimator. A graphical representation of the $\Delta \chi^2$ map can be found in Fig.~\ref{fig:pep_vs_CNO}. }
\label{MegaTable}
\end{table*}

Figure~\ref{fig:pep_vs_CNO} presents the best--fit value for the $pep$--$\nu$ interaction rate when the CNO--$\nu$ interaction rate is fixed at different values and shows the $\Delta \chi^2$ map as a function of both $pep$ and CNO neutrino interaction rates. The numerical values of this map are given in Table~\ref{MegaTable}.

The probability that the data arises from the background--only hypothesis, excluding the signals from $pep$ and CNO neutrinos, is estimated to be $3 \times 10^{-7}$. A comparison of the energy spectrum of the background--only hypothesis with the best--fit result in the $pep$--shoulder energy region is given in Fig.~\ref{fig:AlvaroFig9.5}.

As may be observed in Figs.~\ref{fig:pep_Ene_nhits} and ~\ref{fig:pep_Ene_npe}, the $^{11}$C, $^{10}$C, and $^6$He rates are much smaller in the spectrum of events after the TFC veto, as expected. Using the fit results we can measure the residual fraction of $^{11}$C background after the TFC veto to be $0.094\pm0.009$. The measured production rates of cosmogenic isotopes $^{10}$C and $^6$He are also consistent with other Borexino analyses~\cite{bxB8}, \cite{cosmogenics} and those obtained by extrapolating KamLAND data~\cite{KamlandCosmogenics}.

The results obtained on the $^7$Be neutrino interaction rate and the $^{85}$Kr activity are consistent with our measurement reported in Section
\ref{sec:Be7results}. The rate of the $^{210}$Bi  obtained (about $35 \%$ higher than that reported in the context of the measurement of the  $^7$Be neutrino interaction) 
is due to the different choice of the data set used for the $pep$ and $^7$Be 
analysis and to the change of the $^{210}$Bi rate with time (see Fig.~\ref{fig:Birate}). The 2007 data set, corresponding to the lower $^{210}$Bi
rate, is not used for the $pep$ analysis and this leads to a mean value of $^{210}$Bi rate higher than the one obtained in Section~\ref{sec:Be7results}.
In addition, in  the $^7$Be--$\nu$ analysis the CNO contribution was fixed to the high--metallicity solar model prediction (possible variations of the CNO--$\nu$ rate were included in the systematic uncertainty, see Section~\ref{sec:Be7results}) while it is a free parameter in the $pep$ analysis: the fit prefers a value of the CNO interaction rate equal to zero so favoring high values of the $^{210}$Bi rate.

The fit method, the reliability of the fit results and the interpretation of the likelihood ratio test as a $\Delta \chi^2$  test have been validated with the use of data--like samples of known input composition obtained with the Monte-Carlo.

The dominant sources of systematic uncertainties for the $pep$--neutrino interaction rate are given in 
Table~\ref{tab:pepCNO_syst} with their estimated values.
These systematics increase the upper limit in the CNO
neutrino interaction rate by 0.8\,cpd/100\,ton.
The evaluation of the systematic uncertainty due to the energy scale, fit procedures, and live--time prior to the TFC veto
 has been performed as described for the $^7$Be--$\nu$ analysis in Section~\ref{sec:Be7results}.
The contribution of FV--uncertainty has been evaluated with the same method of $^7$Be--$\nu$ analysis, but using a higher energy range that also includes electron recoils from $pep$ and CNO--neutrino interactions (300 -- 1600 $N_{pe}^d$).

We evaluated, in addition, the contribution of the live--time uncertainty due to the TFC veto. The
statistical uncertainty related to the counting method (Section~\ref{subsec:tfc}) used to estimate
the relative exposure after the TFC vetoes is $<$0.5\% thanks to the large
number of events considered ($^{210}$Po or simulated). We include as systematic uncertainty the discrepancy between the two methods described in Section
\ref{sec:c11} which is $<$1\%.
The overall uncertainty in the exposure introduced by the live--time estimation and the TFC
veto  is less than 1\%.

To test the robustness of the fit against the inclusion of a small fraction of $\gamma$ rays in both the sample of events used to construct the PDF for electrons and the data to fit, we decreased the energy end point of the PS--BDT fit from 900 $N^d_{pe}$ to 700 $N^d_{pe}$. As the $\gamma$--ray contribution in both cases increases with increasing energy (Fig.~\ref{fig:214BiGammaEnergy} and Fig.~\ref{fig:pep_Ene_npe}, respectively), this restriction mitigates any possible systematic effect associated with the presence of a small number of $\gamma$ rays in the electron data. The fit performed with the lower--energy end point of the PS--BDT distributions (PDF and data) returned a central value of the $pep$ interaction rate increased by 2.7\%. This increase has been taken as the systematic uncertainty due to possible $\gamma$--rays contamination in the test sample used to build the PS--BDT distributions.

The $^{85}$Kr value returned by the fit is $2 \sigma$ away (lower) from the independent measure obtained with
the coincidence analysis. We have included in the study of the systematic uncertainty the variation of the $pep$--neutrino interaction rate
obtained including in the likelihood a constraint describing the information about the $^{85}$Kr. This contribution is
a Gaussian--approximated term:
\begin{equation}
 -ln L_G = \frac {(R-R_0)^2} {2 \sigma_0^2},
\label{Lkr}
\end{equation}
where $R$ is the $^{85}$Kr rate in the fit and $R_0$ and $\sigma_0$ are the central value and the standard deviation of the independent constraint.
As reported in Table~\ref{tab:pepCNO_syst}, the $pep$ central value increases by 3.9\%.

\begin{table}[h]
\begin{center}
\begin{tabular}{lc} \hline \hline
Source                                         & Uncertainty (\%)           \\ \hline
Fiducial exposure                              & $^{+0.6}_{-1.1}$    \\
Energy response                                & $\pm$ 4.1           \\
$^{210}$Bi  spectral shape                     & $^{+1.0}_{-5.0}$        \\
Fit methods                                    & $\pm$ 5.7           \\ 
Inclusion of independent  $^{85}$Kr estimate & $^{+3.9}_{-0}$ \\
$\gamma$ rays in pulse--shape distribution      & $\pm$2.7    \\
Stat. unc. in pulse--shape distributions        & $\pm 5$   \\
\hline 
Total systematic uncertainty  &            $\pm$10.0 \\
\hline \hline
\end{tabular}
\end{center}
\caption{Systematic uncertainties of the $pep$--$\nu$ interaction rate. These systematics increase the upper limit in the CNO--neutrino interaction rate by 0.8\,cpd/100\,ton.}
\label{tab:pepCNO_syst}
\end{table}

The number of events used to train the PS--BDT and to build its PDF is not much larger than the number of fitted events in the energy region of $^{11}$C. Because the former number of events is low, the statistical uncertainty in the PS--BDT PDFs may be not negligible. In order to estimate such uncertainty, we have created 100 more PS--BDT PDFs for e$^{+}$ and e$^{-}$; the bin content of these PDFs has been extracted according to Poisson statistics, using the original PS--BDT PDF bin content as the expected $\mu$ value.
We have performed 100 mutivariate fits using these 100 simulated PS--BDT PDFs; the standard deviation of the distribution of the corresponding 100 central values of $pep$--$\nu$ interaction rates, amounting to 5\% of the mean value, has been used as an estimate of the systematic uncertainty due to the use of PS--BDT PDFs with limited statistics (listed in Table~\ref{tab:pepCNO_syst} as statistical uncertainties in pulse--shape distribution).

\begin{figure}
\begin{center}
\includegraphics[width = 0.48 \textwidth]{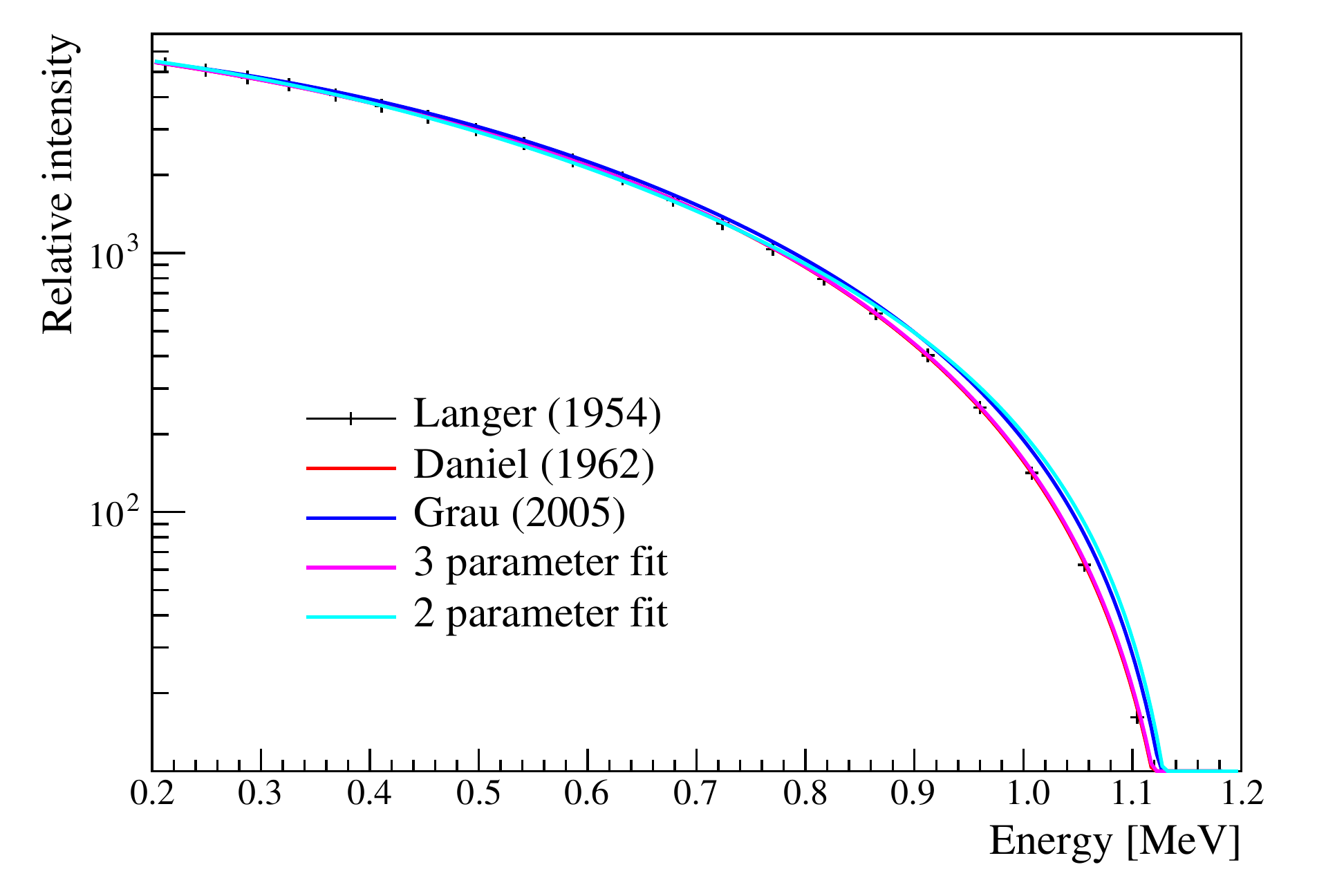}
\caption{Comparison between the $^{210}$Bi energy spectrum measured using the magnetic spectrometer 
\cite{bib:LangerBi210} (crosses, labeled as Langer (1954)) with spectra
obtained using different correction factors: those calculated in \cite{bib:DanielBi210} (red curve labeled as Daniel (1962))  and in \cite{bib:Grau2} (blues curve labeled as Grau (2005)). Also shown are the fit to the Langer data according to Eq.\ref{eq:Bi210Fit} with all the three parameters and with b=0.}
\label{fig:CompareBi}
\end{center}
\end{figure}

The shape of the $^{210}$Bi $\beta$--decay spectrum was measured  using a magnetic spectrometer,
where the electron energy $E$ has an uncertainty $<$0.1\% and the statistical uncertainty in the relative intensity
is $<$1\% for E $<$ 960\,keV~\cite{bib:LangerBi210}.
Figure~\ref{fig:CompareBi} shows the comparison between this measurement and the spectra obtained using different correction factors as it results in \cite{bib:DanielBi210} and \cite{bib:Grau2}. In addition, the fit of the magnetic spectrometer data with a function:
\begin{equation}
C(W)= 1+ a W + b/W +c W^2,
\label{eq:Bi210Fit}
\end{equation}
where $W=1+ E/m_e c^2$ is shown.
The results agree in relative intensity at 1$\%$ level.
More recent measurements of the $^{210}$Bi  spectrum have been performed using Cherenkov and scintillator light detectors~\cite{bib:Grau1}, \cite{bib:Grau2}.
The discrepancy between the $\beta$ spectrum obtained from the scintillator measurement, also shown in Fig.~\ref{fig:CompareBi}, and the one from the magnetic spectrometer becomes $>$10\% for E $>$ 900\,keV and as large
as 30\% near the end--point, the region that is most influential in the determination of the neutrino rates.
These effects are included in the systematic uncertainty shown in Table~\ref{tab:pepCNO_syst}.

\section{Neutrino oscillation analysis with the Borexino results}
\label{sec:GbAn}
%%%%%%%%%%%%%%%%%%%%%%%%%%%%%%%% SHORT-CUTS %%%%%%%%%%%%%%%%%%%%%%%%%%%%%%%%%%%%
\newcommand{\D} 	{$\Delta m^2_{21}$}
\newcommand{\ta} 	{$\theta_{12}$}
\newcommand{\n}		{$\nu_e$}
\newcommand{\C}		{$\chi^2$}
\newcommand{\Mr}[1]	{\mathrm{#1}}
%%%%%%%%%%%%%%%%%%%%%%%%%%%%%%%%%%%%%%%%%%%%%%%%%%%%%%%%%%%%%%%%%%%%%%%%%%%%%%%%%

In this Section we discuss the physical implications of the Borexino results in the context of the neutrino oscillations and of the solar models.

We show the regions of the oscillation parameters $\Delta m^2_{21}$ and $\tan^2\theta_{12}$ determined by the Borexino data alone and by the Borexino data combined with  that of the others solar neutrino  experiments. Particularly interesting is the fact that the solar neutrino results, once combined, single out the LMA region even without including in the global analysis the results of the KamLAND experiment about reactor antineutrinos. The LMA solution is thus obtained without assuming the validity of CPT symmetry.
We also show that, despite  the Borexino results  have significantly contributed  in the determination of the $^7$Be--$\nu$ flux, both the theoretical and experimental uncertainties are still too large  to distinguish between high-- and low--metallicity solar models. 
Finally, one of the most interesting outcomes of the low--energy solar neutrino measurement of Borexino is the experimental knowledge of the neutrino survival probability as a function of energy: these results are discussed in Section~\ref{sec:Pee}.

%%%%%%%%%%%%%%%%%%%%%%%%%%%%%%%%
\begin{figure*}[!t]
\centering
\begin{minipage}[c]{\textwidth}
 \centering{\includegraphics[width=0.8\textwidth]{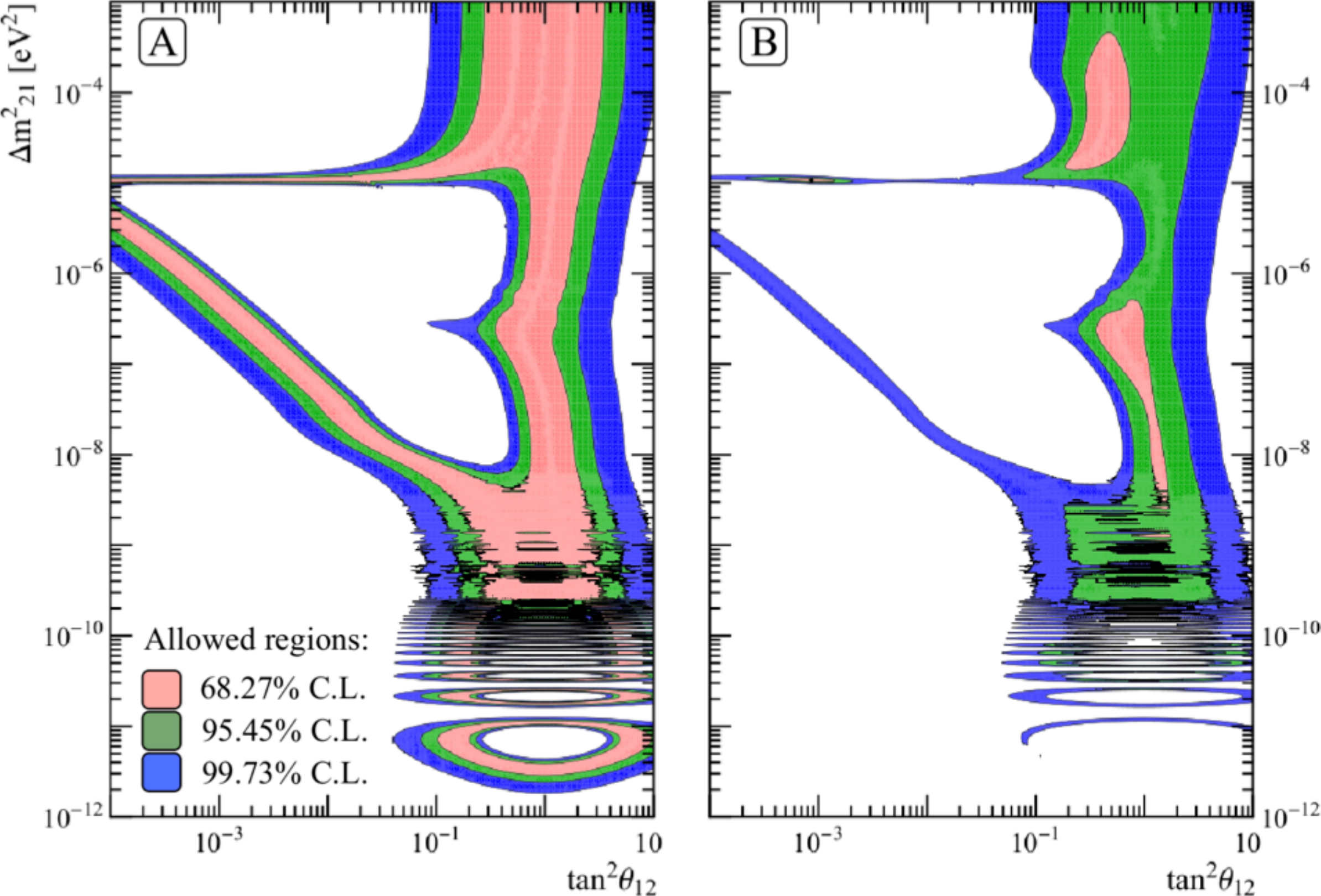}}
 \vspace{1em}
\end{minipage}
\begin{minipage}[c]{\textwidth}
 \centering{\includegraphics[width=0.8\textwidth]{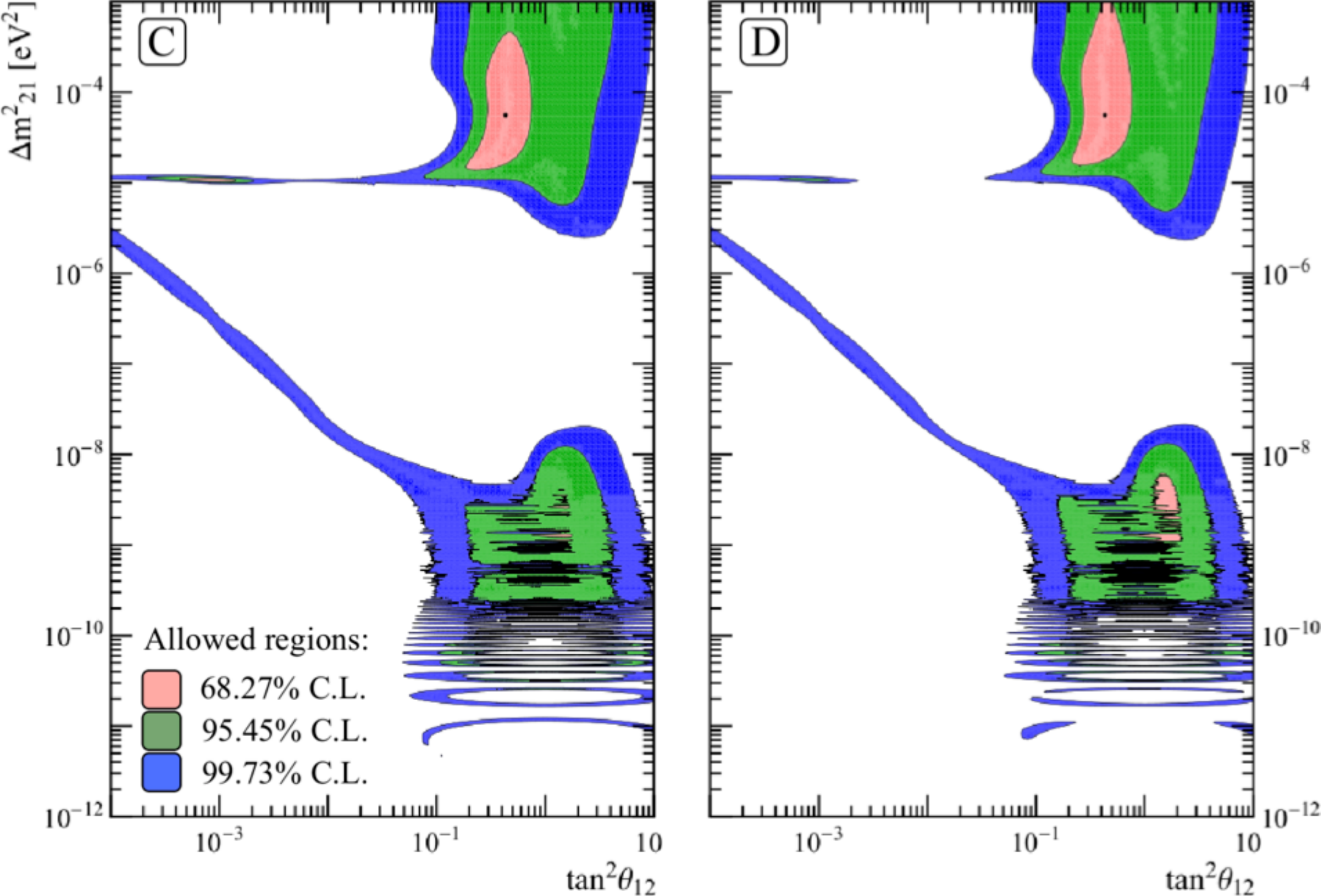}}
\end{minipage}
 \caption{The Borexino data analysis in the $\tan^2\theta_{12}\,-\,$\D~space. Allowed regions
(NDF = 2) at 68.27\% C.L. (pink), 95.45\% C.L. (green), and 99.73\% C.L. (blue). Panel A: impact of the
Borexino $^7$Be--$\nu$ rate measurement. Panel B: the combined analysis of Borexino measurements of $^7$Be- and $^8$B--$\nu$ rates. Panel C: the impact of $^7$Be- and $^8$B--$\nu$ rates together with the $^7$Be--$\nu$ day--night asymmetry results.
Panel D: the global impact of all the Borexino measurements to date, including the $pep$--$\nu$
rate.}
 \label{fig:GbAn:BX}
\end{figure*}
%%%%%%%%%%%%%%%%%%%%%%%%%%%%%%%%

Assuming the standard three--neutrino framework and the energy range of solar 
neutrinos, it is possible to perform an effective three--flavor analysis by reducing 
the Hamiltonian which describes the oscillations phenomena to a $2 \times 2$ matrix, 
the so--called \textit{effective Hamiltonian}, $\mathcal H_{\Mr{eff}}$ \cite{Glob1, Glob2}. 
This yields the survival probability of an electron neutrino to be defined as:
\begin{equation}\label{eq:GbAn:Pee_step1}
P^{3\nu}_{ee}= \sin^4 \theta_{13}+\cos^4 \theta_{13}\,P^{2\nu}_{ee}\:,
\end{equation}
where $P^{2\nu}_{ee}=|\langle \nu_e | \mathcal H_{\rm eff} | \nu_e\rangle |^2$.

In the two--neutrino mixing case, the survival probability for a solar electron neutrino of 
given energy can be written as \cite{Glob3}:
\begin{equation}
\begin{array}{rl}
P^{2\nu}_{ee}=&P^S\,(1-P^E) + (1-P^S)\,P^E \\
 & + 2\,\cos \xi \,\sqrt{P^S\,(1-P^E)\,(1-P^S)\,P^E}\:,
\end{array}
\label{eq:GbAn:Pee_step3}
\end{equation}
where $P^S$ is the probability that a \n~produced in the Sun becomes a neutrino mass 
eigenstate $\nu_{1}$, $P^E$ is the probability
that a neutrino propagating in vacuum as mass eigenstate $\nu_{2}$ is detected on Earth
as a \n, and the factor $\xi$ is defined as: 
\begin{equation}
\xi = \frac{\Delta m^2_{12}}{2E} (L-r)\:,
\label{eq:xi}
\end{equation}
where $L$ is the average distance between the center of the Sun and the surface of the Earth and $r$ is the distance between
the neutrino production point and the center of the Sun.

The survival probability is computed by 
dealing separately $P^S$  and $P^E$: these two quantities are 
calculated for each set of parameters \D$/4E$, $\tan^2\theta_{12}$, 
$\sin^2 \theta_{13}$, according to the indications of the standard solar model~\cite{HighMet_GS98SSM}. The propagation of neutrinos inside the Earth has been evaluated by selecting shells with uniform density
according to the Earth model described in~\cite{PREM}.

For all experiments except Borexino we use the ideal zenith exposure.
In the Borexino case, it was possible to use the experimental exposure function 
weighted by the real live--time.

The parameter estimation is obtained by finding the minimum of the $\chi^2$ function and by 
tracing the iso--$\Delta \chi^2$ contours around it.

If $\Mr{R^{i,\,\scriptscriptstyle A}_{\scriptscriptstyle{EXP}}}$
is the set of results of the measurement $i$ actually obtained by the $A$ experiment, 
and $\Mr{R^{i,\,\scriptscriptstyle A}_{\scriptscriptstyle{THEO}}}(\Delta \Mr{m}^2_{21},\:
\tan^2\theta_{12},\:
\sin^2\theta_{13},\:\Mr{\Phi_{\nu,\,\scriptscriptstyle A}})$ is the corresponding set of 
theoretical predictions, then the \C~of the $A$ experiment is defined as:
\begin{equation}
\begin{array}{rl}
\Mr{\chi^2_{\scriptscriptstyle A}=} & \Mr{\left[R^{i,\,\scriptscriptstyle A}_
{\scriptscriptstyle{EXP}}-R^{i,\,
\scriptscriptstyle A}_{\scriptscriptstyle{THEO}}(\Delta m^2_{21},\:\theta_{12},\:\theta_{13},\:
\Phi_{\nu,\,\scriptscriptstyle A})\right]\:\sigma^{-2}_{ij}\:\left[R^{j,\,\scriptscriptstyle A}
_{\scriptscriptstyle{EXP}}\right.} \\
 & \Mr{\left. -R^{j,\,\scriptscriptstyle A}_{\scriptscriptstyle{THEO}}
(\Delta m^2_{21},\:\theta_{12},\:\theta_{13},\:\Phi_{\nu,\,\scriptscriptstyle A})\right]}
\end{array}
\end{equation}
The error matrix $\sigma_{ij}$ includes both the theoretical and experimental uncertainties as well 
as the cross--correlations between errors on the different parameters.

The \C--projections for each parameter of the fit are then obtained by marginalizing over \D, 
$\tan^2\theta_{12}$, and $\sin^2\theta_{13}$. Unless otherwise stated, the uncertainties we quote
correspond to 1$\sigma$.

\subsection{Analysis of the Borexino data}

In this section we report the results on the neutrino oscillation parameters obtained considering the Borexino data alone.
We assume the high--metallicity SSM. 

The theoretical correlation 
factor between $^7$Be and $^8$B neutrino fluxes is taken from \cite{BeBoCorr}.
After computing the survival probabilities $\Mr{P^S}$ and $\Mr{P^E}$, the
expected rates are evaluated  taking into account the cross sections of
the processes convolved with the detector resolution at the particular investigated energies.

The panels in Fig.~\ref{fig:GbAn:BX} show the effects of the 
analysis of the Borexino $^7$Be--$\nu$ interaction rate as in~\cite{be7-2011} (panel A), 
and the combination of $^7$Be plus $^8$B ($T > 3000$\,keV) neutrino rates and $^8$B spectral 
shape as in \cite{bxB8} (panel B), plus the measurement of a null $^7$Be--$\nu$ day--night 
asymmetry as in~\cite{DNLetter} (panel C), plus the $pep$--$\nu$ total count rate of~\cite{PepBorex} (panel D).
Although Borexino, like all other solar neutrino experiments, does not have significant sensitivity
to $\theta_{13}$, we directly report the results obtained by assuming $\sin^2\theta_{13}$ = 0.0241~\cite{OscParam}.

\begin{figure}[t]
\hspace{-0.8pc}
\centering{\includegraphics[width=0.45\textwidth]{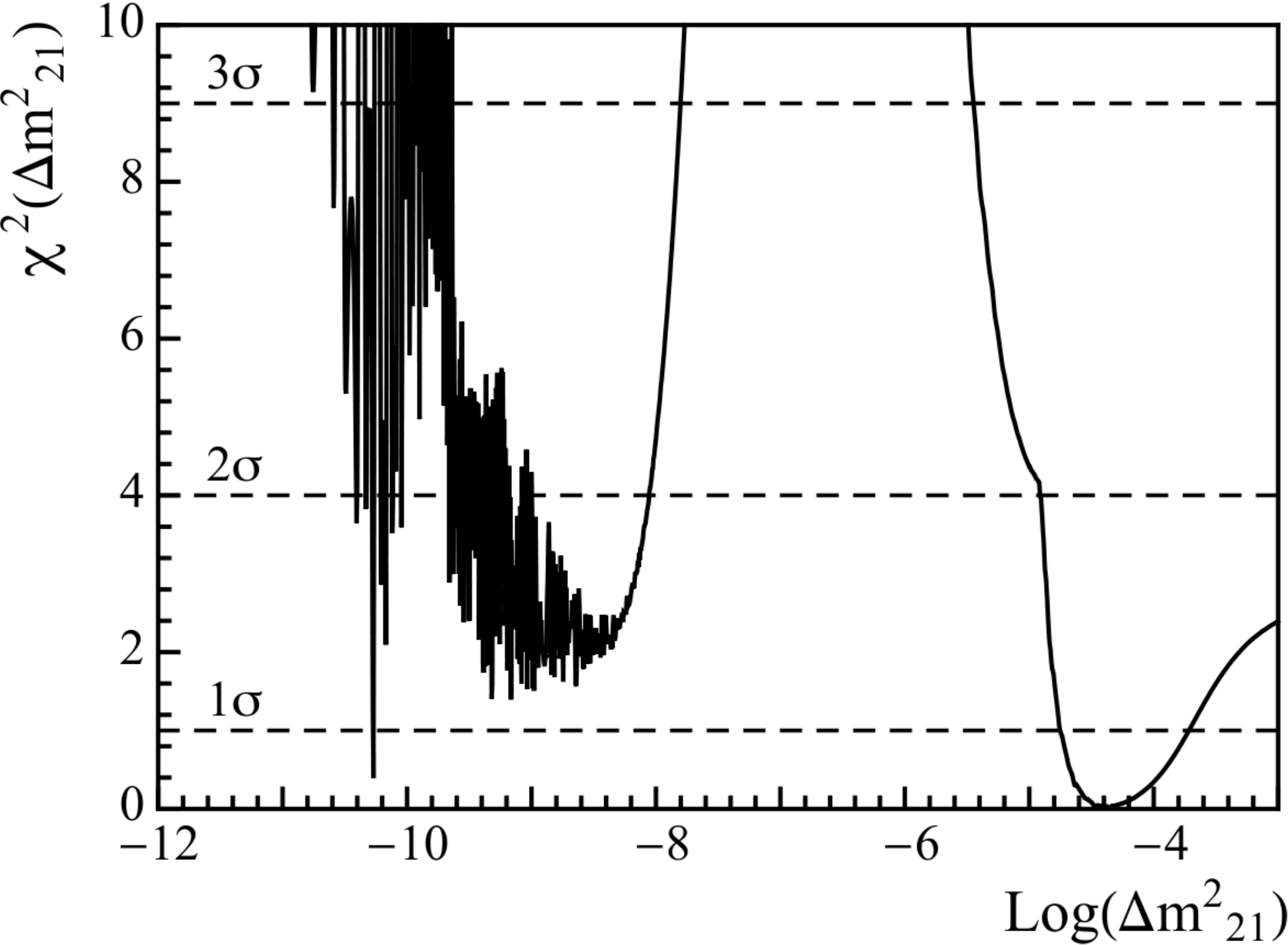}}
\caption{The \C-profile of the \D~parameter in the Borexino results analysis. 
The dashed lines indicate the 1$\sigma$ [$\,$\C(\D)$ = 1\,$], 
2$\sigma$ [$\,$\C(\D)$ = 4\,$], and 3$\sigma$ [$\,$\C(\D)$ = 9\,$] levels. The MSW--LOW region ($10^{-8}\,\mathrm{eV^2} < \Delta m^2_{21} < 10^{-6}\,\mathrm{eV^2}$)
is ruled out at more than 8.5$\sigma$.}
\label{fig:GbAn:BXDm}
\end{figure}

Figure~\ref{fig:GbAn:BXDm} shows a clear output of this study, the rejection of 
the LOW solution ($10^{-8}\,\mathrm{eV^2} < \Delta m^2_{21} < 10^{-6}\,\mathrm{eV^2}$) in the MSW scenario: 
the Borexino experiment alone is able to rule out the LOW mass regime at more than 8.5$\sigma$.

\subsection{Combined analysis of  solar neutrino experiments results}

In this section we present (as before, in the framework of the high--metallicity standard solar model)
the results on the oscillation parameters obtained by a combined analysis of all the solar neutrino experiments with and without Borexino.
We do not include here the results on reactor antineutrinos obtained with the KamLAND experiment. 
We first analyzed the impact of the results excluding Borexino and including the data from
the Homestake \cite{Homestake}, GALLEX/GNO--\cite{GALLEX}, SAGE \cite{SAGE}, 
SNO \cite{SNOa, SNOb}, and Super-Kamiokande \cite{SK1, SK3}.

%%%%%%%%%%%%%%%%%%%%%%%%%%%%%%%%

\begin{figure}[b]
\vspace{-2 mm}
\centering
 \includegraphics[width=0.48\textwidth]{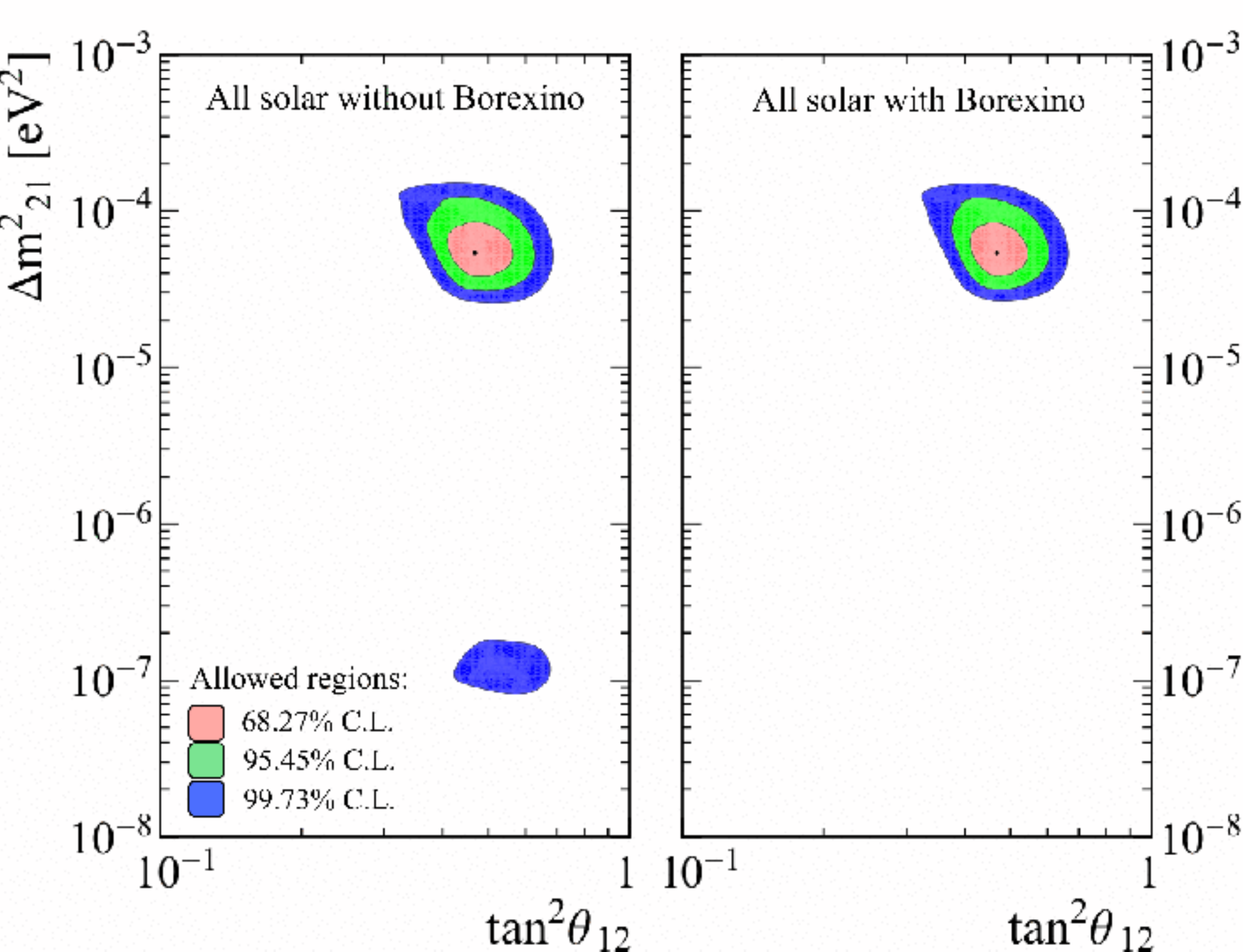}
 \caption{Allowed regions (NDF = 3) of the space of parameters
at 68.27\% C.L. (pink), 95.45\% C.L. (green), and 99.73\% C.L. (blue) by the
solar-without Borexino (left panel) and solar-with Borexino (right panel) data set.}
 \label{fig:GbAn:SOLAR}
\end{figure}

\begin{figure*}[ht]
\centering
\hspace{-0.6pc}
 \centering{\includegraphics[width=0.98\textwidth]{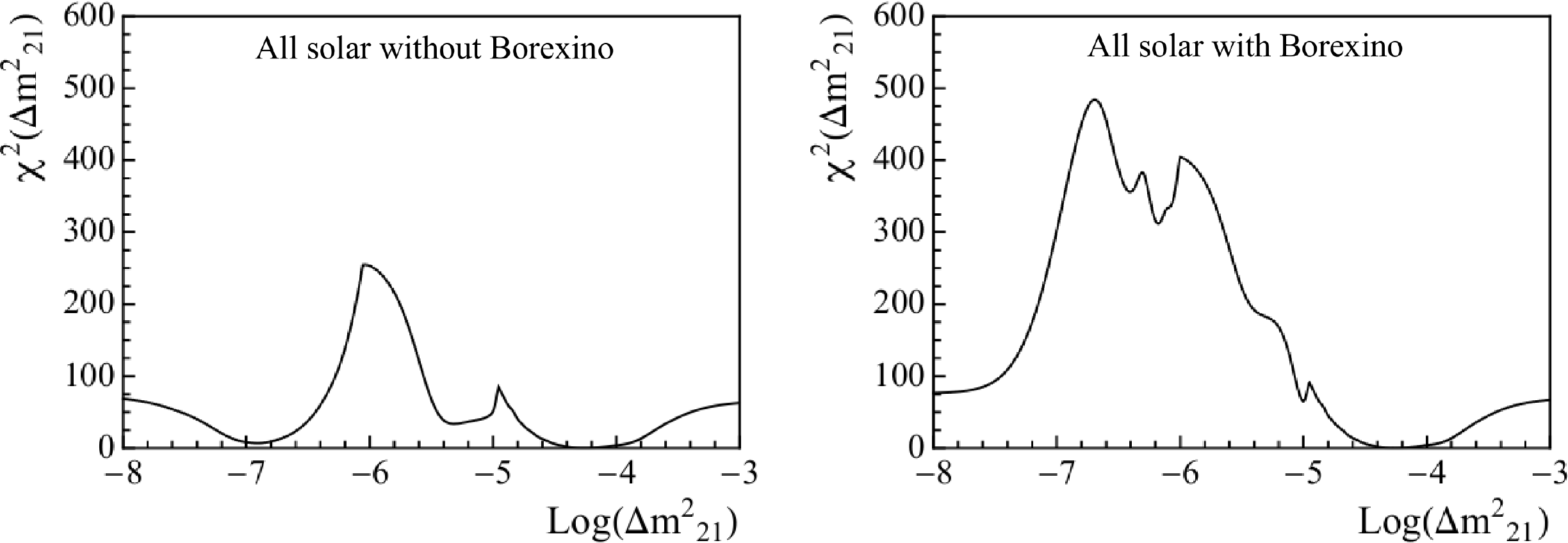}}
 \caption{Comparison of the \C-profile for \D~obtained by the analysis of all
available solar data without (left) and with (right) the Borexino contribution, after marginalization
over $\tan^2\theta_{12}$ and $\sin^2\theta_{13}$.}
 \label{fig:GbAn:DmComp}
\end{figure*}

The left panel of Fig.~\ref{fig:GbAn:SOLAR} shows the resulting  allowed regions for the oscillation 
parameters. 
In this case, the best--fit point ($\Mr{\Delta m^2_{21} =5.4^{+1.7}_{-1.1} \times 10^{-5} \, eV^2}$, 
$\tan^2 \theta_{12}  = 0.479^{+0.035}_{-0.042}$, $\sin^2 \theta_{13} < 0.029$) belongs 
to the LMA region but a small portion of LOW region is still allowed at 
$\Delta\chi^2=11.83$. The right panel of Fig.~\ref{fig:GbAn:SOLAR}
shows the same allowed regions once the Borexino data are included.
The best fit ($\Mr{\Delta m^2_{21} =5.4^{+1.7}_{-1.1} \times 10^{-5} 
\, eV^2}$, $\tan^2 \theta_{12}  = 0.468^{+0.031}_{-0.044}$, $\sin^2 \theta_{13} < 0.030$)
is slightly modified while the LOW region is strongly excluded at 
$\Delta\chi^2 > 190$. Therefore, after the inclusion of the
Borexino data, solar neutrino data alone can single out the LMA solution
with very high confidence (Fig.~\ref{fig:GbAn:DmComp}), without using the Kamland antineutrinos data and thus
without relying on CPT symmetry.

%%%%%%%%%%%%%%%%%%%%%%%%%%%%%%%%

%%%%%%%%%%%%%%%%%%%%%%%%%%%%%%%%
%%%%%%%%%%%%%%%%%%%%%%%%%%%%%%%%

%%%%%%%%%%%%%%%%%%%%%%%%%%%%%%%%
%%%%%%%%%%%%%%%%%%%%%%%%%%%%%%%%

\subsection{Combined analysis of solar plus KamLAND experimental results}\label{sec:GbAn:SOLKL}

The KamLAND contribution in the neutrino oscillation scenario is taken into account 
according to \cite{KL} where a parametric expression of the survival probability, as 
well as the observed values with the relative uncertainties, are reported. 

Assuming the CPT invariance, the analysis of the KamLAND measurements singles out the LMA 
oscillation solution at more than 99.73\% C.L. The best fit obtained 
after marginalizing over the three oscillation parameters is 
$\Mr{\Delta m^2_{21} =7.50^{+0.19}_{-0.20} \times 10^{-5} \: eV^2}$, 
$\tan^2 \theta_{12}  = 0.437^{+0.073}_{-0.060}$, and $\sin^2 \theta_{13} < 0.034$.

After having analyzed both the solar and KamLAND experiments, the next logical step
is a combined analysis of the solar plus KamLAND data.
Since there are no correlations between those sets of data, this study is accomplished 
by directly summing the two \C-outcomes and it results with the best--fit point still 
belonging to the LMA regime: $\Mr{\Delta m^2_{21} =7.50^{+0.18}_{-0.21} \times 10^{-5} 
\: eV^2}$, $\tan^2 \theta_{12}  = 0.457^{+0.038}_{-0.025}$, $\sin^2 \theta_{13} = 
0.023^{+0.014}_{-0.018}$.

\subsection{The solar metallicity controversy}
The analysis so far described were performed under the assumption that the expected 
neutrino fluxes, including their estimated (and correlated)
uncertainties, are predicted by the high--metallicity hypothesis of the 
standard solar model.

The best way to approach the study of the SSM parameters and to 
look deeper into the low/high--metallicity controversy is to analyze the data 
leaving the neutrino fluxes as free parameters of the fit.
We define the reduced fluxes
(or astrophysical factors) $f_{\Mr{Be}}$ and $f_{\Mr{B}}$ where $f_i$ is the
ratio of the true flux to the flux $\Phi^{\Mr{HIGH}}_{\Mr{SSM}}$ predicted by the high--metallicity standard solar model. 
Thus, in the beryllium and boron case, the reduced fluxes are:
\begin{equation}\label{eq:GbAn:RedFlux}
f_{\Mr{Be}}=\cfrac{\Phi(^7\Mr{Be})}{\Phi(^7\Mr{Be})^{\Mr{HIGH}}_{\Mr{SSM}}} \quad \Mr{and} \quad
f_{\Mr{B}}=\cfrac{\Phi(^8\Mr{B})}{\Phi(^8\Mr{B})^{\Mr{HIGH}}_{\Mr{SSM}}}\:.
\end{equation}
By construction, the theoretical beryllium and boron reduced fluxes in the high--metallicity hypothesis result $f_{\Mr{Be}}=1.00 \pm 0.07$ and $f_{\Mr{B}}=1.00 \pm 0.14$.
Instead, in the case of low--metallicity hypothesis, the expected fluxes are 
$f_{\Mr{Be}}=0.91 \pm 0.06$ and $f_{\Mr{B}}=0.82 \pm 0.11$.

Conservation of energy during the solar fusion is 
implemented by imposing the luminosity constraint \cite{LUMIN, ROAD}.

If the global fit is performed on the solar--without Borexino plus KamLAND data set, the 
constraint on beryllium is very weak and the best values for $f_{\Mr{Be}}$ and 
$f_{\Mr{B}}$ are found to be $f_{\Mr{Be}}=0.76^{+ 0.22}_{-0.21}$ and 
$f_{\Mr{B}}=0.90^{+ 0.02}_{-0.02}$. This is due to the fact that $^7$Be flux is 
very poorly constrained by any solar experiment other than Borexino.

Once the Borexino current measurements are included, the 
situation significantly improves and the best fit are 
$f_{\Mr{Be}}=0.95^{+ 0.05}_{-0.04}$, and $f_{\Mr{B}}=0.90^{+ 0.02}_{-0.02}$
corresponding to the neutrino fluxes $\Phi_{\Mr{Be}}= (4.75^{+ 0.26}_{-0.22}) 
\times 10^{9}$\,cm$^{-2}\,$s$^{-1}$, and
$\Phi_{\Mr{B}}= (5.02^{+ 0.17}_{-0.19}) \times 
10^{6}$\,cm$^{-2}\,$s$^{-1}$ respectively.

For $f_{\Mr{B}}$, the best fit value obtained with the two data sets
does not change significantly since the $^8$B flux is mainly determined by the 
results of the SNO and Super-Kamiokande experiments.

The best fit for the oscillation parameters are found to be 
$\Mr{\Delta m^2_{21} =7.50^{+0.17}_{-0.23} \times 10^{-5}~eV^2}$, and 
$\tan^2 \theta_{12}  = 0.452^{+0.029}_{-0.034}$, fully compatible with those 
obtained by fixing all the fluxes to the  standard solar model predictions 
(Section~\ref{sec:GbAn:SOLKL}). In this specific analysis, $\theta_{13}$ is assumed equal to 0.

%%%%%%%%%%%%%%%%%%%%%%%%%%%%%%%%

\begin{figure}[t]
%\vspace{-3 cm}
\hspace{-1.3pc}
\centering{\includegraphics[width=0.49\textwidth]{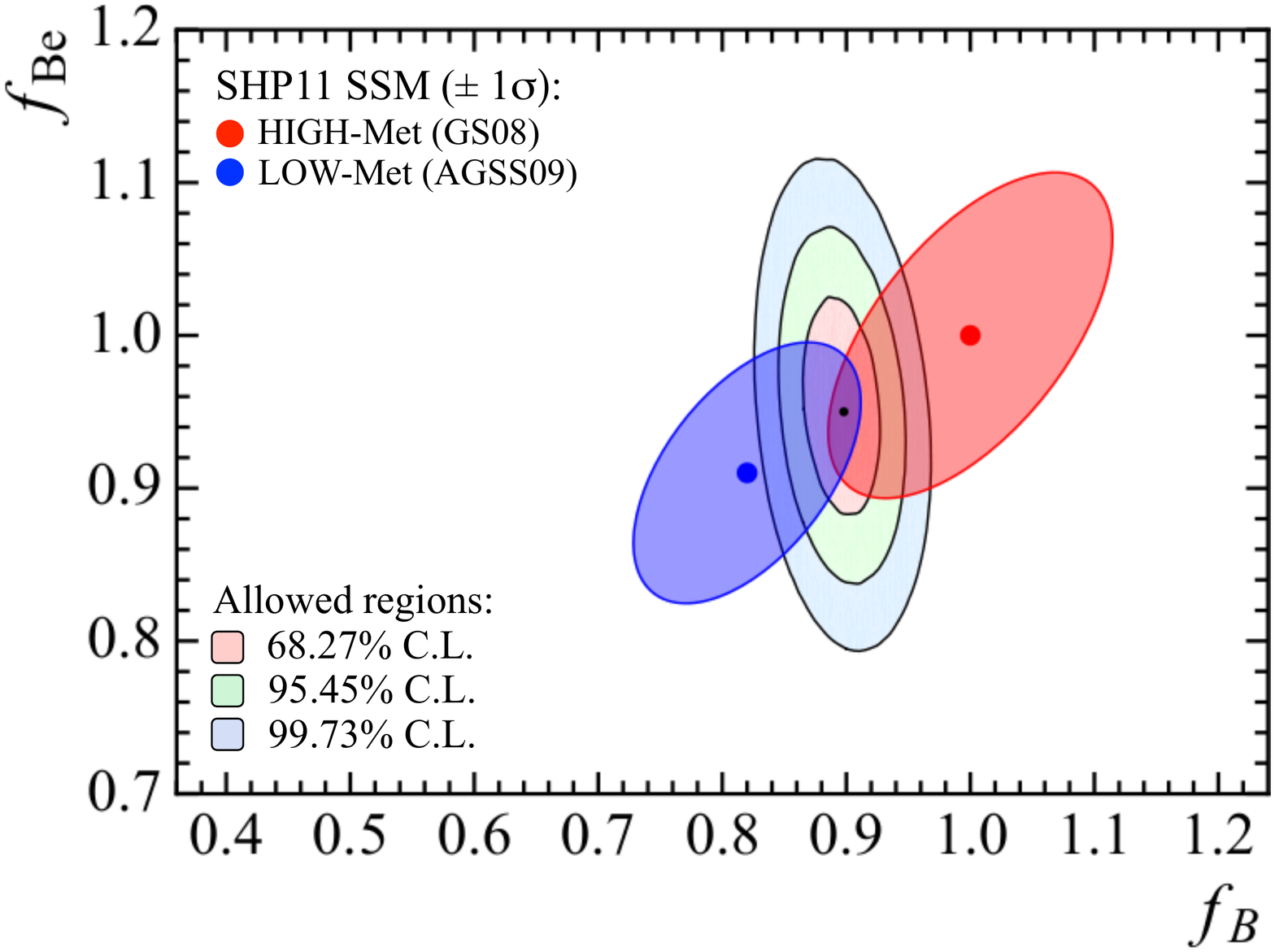}}
 \caption{
The 1$\sigma$ theoretical range of high (red) and low (blue) 
metallicity Standard Solar Model for $f_{\Mr{Be}}$ and $f_{\Mr{B}}$, compared to the 1$\sigma$ 
(light pink), 2$\sigma$ (light green), and 3$\sigma$ (light blue) allowed regions by 
the global analysis of solar-with Borexino plus KamLAND results. The theoretical
correlation factors are taken from \cite{BeBoCorr}.}
 \label{fig:GbAn:BeBoMet}
\end{figure}
%%%%%%%%%%%%%%%%%%%%%%%%%%%%%%%%

It is interesting to compare the result of the global analysis on solar--with
Borexino plus KamLAND results, with the theoretical expectations for
$f_{\Mr{Be}}$ and $f_{\Mr{B}}$.
From Fig.~\ref{fig:GbAn:BeBoMet} it is clear that the actual neutrino data cannot 
discriminate between the low/high--metallicity hypotheses in the solar model: both 
the 1$\sigma$ theoretical range of low/high--metallicity models lies in the 
3$\sigma$ allowed region by the current solar plus KamLAND data. 

At present, no experimental results help to disentangle between the two metallicity
scenarios: the theoretical uncertainty on $^7$Be and $^8$B neutrinos is of the order of their
experimental precision. An improvement in the determination of the different solar 
parameters is needed.

\section{The neutrino survival probability}
\label{sec:Pee}

\begin{figure}[h]
\begin{center}
\includegraphics[width = 0.48\textwidth]{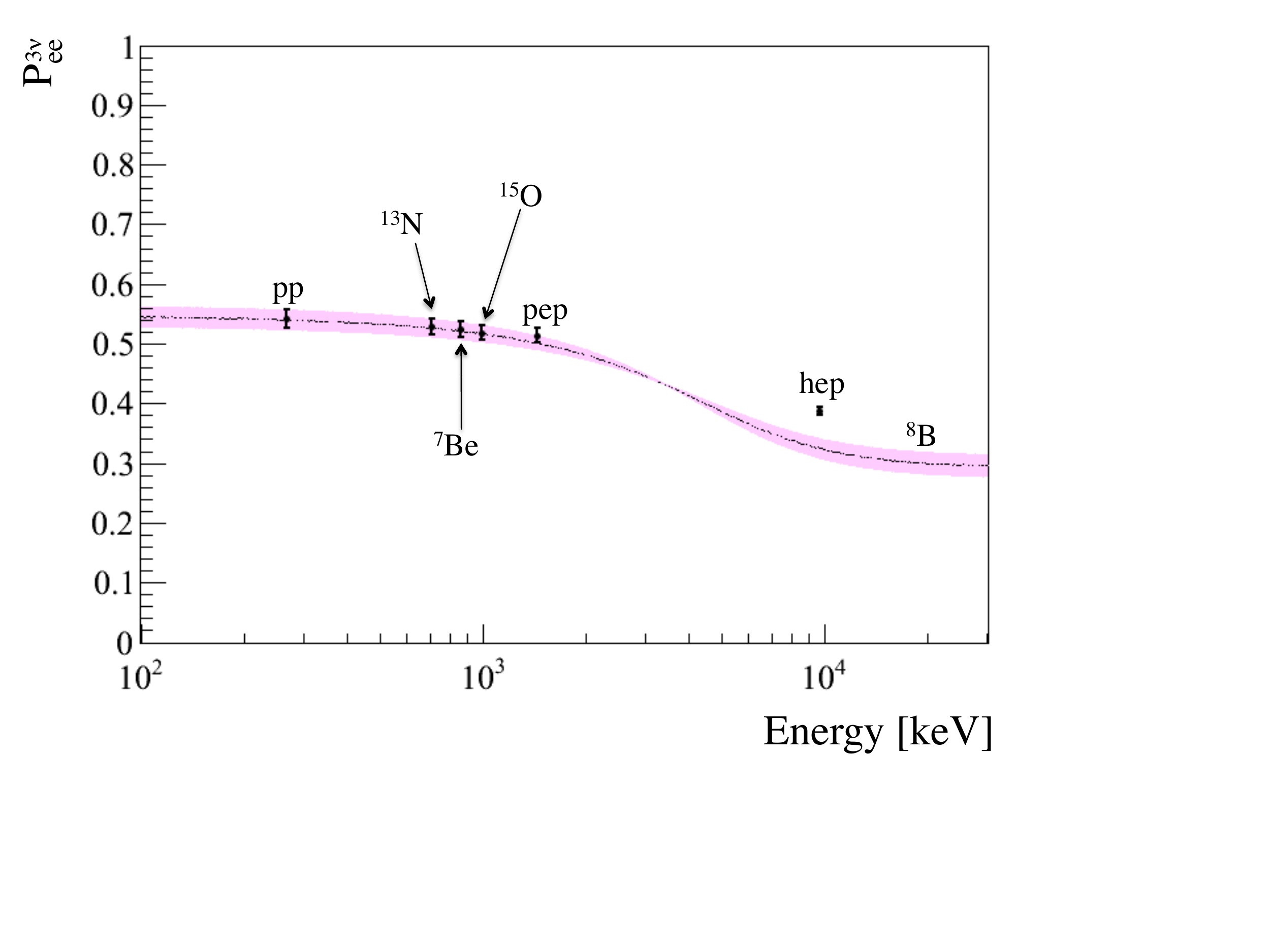}
\caption{Electron neutrino survival probability as a function of neutrino energy according to MSW--LMA model, see Eq.~\ref{eq:ApproxPee}. The pink band is for $^8$B solar neutrinos, considering their production region in the Sun.  The other points represent other solar neutrino fluxes, considering their proper production regions, and reported for mono--energetic $^7$Be (862\,keV) and $pep$ (1444\,keV) neutrinos and for the mean energies of fluxes with continuous energy spectrum: $pp$ (267\,keV), $^{13}$N (707\,keV), $^{15}$O (997\,keV), and $hep$ (9625\,keV).   }
\label{fig:PeeT}
\end{center}
\end{figure}

Solar neutrino oscillations are characterized by the survival probability $P^{3\nu}_{ee}$ (defined in Section~\ref{sec:GbAn} with the relation
\ref{eq:GbAn:Pee_step1}) of electron neutrinos produced in the Sun reaching the detector on Earth.
$P^{3\nu}_{ee}$ depends on the oscillation parameters and on the neutrino energy. In the MSW--LMA model it shows specific features related to the
matter effects taking place while the neutrinos travel inside the Sun (MSW). These  effects influence 
the propagation of $\nu_e$  and $\nu_x$  differently, as the scattering probability of $\nu_e$ off electrons is larger than that of $\nu_x$ due to CC interactions. The effective Hamiltonian depends on the electron density $n_e$ in the Sun and,
considering the case in which the propagation of neutrinos in the Sun satisfies proper hypothesis of adiabaticity, the resulting
survival probability (formula \ref{eq:GbAn:Pee_step1}) does not depend on details of the Sun density profile and is well approximated by the following simple form \cite{bib:Bah2004}:

\begin{equation}
%\begin{array}{rl}
P^{3\nu}_{ee} = 
\frac {1} {2} \cos^4 \theta_{13}
\left( 1+ \cos 2\theta^M_{12} \cos 2\theta_{12} \right), 
%\end{array}
\label{eq:ApproxPee}
\end{equation}
where  $ \theta^M_{12}$ is called mixing angle in matter 
\begin{equation}
\cos 2\theta^M_{12}=\frac{\cos 2\theta_{12}-\beta}{\sqrt{(\cos2\theta_{12}-\beta)^2 + \sin^2 2 \theta_{12}}},
\label{eq:thetaM}
\end{equation}
with 
\begin{equation}
\beta= \frac{2 \sqrt{2} G_F cos^2\theta_{13} n_e E_\nu}{\Delta m^2_{12}},
\label{eq:betaMSW}
\end{equation}
where $G_F$ is the Fermi coupling constant, $n_e$ is the electron density in the Sun calculated at the neutrino production point, and $E_\nu$ is the neutrino energy.

Neutrinos from different reactions are produced in the Sun at different radii 
\cite{bib:Bah10000} and the electron density in the Sun decreases with increasing radius. This means that $P^{3\nu}_{ee}$ for a given neutrino energy is expected to depend on the neutrino species under consideration. Figure~\ref{fig:PeeT} shows $P^{3\nu}_{ee}$ calculated according to the MSW--LMA. The band refers to the $^8$B neutrinos and was obtained averaging the value of $P^{3\nu}_{ee}$ for each energy calculated for different radii (that is different $n_e$) in the Sun according to the proper radial distribution of the production point of the $^8$B--$\nu$'s. The width of the curve is due to the uncertainties (1$\sigma$) associated with the mixing angles and $\Delta m^2_{12}$. The plot also shows the value of the $P^{3\nu}_{ee}$ calculated for the mono--energetic $^7$Be and $pep$ neutrinos, considering their proper production regions in the Sun. Similarly, points at mean energies of $pp$, CNO, and $hep$ neutrino fluxes (having continuous energy spectra) are also included. From this figure we see that the dependence of the survival probability from  the neutrino production region in the Sun is small and it is masked by current uncertainties. The curve calculated for the $^8$B neutrinos matches well the prediction of the MSW--LMA model for $P^{3\nu}_{ee}$ versus energy.

The relative importance of the MSW matter term and the kinematic vacuum oscillation is described by the quantity $\beta(E_{\nu}$), defined in Eq.~\ref{eq:betaMSW}. For $\beta < \cos 2\theta_{12} \simeq 0.4 $, the survival probability reaches the value corresponding to vacuum-averaged oscillations ($\sim$0.55), while for $\beta > 1$, it corresponds to matter-dominated oscillations ($\sim$0.30). The $P^{3\nu}_{ee}$ in the MSW--LMA model exhibits a strong energy dependence only in the region around 2\,MeV, where $P^{3\nu}_{ee}$ is characterized by a transition between the values corresponding to these two limiting regimes, see Fig.~\ref{fig:PeeT}.
The measurement of the low-energy solar neutrinos spectrum with Borexino offers the perfect frame to test this prediction of the MSW--LMA oscillation model. Different oscillation models, including the possibility that neutrinos
undergo non standard interactions, predict survival probabilities with a significantly different energy dependence~\cite{NSI}.

The value of $P^{3\nu}_{ee}$ for the mono--energetic $^7$Be neutrinos is obtained from the Borexino measurement of the interaction rate $R(^7\rm{Be})$ using the relation
\begin{equation}
R(^7\rm{Be})= \Phi(^7\rm{Be}) \left( P^{3\nu}_{ee} \sigma_{\nu e} + (1-P^{3\nu}_{ee})\sigma_{\nu x} \right ) N_{e^-}, 
\label{expPee}
\end{equation}
where $N_{e^-}$ is the number of target electrons (reported in Table~\ref{tab:FV}) and $\Phi(^7\rm{Be})$ is the flux of neutrinos produced in the Sun,  listed in Table~\ref{table:Rate}.
$P^{3\nu}_{ee}$ for $pep$ neutrinos is obtained in the same way. 

\begin{figure}[t]
\begin{center}
\includegraphics[width = 0.5\textwidth]{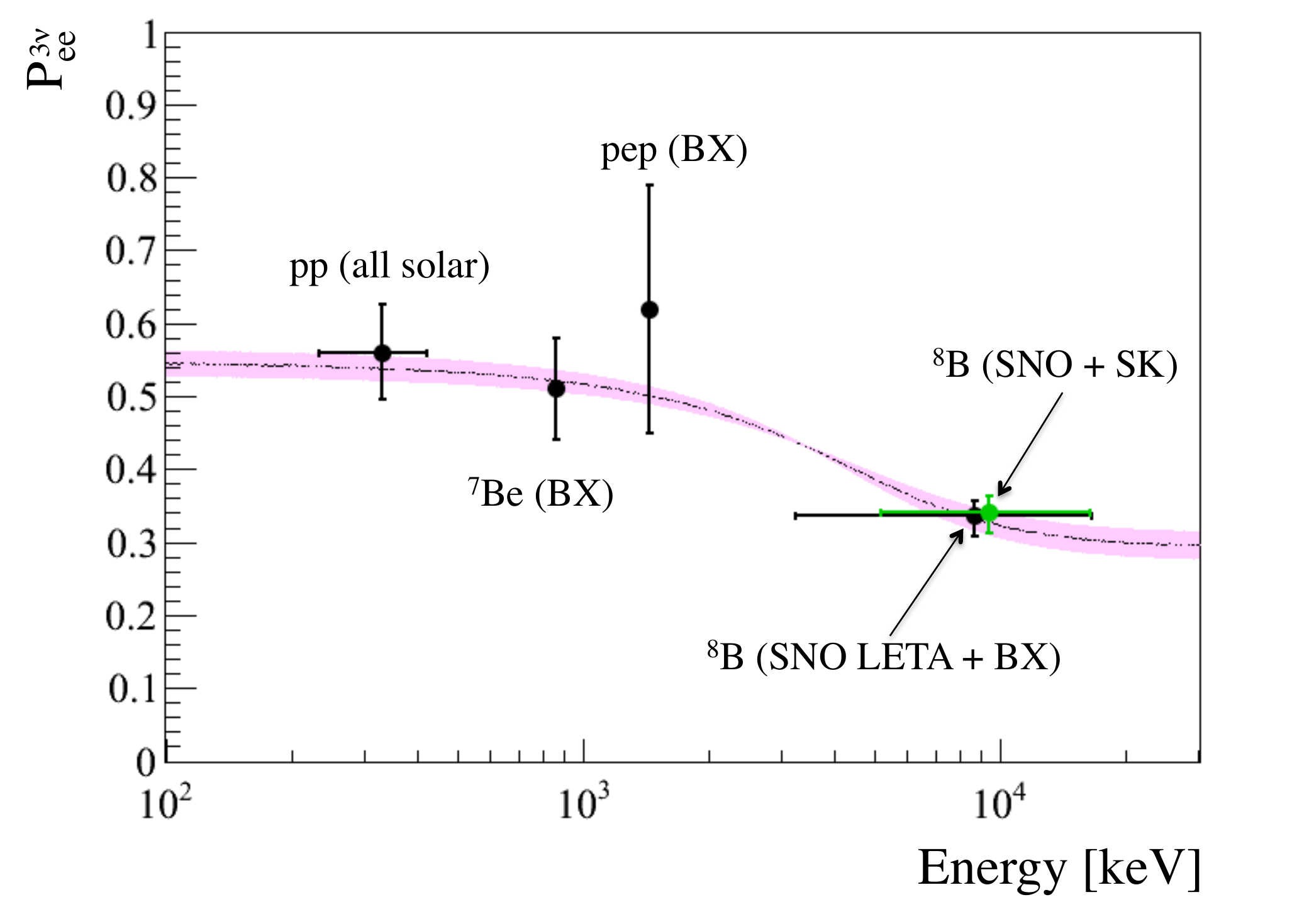}
\caption{Electron neutrino survival probability as a function of neutrino energy according to MSW--LMA model. The band is the same as in Fig.~\ref{fig:PeeT}, calculated for the production region of $^8$B solar neutrinos which represents well also other species of solar neutrinos. The points represent the solar neutrino experimental data for $^7$Be and $pep$ mono--energetic neutrinos (Borexino data), for $^8$B neutrinos detected above 5000\,keV of scattered-electron energy $T$ (SNO and Super-Kamiokande data) and for $T > 3000$\,keV (SNO LETA + Borexino data), and for $pp$ neutrinos considering all solar neutrino data, including radiochemical experiments. }
\label{fig:ExpPee}
\end{center}
\end{figure}

Using the fluxes of the high--metallicity solar model GS98~\cite{HighMet_GS98SSM}, we get $P^{3\nu}_{ee}(E_\nu = 862$\,keV) = 0.51 $\pm$ 0.07 including both the experimental and theoretical (solar model)  uncertainties and $P^{3\nu}_{ee}(E_\nu$ = 1440\,keV) = 0.62 $\pm$ 0.17.
A combined analysis of the Borexino data together with those of other solar experiments allows to obtain also the values of survival probability for the $pp$ and $^8$B neutrinos. Figure~\ref{fig:ExpPee} reports the results.

\section{Conclusions and perspectives}
\label{sec:concl}

 The rich scientific harvest of the Borexino Phase-I was made possible by the extreme radio--purity of the detector and of its liquid scintillator core in particular. Challenging design purity levels have been mostly met, and, in some cases, surpassed by a few orders of magnitude. 

The central physics goal was achieved with the 5\% measurement of the $^7$Be solar neutrino rate.  
Three more measurements beyond the scope of the original proposal were made as well:  the first observation of the solar $pep$ neutrinos, the most stringent experimental constraint on the flux of CNO neutrinos, and the low-threshold measurement of the $^8$B solar neutrino interaction rate.
The latter measurement was possible thanks to the extremely low background rate above natural radioactivity, while the first two exploited the superior particle identification capability of the scintillator and an efficient cosmogenic background subtraction. All measurements benefit from an extensive calibration campaign with radioactive sources that preserved scintillator radio--purity.

In this paper we have described the sources of background and the data analysis methods that led to the published solar neutrinos results.
We also reported, for the first time, the detection of the annual modulation of the $^7$Be solar neutrino rate, consistent with their solar origin.
The implications of Borexino solar neutrino results for neutrino and solar physics were also discussed, both stand--alone and in combination with other solar neutrino data.

Additional important scientific results (not discussed in this paper) were the detection of geo--neutrinos \cite{geonu} and state-of-the art upper limits on many rare and exotic processes \cite{BxRare}.

Borexino has performed several purification cycles in 2010 and 2011 by means of water 
extraction~\cite{BxFluidHandling} in batch mode, reducing even further several background components, among which $^{85}$Kr, $^{210}$Bi, and the $^{238}$U and $^{232}$Th chains. After these purification cycles, the Borexino Phase-II has started at the beginning of 2012, with the goal of improving all solar neutrino measurements.
Borexino is also an ideal apparatus to look for short baseline neutrino oscillations into sterile species using strong artificial neutrino and anti--neutrino sources \cite{bib:sterile}. An experimental program, called SOX (Source Oscillation eXperiment), was approved and it is now in progress.

\par
\par
The Borexino program is made possible by funding from INFN (Italy), NSF (USA), BMBF, DFG and MPG (Germany), NRC Kurchatov Institute
(Russia) and NCN (Poland). We acknowledge the generous support of the Laboratory Nazionali del Gran Sasso (Italy).

 \end{document}